\documentclass[prd,aps,nofootinbib,showpacs,preprintnumbers,amsmath,amssymb,floatfix]{revtex4}

\usepackage{times}
\usepackage{hyperref}

\usepackage{graphicx}
\usepackage{dcolumn}
\usepackage{bm}
\usepackage{slashed}
\usepackage{color}

\begin{document}

\title{Baryon magnetic moment in large-$N_c$ chiral perturbation theory: Complete analysis for $N_c=3$}

\author{
Rub\'en Flores-Mendieta
}
\affiliation{
Instituto de F{\'\i}sica, Universidad Aut\'onoma de San Luis Potos{\'\i}, \'Alvaro Obreg\'on 64, Zona Centro, San Luis Potos{\'\i}, S.L.P.\ 78000, Mexico
}

\author{
Carlos Isaac Garc{\'\i}a
}
\affiliation{
Instituto de F{\'\i}sica, Universidad Aut\'onoma de San Luis Potos{\'\i}, \'Alvaro Obreg\'on 64, Zona Centro, San Luis Potos{\'\i}, S.L.P.\ 78000, Mexico
}

\author{
Johann Hern\'andez
}
\affiliation{
Instituto de F{\'\i}sica, Universidad Aut\'onoma de San Luis Potos{\'\i}, \'Alvaro Obreg\'on 64, Zona Centro, San Luis Potos{\'\i}, S.L.P.\ 78000, Mexico
}

\author{
Mar{\'\i}a Anabel Trejo
}
\affiliation{
Instituto de F{\'\i}sica, Universidad Aut\'onoma de San Luis Potos{\'\i}, \'Alvaro Obreg\'on 64, Zona Centro, San Luis Potos{\'\i}, S.L.P.\ 78000, Mexico
}

\date{}

\begin{abstract}
Baryon magnetic moments are computed in baryon chiral perturbation theory in the large-$N_c$ limit at one-loop order, where $N_c$ is the number of color charges. Orders $\mathcal{O}(m_q^{1/2})$ and $\mathcal{O}(m_q \ln m_q)$ corrections are both evaluated including all the operator structures that participate at the physical value $N_c=3$. The complete expressions for octet and decuplet baryon magnetic moments in addition to octet-octet and decuplet-octet baryon transition moments are thus compared to their available counterparts obtained in heavy baryon chiral perturbation theory for degenerate intermediate baryons in the loops. Theoretical expressions fully agree at the physical values $N_c=3$ and $N_f=3$ flavors of light quarks. Some numerical evaluations are produced via a least-squares fit to explore the free parameters in the analysis. Results point out the necessity of incorporating the effects of nondegenerate intermediate baryons in the loops for a consistent determination of these free parameters.
\end{abstract}

\maketitle

\section{\label{sec:intro}Introduction}

In the limit of exact $SU(3)$ flavor symmetry, Coleman and Glashow \cite{coleman} first derived a set of relations among magnetic moments of the octet baryons. Their celebrated relations read
\begin{eqnarray}
\begin{array}{lcl}
\mu_{\Sigma^+}^{SU(3)} = \mu_p^{SU(3)}, & \qquad \qquad & \mu_{\Sigma^-}^{SU(3)} + \mu_n^{SU(3)} = -\mu_p^{SU(3)}, \\[4mm]
2\mu_\Lambda^{SU(3)} = \mu_n^{SU(3)}, & \qquad \qquad & \mu_{\Xi^-}^{SU(3)} = \mu_{\Sigma^-}^{SU(3)}, \\[4mm]
\mu_{\Xi^0}^{SU(3)} = \mu_n^{SU(3)}, & \qquad \qquad & 2\mu_{\Sigma^0\Lambda}^{SU(3)} = -\sqrt{3}\mu_n^{SU(3)}, \label{eq:cg}
\end{array}
\label{eq:treeval}
\end{eqnarray}
along with the isospin relation
\begin{equation}
\mu_{\Sigma^+}^{SU(3)} - 2 \mu_{\Sigma^0}^{SU(3)} + \mu_{\Sigma^-}^{SU(3)} = 0, \label{eq:isos}
\end{equation}
where $\mu_B^{SU(3)}$ represents the magnetic moment of the octet baryon $B$ in the $SU(3)$ symmetry limit.

Beyond the symmetry limit, various methods have been implemented for the evaluation of baryon magnetic moments. An important selection of such methods prior 2009 can be found in Ref.~\cite{rfm09}; a more recent analysis in the context of covariant chiral perturbation theory was presented in Ref.~\cite{xiao}. One of the earliest methods is chiral perturbation theory. Caldi and Pagels pointed out that nonanalytical corrections of orders $\mathcal{O}(m_q^{1/2})$ and $\mathcal{O}(m_q \ln m_q)$ in the perturbative series are calculable \cite{caldi}. They tackled the former and found them to be as large as the lowest-order values, which would indicate a breakdown of the perturbative expansion. It was not until the arrival of heavy baryon chiral perturbation theory (HBCHPT) first introduced by Jenkins and Manohar \cite{jm91a,jm91b} that some aspects of the theory were properly understood. When the method was applied to the renormalization of the baryon axial current, chiral logarithmic corrections to the axial couplings in hyperon semileptonic decays were found to be as large as the lowest order values when \textit{only intermediate octet baryons were included in the loops} \cite{jm91a}. The inclusion of both octet and decuplet baryons in the loops reduced considerably the corrections with respect to the case with the inclusion of octet states alone \cite{jm91b}. The cancellation pointed out phenomenologically in Refs.~\cite{jm91a,jm91b} was later proved in the context of the $1/N_c$ expansion of QCD in Refs.~\cite{dm91a,dm91b,djm94,djm95,dai}, where $N_c$ is the number of quark charges.

The earliest analysis of the magnetic moments of octet baryons in HBCHPT to orders $\mathcal{O}(m_q^{1/2})$ and $\mathcal{O}(m_q \ln m_q)$ was presented in Ref.~\cite{jen92}. There, it was concluded that, unlike the axial current case, the inclusion of intermediate decuplet baryons in the loops does not partially cancel the contribution from intermediate octet baryons. The use of the combined formalism in $1/N_c$ and chiral corrections \cite{march,jen96} has shed light on the subject \cite{rfm09,rfm14} by allowing one to perform a rigorous analytical evaluation of the cancellations that follow from the large-$N_c$ spin-flavor symmetry of baryons. In Ref.~\cite{rfm09}, one-loop corrections to magnetic moments to relative order $1/N_c^3$ in the $1/N_c$ expansion were carried out under the limit $\Delta\to 0$, where $\Delta \equiv M_T-M_B$ is the average decuplet-octet mass difference. A more refined analysis was later presented in Ref.~\cite{rfm14}, where the assumption of degenerate intermediate baryons was lifted and explicit $SU(3)$ symmetry breaking (SB) effects were also included.

The aim of the present paper is to improve the analyses of Refs.~\cite{rfm09,rfm14} in a few aspects. Mainly, all $1/N_c$ corrections to the baryon magnetic moment allowed for $N_c=3$ will be evaluated, motivated by a recent calculation of the baryon axial coupling \cite{rfm21}. While corrections of order $\mathcal{O}(m_q^{1/2})$ will be carried out for nonzero $\Delta$, corrections of order $\mathcal{O}(m_q \ln m_q)$ will keep the $\Delta=0$ assumption for reasons that will become apparent later. Complete expressions for all 27 magnetic moments of octet and decuplet baryons and decuplet-octet transition moments are provided. Despite their lengths, the analytical forms are basically simple and organized in a way that are easy to handle. Their main usefulness lies in that they can be used to perform an \textit{analytical} comparison to the available expressions obtained in conventional HBCHPT (the effective field theory with no $1/N_c$ expansion) of Ref.~\cite{jen92}. Therefore, the main contribution of this paper is to explicitly show that baryon chiral perturbation theory in the large-$N_c$ limit and HBCHPT analyses of baryon octet magnetic moments fully agree at the physical value $N_c=3$ for $N_f=3$ flavors of light quarks.

The organization of the paper is as follows. Some introductory aspects of large-$N_c$ chiral perturbation theory are provided in Sec.~\ref{sec:introln}; in passing, notation and conventions are introduced. After a brief review of baryon magnetic moments at tree level in Sec.~\ref{sec:tree}, the discussion is focused on the computation of one-loop corrections in Sec.~\ref{sec:1l}; because of their different group theoretical properties, corrections of orders $\mathcal{O}(m_q^{1/2})$ and $\mathcal{O}(m_q\ln m_q)$ are studied separately in Secs.~\ref{sec:mq} and \ref{sec:mqlnmq}, respectively, followed by their corresponding analytical comparisons with HBCHPT results. The issue of explicit SB is reviewed in Sec.~\ref{sec:sb}, based on the analysis of Ref.~\cite{rfm14}. Gathering together all partial results allows one to carry out a numerical analysis to determine the free parameters of the theory, by making a least-squares fit to the available data \cite{part}. Results are presented in Sec.~\ref{sec:num} and some closing remarks are provided in Sec.~\ref{sec:con}. The paper is complemented by five appendixes where the complete although lengthy formulas of baryon magnetic moments are relegated.

\section{\label{sec:introln}Operator analysis in the $1/N_c$ expansion}

The $1/N_c$ expansion has been very useful in understanding the spin-flavor structure of baryons in QCD \cite{dm91a,dm91b,djm94,djm95}.
For the physically interesting case of three light flavors, $N_f=3$, the lowest-lying baryon states fall into a representation of the spin-flavor group $SU(6)$. At the physical value $N_c=3$, this is the usual $\mathbf{56}$ dimensional representation of $SU(6)$. The $J^P = 1/2^+$ octet containing the nucleon and the $J^P = 3/2^+$ decuplet containing the $\Delta(1232)$ together make up the ground-state 56-plet, in which the orbital angular momenta between the quark pairs are zero and the spatial part of the state function is symmetric.

The present analysis builds on the $1/N_c$ baryon chiral Lagrangian $\mathcal{L}_{\text{baryon}}$ introduced in Ref.~\cite{jen96}. This Lagrangian incorporates nonet symmetry and the contracted spin-flavor symmetry for baryons in the large-$N_c$ limit; its definite form reads
\begin{equation}
\mathcal{L}_{\text{baryon}} = i \mathcal{D}^0 - \mathcal{M}_{\text{hyperfine}} + \text{Tr} \left(\mathcal{A}^k \lambda^c \right) A^{kc} + \frac{1}{N_c} \text{Tr} \left(\mathcal{A}^k \frac{2I}{\sqrt 6}\right) A^k + \ldots, \label{eq:ncch}
\end{equation}
with
\begin{equation}
\mathcal{D}^0 = \partial^0 \openone + \text{Tr} \left(\mathcal{V}^0 \lambda^c\right) T^c. \label{eq:kin}
\end{equation}

The ellipses in Eq.~(\ref{eq:ncch}) represent higher partial wave meson couplings which occur at subleading orders in the $1/N_c$ expansion for $N_c > 3$. In the large-$N_c$ limit, all of these higher partial waves vanish so the meson coupling to baryons is purely $p$ wave.

Meson fields participate in $\mathcal{L}_{\text{baryon}}$ through the vector and axial-vector combinations
\begin{equation}
\mathcal{V}^0 = \frac12 \left[ \xi \partial^0 \xi^\dagger + \xi^\dagger \partial^0 \xi \right], \qquad
\mathcal{A}^k = \frac{i}{2} \left(\xi \nabla^k \xi^\dagger - \xi^\dagger \nabla^k \xi\right), \qquad \qquad
\xi(x)=\exp[i\Pi(x)/f],
\end{equation}
where $\Pi(x)$ represents the nonet of Goldstone boson fields and $f \approx 93$ $\mathrm{MeV}/c^2$ is the pion decay constant.

Each term in $\mathcal{L}_{\text{baryon}}$ is made up by a baryon operator. The baryon kinetic energy term involves the spin-flavor identity, $\mathcal{M}_{\text{hyperfine}}$ represents the hyperfine baryon mass operator which incorporates the spin splittings of the tower of baryon states with spins $1/2,\ldots, N_c/2$ in the flavor representations, and $A^k$ and $A^{kc}$ stand for the flavor singlet and flavor octet axial current operators, respectively. All these baryon operators have an expansion in operators whose coefficients are inverse powers of $N_c$ \cite{djm95}. To a given order in $1/N_c$, the expansions can be truncated and linked to physics by evaluating their matrix elements between $SU(6)$ symmetric states at $N_c = 3$.

For any representation of $SU(6)$, polynomials in the generators
\begin{equation}
J^k = q^\dagger \frac{\sigma^k}{2} q, \qquad T^c = q^\dagger \frac{\lambda^c}{2} q, \qquad G^{kc} = q^\dagger
\frac{\sigma^k}{2}\frac{\lambda^c}{2} q, \label{eq:su6gen}
\end{equation}
form a complete set of operators \cite{djm95}. In the above relations, $q^\dagger$ and $q$ represent $SU(6)$ operators that create and annihilate states in the fundamental representation of $SU(6)$, and $\sigma^k$ and $\lambda^c$ are the Pauli spin and Gell-Mann flavor matrices, respectively. The spin-flavor generators satisfy the commutation relations listed in Table \ref{tab:surel}.
\begingroup
\begin{table}
\caption{\label{tab:surel}$SU(2N_f)$ commutation relations.}
\bigskip
\label{tab:su2fcomm}
\centerline{\vbox{ \tabskip=0pt \offinterlineskip
\halign{
\strut\quad $ # $\quad\hfil&\strut\quad $ # $\quad \hfil\cr
\multispan2\hfil $\left[J^i,T^a\right]=0,$ \hfil \cr
\noalign{\medskip}
\left[J^i,J^j\right]=i\epsilon^{ijk} J^k,
&\left[T^a,T^b\right]=i f^{abc} T^c,\cr
\noalign{\medskip}
\left[J^i,G^{ja}\right]=i\epsilon^{ijk} G^{ka},
&\left[T^a,G^{ib}\right]=i f^{abc} G^{ic},\cr
\noalign{\medskip}
\multispan2\hfil$\displaystyle [G^{ia},G^{jb}] = \frac{i}{4}\delta^{ij}
f^{abc} T^c + \frac{i}{2N_f} \delta^{ab} \epsilon^{ijk} J^k + \frac{i}{2} \epsilon^{ijk} d^{abc} G^{kc}.$ \hfill\cr
}}}
\end{table}
\endgroup

The way in which large-$N_c$ dynamics enters can best be seen through some examples. The $1/N_c$ expansion of the baryon mass operator $\mathcal{M}$ can be written as \cite{djm94,djm95}
\begin{eqnarray}
\mathcal{M} = \tilde{m}_0 N_c \openone + \sum_{n=2,4}^{N_c-1} \tilde{m}_{n} \frac{1}{N_c^{n-1}} J^n, \label{eq:mop}
\end{eqnarray}
where $\tilde{m}_n$ are unknown coefficients. While the first summand on the right-hand side of Eq.~(\ref{eq:mop}) is the overall spin-independent mass of the baryon multiplet and is removed from the chiral Lagrangian by the heavy baryon field redefinition~\cite{jm91a}, the spin-dependent ones define $\mathcal{M}_{\text{hyperfine}}$ introduced in the chiral Lagrangian (\ref{eq:ncch}). In the large-$N_c$ limit, $\Delta=\langle \mathcal{M}\rangle_{\frac32}-\langle \mathcal{M}\rangle_{\frac12} \propto 1/N_c$, so decuplet and octet baryons become degenerate and form a single irreducible representation of the contracted spin-flavor symmetry of baryons in large-$N_c$ QCD \cite{djm95}.

At the physical value $N_c=3$ the hyperfine mass expansion reduces to
\begin{eqnarray}
\mathcal{M} _{\text{hyperfine}} = \frac{\tilde{m}_2}{N_c} J^2, \label{eq:smop}
\end{eqnarray}
so $\Delta$ becomes $\tilde{m}_2$.

The baryon flavor singlet axial current $A^k$ is a spin-1 object and a singlet under $SU(3)$; its $1/N_c$ expansion reads \cite{djm95}
\begin{equation}
A^k = \sum_{n=1,3}^{N_c} b_n^{1,1} \frac{1}{N_c^{n-1}} \mathcal{D}_n^k, \label{eq:asin}
\end{equation}
where $\mathcal{D}_1^k = J^k$ and $\mathcal{D}_{2m+1}^k = \{J^2,\mathcal{D}_{2m-1}^k\}$ for $m\geq 1$. The superscript on the operator coefficients of $A^k$ denotes that they refer to the baryon singlet current. At $N_c=3$, Eq.~(\ref{eq:asin}) becomes
\begin{equation}
A^k = b_1^{1,1} J^k + b_3^{1,1} \frac{1}{N_c^2} \{J^2,J^k\}.
\end{equation}

The baryon flavor octet axial current $A^{kc}$ is a spin-1 object, an octet under $SU(3)$ and odd under time reversal; its $1/N_c$ expansion can be written as \cite{djm94,djm95}
\begin{equation}
A^{kc} = a_1 G^{kc} + \sum_{n=2,3}^{N_c} b_n \frac{1}{N_c^{n-1}} \mathcal{D}_n^{kc} + \sum_{n=3,5}^{N_c} c_n
\frac{1}{N_c^{n-1}} \mathcal{O}_n^{kc}, \label{eq:akcfull}
\end{equation}
where the unknown coefficients $a_1$, $b_n$, and $c_n$ have expansions in powers of $1/N_c$ and are order unity at leading order in the $1/N_c$ expansion. The basic operators in expansion (\ref{eq:akcfull}) are
\begin{eqnarray}
\mathcal{D}_2^{kc} & = & J^kT^c, \label{eq:d2kc} \\
\mathcal{D}_3^{kc} & = & \{J^k,\{J^r,G^{rc}\}\}, \label{eq:d3kc} \\
\mathcal{O}_3^{kc} & = & \{J^2,G^{kc}\} - \frac12 \{J^k,\{J^r,G^{rc}\}\}, \label{eq:o3kc}
\end{eqnarray}
so that $\mathcal{D}_n^{kc}=\{J^2,\mathcal{D}_{n-2}^{kc}\}$ and $\mathcal{O}_n^{kc}=\{J^2,\mathcal{O}_{n-2}^{kc}\}$ for $n\geq 4$. Notice that $\mathcal{D}_n^{kc}$ are diagonal operators with nonzero matrix elements only between states with the same spin, and the $\mathcal{O}_n^{kc}$ are purely off-diagonal operators with nonzero matrix elements only between states with different spin. At $N_c = 3$ the series (\ref{eq:akcfull}) can be truncated as
\begin{equation}
A^{kc} = a_1 G^{kc} + b_2 \frac{1}{N_c} \mathcal{D}_2^{kc} + b_3 \frac{1}{N_c^2} \mathcal{D}_3^{kc} + c_3 \frac{1}{N_c^2} \mathcal{O}_3^{kc}. \label{eq:akc}
\end{equation}
At leading order in the $1/N_c$ expansion, $A^{kc}$ is order $\mathcal{O}(N_c)$.

It should be emphasized that keeping all four terms in Eq.~(\ref{eq:akc}) allows for arbitrary values of the four possible $SU(3)$ symmetric couplings of pseudoscalar mesons to the octet and decuplet baryons $D$, $F$, $\mathcal{C}$, and $\mathcal{H}$ introduced in Refs.~\cite{jm91a,jm91b}. This is the reason why for $N_c=3$ it is not necessary to go beyond operator products of third order in the spin-flavor generators.

\section{\label{sec:tree}Baryon magnetic moment at tree level}

The starting point in the present analysis is the fact that in the large-$N_c$ limit, the baryon magnetic moments have the same kinematic properties as the baryon axial couplings so they can be expressed in terms of the very same operators \cite{dai}. Since much of the work has already been advanced in Refs.~\cite{rfm09,rfm14,rfm21}, some partial results presented in these references will be borrowed.

Accordingly, the $1/N_c$ expansion of the operator that yields the baryon magnetic moment operator becomes \cite{rfm09}
\begin{equation}
M^{kc} = m_1 G^{kc} + \frac{1}{N_c} m_2 \mathcal{D}_2^{kc} + \frac{1}{N_c^2} m_3 \mathcal{D}_3^{kc} + \frac{1}{N_c^2} m_4 \mathcal{O}_3^{kc}, \label{eq:mkc}
\end{equation}
which is also order $\mathcal{O}(N_c)$ at leading order in the $1/N_c$ expansion; $m_i$ are unknown coefficients which also possess a $1/N_c$ expansion starting at order 1. Under the assumption of $SU(3)$ symmetry, the unknown coefficients $m_i$ are independent of $k$ so they are unrelated to $a_1$, $b_2$, $b_3$, or $c_3$ \cite{rfm09}.

The baryon magnetic moment operator is thus defined as
\begin{equation}
M^k \equiv M^{kQ} = M^{k3} + \frac{1}{\sqrt{3}}M^{k8}, \label{eq:mQ}
\end{equation}
where the spin index will be fixed to 3 and the flavor index becomes $Q=3+(1/\sqrt{3})8$. Hereafter, any operators of the form $X^Q$ and $X^{\overline{Q}}$ should be understood as $X^3+(1/\sqrt{3})X^8$ and $X^3-(1/\sqrt{3})X^8$, respectively. The magnetic moments are proportional to the quark charge matrix $\mathrm{diag}(2/3,-1/3,-1/3)$, so they can be separated into isovector and isoscalar components, $M^{33}$ and $M^{38}$, respectively.

The baryon magnetic moments at tree level can be straightforwardly obtained by evaluating the matrix elements of the operators that appear in (\ref{eq:mQ}) between $SU(6)$ baryon symmetric states. The universality of operator (\ref{eq:mQ}) is such that it allows one to compute all possible $27$ magnetic moments: Eight magnetic moments for the octet baryons, ten more for the decuplet baryons and one for the octet-octet and eight for the decuplet-octet transition moments. At tree level they will be denoted by $\mu_{B}^{(0)} = \langle B|M^Q|B \rangle$, $\mu_{T}^{(0)} = \langle T|M^Q|T\rangle$, $\mu_{BB^\prime}^{(0)} = \langle B|M^Q|B^\prime \rangle$, and $\mu_{TB}^{(0)} = \langle T|M^Q|B \rangle$, where $B$ and $T$ stand for octet and decuplet baryons, respectively. The theoretical expressions can generically be given by
\begin{equation}
\mu_{B}^{(0)} = \sum_{j=1}^4 \mu_j \langle B|S_j^{3Q}|B\rangle, \label{eq:mmtre}
\end{equation}
where the coefficients $\mu_j$ can be easily read off from Eq.~(\ref{eq:mQ}) and the operator basis $\{S_i\}$ used to describe tree-level (and the singlet contribution of) magnetic moments reads
\begin{eqnarray}
\label{eq:basis1}
\begin{tabular}{lllll}
$S_1^{kc} = G^{kc}$, &
$S_2^{kc} = \mathcal{D}_2^{kc}$, &
$S_3^{kc} = \mathcal{D}_3^{ke}$, &
$S_4^{kc} = \mathcal{O}_3^{kc}$, &
$S_5^{kc} = \mathcal{D}_4^{kc}$, \\
$S_6^{kc} = \mathcal{D}_5^{ke}$, &
$S_7^{kc} = \mathcal{O}_5^{ke}$, &
$S_8^{kc} = \mathcal{D}_6^{kc}$, &
$S_9^{kc} = \mathcal{D}_7^{kc}$, &
$S_{10}^{kc} = \mathcal{O}_7^{kc}$.
\end{tabular}
\end{eqnarray}
Of course, it should be recalled that $\mu^{(0)}$ also define $\mu^{SU(3)}$; both quantities will be used interchangeably hereafter.

Nontrivial matrix elements\footnote{A baryon operator $X_j^{kc}$ yields a trivial matrix element in two possible ways: Either by definition $\langle X_j^{3c} \rangle = 0$ or $\langle X_j^{3c} \rangle = \langle\{J^2,X_{j-2}^{3c}\} \rangle$ for $c=3,8$. Hereafter, trivial matrix elements will not be listed in tables.} of the baryon operators that constitute the basis (\ref{eq:basis1}) are listed in Tables \ref{t:mm1O}, \ref{t:mm1T}, and \ref{t:mm1TO}. The resultant expressions for the magnetic moments at tree level are thus listed in the column labeled (a) in Table \ref{t:treeandnum}.

\begin{table*}
\caption{\label{t:mm1O}Nontrivial matrix elements of the operators involved in the magnetic moments of octet baryons at tree level. The entries for isoscalar components correspond to $\sqrt{3} \langle S_i^{38} \rangle$.}
\begin{ruledtabular}
\begin{tabular}{lccccccccc}
& $\displaystyle n$ & $\displaystyle p$ & $\displaystyle \Sigma^-$ & $\displaystyle \Sigma^0$ & $\displaystyle \Sigma^+$ & $\displaystyle \Xi^-$ & $\displaystyle \Xi^0$ & $\displaystyle \Lambda$ & $\displaystyle \Sigma^0\Lambda$ \\
\hline
$\langle S_1^{33} \rangle$ & $-\frac{5}{12}$ & $\frac{5}{12}$ & $-\frac13$ & $0$ & $\frac13$ & $\frac{1}{12}$ & $-\frac{1}{12}$ & $0$ & $\frac{1}{2 \sqrt{3}}$ \\
$\langle S_2^{33} \rangle$ & $-\frac14$ & $\frac14$ & $-\frac12$ & $0$ & $\frac12$ & $-\frac14$ & $\frac14$ & $0$ & $0$ \\
$\langle S_3^{33} \rangle$ & $-\frac54$ & $\frac54$ & $-1$ & $0$ & $1$ & $\frac14$ & $-\frac14$ & $0$ & $\frac{\sqrt{3}}{2}$ \\
\hline
$\langle S_1^{38} \rangle$ & $\frac14$ & $\frac14$ & $\frac12$ & $\frac12$ & $\frac12$ & $-\frac34$ & $-\frac34$ & $-\frac12$ & $0$ \\
$\langle S_2^{38} \rangle$ & $\frac34$ & $\frac34$ & $0$ & $0$ & $0$ & $-\frac34$ & $-\frac34$ & $0$ & $0$ \\
$\langle S_3^{38} \rangle$ & $\frac34$ & $\frac34$ & $\frac32$ & $\frac32$ & $\frac32$ & $-\frac94$ & $-\frac94$ & $-\frac32$ & $0$ \\
\end{tabular}
\end{ruledtabular}
\end{table*}

\begin{table*}
\caption{\label{t:mm1T}Nontrivial matrix elements of the operators involved in the magnetic moments of decuplet baryons at tree level. The entries for isoscalar components correspond to $\sqrt{3} \langle S_i^{38} \rangle$.}
\begin{ruledtabular}
\begin{tabular}{lcccccccccc}
& $\displaystyle \Delta^{++}$ & $\displaystyle \Delta^+$ & $\displaystyle \Delta^0$ & $\displaystyle \Delta^-$ & $\displaystyle {\Sigma^*}^+$ & $\displaystyle {\Sigma^*}^0$ & $\displaystyle {\Sigma^*}^-$ & $\displaystyle {\Xi^*}^0$ & $\displaystyle {\Xi^*}^-$ & $\displaystyle \Omega^-$ \\
\hline
$\langle S_1^{33} \rangle$ & $\frac34$ & $\frac14$ & $-\frac14$ & $-\frac34$ & $\frac12$ & $0$ & $-\frac12$ & $\frac14$ & $-\frac14$ & $0$ \\
$\langle S_2^{33} \rangle$ & $\frac94$ & $\frac34$ & $-\frac34$ & $-\frac94$ & $\frac32$ & $0$ & $-\frac32$ & $\frac34$ & $-\frac34$ & $0$ \\
$\langle S_3^{33} \rangle$ & $\frac{45}{4}$ & $\frac{15}{4}$ & $-\frac{15}{4}$ & $-\frac{45}{4}$ & $\frac{15}{2}$ & $0$ & $-\frac{15}{2}$ & $\frac{15}{4}$ & $-\frac{15}{4}$ & $0$ \\
\hline
$\langle S_1^{38} \rangle$ & $\frac34$ & $\frac34$ & $\frac34$ & $\frac34$ & $0$ & $0$ & $0$ & $-\frac34$ & $-\frac34$ & $-\frac32$ \\
$\langle S_2^{38} \rangle$ & $\frac94$ & $\frac94$ & $\frac94$ & $\frac94$ & $0$ & $0$ & $0$ & $-\frac94$ & $-\frac94$ & $-\frac92$ \\
$\langle S_3^{38} \rangle$ & $\frac{45}{4}$ & $\frac{45}{4}$ & $\frac{45}{4}$ & $\frac{45}{4}$ & $0$ & $0$ & $0$ & $-\frac{45}{4}$ & $-\frac{45}{4}$ & $-\frac{45}{2}$ \\
\end{tabular}
\end{ruledtabular}
\end{table*}

\begin{table*}
\caption{\label{t:mm1TO}Nontrivial matrix elements of the operators involved in the decuplet to octet transition moments at tree level. The entries for isovector and isoscalar components correspond to $\sqrt{2} \langle S_i^{33} \rangle$ and $\sqrt{6} \langle S_j^{38} \rangle$, respectively.}
\begin{ruledtabular}
\begin{tabular}{lcccccccc}
& $\displaystyle \Delta^+p$ & $\displaystyle \Delta^0n$ & $\displaystyle {\Sigma^*}^0\Lambda$ & $\displaystyle {\Sigma^*}^0\Sigma^0$ & $\displaystyle {\Sigma^*}^+\Sigma^+$ & $\displaystyle {\Sigma^*}^-\Sigma^-$ & $\displaystyle {\Xi^*}^0\Xi^0$ & $\displaystyle {\Xi^*}^-\Xi^-$ \\
\hline
$\langle S_1^{33} \rangle$ & $\frac23$ & $\frac23$ & $\frac{1}{\sqrt{3}}$ & $0$ & $\frac13$ & $-\frac13$ & $\frac13$ & $-\frac13$ \\
$\langle S_4^{33} \rangle$ & $3$ & $3$ & $\frac{3 \sqrt{3}}{2}$ & $0$ & $\frac32$ & $-\frac32$ & $\frac32$ & $-\frac32$ \\
\hline
$\langle S_1^{38} \rangle$ & $0$ & $0$ & $0$ & $1$ & $1$ & $1$ & $1$ & $1$ \\
$\langle S_4^{38} \rangle$ & $0$ & $0$ & $0$ & $\frac92$ & $\frac92$ & $\frac92$ & $\frac92$ & $\frac92$ \\
\end{tabular}
\end{ruledtabular}
\end{table*}

The main goal of the present analysis is to carry out an \textit{analytical} comparison with HBCHPT results of Ref.~\cite{jen92}. The comparison can be made following a simple procedure. First, it is convenient to introduce the relations between the operator coefficients $m_i$ of Eq.~(\ref{eq:mkc}) and the $SU(3)$ invariants $\mu_D$, $\mu_F$, $\mu_C$, and $\mu_T$ used to parametrize the baryon magnetic moments in HBCHPT \cite{jen92}. At $N_c=3$, the relations read \cite{rfm09}
\begin{subequations}
\begin{eqnarray}
\mu_D & = & \frac12 m_1 + \frac16 m_3, \\
\mu_F & = & \frac13 m_1 + \frac16 m_2 + \frac19 m_3, \\
\mu_C & = & \frac12 m_1 + \frac12 m_2 + \frac56 m_3, \\
\mu_T & = & -2 m_1 - m_4,
\end{eqnarray}
\end{subequations}
so the inverse relations become
\begin{subequations}
\label{eq:su3inv}
\begin{eqnarray}
m_1 & = & \frac32 \mu_D + \frac32 \mu_F - \frac12 \mu_C, \\
m_2 & = & -4 \mu_D + 6 \mu_F, \\
m_3 & = & \frac32 \mu_D - \frac92 \mu_F + \frac32 \mu_C, \\
m_4 & = & -3 \mu_D - 3 \mu_F + \mu_C - \mu_T.
\end{eqnarray}
\end{subequations}
Second, by using the inverse relations (\ref{eq:su3inv}), the tree-level magnetic moments can be rewritten in terms of the $SU(3)$ invariants $\mu_D$, $\mu_F$, $\mu_C$, and $\mu_T$, which yields the expressions listed in the column labeled (b) in Table \ref{t:treeandnum}. These last expressions are the ones suitable for comparison with HBCHPT. For octet and decuplet baryons these expressions fully agree with the ones reported in Ref.~\cite{jen92}. Tree-level magnetic moments for octet baryons are given in terms of $\alpha_B$ of Eq.~(23) of this reference, whereas for decuplet baryons, they are normalized to be $\mu_C$ times the electric charge of the corresponding baryon. For decuplet-octet transition moments, no explicit theoretical expressions in the context of HBCHPT are available so no direct comparison is possible.

\begin{table*}
\caption{\label{t:treeandnum} Tree-level expressions of baryon magnetic moments. Expressions in (a) are evaluated in the context of the $1/N_c$ expansion; expressions in (b) follow from the ones given in (a) by using relations (\ref{eq:su3inv}) to compare with heavy baryon chiral perturbation theory results.}
\begin{ruledtabular}
\begin{tabular}{lcc}
\textrm{Baryon} & \multicolumn{2}{c}{Tree-level values, $\mu_B^{(0)}$} \\
& (a) & (b) \\
\hline
$n$ & $-\frac13 m_1-\frac19 m_3$ & $-\frac23 \mu_D$ \\
$p$ & $\frac12 m_1+\frac16 m_2+\frac16 m_3$ & $\frac13 \mu_D+\mu_F$ \\
$\Sigma^-$ & $-\frac16 m_1-\frac16 m_2-\frac{1}{18} m_3$ & $\frac13 \mu_D-\mu_F$ \\
$\Sigma^0$ & $\frac16 m_1+\frac{1}{18} m_3$ & $\frac13 \mu_D$ \\
$\Sigma^+$ & $\frac12 m_1+\frac16 m_2+\frac16 m_3$ & $\frac13 \mu_D+\mu_F$ \\
$\Xi^-$ & $-\frac16 m_1-\frac16 m_2-\frac{1}{18} m_3$ & $\frac13 \mu_D-\mu_F$ \\
$\Xi^0$ & $-\frac13 m_1-\frac19 m_3$ & $-\frac23 \mu_D$ \\
$\Lambda$ & $-\frac16 m_1-\frac{1}{18} m_3$ & $-\frac13 \mu_D$ \\
$\Sigma^0\Lambda$ & $\frac{1}{2\sqrt{3}} m_1+\frac{1}{6\sqrt{3}} m_3$ & $\frac{1}{\sqrt{3}} \mu_D$ \\
$\Delta^{++}$ & $m_1+m_2+\frac53 m_3$ & $2 \mu_C$ \\
$\Delta^+$ & $\frac12 m_1+\frac12 m_2+\frac56 m_3$ & $\mu_C$ \\
$\Delta^0$ & $0$ & $0$ \\
$\Delta^-$ & $-\frac12 m_1-\frac12 m_2-\frac56 m_3$ & $-\mu_C$ \\
${\Sigma^*}^+$ & $\frac12 m_1+\frac12 m_2+\frac56 m_3$ & $\mu_C$ \\
${\Sigma^*}^0$ & $0$ & $0$ \\
${\Sigma^*}^-$ & $-\frac12 m_1-\frac12 m_2-\frac56 m_3$ & $-\mu_C$ \\
${\Xi^*}^0$ & $0$ & $0$ \\
${\Xi^*}^-$ & $-\frac12 m_1-\frac12 m_2-\frac56 m_3$ & $-\mu_C$ \\
$\Omega^-$ & $-\frac12 m_1-\frac12 m_2-\frac56 m_3$ & $-\mu_C$ \\
$\Delta^+p$ & $\frac{1}{3\sqrt{2}}(2 m_1+m_4)$ & $-\frac{1}{3\sqrt{2}} \mu_T$ \\
$\Delta^0n$ & $\frac{1}{3\sqrt{2}}(2 m_1+m_4)$ & $-\frac{1}{3\sqrt{2}} \mu_T$ \\
${\Sigma^*}^0\Lambda$ & $\frac{1}{2\sqrt{6}}(2 m_1+m_4)$ & $-\frac{1}{2\sqrt{6}} \mu_T$ \\
${\Sigma^*}^0\Sigma^0$ & $\frac{1}{6\sqrt{2}}(2 m_1+m_4)$ & $-\frac{1}{6\sqrt{2}} \mu_T$ \\
${\Sigma^*}^+\Sigma^+$ & $\frac{1}{3\sqrt{2}}(2 m_1+m_4)$ & $-\frac{1}{3\sqrt{2}} \mu_T$ \\
${\Sigma^*}^-\Sigma^-$ & $0$ & $0$ \\
${\Xi^*}^0\Xi^0$ & $\frac{1}{3\sqrt{2}}(2 m_1+m_4)$ & $-\frac{1}{3\sqrt{2}} \mu_T$ \\
${\Xi^*}^-\Xi^-$ & $0$ & $0$ \\
\end{tabular}
\end{ruledtabular}
\end{table*}

Once tree-level values of baryon magnetic moments are obtained, one-loop corrections are discussed in the next sections.

\section{\label{sec:1l}One-loop corrections to baryon magnetic moments}

\begin{figure}[ht]
\scalebox{0.32}{\includegraphics{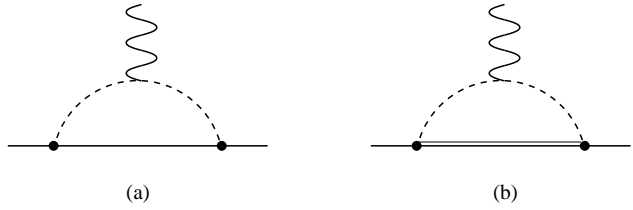}}
\caption{\label{fig:mmloop1}Feynman diagrams that yield order $\mathcal{O}(m_q^{1/2})$ corrections to the magnetic moments of octet baryons. Dashed lines denote mesons and single and double solid lines denote octet and decuplet baryons, respectively. Similar diagrams arise for the magnetic moment of decuplet baryons and for decuplet-octet transition moments.}
\end{figure}

\begin{figure}[ht]
\scalebox{0.32}{\includegraphics{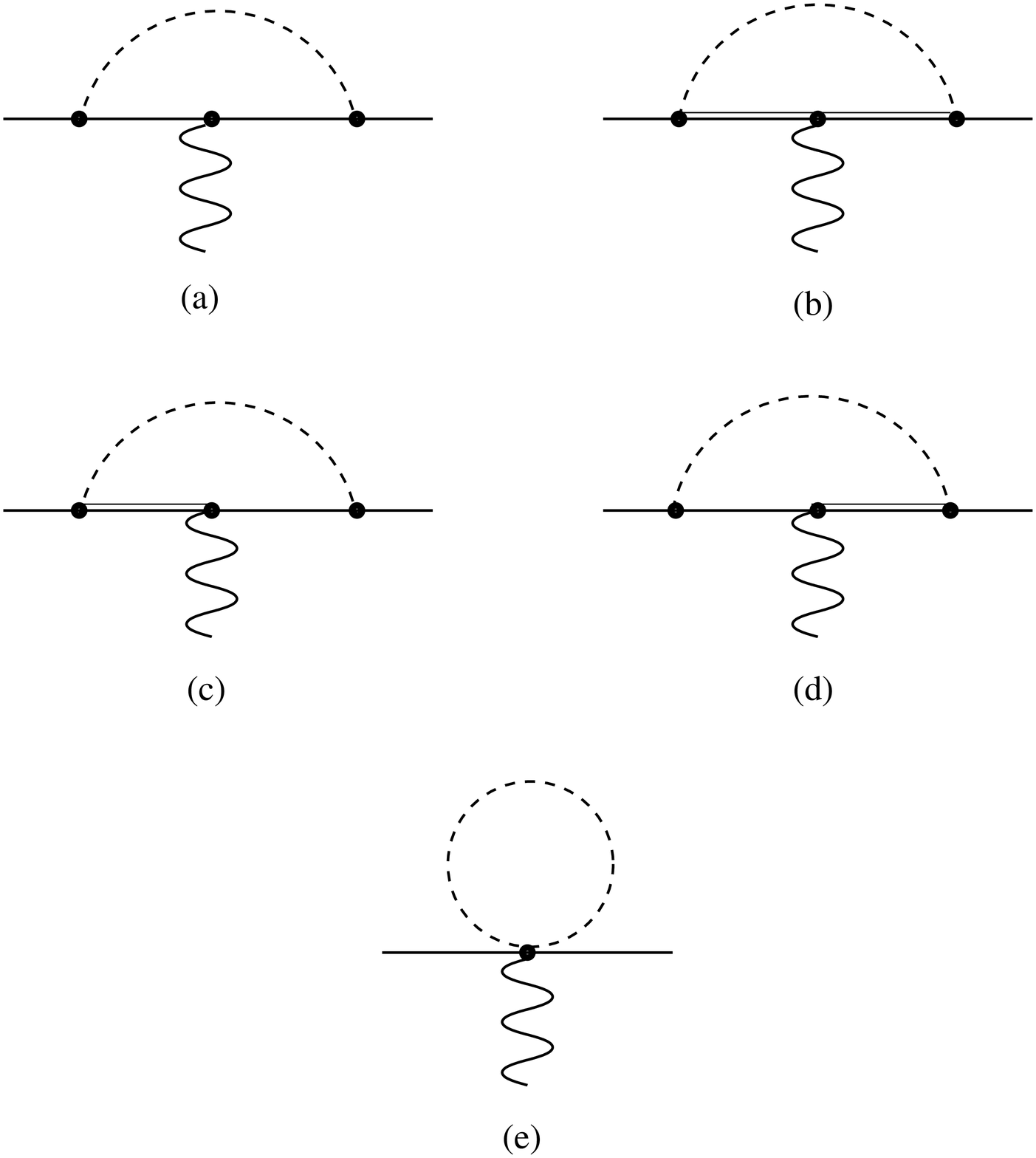}}
\caption{\label{fig:mmloop2}Feynman diagrams that yield order $\mathcal{O}(m_q \ln m_q)$ corrections to the magnetic moments of octet baryons. Dashed lines denote mesons and single and double solid lines denote octet and decuplet baryons, respectively. The wave function renormalization graphs are omitted in the figure but are nevertheless considered in the analysis. Similar diagrams arise for the magnetic moment of decuplet baryons and for decuplet-octet transition moments.}
\end{figure}

Baryon magnetic moments get corrections at one-loop order from the diagrams displayed in Figs.~\ref{fig:mmloop1} and \ref{fig:mmloop2}, which contribute to orders $\mathcal{O}(m_q^{1/2})$ and $\mathcal{O}(m_q \ln m_q)$, respectively. The group theoretical properties of these diagrams have been discussed in detail in previous works \cite{rfm09,rfm14} to a certain order in the $1/N_c$ expansion, so some partial results will be borrowed. A useful $1/N_c$ power counting scheme introduced in Ref.~\cite{rfm00} becomes handy for the purposes of the present analysis. On general grounds, the meson-baryon vertex is proportional to $g_A/f$; in the large-$N_c$ limit, $g_A \propto N_c$ and $f\propto \sqrt{N_c}$, so that the meson-baryon vertex is of order $\mathcal{O}(\sqrt{N_c})$ and grows with $N_c$. The baryon propagator is $i/(k\cdot v)$ and is $N_c$ independent and so is the meson propagator. Besides, in the $\overline{\mathrm{MS}}$ scheme, loop integrals are given by the pole structure of the propagators, so loop integrals are $N_c$ independent too. The tree-level matrix element of the baryon magnetic moment is thus is of order $\mathcal{O}(N_c)$.

In this section, one-loop corrections will be evaluated to all orders allowed for $N_c=3$ in the $1/N_c$ expansion. Each correction is dealt with separately due to its inherent complexity. 

\subsection{\label{sec:mq}Order ${\mathcal O}(m_q^{1/2})$ correction}

The one-loop correction of order ${\mathcal O}(m_q^{1/2})$ to baryon magnetic moments arising from Fig.~\ref{fig:mmloop1} can be expressed as \cite{rfm09}
\begin{equation}
\delta M_{\mathrm{loop\, 1}}^k = \sum_{\textsf{j}} \epsilon^{ijk} A^{ia} \mathcal{P}_{\textsf{j}} A^{jb} \Gamma^{ab}(\Delta_{\textsf{j}}). \label{eq:corrloop1}
\end{equation}
This correction has been studied in Refs.~\cite{rfm09} and \cite{rfm14} to relative order $1/N_c^3$ in the $1/N_c$ expansion for $\Delta=0$ and $\Delta \neq 0$, respectively. For definiteness, in Eq.~(\ref{eq:corrloop1}), the explicit sum over spin $\mathsf{j}$ is indicated but the sums over spin and flavor indices are understood, the baryon axial current operators $A^{ia}$ and $A^{jb}$, Eq.~(\ref{eq:akc}), are used at the meson-baryon vertices, $\mathcal{P}_{\mathsf{j}}$ is a spin projection operator for spin $J=\mathsf{j}$, and $\Gamma^{ab}(\Delta_{\textsf{j}})$ is an antisymmetric tensor which depends on the difference of the hyperfine mass splitting for spin $J=\mathsf{j}$ and the external baryon. The most general form of $\mathcal{P}_{\mathsf{j}}$ for arbitrary $N_c$ can be found in Ref.~\cite{jen96}. The spin-$\frac12$ and spin-$\frac32$ projectors for $N_c=3$ required here reduce to
\begin{subequations}
\label{eq:projnc3}
\begin{eqnarray}
\mathcal{P}_\frac12 & = & -\frac13 \left[ J^2 - \frac{15}{4} \right], \\
\mathcal{P}_\frac32 & = & \frac13 \left[ J^2 - \frac34 \right],
\end{eqnarray}
\end{subequations}
with
\begin{subequations}
\begin{equation}
\Delta_\frac12 = \left\{
\begin{array}{ll}
\displaystyle 0, & \mathsf{j}_{\mathrm{ext}}=\frac12, \\[2mm]
\displaystyle -\Delta, & \mathsf{j}_{\mathrm{ext}}=\frac32,
\end{array}
\right.
\end{equation}
\begin{equation}
\Delta_\frac32 = \left\{
\begin{array}{ll}
\displaystyle \Delta, & \mathsf{j}_{\mathrm{ext}}=\frac12, \\[2mm]
\displaystyle 0, & \mathsf{j}_{\mathrm{ext}}=\frac32. \\[2mm]
\end{array}
\right.
\end{equation}
\end{subequations}

The $\Gamma^{ab}(\Delta_{\textsf{j}})$ tensor, in turn, can be decomposed as \cite{rfm14}
\begin{equation}
\Gamma^{ab}(\Delta_{\textsf{j}}) = A_0(\Delta_{\textsf{j}}) \Gamma_0^{ab} + A_1(\Delta_{\textsf{j}}) \Gamma_1^{ab} + A_2(\Delta_{\textsf{j}}) \Gamma_2^{ab},
\end{equation}
where the tensors $\Gamma_i^{ab}$ are written as \cite{dai}
\begin{subequations}
\begin{eqnarray}
& & \Gamma_0^{ab} = f^{abQ}, \\
& & \Gamma_1^{ab} = f^{ab\overline{Q}}, \\
& & \Gamma_2^{ab} = f^{aeQ}d^{be8} - f^{beQ}d^{ae8} - f^{abe}d^{eQ8}. \label{eq:tens}
\end{eqnarray}
\end{subequations}
$\Gamma_0^{ab}$ and $\Gamma_1^{ab}$ are both $SU(3)$ octets and transform as the electric charge, except that the latter is rotated by $\pi$ in isospin space. $\Gamma_2^{ab}$ breaks $SU(3)$ as $\mathbf{10}+\overline{\mathbf{10}}$ \cite{dai}.

The $A_i(\Delta_{\textsf{j}})$ coefficients, on the other hand, read
\begin{subequations}
\label{eq:ais}
\begin{eqnarray}
A_0(\Delta_{\textsf{j}}) & = & \frac13 [ I_1(m_\pi,\Delta_{\textsf{j}},\mu)+2I_1(m_K,\Delta_{\textsf{j}},\mu) ], \\
A_1(\Delta_{\textsf{j}}) & = & \frac13 [ I_1(m_\pi,\Delta_{\textsf{j}},\mu)-I_1(m_K,\Delta_{\textsf{j}},\mu) ], \\
A_2(\Delta_{\textsf{j}}) & = & \frac{1}{\sqrt{3}}[ I_1(m_\pi,\Delta_{\textsf{j}},\mu)-I_1(m_K,\Delta_{\textsf{j}},\mu) ],
\end{eqnarray}
\end{subequations}
which are expressed in terms of the loop integral \cite{jen92}
\begin{eqnarray}
\frac{8\pi^2 f^2}{M_N} I_1(m,\Delta,\mu) = -\Delta \ln \frac{m^2}{\mu^2} + \left\{ \begin{array}{ll} \displaystyle 2\sqrt{m^2-\Delta^2}\left[\frac{\pi}{2}-\tan^{-1} \frac{\Delta}{\sqrt{m^2-\Delta^2}} \right], & |\Delta| \leq m, \\ [6mm]
\displaystyle \sqrt{\Delta^2-m^2} \left[-2i\pi + \ln{\frac{\Delta-\sqrt{\Delta^2-m^2}}{\Delta+\sqrt{\Delta^2-m^2}}} \right], & |\Delta| > m, \end{array} \right. \label{eq:loopi}
\end{eqnarray}
where $M_N$ and $m$ denote the nucleon and meson masses, respectively, and $\mu$ is the scale of dimensional regularization. In the limit of vanishing $\Delta$, the integral reduces to
\begin{equation}
I_1(m,0,\mu) = \frac{1}{8\pi f^2} M_N m,
\end{equation}
where the order $\mathcal{O}(m_q^{1/2})$ now becomes evident. A close inspection to Eq.~(\ref{eq:corrloop1}) reveals that, according to the $1/N_c$ power counting scheme reviewed above, the diagram is actually $\mathcal{O}(m_q^{1/2}N_c)$, so it is leading order in $N_c$. In the limit of small $m_q$, this diagram should be the dominant source of SB.

Collecting all partial contributions, $\delta M_{\mathrm{loop\, 1}}^k$ can be expressed as \cite{rfm14}
\begin{equation}
\delta M_{\mathrm{loop\, 1}}^k = \sum_{\mathsf{j}} \left[A_0(\Delta_{\mathsf{j}}) M_{\mathbf{8},\mathrm{loop\, 1}}^{kQ}(\mathcal{P}_{\mathsf{j}}) + A_1(\Delta_{\mathsf{j}}) M_{\mathbf{8},\mathrm{loop\, 1}}^{k\overline{Q}}(\mathcal{P}_{\mathsf{j}}) + A_2(\Delta_{\mathsf{j}}) M_{\mathbf{10}+\overline{\mathbf{10}},\mathrm{loop\, 1}}^{kQ}(\mathcal{P}_{\mathsf{j}}) \right], \label{eq:loop1}
\end{equation}
where the flavor contributions $M_{\mathbf{rep},\mathrm{loop\, 1}}^{kc}$ transforming under representation $\mathbf{rep}$ of $SU(3)$ read
\begin{equation}
M_{\mathbf{8},\mathrm{loop\, 1}}^{kc}(\mathcal{P}_{\mathsf{j}}) = \epsilon^{ijk} f^{abc} A^{ia}\mathcal{P}_{\mathsf{j}}A^{jb}, \label{eq:m8l1}
\end{equation}
and
\begin{equation}
M_{\mathbf{10}+\overline{\mathbf{10}},\mathrm{loop\, 1}}^{kc}(\mathcal{P}_{\mathsf{j}}) = \epsilon^{ijk}(f^{aec}d^{be8} - f^{bec}d^{ae8} - f^{abe}d^{ec8})A^{ia} \mathcal{P}_{\mathsf{j}} A^{jb}. \label{eq:m10l1}
\end{equation}
Terms up to relative order $1/N_c^3$ in the $1/N_c$ expansion from the above expressions have been evaluated for spin-independent and spin-dependent contributions in Refs.~\cite{rfm09} and \cite{rfm14}, respectively. Terms that participate to the next relative order, $1/N_c^4$, for instance $\mathcal{D}_3^{ia} \mathcal{O}_3^{ia}$ or $\mathcal{D}_3^{ia} J^2 \mathcal{D}_3^{ia}$, would complete the calculation for $N_c=3$ so they are evaluated and listed in Appendix \ref{app:rloop1} for the sake of completeness.

Order $\mathcal{O}(m_q^{1/2})$ corrections to baryon magnetic moments can be cast into the generic form
\begin{equation}
\delta \mu_{B}^{\mathrm{(loop\, 1)}} = \sum_{j=1}^{41} \mu_j^{\mathrm{(loop\, 1)}} \langle B|O_j^{3Q}|B\rangle, \label{eq:mml1}
\end{equation}
where $\mu_j^{\mathrm{(loop\, 1)}}$ are some coefficients and the operator basis $\{O_j\}$ reads
\begin{eqnarray}
\label{eq:basis8}
\begin{array}{ll}
O_{1}^{kc} = d^{c8e} G^{ke}, & 
O_{2}^{kc} = \delta^{c8} J^k, \\
O_{3}^{kc} = d^{c8e} \mathcal{D}_2^{ke}, & 
O_{4}^{kc} = \{G^{kc},T^8\}, \\
O_{5}^{kc} = \{G^{k8},T^c\}, & 
O_{6}^{kc} = i f^{c8e} [J^2,G^{ke}], \\ 
O_{7}^{kc} = d^{c8e} \mathcal{D}_3^{ke}, & 
O_{8}^{kc} = d^{c8e} \mathcal{O}_3^{ke}, \\
O_{9}^{kc} = \{G^{kc},\{J^r,G^{r8}\}\}, & 
O_{10}^{kc} = \{G^{k8},\{J^r,G^{rc}\}\}, \\
O_{11}^{kc} = \{J^k,\{T^c,T^8\}\}, & 
O_{12}^{kc} = \{J^k,\{G^{rc},G^{r8}\}\}, \\
O_{13}^{kc} = \delta^{c8} \{J^2,J^k\}, & 
O_{14}^{kc} = d^{c8e} \mathcal{D}_4^{ke}, \\
O_{15}^{kc} = \{\mathcal{D}_2^{kc},\{J^r,G^{r8}\}\}, & 
O_{16}^{kc} = \{\mathcal{D}_2^{k8},\{J^r,G^{rc}\}\}, \\
O_{17}^{kc} = \{J^2,\{G^{kc},T^8\}\}, & 
O_{18}^{kc} = \{J^2,\{G^{k8},T^c\}\}, \\
O_{19}^{kc} = i f^{c8e} \{J^2,[J^2,G^{ke}]\}, & 
O_{20}^{kc} = d^{c8e} \mathcal{D}_5^{ke}, \\
O_{21}^{kc} = d^{c8e} \mathcal{O}_5^{ke}, &
O_{22}^{kc} = \{J^2,\{G^{kc},\{J^r,G^{r8}\}\}\}, \\
O_{23}^{kc} = \{J^2,\{G^{k8},\{J^r,G^{rc}\}\}\}, &
O_{24}^{kc} = \{J^2,\{J^k,\{T^c,T^8\}\}\}, \\
O_{25}^{kc} = \{J^2,\{J^k,\{G^{rc},G^{r8}\}\}\}, &
O_{26}^{kc} = \{J^k,\{\{J^m,G^{mc}\},\{J^r,G^{r8}\}\}\}, \\
O_{27}^{kc} = \delta^{c8} \{J^2,\{J^2,J^k\}\}, &
O_{28}^{kc} = d^{c8e} \mathcal{D}_6^{ke}, \\
O_{29}^{kc} = \{J^2,\{\mathcal{D}_2^{kc},\{J^r,G^{r8}\}\}\}, &
O_{30}^{kc} = \{J^2,\{\mathcal{D}_2^{k8},\{J^r,G^{rc}\}\}\}, \\ 
O_{31}^{kc} = \{J^2,\{J^2,\{G^{kc},T^8\}\}\}, &
O_{32}^{kc} = \{J^2,\{J^2,\{G^{k8},T^c\}\}\}, \\ 
O_{33}^{kc} = i f^{c8e} \{J^2,\{J^2,[J^2,G^{ke}]\}\}, &
O_{34}^{kc} = d^{c8e} \mathcal{D}_7^{ke}, \\
O_{35}^{kc} = d^{c8e} \mathcal{O}_7^{ke}, & 
O_{36}^{kc} = \{J^2,\{J^2,\{G^{kc},\{J^r,G^{r8}\}\}\}\}, \\
O_{37}^{kc} = \{J^2,\{J^2,\{G^{k8},\{J^r,G^{rc}\}\}\}\}, & 
O_{38}^{kc} = \{J^2,\{J^2,\{J^k,\{T^c,T^8\}\}\}\}, \\
O_{39}^{kc} = \{J^2,\{J^2,\{J^k,\{G^{rc},G^{r8}\}\}\}\}, & 
O_{40}^{kc} = \{J^2,\{J^k,\{\{J^m,G^{mc}\},\{J^r,G^{r8}\}\}\}\}, \\
O_{41}^{kc} = \delta^{c8} \{J^2,\{J^2,\{J^2,J^k\}\}\}. & \\
\end{array}
\end{eqnarray}
Nontrivial matrix elements for the baryon operators contained in the operator basis (\ref{eq:basis8}) are listed in Tables \ref{t:mm8O}, \ref{t:mm8T}, and \ref{t:mm8TO}.

\begin{table*}
\caption{\label{t:mm8O}Nontrivial matrix elements of the operators involved in the magnetic moments of octet baryons: flavor octet representation.
The entries for isovector components correspond to $\sqrt{3} \langle O_i^{33} \rangle$.}
\begin{ruledtabular}
\begin{tabular}{lccccccccc}
& $\displaystyle n$ & $\displaystyle p$ & $\displaystyle \Sigma^-$ & $\displaystyle \Sigma^0$ & $\displaystyle \Sigma^+$ & $\displaystyle \Xi^-$ & $\displaystyle \Xi^0$ & $\displaystyle \Lambda$ & $\displaystyle \Sigma^0\Lambda$ \\[2mm]
\hline
$\langle O_{1}^{33} \rangle$ & $-\frac{5}{12}$ & $\frac{5}{12}$ & $-\frac13$ & $0$ & $\frac13$ & $\frac{1}{12}$ & $-\frac{1}{12}$ & $0$ & $\frac{1}{2 \sqrt{3}}$ \\
$\langle O_{2}^{33} \rangle$ & $0$ & $0$ & $0$ & $0$ & $0$ & $0$ & $0$ & $0$ & $0$ \\
$\langle O_{3}^{33} \rangle$ & $-\frac14$ & $\frac14$ & $-\frac12$ & $0$ & $\frac12$ & $-\frac14$ & $\frac14$ & $0$ & $0$ \\
$\langle O_{4}^{33} \rangle$ & $-\frac54$ & $\frac54$ & $0$ & $0$ & $0$ & $-\frac14$ & $\frac14$ & $0$ & $0$ \\
$\langle O_{5}^{33} \rangle$ & $-\frac14$ & $\frac14$ & $-1$ & $0$ & $1$ & $\frac34$ & $-\frac34$ & $0$ & $0$ \\
$\langle O_{7}^{33} \rangle$ & $-\frac54$ & $\frac54$ & $-1$ & $0$ & $1$ & $\frac14$ & $-\frac14$ & $0$ & $\frac{\sqrt{3}}{2}$ \\
$\langle O_{9}^{33} \rangle$ & $-\frac58$ & $\frac58$ & $-1$ & $0$ & $1$ & $-\frac38$ & $\frac38$ & $0$ & $0$ \\
$\langle O_{10}^{33} \rangle$ & $-\frac58$ & $\frac58$ & $-1$ & $0$ & $1$ & $-\frac38$ & $\frac38$ & $0$ & $0$ \\
$\langle O_{11}^{33} \rangle$ & $-\frac32$ & $\frac32$ & $0$ & $0$ & $0$ & $\frac32$ & $-\frac32$ & $0$ & $0$ \\
$\langle O_{12}^{33} \rangle$ & $-\frac58$ & $\frac58$ & $-2$ & $0$ & $2$ & $-\frac{11}{8}$ & $\frac{11}{8}$ & $0$ & $-\frac{\sqrt{3}}{2}$ \\
$\langle O_{15}^{33} \rangle$ & $-\frac38$ & $\frac38$ & $-\frac32$ & $0$ & $\frac32$ & $\frac98$ & $-\frac98$ & $0$ & $0$ \\
$\langle O_{16}^{33} \rangle$ & $-\frac{15}{8}$ & $\frac{15}{8}$ & $0$ & $0$ & $0$ & $-\frac38$ & $\frac38$ & $0$ & $0$ \\
$\langle O_{26}^{33} \rangle$ & $-\frac{15}{8}$ & $\frac{15}{8}$ & $-3$ & $0$ & $3$ & $-\frac98$ & $\frac98$ & $0$ & $0$ \\
\hline
$\langle O_{1}^{38} \rangle$ & $-\frac{1}{12}$ & $-\frac{1}{12}$ & $-\frac16$ & $-\frac16$ & $-\frac16$ & $\frac14$ & $\frac14$ & $\frac16$ & $0$ \\
$\langle O_{2}^{38} \rangle$ & $\frac12$ & $\frac12$ & $\frac12$ & $\frac12$ & $\frac12$ & $\frac12$ & $\frac12$ & $\frac12$ & $0$ \\
$\langle O_{3}^{38} \rangle$ & $-\frac14$ & $-\frac14$ & $0$ & $0$ & $0$ & $\frac14$ & $\frac14$ & $0$ & $0$ \\
$\langle O_{4}^{38} \rangle$ & $\frac14$ & $\frac14$ & $0$ & $0$ & $0$ & $\frac34$ & $\frac34$ & $0$ & $0$ \\
$\langle O_{5}^{38} \rangle$ & $\frac14$ & $\frac14$ & $0$ & $0$ & $0$ & $\frac34$ & $\frac34$ & $0$ & $0$ \\
$\langle O_{7}^{38} \rangle$ & $-\frac14$ & $-\frac14$ & $-\frac12$ & $-\frac12$ & $-\frac12$ & $\frac34$ & $\frac34$ & $\frac12$ & $0$ \\
$\langle O_{9}^{38} \rangle$ & $\frac18$ & $\frac18$ & $\frac12$ & $\frac12$ & $\frac12$ & $\frac98$ & $\frac98$ & $\frac12$ & $0$ \\
$\langle O_{10}^{38} \rangle$ & $\frac18$ & $\frac18$ & $\frac12$ & $\frac12$ & $\frac12$ & $\frac98$ & $\frac98$ & $\frac12$ & $0$ \\
$\langle O_{11}^{38} \rangle$ & $\frac32$ & $\frac32$ & $0$ & $0$ & $0$ & $\frac32$ & $\frac32$ & $0$ & $0$ \\
$\langle O_{12}^{38} \rangle$ & $\frac18$ & $\frac18$ & $\frac32$ & $\frac32$ & $\frac32$ & $\frac{17}{8}$ & $\frac{17}{8}$ & $\frac12$ & $0$ \\
$\langle O_{15}^{38} \rangle$ & $\frac38$ & $\frac38$ & $0$ & $0$ & $0$ & $\frac98$ & $\frac98$ & $0$ & $0$ \\
$\langle O_{16}^{38} \rangle$ & $\frac38$ & $\frac38$ & $0$ & $0$ & $0$ & $\frac98$ & $\frac98$ & $0$ & $0$ \\
$\langle O_{26}^{38} \rangle$ & $\frac38$ & $\frac38$ & $\frac32$ & $\frac32$ & $\frac32$ & $\frac{27}{8}$ & $\frac{27}{8}$ & $\frac32$ & $0$ \\
\end{tabular}
\end{ruledtabular}
\end{table*}

\begin{table*}
\caption{\label{t:mm8T}Nontrivial matrix elements of the operators involved in the magnetic moments of decuplet baryons: flavor octet representation. The entries for isovector components correspond to $\sqrt{3} \langle O_i^{33} \rangle$.}
\begin{ruledtabular}
\begin{tabular}{lcccccccccc}
& $\displaystyle \Delta^{++}$ & $\displaystyle \Delta^+$ & $\displaystyle \Delta^0$ & $\displaystyle \Delta^-$ & $\displaystyle {\Sigma^*}^+$ & $\displaystyle {\Sigma^*}^0$ & $\displaystyle {\Sigma^*}^-$ & $\displaystyle {\Xi^*}^0$ & $\displaystyle {\Xi^*}^-$ & $\displaystyle \Omega^-$ \\[2mm]
\hline
$\langle O_{1}^{33} \rangle$ & $\frac34$ & $\frac14$ & $-\frac14$ & $-\frac34$ & $\frac12$ & $0$ & $-\frac12$ & $\frac14$ & $-\frac14$ & $0$ \\
$\langle O_{2}^{33} \rangle$ & $0$ & $0$ & $0$ & $0$ & $0$ & $0$ & $0$ & $0$ & $0$ & $0$ \\
$\langle O_{3}^{33} \rangle$ & $\frac94$ & $\frac34$ & $-\frac34$ & $-\frac94$ & $\frac32$ & $0$ & $-\frac32$ & $\frac34$ & $-\frac34$ & $0$ \\
$\langle O_{4}^{33} \rangle$ & $\frac94$ & $\frac34$ & $-\frac34$ & $-\frac94$ & $0$ & $0$ & $0$ & $-\frac34$ & $\frac34$ & $0$ \\
$\langle O_{5}^{33} \rangle$ & $\frac94$ & $\frac34$ & $-\frac34$ & $-\frac94$ & $0$ & $0$ & $0$ & $-\frac34$ & $\frac34$ & $0$ \\
$\langle O_{7}^{33} \rangle$ & $\frac{45}{4}$ & $\frac{15}{4}$ & $-\frac{15}{4}$ & $-\frac{45}{4}$ & $\frac{15}{2}$ & $0$ & $-\frac{15}{2}$ & $\frac{15}{4}$ & $-\frac{15}{4}$ & $0$ \\
$\langle O_{9}^{33} \rangle$ & $\frac{45}{8}$ & $\frac{15}{8}$ & $-\frac{15}{8}$ & $-\frac{45}{8}$ & $0$ & $0$ & $0$ & $-\frac{15}{8}$ & $\frac{15}{8}$ & $0$ \\
$\langle O_{10}^{33} \rangle$ & $\frac{45}{8}$ & $\frac{15}{8}$ & $-\frac{15}{8}$ & $-\frac{45}{8}$ & $0$ & $0$ & $0$ & $-\frac{15}{8}$ & $\frac{15}{8}$ & $0$ \\
$\langle O_{11}^{33} \rangle$ & $\frac{27}{2}$ & $\frac92$ & $-\frac92$ & $-\frac{27}{2}$ & $0$ & $0$ & $0$ & $-\frac92$ & $\frac92$ & $0$ \\
$\langle O_{12}^{33} \rangle$ & $\frac{45}{8}$ & $\frac{15}{8}$ & $-\frac{15}{8}$ & $-\frac{45}{8}$ & $\frac32$ & $0$ & $-\frac32$ & $-\frac38$ & $\frac38$ & $0$ \\
$\langle O_{15}^{33} \rangle$ & $\frac{135}{8}$ & $\frac{45}{8}$ & $-\frac{45}{8}$ & $-\frac{135}{8}$ & $0$ & $0$ & $0$ & $-\frac{45}{8}$ & $\frac{45}{8}$ & $0$ \\
$\langle O_{16}^{33} \rangle$ & $\frac{135}{8}$ & $\frac{45}{8}$ & $-\frac{45}{8}$ & $-\frac{135}{8}$ & $0$ & $0$ & $0$ & $-\frac{45}{8}$ & $\frac{45}{8}$ & $0$ \\
$\langle O_{26}^{33} \rangle$ & $\frac{675}{8}$ & $\frac{225}{8}$ & $-\frac{225}{8}$ & $-\frac{675}{8}$ & $0$ & $0$ & $0$ & $-\frac{225}{8}$ & $\frac{225}{8}$ & $0$ \\
\hline
$\langle O_{1}^{38} \rangle$ & $-\frac14$ & $-\frac14$ & $-\frac14$ & $-\frac14$ & $0$ & $0$ & $0$ & $\frac14$ & $\frac14$ & $\frac12$ \\
$\langle O_{2}^{38} \rangle$ & $\frac32$ & $\frac32$ & $\frac32$ & $\frac32$ & $\frac32$ & $\frac32$ & $\frac32$ & $\frac32$ & $\frac32$ & $\frac32$
\\
$\langle O_{3}^{38} \rangle$ & $-\frac34$ & $-\frac34$ & $-\frac34$ & $-\frac34$ & $0$ & $0$ & $0$ & $\frac34$ & $\frac34$ & $\frac32$ \\
$\langle O_{4}^{38} \rangle$ & $\frac34$ & $\frac34$ & $\frac34$ & $\frac34$ & $0$ & $0$ & $0$ & $\frac34$ & $\frac34$ & $3$ \\
$\langle O_{5}^{38} \rangle$ & $\frac34$ & $\frac34$ & $\frac34$ & $\frac34$ & $0$ & $0$ & $0$ & $\frac34$ & $\frac34$ & $3$ \\
$\langle O_{7}^{38} \rangle$ & $-\frac{15}{4}$ & $-\frac{15}{4}$ & $-\frac{15}{4}$ & $-\frac{15}{4}$ & $0$ & $0$ & $0$ & $\frac{15}{4}$ & $\frac{15}{4}$ & $\frac{15}{2}$ \\
$\langle O_{9}^{38} \rangle$ & $\frac{15}{8}$ & $\frac{15}{8}$ & $\frac{15}{8}$ & $\frac{15}{8}$ & $0$ & $0$ & $0$ & $\frac{15}{8}$ & $\frac{15}{8}$ & $\frac{15}{2}$ \\
$\langle O_{10}^{38} \rangle$ & $\frac{15}{8}$ & $\frac{15}{8}$ & $\frac{15}{8}$ & $\frac{15}{8}$ & $0$ & $0$ & $0$ & $\frac{15}{8}$ & $\frac{15}{8}$ & $\frac{15}{2}$ \\
$\langle O_{11}^{38} \rangle$ & $\frac92$ & $\frac92$ & $\frac92$ & $\frac92$ & $0$ & $0$ & $0$ & $\frac92$ & $\frac92$ & $18$ \\
$\langle O_{12}^{38} \rangle$ & $\frac{15}{8}$ & $\frac{15}{8}$ & $\frac{15}{8}$ & $\frac{15}{8}$ & $\frac32$ & $\frac32$ & $\frac32$ & $\frac{27}{8}$ & $\frac{27}{8}$ & $\frac{15}{2}$ \\
$\langle O_{15}^{38} \rangle$ & $\frac{45}{8}$ & $\frac{45}{8}$ & $\frac{45}{8}$ & $\frac{45}{8}$ & $0$ & $0$ & $0$ & $\frac{45}{8}$ & $\frac{45}{8}$ & $\frac{45}{2}$ \\
$\langle O_{16}^{38} \rangle$ & $\frac{45}{8}$ & $\frac{45}{8}$ & $\frac{45}{8}$ & $\frac{45}{8}$ & $0$ & $0$ & $0$ & $\frac{45}{8}$ & $\frac{45}{8}$ & $\frac{45}{2}$ \\
$\langle O_{26}^{38} \rangle$ & $\frac{225}{8}$ & $\frac{225}{8}$ & $\frac{225}{8}$ & $\frac{225}{8}$ & $0$ & $0$ & $0$ & $\frac{225}{8}$ & $\frac{225}{8}$ & $\frac{225}{2}$ \\
\end{tabular}
\end{ruledtabular}
\end{table*}

\begin{table*}
\caption{\label{t:mm8TO}Nontrivial matrix elements of the operators involved in the decuplet to octet transition magnetic moments: Flavor octet representation. The entries for isovector and isoscalar components correspond to $\sqrt{6} \langle O_i^{33} \rangle$ and $\sqrt{2} \langle O_j^{38} \rangle$, respectively.}
\begin{ruledtabular}
\begin{tabular}{lcccccccc}
& $\displaystyle \Delta^+p$ & $\displaystyle \Delta^0n$ & $\displaystyle {\Sigma^*}^0\Lambda$ & $\displaystyle {\Sigma^*}^0\Sigma^0$ & $\displaystyle {\Sigma^*}^+\Sigma^+$ & $\displaystyle {\Sigma^*}^-\Sigma^-$ & $\displaystyle {\Xi^*}^0\Xi^0$ & $\displaystyle {\Xi^*}^-\Xi^-$ \\[2mm]
\hline
$\langle O_{1}^{33} \rangle$ & $\frac23$ & $\frac23$ & $\frac{1}{\sqrt{3}}$ & $0$ & $\frac13$ & $-\frac13$ & $\frac13$ & $-\frac13$ \\
$\langle O_{4}^{33} \rangle$ & $2$ & $2$ & $0$ & $0$ & $0$ & $0$ & $-1$ & $1$ \\
$\langle O_{5}^{33} \rangle$ & $0$ & $0$ & $0$ & $0$ & $2$ & $-2$ & $1$ & $-1$ \\
$\langle O_{8}^{33} \rangle$ & $3$ & $3$ & $\frac{3 \sqrt{3}}{2}$ & $0$ & $\frac32$ & $-\frac32$ & $\frac32$ & $-\frac32$ \\
$\langle O_{9}^{33} \rangle$ & $3$ & $3$ & $-\frac{\sqrt{3}}{2}$ & $0$ & $\frac12$ & $-\frac12$ & $-2$ & $2$ \\
$\langle O_{10}^{33} \rangle$ & $0$ & $0$ & $-\frac{\sqrt{3}}{2}$ & $0$ & $\frac72$ & $-\frac72$ & $1$ & $-1$ \\
\hline
$\langle O_{1}^{38} \rangle$ & $0$ & $0$ & $0$ & $-\frac13$ & $-\frac13$ & $-\frac13$ & $-\frac13$ & $-\frac13$ \\
$\langle O_{4}^{38} \rangle$ & $0$ & $0$ & $0$ & $0$ & $0$ & $0$ & $-1$ & $-1$ \\
$\langle O_{5}^{38} \rangle$ & $0$ & $0$ & $0$ & $0$ & $0$ & $0$ & $-1$ & $-1$ \\
$\langle O_{8}^{38} \rangle$ & $0$ & $0$ & $0$ & $-\frac32$ & $-\frac32$ & $-\frac32$ & $-\frac32$ & $-\frac32$ \\
$\langle O_{9}^{38} \rangle$ & $0$ & $0$ & $0$ & $\frac12$ & $\frac12$ & $\frac12$ & $-2$ & $-2$ \\
$\langle O_{10}^{38} \rangle$ & $0$ & $0$ & $0$ & $\frac12$ & $\frac12$ & $\frac12$ & $-2$ & $-2$ \\
\end{tabular}
\end{ruledtabular}
\end{table*}

Resultant expressions are, for instance,
\begin{eqnarray}
\delta \mu_{\Sigma^-}^{\mathrm{(loop\, 1)}} & = & \left[ \frac{7}{18} a_1^2 + \frac29 a_1b_2 + \frac{1}{18} b_2^2 + \frac{7}{27} a_1b_3 + \frac{2}{27} b_2b_3 + \frac{7}{162} b_3^2 \right] I_1(m_\pi,0,\mu) \nonumber \\
& & \mbox{} + \left[ \frac{1}{36} a_1^2 - \frac{1}{18} a_1b_2 + \frac{1}{36} b_2^2 + \frac{1}{54} a_1b_3 - \frac{1}{54} b_2b_3 + \frac{1}{324} b_3^2 \right] I_1(m_K,0,\mu) \nonumber \\
& & \mbox{} + \left[ -\frac{1}{18} a_1^2 - \frac{1}{18} a_1c_3 - \frac{1}{72} c_3^2 \right] I_1(m_\pi,\Delta,\mu) + \left[ -\frac19 a_1^2 - \frac19 a_1c_3 - \frac{1}{36} c_3^2 \right] I_1(m_K,\Delta,\mu), \label{eq:case1}
\end{eqnarray}
and
\begin{eqnarray}
\delta \mu_{{\Sigma^*}^-}^{\mathrm{(loop\, 1)}} & = & \left[ \frac16 a_1^2 + \frac13 a_1b_2 + \frac16 b_2^2 + \frac59 a_1b_3 + \frac59 b_2b_3 + \frac{25}{54} b_3^2 \right] I_1(m_\pi,0,\mu) \nonumber \\
& & \mbox{} + \left[ \frac{1}{12} a_1^2 + \frac16 a_1b_2 + \frac{1}{12} b_2^2 + \frac{5}{18} a_1b_3 + \frac{5}{18} b_2b_3 + \frac{25}{108} b_3^2 \right] I_1(m_K,0,\mu) \nonumber \\
& & \mbox{} + \left[ \frac13 a_1^2 + \frac13 a_1c_3 + \frac{1}{12} c_3^2 \right] I_1(m_\pi,-\Delta,\mu) + \left[ \frac16 a_1^2 + \frac16 a_1c_3 + \frac{1}{24} c_3^2 \right] I_1(m_K,-\Delta,\mu). \label{eq:case2}
\end{eqnarray}
All 27 resultant expressions are listed in full in Appendix \ref{app:Loop1}.

It can be easily verified that Coleman and Glashow relations are satisfied when order $\mathcal{O}(m_q^{1/2})$ corrections are included to baryon magnetic moments, even for $\Delta\neq 0$. For decuplet baryons the $I=2$ and $I=3$ sum rules introduced in Ref.~\cite{lebed} are also satisfied. For $I=2$
\begin{equation}
\mu_{\Delta^{++}}^{\mathrm{(loop\, 1)}} - \mu_{\Delta^+}^{\mathrm{(loop\, 1)}} - \mu_{\Delta^0}^{\mathrm{(loop\, 1)}} + \mu_{\Delta^-}^{\mathrm{(loop\, 1)}} = 0,
\end{equation}
\begin{equation}
\mu_{{\Sigma^*}^+}^{\mathrm{(loop\, 1)}} - 2 \mu_{{\Sigma^*}^0}^{\mathrm{(loop\, 1)}} + \mu_{{\Sigma^*}^-}^{\mathrm{(loop\, 1)}} = 0,
\end{equation}
whereas for $I=3$
\begin{equation}
\mu_{\Delta^{++}}^{\mathrm{(loop\, 1)}} - 3 \mu_{\Delta^+}^{\mathrm{(loop\, 1)}} + 3 \mu_{\Delta^0}^{\mathrm{(loop\, 1)}} - \mu_{\Delta^-}^{\mathrm{(loop\, 1)}} = 0.
\end{equation}

For transition magnetic moments, the isotensor combinations for $I=2$ read \cite{lebed}
\begin{equation}
\mu_{\Delta^{+}p}^{\mathrm{(loop\, 1)}} - \mu_{\Delta^{0}n}^{\mathrm{(loop\, 1)}} = 0,
\end{equation}
and
\begin{equation}
\mu_{{\Sigma^{*}}^+\Sigma^+}^{\mathrm{(loop\, 1)}} - 2 \mu_{{\Sigma^{*}}^0\Sigma^0}^{\mathrm{(loop\, 1)}} + \mu_{{\Sigma^{*}}^-\Sigma^-}^{\mathrm{(loop\, 1)}} = 0, \label{eq:is6}
\end{equation}
where $\mu_X^{\mathrm{(loop\, 1)}}$ should be understood as $\mu_X + \delta \mu_X^{\mathrm{(loop\, 1)}}$ for baryon $X$.	

\subsubsection{\label{sec:comL1}Comparison with heavy chiral perturbation theory results}

The full expressions (\ref{eq:mun}) to (\ref{eq:muxixi}) can be rewritten in terms of the flavor octet baryon-meson couplings $D$, $F$, $\mathcal{C}$, and $\mathcal{H}$ introduced in Refs.~\cite{jm91a,jm91b}, which are related to the coefficients of the $1/N_c$ expansion $a_1$, $b_2$, $b_3$, and $c_3$ at $N_c=3$. The relations are \cite{jen96}
\begin{subequations}
\label{eq:rel1}
\begin{eqnarray}
& & D = \frac12 a_1 + \frac16 b_3, \\
& & F = \frac13 a_1 + \frac16 b_2 + \frac19 b_3, \\
& & \mathcal{C} = - a_1 - \frac12 c_3, \\
& & \mathcal{H} = - \frac32 a_1 - \frac32 b_2 - \frac52 b_3.
\end{eqnarray}
\end{subequations}
so the inverse relations become
\begin{subequations}
\label{eq:rel1inv}
\begin{eqnarray}
& & a_1 = \frac32 D + \frac32 F + \frac16 \mathcal{H}, \\
& & b_2 = -4D + 6F, \\
& & b_3 = \frac32 D - \frac92 F - \frac12 \mathcal{H}, \\
& & c_3 = - 3D - 3F -2 \mathcal{C} - \frac13 \mathcal{H}.
\end{eqnarray}
\end{subequations}

Using the inverse relations (\ref{eq:rel1inv}), expressions (\ref{eq:mun}) to (\ref{eq:muxixi}) now become (\ref{eq:munch}) to (\ref{eq:muxixich}), respectively. In particular, for magnetic moments in the case study, Eqs.~(\ref{eq:case1}) and (\ref{eq:case2}) can be rewritten as 
\begin{equation}
\delta \mu_{\Sigma^-}^{\mathrm{(loop\, 1)}} = \frac23(D^2+3F^2) I_1(m_\pi,0,\mu) + (D-F)^2 I_1(m_K,0,\mu) - \frac{1}{18} \mathcal{C}^2 I_1(m_\pi,\Delta,\mu) - \frac19 \mathcal{C}^2 I_1(m_K,\Delta,\mu),
\end{equation}
and
\begin{equation}
\delta \mu_{{\Sigma^*}^-}^{\mathrm{(loop\, 1)}} = \frac{2}{27} \mathcal{H}^2 I_1(m_\pi,0,\mu) + \frac{1}{27} \mathcal{H}^2 I_1(m_K,0,\mu) + \frac13 \mathcal{C}^2 I_1(m_\pi,-\Delta,\mu) + \frac16 \mathcal{C}^2 I_1(m_K,-\Delta,\mu).
\end{equation}

In the context of HBCHPT, order $\mathcal{O}(m_q^{1/2})$ corrections to the magnetic moments of octet baryons can be organized as \cite{jen92},
\begin{equation}
\delta \mu_i^{\mathrm{(loop\, 1)}} = \sum_{P=\pi, K}\beta_i^{(P)}I_1(m_P,0,\mu) + \sum_{P=\pi, K}\beta_i^{\prime (P)}I_1(m_P,\Delta,\mu), \label{eq:l1ch}
\end{equation}
where $\beta_i^{(P)}$ and $\beta_i^{\prime (P)}$ are the contributions arising from loop graphs of Fig.~\ref{fig:mmloop1} with intermediate octet and decuplet baryons, respectively. In the limit of vanishing $\Delta$, expressions (\ref{eq:munch}) to (\ref{eq:mul}) and (\ref{eq:musl} agree in full with the corresponding ones attainable from Eq.~(\ref{eq:l1ch}).

\subsection{\label{sec:mqlnmq}Order $\mathcal{O}(m_q \ln m_q)$ correction}

The one-loop corrections to baryon magnetic moments from the Feynman diagrams depicted in Fig.~\ref{fig:mmloop2} have a nonanalytic dependence on the quark mass of the form $m_q \ln m_q$. The computation of these diagrams requires a rather formidable effort to reduce the operator structures involved. In Refs.~\cite{rfm09,rfm14}, relative corrections to order $1/N_c^4$ in the $1/N_c$ expansion were included. The incorporation of all the structures present for $N_c=3$ needs the inclusion of relative terms of up to order $1/N_c^6$. Again, a great deal of computational ease is gained by using some of the operator structures already reduced in the renormalized baryon axial current computed in Ref.~\cite{rfm21}. Other structures appear for the first time and need to be reduced.

Diagrams \ref{fig:mmloop2}(a-d) present a few interesting features so they are studied first.

\subsubsection{Diagrams \ref{fig:mmloop2}(a)-\ref{fig:mmloop2}(d)}

Feynman diagrams depicted in Fig.~\ref{fig:mmloop2}(a)-~\ref{fig:mmloop2}(d) contribute to the baryon magnetic moment operator, for $\Delta=0$, as \cite{rfm09,rfm14}
\begin{equation}
\delta M_{\textrm{loop 2ad}}^k = \frac12 \left[A^{ja},\left[A^{jb},M^k \right] \right] \Pi^{ab}, \label{eq:corrloop2}
\end{equation}
The double commutator structure in Eq.~(\ref{eq:corrloop2}) involves three axial current operators, so naively this structure should be order $\mathcal{O}(N_c^3)$. However, it has been explicitly shown \cite{rfm00} that there are large-$N_c$ cancellations in the sum over intermediate baryon states in the loop. The cancellations are a consequence of the spin-flavor symmetry of large-$N_c$ QCD \cite{dm91a,dm91b,djm95} and only occur when the ratios of $F$, $D$, $\mathcal{C}$, and $\mathcal{H}$ are close to their $SU(6)$ values. Therefore, the double commutator structure is at most of order $\mathcal{O}(N_c)$.

On the other hand, $\Pi^{ab}$ is a symmetric tensor which contains meson-loop integrals and decomposes into flavor singlet, flavor $\mathbf{8}$, and flavor $\mathbf{27}$ representations as \cite{jen96}
\begin{equation}
\Pi^{ab} = F_\mathbf{1} \delta^{ab} + F_\mathbf{8} d^{ab8} + F_\mathbf{27} \left[ \delta^{a8} \delta^{b8} - \frac18 \delta^{ab} - \frac35 d^{ab8} d^{888}\right], \label{eq:pisym}
\end{equation}
where
\begin{equation}
F_\mathbf{1} = \frac18 \left[3I_2(m_\pi,0,\mu) + 4I_2(m_K,0,\mu) + I_2(m_\eta,0,\mu) \right], \label{eq:F1}
\end{equation}
\begin{equation}
F_\mathbf{8} = \frac{2\sqrt 3}{5} \left[\frac32 I_2(m_\pi,0,\mu) - I_2(m_K,0,\mu) - \frac12 I_2(m_\eta,0,\mu) \right], \label{eq:F8}
\end{equation}
and
\begin{equation}
F_\mathbf{27} = \frac13 I_2(m_\pi,0,\mu) - \frac43 I_2(m_K,0,\mu) + I_2(m_\eta,0,\mu). \label{eq:F27}
\end{equation}
Equations (\ref{eq:F1})-(\ref{eq:F27}) are linear combinations of $I_2(m_\pi,0,\mu)$, $I_2(m_K,0,\mu)$, and $I_2(m_\eta,0,\mu)$, where $I_2(m,\Delta,\mu)$ represents the loop integral, which can be found in Ref.~\cite{rfm14}. In the degeneracy limit $\Delta\to 0$, this function reduces to
\begin{equation}
I_2(m,0,\mu) = - \frac{m^2}{16\pi^2f^2} \ln{\frac{m^2}{\mu^2}}, \label{eq:fprime}
\end{equation}
where $\mu$ is the scale of dimensional regularization and only nonanalytic terms in $m$ have been retained.

Expression (\ref{eq:corrloop2}) can be organized in terms of the flavor $\mathbf{1}$, $\mathbf{8}$, and $\mathbf{27}$ contributions as \cite{rfm09}
\begin{equation}
\delta M_{\textrm{loop 2ad}}^k = F_\mathbf{1} M_{\mathbf{1},\textrm{loop 2ad}}^{kQ} + F_\mathbf{8} M_{\mathbf{8},\textrm{loop 2ad}}^{kQ} + F_\mathbf{27} M_{\mathbf{27},\textrm{loop 2ad}}^{kQ}. \label{eq:loop2ad}
\end{equation}
The matrix elements of the operator structures $M_{\mathbf{rep},\textrm{loop 2ad}}^{kQ}$ have the generic forms
\begin{equation}
\delta \mu_{j,\mathbf{1}}^{(\mathrm{loop\, 2ad})} = \sum_{j=1}^{10} \mu_{j,\mathrm{1}}^{(\mathrm{loop\, 2ad})} \langle B|S_j^{3Q}|B\rangle, \label{eq:mmsl2}
\end{equation}
\begin{equation}
\delta \mu_{j,\mathbf{8}}^{(\mathrm{loop\, 2ad})} = \sum_{j=1}^{41} \mu_{j,\mathrm{8}}^{(\mathrm{loop\, 2ad})} \langle B|O_j^{3Q}|B\rangle, \label{eq:mmol2}
\end{equation}
\begin{equation}
\delta \mu_{j,\mathbf{27}}^{(\mathrm{loop\, 2ad})} = \sum_{j=1}^{167} \mu_{j,\mathrm{27}}^{(\mathrm{loop\, 2ad})} \langle B|T_j^{3Q}|B\rangle, \label{eq:mmtl2}
\end{equation}
where as before $\mu_{j,\mathrm{rep}}^{(\mathrm{loop\, 2ad})}$ are some coefficients, the operator bases $\{S_i\}$ and $\{O_j\}$ are listed in (\ref{eq:basis1}) and (\ref{eq:basis8}), respectively, and the operator basis $\{T_k\}$ is given by
\begin{eqnarray}
\nonumber
\begin{array}{ll}
T_{1}^{kc} = f^{c8e} f^{8eg} G^{kg}, &
T_{2}^{kc} = d^{c8e} d^{8eg} G^{kg}, \\
T_{3}^{kc} = \delta^{c8} G^{k8}, &
T_{4}^{kc} = d^{c88} J^k, \\
T_{5}^{kc} = f^{c8e} f^{8eg} \mathcal{D}_2^{kg}, &
T_{6}^{kc} = d^{c8e} d^{8eg} \mathcal{D}_2^{kg}, \\
T_{7}^{kc} = d^{ceg} d^{88e} \mathcal{D}_2^{kg}, &
T_{8}^{kc} = \delta^{c8} \mathcal{D}_2^{k8}, \\
T_{9}^{kc} = d^{c8e} \{G^{ke},T^8\}, &
T_{10}^{kc} = d^{88e} \{G^{ke},T^c\}, \\
T_{11}^{kc} = i \epsilon^{kim} f^{c8e} f^{8eg} \{J^i,G^{mg}\}, &
T_{12}^{kc} = i f^{c8e} [G^{ke},\{J^r,G^{r8}\}], \\
T_{13}^{kc} = i f^{c8e} [G^{k8},\{J^r,G^{re}\}], &
T_{14}^{kc} = f^{c8e} f^{8eg} \mathcal{D}_3^{kg}, \\
T_{15}^{kc} = d^{c8e} d^{8eg} \mathcal{D}_3^{kg}, &
T_{16}^{kc} = d^{ceg} d^{88e} \mathcal{D}_3^{kg}, \\
T_{17}^{kc} = i f^{c8e} d^{8eg} \mathcal{D}_3^{kg}, &
T_{18}^{kc} = i d^{c8e} f^{8eg} \mathcal{D}_3^{kg}, \\
T_{19}^{kc} = \delta^{c8} \mathcal{D}_3^{k8}, &
T_{20}^{kc} = f^{c8e} f^{8eg} \mathcal{O}_3^{kg}, \\
T_{21}^{kc} = d^{c8e} d^{8eg} \mathcal{O}_3^{kg}, &
T_{22}^{kc} = d^{ceg} d^{88e} \mathcal{O}_3^{kg}, \\
T_{23}^{kc} = \delta^{c8} \mathcal{O}_3^{k8}, &
T_{24}^{kc} = d^{c88} \{J^2,J^k\}, \\
T_{25}^{kc} = \{G^{kc},\{T^8,T^8\}\}, &
T_{26}^{kc} = \{G^{k8},\{T^c,T^8\}\}, \\
T_{27}^{kc} = \{G^{kc},\{G^{r8},G^{r8}\}\}, &
T_{28}^{kc} = \{G^{k8},\{G^{rc},G^{r8}\}\}, \\
T_{29}^{kc} = d^{c8e} \{J^k,\{G^{re},G^{r8}\}\}, &
T_{30}^{kc} = d^{88e} \{J^k,\{G^{rc},G^{re}\}\}, \\
T_{31}^{kc} = d^{c8e} \{G^{ke},\{J^r,G^{r8}\}\}, &
T_{32}^{kc} = d^{c8e} \{G^{k8},\{J^r,G^{re}\}\}, \\
T_{33}^{kc} = d^{88e} \{G^{kc},\{J^r,G^{re}\}\}, &
T_{34}^{kc} = d^{88e} \{G^{ke},\{J^r,G^{rc}\}\}, \\
T_{35}^{kc} = \epsilon^{kim} f^{c8e} \{T^e,\{J^i,G^{m8}\}\}, &
T_{36}^{kc} = \epsilon^{kim} f^{c8e} \{T^8,\{J^i,G^{me}\}\}, \\
T_{37}^{kc} = f^{c8e} f^{8eg} \mathcal{D}_4^{kg}, &
T_{38}^{kc} = d^{c8e} d^{8eg} \mathcal{D}_4^{kg}, \\
T_{39}^{kc} = d^{ceg} d^{88e} \mathcal{D}_4^{kg}, &
T_{40}^{kc} = i f^{c8e} d^{8eg} \mathcal{D}_4^{kg}, \\
T_{41}^{kc} = \delta^{c8} \mathcal{D}_4^{k8}, &
T_{42}^{kc} = d^{c8e} \{J^2,\{G^{ke},T^8\}\}, \\
T_{43}^{kc} = d^{88e} \{J^2,\{G^{ke},T^c\}\}, &
T_{44}^{kc} = i \epsilon^{kim} f^{c8e} f^{8eg} \{J^2,\{J^i,G^{mg}\}\}, \\
T_{45}^{kc} = i \epsilon^{kim} \delta^{c8} \{J^2,\{J^i,G^{m8}\}\}, &
T_{46}^{kc} = \{\mathcal{D}_2^{kc},\{T^8,T^8\}\}, \\
T_{47}^{kc} = \{\mathcal{D}_2^{kc},\{G^{r8},G^{r8}\}\}, &
T_{48}^{kc} = \{\mathcal{D}_2^{k8},\{G^{rc},G^{r8}\}\}, \\
T_{49}^{kc} = d^{c8e} \{\mathcal{D}_2^{k8},\{J^r,G^{re}\}\}, &
T_{50}^{kc} = d^{88e} \{\mathcal{D}_2^{kc},\{J^r,G^{re}\}\}, \\
T_{51}^{kc} = i f^{c8e} \{\mathcal{D}_2^{ke},\{J^r,G^{r8}\}\}, &
T_{52}^{kc} = \{\{J^r,G^{rc}\},\{G^{k8},T^8\}\}, \\
T_{53}^{kc} = \{\{J^r,G^{r8}\},\{G^{kc},T^8\}\}, &
T_{54}^{kc} = \{\{J^r,G^{r8}\},\{G^{k8},T^c\}\}, \\
T_{55}^{kc} = i \epsilon^{kim} \{\{J^i,G^{m8}\},\{G^{r8},G^{rc}\}\}, &
T_{56}^{kc} = i \epsilon^{kim} \{\{J^i,G^{mc}\},\{G^{r8},G^{r8}\}\}, \\
T_{57}^{kc} = i \epsilon^{rim} \{G^{k8},\{J^r,\{G^{ic},G^{m8}\}\}\}, &
T_{58}^{kc} = i \epsilon^{rim} d^{c8e} \{J^k,\{J^r,\{G^{i8},G^{me}\}\}\}, \\
T_{59}^{kc} = i \epsilon^{kim} f^{cae} f^{8eb} \{\{J^i,G^{m8}\},\{T^a,T^b\}\}, &
T_{60}^{kc} = i f^{c8e} \{J^k,[\{J^i,G^{ie}\},\{J^r,G^{r8}\}]\}, \\
T_{61}^{kc} = i f^{c8e} \{\{J^r,G^{re}\},[J^2,G^{k8}]\}, &
T_{62}^{kc} = i f^{c8e} \{\{J^r,G^{r8}\},[J^2,G^{ke}]\}, \\
T_{63}^{kc} = i f^{c8e} \{J^2,[G^{ke},\{J^r,G^{r8}\}]\}, &
T_{64}^{kc} = i f^{c8e} \{J^2,[G^{k8},\{J^r,G^{re}\}]\}, \\
T_{65}^{kc} = d^{c8e} \{J^2,[G^{ke},\{J^r,G^{r8}\}]\}, &
T_{66}^{kc} = d^{c8e} \{J^2,[G^{k8},\{J^r,G^{re}\}]\}, \\
T_{67}^{kc} = [G^{kc},\{\{J^m,G^{m8}\},\{J^r,G^{r8}\}\}], &
T_{68}^{kc} = [G^{k8},\{\{J^m,G^{m8}\},\{J^r,G^{rc}\}\}], \\
T_{69}^{kc} = \{\{J^m,G^{mc}\},[G^{k8},\{J^r,G^{r8}\}]\}, &
T_{70}^{kc} = i \epsilon^{kim} f^{cea} f^{e8b} \{\{J^i,G^{m8}\},\{G^{ra},G^{rb}\}\}, \\
T_{71}^{kc} = f^{c8e} f^{8eg} \mathcal{D}_5^{kg}, &
T_{72}^{kc} = d^{c8e} d^{8eg} \mathcal{D}_5^{kg}, \\
T_{73}^{kc} = d^{ceg} d^{88e} \mathcal{D}_5^{kg}, &
T_{74}^{kc} = i f^{c8e} d^{8eg} \mathcal{D}_5^{kg}, \\
T_{75}^{kc} = i d^{c8e} f^{8eg} \mathcal{D}_5^{kg}, &
T_{76}^{kc} = \delta^{c8} \mathcal{D}_5^{k8}, \\
T_{77}^{kc} = f^{c8e} f^{8eg} \mathcal{O}_5^{kg}, &
T_{78}^{kc} = d^{c8e} d^{8eg} \mathcal{O}_5^{kg}, \\
T_{79}^{kc} = d^{ceg} d^{88e} \mathcal{O}_5^{kg}, &
T_{80}^{kc} = \delta^{c8} \mathcal{O}_5^{k8}, \\
T_{81}^{kc} = d^{c88} \{J^2,\{J^2,J^k\}\}, &
T_{82}^{kc} = \{J^2,\{G^{kc},\{T^8,T^8\}\}\}, \\
T_{83}^{kc} = \{J^2,\{G^{k8},\{T^c,T^8\}\}\}, &
T_{84}^{kc} = \{J^2,\{G^{kc},\{G^{r8},G^{r8}\}\}\}, \\
T_{85}^{kc} = \{J^2,\{G^{k8},\{G^{rc},G^{r8}\}\}\}, &
T_{86}^{kc} = d^{c8e} \{J^2,\{J^k,\{G^{re},G^{r8}\}\}\}, \\
T_{87}^{kc} = d^{88e} \{J^2,\{J^k,\{G^{rc},G^{re}\}\}\}, &
T_{88}^{kc} = d^{c8e} \{J^2,\{G^{ke},\{J^r,G^{r8}\}\}\}, \\
\end{array}
\end{eqnarray}
\begin{eqnarray}
\label{eq:basis27}
\begin{array}{ll}
T_{89}^{kc} = d^{c8e} \{J^2,\{G^{k8},\{J^r,G^{re}\}\}\}, &
T_{90}^{kc} = d^{88e} \{J^2,\{G^{kc},\{J^r,G^{re}\}\}\}, \\
T_{91}^{kc} = d^{88e} \{J^2,\{G^{ke},\{J^r,G^{rc}\}\}\}, &
T_{92}^{kc} = \epsilon^{kim} f^{c8e} \{J^2,\{T^e,\{J^i,G^{m8}\}\}\}, \\
T_{93}^{kc} = \epsilon^{kim} f^{c8e} \{J^2,\{T^8,\{J^i,G^{me}\}\}\}, &
T_{94}^{kc} = \{G^{kc},\{\{J^m,G^{m8}\},\{J^r,G^{r8}\}\}\}, \\
T_{95}^{kc} = \{G^{k8},\{\{J^m,G^{m8}\},\{J^r,G^{rc}\}\}\}, &
T_{96}^{kc} = \{J^k,\{\{J^m,G^{mc}\},\{G^{r8},G^{r8}\}\}\}, \\
T_{97}^{kc} = \{J^k,\{\{J^m,G^{m8}\},\{G^{r8},G^{rc}\}\}\}, &
T_{98}^{kc} = \{\mathcal{D}_2^{kc},\{T^8,\{J^r,G^{r8}\}\}\}, \\
T_{99}^{kc} = \{\mathcal{D}_2^{k8},\{T^8,\{J^r,G^{rc}\}\}\}, &
T_{100}^{kc} = d^{c8e} \{\mathcal{D}_3^{ke},\{J^r,G^{r8}\}\}, \\
T_{101}^{kc} = d^{88e} \{\mathcal{D}_3^{kc},\{J^r,G^{re}\}\}, &
T_{102}^{kc} = \epsilon^{kim} f^{ab8} \{\{J^i,G^{m8}\},\{T^a,\{G^{rb},G^{rc}\}\}\}, \\
T_{103}^{kc} = i \epsilon^{kim} d^{c8e} \{J^2,\{T^e,\{J^i,G^{m8}\}\}\}, &
T_{104}^{kc} = i \epsilon^{kil} [\{J^i,G^{l8}\},\{\{J^m,G^{m8}\},\{J^r,G^{rc}\}\}], \\
T_{105}^{kc} = f^{c8e} f^{8eg} \mathcal{D}_6^{kg}, &
T_{106}^{kc} = d^{c8e} d^{8eg} \mathcal{D}_6^{kg}, \\
T_{107}^{kc} = d^{ceg} d^{88e} \mathcal{D}_6^{kg}, &
T_{108}^{kc} = i f^{c8e} d^{8eg} \mathcal{D}_6^{kg}, \\
T_{109}^{kc} = \delta^{c8} \mathcal{D}_6^{k8}, &
T_{110}^{kc} = d^{c8e} \{J^2,\{J^2,\{G^{ke},T^8\}\}\}, \\
T_{111}^{kc} = d^{88e} \{J^2,\{J^2,\{G^{ke},T^c\}\}\}, &
T_{112}^{kc} = i \epsilon^{kim} \delta^{c8} \{J^2,\{J^2,\{J^i,G^{m8}\}\}\}, \\
T_{113}^{kc} = \{J^2,\{\mathcal{D}_2^{kc},\{G^{r8},G^{r8}\}\}\}, &
T_{114}^{kc} = \{J^2,\{\mathcal{D}_2^{k8},\{G^{rc},G^{r8}\}\}\}, \\
T_{115}^{kc} = d^{c8e} \{J^2,\{\mathcal{D}_2^{k8},\{J^r,G^{re}\}\}\}, &
T_{116}^{kc} = d^{88e} \{J^2,\{\mathcal{D}_2^{kc},\{J^r,G^{re}\}\}\}, \\
T_{117}^{kc} = i f^{c8e} \{J^2,\{\mathcal{D}_2^{ke},\{J^r,G^{r8}\}\}\}, &
T_{118}^{kc} = \{J^2,\{\{J^r,G^{rc}\},\{G^{k8},T^8\}\}\}, \\
T_{119}^{kc} = \{J^2,\{\{J^r,G^{r8}\},\{G^{kc},T^8\}\}\}, &
T_{120}^{kc} = \{J^2,\{\{J^r,G^{r8}\},\{G^{k8},T^c\}\}\}, \\
T_{121}^{kc} = i \epsilon^{kim} \{J^2,\{\{T^c,T^8\},\{J^i,G^{m8}\}\}\}, &
T_{122}^{kc} = i \epsilon^{kim} \{J^2,\{\{G^{rc},G^{r8}\},\{J^i,G^{m8}\}\}\}, \\
T_{123}^{kc} = i \epsilon^{kim} \{J^2,\{\{G^{r8},G^{r8}\},\{J^i,G^{mc}\}\}\}, &
T_{124}^{kc} = i \epsilon^{rim} \{J^2,\{G^{k8},\{J^r,\{G^{ic},G^{m8}\}\}\}\}, \\
T_{125}^{kc} = i \epsilon^{rim} d^{c8e} \{J^2,\{J^k,\{J^r,\{G^{i8},G^{me}\}\}\}\}, &
T_{126}^{kc} = i \epsilon^{kim} f^{cae} f^{8eb} \{J^2,\{\{J^i,G^{m8}\},\{T^a,T^b\}\}\}, \\
T_{127}^{kc} = i f^{c8e} \{J^2,\{J^k,[\{J^i,G^{ie}\},\{J^r,G^{r8}\}]\}\}, &
T_{128}^{kc} = i f^{c8e} \{J^2,\{\{J^r,G^{re}\},[J^2,G^{k8}]\}\}, \\
T_{129}^{kc} = i f^{c8e} \{J^2,\{\{J^r,G^{r8}\},[J^2,G^{ke}]\}\}, &
T_{130}^{kc} = i f^{c8e} \{J^2,\{J^2,[G^{ke},\{J^r,G^{r8}\}]\}\}, \\
T_{131}^{kc} = i f^{c8e} \{J^2,\{J^2,[G^{k8},\{J^r,G^{re}\}]\}\}, &
T_{132}^{kc} = \{\mathcal{D}_2^{kc},\{\{J^m,G^{m8}\},\{J^r,G^{r8}\}\}\}, \\
T_{133}^{kc} = \{\mathcal{D}_2^{k8},\{\{J^m,G^{mc}\},\{J^r,G^{r8}\}\}\}, &
T_{134}^{kc} = i \epsilon^{kim} [\{T^8,\{J^r,G^{r8}\}\},\{J^2,\{J^i,G^{mc}\}\}], \\
T_{135}^{kc} = d^{c8e} \{J^2,\{J^2,[G^{ke},\{J^r,G^{r8}\}]\}\}, &
T_{136}^{kc} = d^{c8e} \{J^2,\{J^2,[G^{k8},\{J^r,G^{re}\}]\}\}, \\
T_{137}^{kc} = \{J^2,[G^{kc},\{\{J^m,G^{m8}\},\{J^r,G^{r8}\}\}]\}, &
T_{138}^{kc} = \{J^2,[G^{k8},\{\{J^m,G^{m8}\},\{J^r,G^{rc}\}\}]\}, \\
T_{139}^{kc} = \{J^2,\{\{J^m,G^{mc}\},[G^{k8},\{J^r,G^{r8}\}]\}\}, &
T_{140}^{kc} = f^{c8e} f^{8eg} \mathcal{D}_7^{kg}, \\
T_{141}^{kc} = d^{c8e} d^{8eg} \mathcal{D}_7^{kg}, &
T_{142}^{kc} = d^{ceg} d^{88e} \mathcal{D}_7^{kg}, \\
T_{143}^{kc} = \delta^{c8} \mathcal{D}_7^{k8}, &
T_{144}^{kc} = f^{c8e} f^{8eg} \mathcal{O}_7^{kg}, \\
T_{145}^{kc} = d^{c8e} d^{8eg} \mathcal{O}_7^{kg}, &
T_{146}^{kc} = d^{ceg} d^{88e} \mathcal{O}_7^{kg}, \\
T_{147}^{kc} = \delta^{c8} \mathcal{O}_7^{k8}, &
T_{148}^{kc} = d^{c88} \{J^2,\{J^2,\{J^2,J^k\}\}\}, \\
T_{149}^{kc} = \{J^2,\{J^2,\{G^{kc},\{G^{r8},G^{r8}\}\}\}\}, &
T_{150}^{kc} = \{J^2,\{J^2,\{G^{k8},\{G^{rc},G^{r8}\}\}\}\}, \\
T_{151}^{kc} = d^{c8e} \{J^2,\{J^2,\{J^k,\{G^{re},G^{r8}\}\}\}\}, &
T_{152}^{kc} = d^{88e} \{J^2,\{J^2,\{J^k,\{G^{rc},G^{re}\}\}\}\}, \\
T_{153}^{kc} = d^{c8e} \{J^2,\{J^2,\{G^{ke},\{J^r,G^{r8}\}\}\}\}, &
T_{154}^{kc} = d^{c8e} \{J^2,\{J^2,\{G^{k8},\{J^r,G^{re}\}\}\}\}, \\
T_{155}^{kc} = d^{88e} \{J^2,\{J^2,\{G^{kc},\{J^r,G^{re}\}\}\}\}, &
T_{156}^{kc} = d^{88e} \{J^2,\{J^2,\{G^{ke},\{J^r,G^{rc}\}\}\}\}, \\
T_{157}^{kc} = \epsilon^{kim} f^{c8e} \{J^2,\{J^2,\{T^e,\{J^i,G^{m8}\}\}\}\}, &
T_{158}^{kc} = \{J^2,\{G^{kc},\{\{J^m,G^{m8}\},\{J^r,G^{r8}\}\}\}\}, \\
T_{159}^{kc} = \{J^2,\{G^{k8},\{\{J^m,G^{m8}\},\{J^r,G^{rc}\}\}\}\}, &
T_{160}^{kc} = \{J^2,\{J^k,\{\{J^m,G^{mc}\},\{G^{r8},G^{r8}\}\}\}\}, \\
T_{161}^{kc} = \{J^2,\{J^k,\{\{J^m,G^{m8}\},\{G^{r8},G^{rc}\}\}\}\}, &
T_{162}^{kc} = d^{c8e} \{J^2,\{\mathcal{D}_3^{ke},\{J^r,G^{r8}\}\}\}, \\
T_{163}^{kc} = d^{88e} \{J^2,\{\mathcal{D}_3^{kc},\{J^r,G^{re}\}\}\}, &
T_{164}^{kc} = \epsilon^{kim} f^{ab8} \{J^2,\{\{J^i,G^{m8}\},\{T^a,\{G^{rb},G^{rc}\}\}\}\}, \\
T_{165}^{kc} = i \epsilon^{kil} \{J^2,[\{J^i,G^{l8}\},\{\{J^m,G^{m8}\},\{J^r,G^{rc}\}\}]\}, &
T_{166}^{kc} = \{\mathcal{D}_3^{kc},\{\{J^m,G^{m8}\},\{J^r,G^{r8}\}\}\}, \\
T_{167}^{kc} = i \epsilon^{kil} \{J^2,\{J^i,\{J^r,[G^{l8},\{G^{r8},\{J^m,G^{mc}\}\}]\}\}\}. &
\end{array}
\end{eqnarray}
The corresponding nontrivial matrix elements of the operators in basis (\ref{eq:basis27}) are listed in Tables \ref{t:mm2733O}--\ref{t:mm2738TO}.

\begin{table*}
\caption{\label{t:mm2733O}Nontrivial matrix elements of the operators involved in the magnetic moments of octet baryons: flavor $\mathbf{27}$ representation.}
\begin{ruledtabular}
\begin{tabular}{lccccccccc}
& $\displaystyle n$ & $\displaystyle p$ & $\displaystyle \Sigma^-$ & $\displaystyle \Sigma^0$ & $\displaystyle \Sigma^+$ & $\displaystyle \Xi^-$ & $\displaystyle \Xi^0$ & $\displaystyle \Lambda$ & $\displaystyle \Lambda\Sigma^0$ \\[2mm]
\hline 
$\langle T_{2}^{33} \rangle$ & $-\frac{5}{36}$ & $\frac{5}{36}$ & $-\frac19$ & $0$ & $\frac19$ & $\frac{1}{36}$ & $-\frac{1}{36}$ & $0$ & $\frac{1}{6 \sqrt{3}}$ \\
$\langle T_{3}^{33} \rangle$ & $0$ & $0$ & $0$ & $0$ & $0$ & $0$ & $0$ & $0$ & $0$ \\
$\langle T_{4}^{33} \rangle$ & $0$ & $0$ & $0$ & $0$ & $0$ & $0$ & $0$ & $0$ & $0$ \\
$\langle T_{6}^{33} \rangle$ & $-\frac{1}{12}$ & $\frac{1}{12}$ & $-\frac16$ & $0$ & $\frac16$ & $-\frac{1}{12}$ & $\frac{1}{12}$ & $0$ & $0$ \\
$\langle T_{7}^{33} \rangle$ & $\frac{1}{12}$ & $-\frac{1}{12}$ & $\frac16$ & $0$ & $-\frac16$ & $\frac{1}{12}$ & $-\frac{1}{12}$ & $0$ & $0$ \\
$\langle T_{8}^{33} \rangle$ & $0$ & $0$ & $0$ & $0$ & $0$ & $0$ & $0$ & $0$ & $0$ \\
$\langle T_{9}^{33} \rangle$ & $-\frac{5}{12}$ & $\frac{5}{12}$ & $0$ & $0$ & $0$ & $-\frac{1}{12}$ & $\frac{1}{12}$ & $0$ & $0$ \\
$\langle T_{10}^{33} \rangle$ & $\frac{1}{12}$ & $-\frac{1}{12}$ & $\frac13$ & $0$ & $-\frac13$ & $-\frac14$ & $\frac14$ & $0$ & $0$ \\
$\langle T_{15}^{33} \rangle$ & $-\frac{5}{12}$ & $\frac{5}{12}$ & $-\frac13$ & $0$ & $\frac13$ & $\frac{1}{12}$ & $-\frac{1}{12}$ & $0$ & $\frac{1}{2 \sqrt{3}}$ \\
$\langle T_{16}^{33} \rangle$ & $\frac{5}{12}$ & $-\frac{5}{12}$ & $\frac13$ & $0$ & $-\frac13$ & $-\frac{1}{12}$ & $\frac{1}{12}$ & $0$ & $-\frac{1}{2 \sqrt{3}}$ \\
$\langle T_{19}^{33} \rangle$ & $0$ & $0$ & $0$ & $0$ & $0$ & $0$ & $0$ & $0$ & $0$ \\
$\langle T_{25}^{33} \rangle$ & $-\frac54$ & $\frac54$ & $0$ & $0$ & $0$ & $\frac14$ & $-\frac14$ & $0$ & $0$ \\
$\langle T_{26}^{33} \rangle$ & $-\frac14$ & $\frac14$ & $0$ & $0$ & $0$ & $-\frac34$ & $\frac34$ & $0$ & $0$ \\
$\langle T_{27}^{33} \rangle$ & $-\frac{5}{48}$ & $\frac{5}{48}$ & $-1$ & $0$ & $1$ & $\frac{17}{48}$ & $-\frac{17}{48}$ & $0$ & $\frac{1}{\sqrt{3}}$ \\
$\langle T_{28}^{33} \rangle$ & $-\frac{5}{48}$ & $\frac{5}{48}$ & $-\frac23$ & $0$ & $\frac23$ & $\frac{11}{16}$ & $-\frac{11}{16}$ & $0$ & $0$ \\
$\langle T_{29}^{33} \rangle$ & $-\frac{5}{24}$ & $\frac{5}{24}$ & $-\frac23$ & $0$ & $\frac23$ & $-\frac{11}{24}$ & $\frac{11}{24}$ & $0$ & $-\frac{1}{2 \sqrt{3}}$ \\
$\langle T_{30}^{33} \rangle$ & $\frac{5}{24}$ & $-\frac{5}{24}$ & $\frac23$ & $0$ & $-\frac23$ & $\frac{11}{24}$ & $-\frac{11}{24}$ & $0$ & $\frac{1}{2 \sqrt{3}}$ \\
$\langle T_{31}^{33} \rangle$ & $-\frac{5}{24}$ & $\frac{5}{24}$ & $-\frac13$ & $0$ & $\frac13$ & $-\frac18$ & $\frac18$ & $0$ & $0$ \\
$\langle T_{32}^{33} \rangle$ & $-\frac{5}{24}$ & $\frac{5}{24}$ & $-\frac13$ & $0$ & $\frac13$ & $-\frac18$ & $\frac18$ & $0$ & $0$ \\
$\langle T_{33}^{33} \rangle$ & $\frac{5}{24}$ & $-\frac{5}{24}$ & $\frac13$ & $0$ & $-\frac13$ & $\frac18$ & $-\frac18$ & $0$ & $0$ \\
$\langle T_{34}^{33} \rangle$ & $\frac{5}{24}$ & $-\frac{5}{24}$ & $\frac13$ & $0$ & $-\frac13$ & $\frac18$ & $-\frac18$ & $0$ & $0$ \\
$\langle T_{46}^{33} \rangle$ & $-\frac34$ & $\frac34$ & $0$ & $0$ & $0$ & $-\frac34$ & $\frac34$ & $0$ & $0$ \\
$\langle T_{47}^{33} \rangle$ & $-\frac{1}{16}$ & $\frac{1}{16}$ & $-\frac32$ & $0$ & $\frac32$ & $-\frac{17}{16}$ & $\frac{17}{16}$ & $0$ & $0$ \\
$\langle T_{48}^{33} \rangle$ & $-\frac{5}{16}$ & $\frac{5}{16}$ & $0$ & $0$ & $0$ & $\frac{11}{16}$ & $-\frac{11}{16}$ & $0$ & $0$ \\
$\langle T_{49}^{33} \rangle$ & $-\frac58$ & $\frac58$ & $0$ & $0$ & $0$ & $-\frac18$ & $\frac18$ & $0$ & $0$ \\
$\langle T_{50}^{33} \rangle$ & $\frac18$ & $-\frac18$ & $\frac12$ & $0$ & $-\frac12$ & $-\frac38$ & $\frac38$ & $0$ & $0$ \\
$\langle T_{52}^{33} \rangle$ & $-\frac58$ & $\frac58$ & $0$ & $0$ & $0$ & $\frac38$ & $-\frac38$ & $0$ & $0$ \\
$\langle T_{53}^{33} \rangle$ & $-\frac58$ & $\frac58$ & $0$ & $0$ & $0$ & $\frac38$ & $-\frac38$ & $0$ & $0$ \\
$\langle T_{54}^{33} \rangle$ & $-\frac18$ & $\frac18$ & $-1$ & $0$ & $1$ & $-\frac98$ & $\frac98$ & $0$ & $0$ \\
$\langle T_{58}^{33} \rangle$ & $0$ & $0$ & $0$ & $0$ & $0$ & $0$ & $0$ & $0$ & $-\frac{\sqrt{3}}{2}$ \\
$\langle T_{65}^{33} \rangle$ & $0$ & $0$ & $0$ & $0$ & $0$ & $0$ & $0$ & $0$ & $-\frac{\sqrt{3}}{4}$ \\
$\langle T_{66}^{33} \rangle$ & $0$ & $0$ & $0$ & $0$ & $0$ & $0$ & $0$ & $0$ & $\frac{\sqrt{3}}{4}$ \\
$\langle T_{94}^{33} \rangle$ & $-\frac{5}{16}$ & $\frac{5}{16}$ & $-1$ & $0$ & $1$ & $\frac{9}{16}$ & $-\frac{9}{16}$ & $0$ & $\frac{\sqrt{3}}{2}$ \\
$\langle T_{95}^{33} \rangle$ & $-\frac{5}{16}$ & $\frac{5}{16}$ & $-1$ & $0$ & $1$ & $\frac{9}{16}$ & $-\frac{9}{16}$ & $0$ & $0$ \\
$\langle T_{96}^{33} \rangle$ & $-\frac{5}{16}$ & $\frac{5}{16}$ & $-3$ & $0$ & $3$ & $\frac{17}{16}$ & $-\frac{17}{16}$ & $0$ & $\sqrt{3}$ \\
$\langle T_{97}^{33} \rangle$ & $-\frac{5}{16}$ & $\frac{5}{16}$ & $-2$ & $0$ & $2$ & $\frac{33}{16}$ & $-\frac{33}{16}$ & $0$ & $0$ \\
$\langle T_{98}^{33} \rangle$ & $-\frac38$ & $\frac38$ & $0$ & $0$ & $0$ & $-\frac98$ & $\frac98$ & $0$ & $0$ \\
$\langle T_{99}^{33} \rangle$ & $-\frac{15}{8}$ & $\frac{15}{8}$ & $0$ & $0$ & $0$ & $\frac38$ & $-\frac38$ & $0$ & $0$ \\
$\langle T_{100}^{33} \rangle$ & $-\frac58$ & $\frac58$ & $-1$ & $0$ & $1$ & $-\frac38$ & $\frac38$ & $0$ & $0$ \\
$\langle T_{101}^{33} \rangle$ & $\frac58$ & $-\frac58$ & $1$ & $0$ & $-1$ & $\frac38$ & $-\frac38$ & $0$ & $0$ \\
$\langle T_{120}^{33} \rangle$ & $-\frac{3}{16}$ & $\frac{3}{16}$ & $-\frac32$ & $0$ & $\frac32$ & $-\frac{27}{16}$ & $\frac{27}{16}$ & $0$ & $0$ \\
$\langle T_{132}^{33} \rangle$ & $-\frac{3}{16}$ & $\frac{3}{16}$ & $-\frac32$ & $0$ & $\frac32$ & $-\frac{27}{16}$ & $\frac{27}{16}$ & $0$ & $0$ \\
$\langle T_{133}^{33} \rangle$ & $-\frac{15}{16}$ & $\frac{15}{16}$ & $0$ & $0$ & $0$ & $\frac{9}{16}$ & $-\frac{9}{16}$ & $0$ & $0$ \\
$\langle T_{166}^{33} \rangle$ & $-\frac{15}{16}$ & $\frac{15}{16}$ & $-3$ & $0$ & $3$ & $\frac{27}{16}$ & $-\frac{27}{16}$ & $0$ & $\frac{3\sqrt{3}}{2}$ \\
\end{tabular}
\end{ruledtabular}
\end{table*}

\begin{table*}
\caption{\label{t:mm2738O}Nontrivial matrix elements of the operators involved in the magnetic moments of octet baryons: flavor $\mathbf{27}$ representation. The entries correspond to $\sqrt{3} \langle T_i^{38} \rangle$.}
\begin{ruledtabular}
\begin{tabular}{lccccccccc}
& $\displaystyle n$ & $\displaystyle p$ & $\displaystyle \Sigma^-$ & $\displaystyle \Sigma^0$ & $\displaystyle \Sigma^+$ & $\displaystyle \Xi^-$ & $\displaystyle \Xi^0$ & $\displaystyle \Lambda$ & $\displaystyle \Lambda\Sigma^0$ \\[2mm]
\hline
$\langle T_{2}^{38} \rangle$ & $\frac{1}{12}$ & $\frac{1}{12}$ & $\frac16$ & $\frac16$ & $\frac16$ & $-\frac14$ & $-\frac14$ & $-\frac16$ & $0$ \\
$\langle T_{3}^{38} \rangle$ & $\frac14$ & $\frac14$ & $\frac12$ & $\frac12$ & $\frac12$ & $-\frac34$ & $-\frac34$ & $-\frac12$ & $0$ \\
$\langle T_{4}^{38} \rangle$ & $-\frac12$ & $-\frac12$ & $-\frac12$ & $-\frac12$ & $-\frac12$ & $-\frac12$ & $-\frac12$ & $-\frac12$ & $0$ \\
$\langle T_{6}^{38} \rangle$ & $\frac14$ & $\frac14$ & $0$ & $0$ & $0$ & $-\frac14$ & $-\frac14$ & $0$ & $0$ \\
$\langle T_{7}^{38} \rangle$ & $\frac14$ & $\frac14$ & $0$ & $0$ & $0$ & $-\frac14$ & $-\frac14$ & $0$ & $0$ \\
$\langle T_{8}^{38} \rangle$ & $\frac34$ & $\frac34$ & $0$ & $0$ & $0$ & $-\frac34$ & $-\frac34$ & $0$ & $0$ \\
$\langle T_{9}^{38} \rangle$ & $-\frac14$ & $-\frac14$ & $0$ & $0$ & $0$ & $-\frac34$ & $-\frac34$ & $0$ & $0$ \\
$\langle T_{10}^{38} \rangle$ & $-\frac14$ & $-\frac14$ & $0$ & $0$ & $0$ & $-\frac34$ & $-\frac34$ & $0$ & $0$ \\
$\langle T_{15}^{38} \rangle$ & $\frac14$ & $\frac14$ & $\frac12$ & $\frac12$ & $\frac12$ & $-\frac34$ & $-\frac34$ & $-\frac12$ & $0$ \\
$\langle T_{16}^{38} \rangle$ & $\frac14$ & $\frac14$ & $\frac12$ & $\frac12$ & $\frac12$ & $-\frac34$ & $-\frac34$ & $-\frac12$ & $0$ \\
$\langle T_{19}^{38} \rangle$ & $\frac34$ & $\frac34$ & $\frac32$ & $\frac32$ & $\frac32$ & $-\frac94$ & $-\frac94$ & $-\frac32$ & $0$ \\
$\langle T_{25}^{38} \rangle$ & $\frac34$ & $\frac34$ & $0$ & $0$ & $0$ & $-\frac94$ & $-\frac94$ & $0$ & $0$ \\
$\langle T_{26}^{38} \rangle$ & $\frac34$ & $\frac34$ & $0$ & $0$ & $0$ & $-\frac94$ & $-\frac94$ & $0$ & $0$ \\
$\langle T_{27}^{38} \rangle$ & $\frac{1}{16}$ & $\frac{1}{16}$ & $\frac32$ & $\frac32$ & $\frac32$ & $-\frac{51}{16}$ & $-\frac{51}{16}$ & $-\frac12$ & $0$ \\
$\langle T_{28}^{38} \rangle$ & $\frac{1}{16}$ & $\frac{1}{16}$ & $\frac32$ & $\frac32$ & $\frac32$ & $-\frac{51}{16}$ & $-\frac{51}{16}$ & $-\frac12$ & $0$ \\
$\langle T_{29}^{38} \rangle$ & $-\frac18$ & $-\frac18$ & $-\frac32$ & $-\frac32$ & $-\frac32$ & $-\frac{17}{8}$ & $-\frac{17}{8}$ & $-\frac12$ & $0$ \\
$\langle T_{30}^{38} \rangle$ & $-\frac18$ & $-\frac18$ & $-\frac32$ & $-\frac32$ & $-\frac32$ & $-\frac{17}{8}$ & $-\frac{17}{8}$ & $-\frac12$ & $0$ \\
$\langle T_{31}^{38} \rangle$ & $-\frac18$ & $-\frac18$ & $-\frac12$ & $-\frac12$ & $-\frac12$ & $-\frac98$ & $-\frac98$ & $-\frac12$ & $0$ \\
$\langle T_{32}^{38} \rangle$ & $-\frac18$ & $-\frac18$ & $-\frac12$ & $-\frac12$ & $-\frac12$ & $-\frac98$ & $-\frac98$ & $-\frac12$ & $0$ \\
$\langle T_{33}^{38} \rangle$ & $-\frac18$ & $-\frac18$ & $-\frac12$ & $-\frac12$ & $-\frac12$ & $-\frac98$ & $-\frac98$ & $-\frac12$ & $0$ \\
$\langle T_{34}^{38} \rangle$ & $-\frac18$ & $-\frac18$ & $-\frac12$ & $-\frac12$ & $-\frac12$ & $-\frac98$ & $-\frac98$ & $-\frac12$ & $0$ \\
$\langle T_{46}^{38} \rangle$ & $\frac94$ & $\frac94$ & $0$ & $0$ & $0$ & $-\frac94$ & $-\frac94$ & $0$ & $0$ \\
$\langle T_{47}^{38} \rangle$ & $\frac{3}{16}$ & $\frac{3}{16}$ & $0$ & $0$ & $0$ & $-\frac{51}{16}$ & $-\frac{51}{16}$ & $0$ & $0$ \\
$\langle T_{48}^{38} \rangle$ & $\frac{3}{16}$ & $\frac{3}{16}$ & $0$ & $0$ & $0$ & $-\frac{51}{16}$ & $-\frac{51}{16}$ & $0$ & $0$ \\
$\langle T_{49}^{38} \rangle$ & $-\frac38$ & $-\frac38$ & $0$ & $0$ & $0$ & $-\frac98$ & $-\frac98$ & $0$ & $0$ \\
$\langle T_{50}^{38} \rangle$ & $-\frac38$ & $-\frac38$ & $0$ & $0$ & $0$ & $-\frac98$ & $-\frac98$ & $0$ & $0$ \\
$\langle T_{52}^{38} \rangle$ & $\frac38$ & $\frac38$ & $0$ & $0$ & $0$ & $-\frac{27}{8}$ & $-\frac{27}{8}$ & $0$ & $0$ \\
$\langle T_{53}^{38} \rangle$ & $\frac38$ & $\frac38$ & $0$ & $0$ & $0$ & $-\frac{27}{8}$ & $-\frac{27}{8}$ & $0$ & $0$ \\
$\langle T_{54}^{38} \rangle$ & $\frac38$ & $\frac38$ & $0$ & $0$ & $0$ & $-\frac{27}{8}$ & $-\frac{27}{8}$ & $0$ & $0$ \\
$\langle T_{58}^{38} \rangle$ & $0$ & $0$ & $0$ & $0$ & $0$ & $0$ & $0$ & $0$ & $0$ \\
$\langle T_{65}^{38} \rangle$ & $0$ & $0$ & $0$ & $0$ & $0$ & $0$ & $0$ & $0$ & $0$ \\
$\langle T_{66}^{38} \rangle$ & $0$ & $0$ & $0$ & $0$ & $0$ & $0$ & $0$ & $0$ & $0$ \\
$\langle T_{94}^{38} \rangle$ & $\frac{3}{16}$ & $\frac{3}{16}$ & $\frac32$ & $\frac32$ & $\frac32$ & $-\frac{81}{16}$ & $-\frac{81}{16}$ & $-\frac32$ & $0$ \\
$\langle T_{95}^{38} \rangle$ & $\frac{3}{16}$ & $\frac{3}{16}$ & $\frac32$ & $\frac32$ & $\frac32$ & $-\frac{81}{16}$ & $-\frac{81}{16}$ & $-\frac32$ & $0$ \\
$\langle T_{96}^{38} \rangle$ & $\frac{3}{16}$ & $\frac{3}{16}$ & $\frac92$ & $\frac92$ & $\frac92$ & $-\frac{153}{16}$ & $-\frac{153}{16}$ & $-\frac32$ & $0$ \\
$\langle T_{97}^{38} \rangle$ & $\frac{3}{16}$ & $\frac{3}{16}$ & $\frac92$ & $\frac92$ & $\frac92$ & $-\frac{153}{16}$ & $-\frac{153}{16}$ & $-\frac32$ & $0$ \\
$\langle T_{98}^{38} \rangle$ & $\frac98$ & $\frac98$ & $0$ & $0$ & $0$ & $-\frac{27}{8}$ & $-\frac{27}{8}$ & $0$ & $0$ \\
$\langle T_{99}^{38} \rangle$ & $\frac98$ & $\frac98$ & $0$ & $0$ & $0$ & $-\frac{27}{8}$ & $-\frac{27}{8}$ & $0$ & $0$ \\
$\langle T_{100}^{38} \rangle$ & $-\frac38$ & $-\frac38$ & $-\frac32$ & $-\frac32$ & $-\frac32$ & $-\frac{27}{8}$ & $-\frac{27}{8}$ & $-\frac32$ & $0$ \\
$\langle T_{101}^{38} \rangle$ & $-\frac38$ & $-\frac38$ & $-\frac32$ & $-\frac32$ & $-\frac32$ & $-\frac{27}{8}$ & $-\frac{27}{8}$ & $-\frac32$ & $0$ \\
$\langle T_{120}^{38} \rangle$ & $\frac{9}{16}$ & $\frac{9}{16}$ & $0$ & $0$ & $0$ & $-\frac{81}{16}$ & $-\frac{81}{16}$ & $0$ & $0$ \\
$\langle T_{132}^{38} \rangle$ & $\frac{9}{16}$ & $\frac{9}{16}$ & $0$ & $0$ & $0$ & $-\frac{81}{16}$ & $-\frac{81}{16}$ & $0$ & $0$ \\
$\langle T_{133}^{38} \rangle$ & $\frac{9}{16}$ & $\frac{9}{16}$ & $0$ & $0$ & $0$ & $-\frac{81}{16}$ & $-\frac{81}{16}$ & $0$ & $0$ \\
$\langle T_{166}^{38} \rangle$ & $\frac{9}{16}$ & $\frac{9}{16}$ & $\frac92$ & $\frac92$ & $\frac92$ & $-\frac{243}{16}$ & $-\frac{243}{16}$ & $-\frac92$ & $0$ \\
\end{tabular}
\end{ruledtabular}
\end{table*}

\begin{table*}
\caption{\label{t:mm2733T}Nontrivial matrix elements of the operators involved in the magnetic moments of decuplet baryons: flavor $\mathbf{27}$ representation.}
\begin{ruledtabular}
\begin{tabular}{lcccccccccc}
& $\displaystyle \Delta^{++}$ & $\displaystyle \Delta^+$ & $\displaystyle \Delta^0$ & $\displaystyle \Delta^-$ & $\displaystyle {\Sigma^*}^+$ & $\displaystyle {\Sigma^*}^0$ & $\displaystyle {\Sigma^*}^-$ & $\displaystyle {\Xi^*}^0$ & $\displaystyle {\Xi^*}^-$ & $\displaystyle \Omega^-$ \\[2mm]
\hline
$\langle T_{2}^{33} \rangle$ & $\frac14$ & $\frac{1}{12}$ & $-\frac{1}{12}$ & $-\frac14$ & $\frac16$ & $0$ & $-\frac16$ & $\frac{1}{12}$ & $-\frac{1}{12}$ & $0$ \\
$\langle T_{3}^{33} \rangle$ & $0$ & $0$ & $0$ & $0$ & $0$ & $0$ & $0$ & $0$ & $0$ & $0$ \\
$\langle T_{4}^{33} \rangle$ & $0$ & $0$ & $0$ & $0$ & $0$ & $0$ & $0$ & $0$ & $0$ & $0$ \\
$\langle T_{6}^{33} \rangle$ & $\frac34$ & $\frac14$ & $-\frac14$ & $-\frac34$ & $\frac12$ & $0$ & $-\frac12$ & $\frac14$ & $-\frac14$ & $0$ \\
$\langle T_{7}^{33} \rangle$ & $-\frac34$ & $-\frac14$ & $\frac14$ & $\frac34$ & $-\frac12$ & $0$ & $\frac12$ & $-\frac14$ & $\frac14$ & $0$ \\
$\langle T_{8}^{33} \rangle$ & $0$ & $0$ & $0$ & $0$ & $0$ & $0$ & $0$ & $0$ & $0$ & $0$ \\
$\langle T_{9}^{33} \rangle$ & $\frac34$ & $\frac14$ & $-\frac14$ & $-\frac34$ & $0$ & $0$ & $0$ & $-\frac14$ & $\frac14$ & $0$ \\
$\langle T_{10}^{33} \rangle$ & $-\frac34$ & $-\frac14$ & $\frac14$ & $\frac34$ & $0$ & $0$ & $0$ & $\frac14$ & $-\frac14$ & $0$ \\
$\langle T_{15}^{33} \rangle$ & $\frac{15}{4}$ & $\frac54$ & $-\frac54$ & $-\frac{15}{4}$ & $\frac52$ & $0$ & $-\frac52$ & $\frac54$ & $-\frac54$ & $0$ \\
$\langle T_{16}^{33} \rangle$ & $-\frac{15}{4}$ & $-\frac54$ & $\frac54$ & $\frac{15}{4}$ & $-\frac52$ & $0$ & $\frac52$ & $-\frac54$ & $\frac54$ & $0$ \\
$\langle T_{19}^{33} \rangle$ & $0$ & $0$ & $0$ & $0$ & $0$ & $0$ & $0$ & $0$ & $0$ & $0$ \\
$\langle T_{25}^{33} \rangle$ & $\frac94$ & $\frac34$ & $-\frac34$ & $-\frac94$ & $0$ & $0$ & $0$ & $\frac34$ & $-\frac34$ & $0$ \\
$\langle T_{26}^{33} \rangle$ & $\frac94$ & $\frac34$ & $-\frac34$ & $-\frac94$ & $0$ & $0$ & $0$ & $\frac34$ & $-\frac34$ & $0$ \\
$\langle T_{27}^{33} \rangle$ & $\frac{15}{16}$ & $\frac{5}{16}$ & $-\frac{5}{16}$ & $-\frac{15}{16}$ & $\frac12$ & $0$ & $-\frac12$ & $\frac{9}{16}$ & $-\frac{9}{16}$ & $0$ \\
$\langle T_{28}^{33} \rangle$ & $\frac{15}{16}$ & $\frac{5}{16}$ & $-\frac{5}{16}$ & $-\frac{15}{16}$ & $0$ & $0$ & $0$ & $\frac{1}{16}$ & $-\frac{1}{16}$ & $0$ \\
$\langle T_{29}^{33} \rangle$ & $\frac{15}{8}$ & $\frac58$ & $-\frac58$ & $-\frac{15}{8}$ & $\frac12$ & $0$ & $-\frac12$ & $-\frac18$ & $\frac18$ & $0$ \\
$\langle T_{30}^{33} \rangle$ & $-\frac{15}{8}$ & $-\frac58$ & $\frac58$ & $\frac{15}{8}$ & $-\frac12$ & $0$ & $\frac12$ & $\frac18$ & $-\frac18$ & $0$ \\
$\langle T_{31}^{33} \rangle$ & $\frac{15}{8}$ & $\frac58$ & $-\frac58$ & $-\frac{15}{8}$ & $0$ & $0$ & $0$ & $-\frac58$ & $\frac58$ & $0$ \\
$\langle T_{32}^{33} \rangle$ & $\frac{15}{8}$ & $\frac58$ & $-\frac58$ & $-\frac{15}{8}$ & $0$ & $0$ & $0$ & $-\frac58$ & $\frac58$ & $0$ \\
$\langle T_{33}^{33} \rangle$ & $-\frac{15}{8}$ & $-\frac58$ & $\frac58$ & $\frac{15}{8}$ & $0$ & $0$ & $0$ & $\frac58$ & $-\frac58$ & $0$ \\
$\langle T_{34}^{33} \rangle$ & $-\frac{15}{8}$ & $-\frac58$ & $\frac58$ & $\frac{15}{8}$ & $0$ & $0$ & $0$ & $\frac58$ & $-\frac58$ & $0$ \\
$\langle T_{46}^{33} \rangle$ & $\frac{27}{4}$ & $\frac94$ & $-\frac94$ & $-\frac{27}{4}$ & $0$ & $0$ & $0$ & $\frac94$ & $-\frac94$ & $0$ \\
$\langle T_{47}^{33} \rangle$ & $\frac{45}{16}$ & $\frac{15}{16}$ & $-\frac{15}{16}$ & $-\frac{45}{16}$ & $\frac32$ & $0$ & $-\frac32$ & $\frac{27}{16}$ & $-\frac{27}{16}$ & $0$ \\
$\langle T_{48}^{33} \rangle$ & $\frac{45}{16}$ & $\frac{15}{16}$ & $-\frac{15}{16}$ & $-\frac{45}{16}$ & $0$ & $0$ & $0$ & $\frac{3}{16}$ & $-\frac{3}{16}$ & $0$ \\
$\langle T_{49}^{33} \rangle$ & $\frac{45}{8}$ & $\frac{15}{8}$ & $-\frac{15}{8}$ & $-\frac{45}{8}$ & $0$ & $0$ & $0$ & $-\frac{15}{8}$ & $\frac{15}{8}$ & $0$ \\
$\langle T_{50}^{33} \rangle$ & $-\frac{45}{8}$ & $-\frac{15}{8}$ & $\frac{15}{8}$ & $\frac{45}{8}$ & $0$ & $0$ & $0$ & $\frac{15}{8}$ & $-\frac{15}{8}$ & $0$ \\
$\langle T_{52}^{33} \rangle$ & $\frac{45}{8}$ & $\frac{15}{8}$ & $-\frac{15}{8}$ & $-\frac{45}{8}$ & $0$ & $0$ & $0$ & $\frac{15}{8}$ & $-\frac{15}{8}$ & $0$ \\
$\langle T_{53}^{33} \rangle$ & $\frac{45}{8}$ & $\frac{15}{8}$ & $-\frac{15}{8}$ & $-\frac{45}{8}$ & $0$ & $0$ & $0$ & $\frac{15}{8}$ & $-\frac{15}{8}$ & $0$ \\
$\langle T_{54}^{33} \rangle$ & $\frac{45}{8}$ & $\frac{15}{8}$ & $-\frac{15}{8}$ & $-\frac{45}{8}$ & $0$ & $0$ & $0$ & $\frac{15}{8}$ & $-\frac{15}{8}$ & $0$ \\
$\langle T_{94}^{33} \rangle$ & $\frac{225}{16}$ & $\frac{75}{16}$ & $-\frac{75}{16}$ & $-\frac{225}{16}$ & $0$ & $0$ & $0$ & $\frac{75}{16}$ & $-\frac{75}{16}$ & $0$ \\
$\langle T_{95}^{33} \rangle$ & $\frac{225}{16}$ & $\frac{75}{16}$ & $-\frac{75}{16}$ & $-\frac{225}{16}$ & $0$ & $0$ & $0$ & $\frac{75}{16}$ & $-\frac{75}{16}$ & $0$ \\
$\langle T_{96}^{33} \rangle$ & $\frac{225}{16}$ & $\frac{75}{16}$ & $-\frac{75}{16}$ & $-\frac{225}{16}$ & $\frac{15}{2}$ & $0$ & $-\frac{15}{2}$ & $\frac{135}{16}$ & $-\frac{135}{16}$ & $0$ \\
$\langle T_{97}^{33} \rangle$ & $\frac{225}{16}$ & $\frac{75}{16}$ & $-\frac{75}{16}$ & $-\frac{225}{16}$ & $0$ & $0$ & $0$ & $\frac{15}{16}$ & $-\frac{15}{16}$ & $0$ \\
$\langle T_{98}^{33} \rangle$ & $\frac{135}{8}$ & $\frac{45}{8}$ & $-\frac{45}{8}$ & $-\frac{135}{8}$ & $0$ & $0$ & $0$ & $\frac{45}{8}$ & $-\frac{45}{8}$ & $0$ \\
$\langle T_{99}^{33} \rangle$ & $\frac{135}{8}$ & $\frac{45}{8}$ & $-\frac{45}{8}$ & $-\frac{135}{8}$ & $0$ & $0$ & $0$ & $\frac{45}{8}$ & $-\frac{45}{8}$ & $0$ \\
$\langle T_{100}^{33} \rangle$ & $\frac{225}{8}$ & $\frac{75}{8}$ & $-\frac{75}{8}$ & $-\frac{225}{8}$ & $0$ & $0$ & $0$ & $-\frac{75}{8}$ & $\frac{75}{8}$ & $0$ \\
$\langle T_{101}^{33} \rangle$ & $-\frac{225}{8}$ & $-\frac{75}{8}$ & $\frac{75}{8}$ & $\frac{225}{8}$ & $0$ & $0$ & $0$ & $\frac{75}{8}$ & $-\frac{75}{8}$ & $0$ \\
$\langle T_{120}^{33} \rangle$ & $\frac{675}{16}$ & $\frac{225}{16}$ & $-\frac{225}{16}$ & $-\frac{675}{16}$ & $0$ & $0$ & $0$ & $\frac{225}{16}$ & $-\frac{225}{16}$ & $0$ \\
$\langle T_{132}^{33} \rangle$ & $\frac{675}{16}$ & $\frac{225}{16}$ & $-\frac{225}{16}$ & $-\frac{675}{16}$ & $0$ & $0$ & $0$ & $\frac{225}{16}$ & $-\frac{225}{16}$ & $0$ \\
$\langle T_{133}^{33} \rangle$ & $\frac{675}{16}$ & $\frac{225}{16}$ & $-\frac{225}{16}$ & $-\frac{675}{16}$ & $0$ & $0$ & $0$ & $\frac{225}{16}$ & $-\frac{225}{16}$ & $0$ \\
$\langle T_{166}^{33} \rangle$ & $\frac{3375}{16}$ & $\frac{1125}{16}$ & $-\frac{1125}{16}$ & $-\frac{3375}{16}$ & $0$ & $0$ & $0$ & $\frac{1125}{16}$ & $-\frac{1125}{16}$ & $0$ \\
\end{tabular}
\end{ruledtabular}
\end{table*}

\begin{table*}
\caption{\label{t:mm2738T}Nontrivial matrix elements of the operators involved in the magnetic moments of decuplet baryons: flavor $\mathbf{27}$ representation. The entries correspond to $\sqrt{3} \langle T_i^{38} \rangle$.}
\begin{ruledtabular}
\begin{tabular}{lcccccccccc}
& $\displaystyle \Delta^{++}$ & $\displaystyle \Delta^+$ & $\displaystyle \Delta^0$ & $\displaystyle \Delta^-$ & $\displaystyle {\Sigma^*}^+$ & $\displaystyle {\Sigma^*}^0$ & $\displaystyle {\Sigma^*}^-$ & $\displaystyle {\Xi^*}^0$ & $\displaystyle {\Xi^*}^-$ & $\displaystyle \Omega^-$ \\[2mm]
\hline
$\langle T_{2}^{38} \rangle$ & $\frac14$ & $\frac14$ & $\frac14$ & $\frac14$ & $0$ & $0$ & $0$ & $-\frac14$ & $-\frac14$ & $-\frac12$ \\
$\langle T_{3}^{38} \rangle$ & $\frac34$ & $\frac34$ & $\frac34$ & $\frac34$ & $0$ & $0$ & $0$ & $-\frac34$ & $-\frac34$ & $-\frac32$ \\
$\langle T_{4}^{38} \rangle$ & $-\frac32$ & $-\frac32$ & $-\frac32$ & $-\frac32$ & $-\frac32$ & $-\frac32$ & $-\frac32$ & $-\frac32$ & $-\frac32$ & $-\frac32$ \\
$\langle T_{6}^{38} \rangle$ & $\frac34$ & $\frac34$ & $\frac34$ & $\frac34$ & $0$ & $0$ & $0$ & $-\frac34$ & $-\frac34$ & $-\frac32$ \\
$\langle T_{7}^{38} \rangle$ & $\frac34$ & $\frac34$ & $\frac34$ & $\frac34$ & $0$ & $0$ & $0$ & $-\frac34$ & $-\frac34$ & $-\frac32$ \\

$\langle T_{8}^{38} \rangle$ & $\frac94$ & $\frac94$ & $\frac94$ & $\frac94$ & $0$ & $0$ & $0$ & $-\frac94$ & $-\frac94$ & $-\frac92$ \\
$\langle T_{9}^{38} \rangle$ & $-\frac34$ & $-\frac34$ & $-\frac34$ & $-\frac34$ & $0$ & $0$ & $0$ & $-\frac34$ & $-\frac34$ & $-3$ \\
$\langle T_{10}^{38} \rangle$ & $-\frac34$ & $-\frac34$ & $-\frac34$ & $-\frac34$ & $0$ & $0$ & $0$ & $-\frac34$ & $-\frac34$ & $-3$ \\
$\langle T_{15}^{38} \rangle$ & $\frac{15}{4}$ & $\frac{15}{4}$ & $\frac{15}{4}$ & $\frac{15}{4}$ & $0$ & $0$ & $0$ & $-\frac{15}{4}$ & $-\frac{15}{4}$ & $-\frac{15}{2}$ \\
$\langle T_{16}^{38} \rangle$ & $\frac{15}{4}$ & $\frac{15}{4}$ & $\frac{15}{4}$ & $\frac{15}{4}$ & $0$ & $0$ & $0$ & $-\frac{15}{4}$ & $-\frac{15}{4}$ & $-\frac{15}{2}$ \\
$\langle T_{19}^{38} \rangle$ & $\frac{45}{4}$ & $\frac{45}{4}$ & $\frac{45}{4}$ & $\frac{45}{4}$ & $0$ & $0$ & $0$ & $-\frac{45}{4}$ & $-\frac{45}{4}$ & $-\frac{45}{2}$ \\
$\langle T_{25}^{38} \rangle$ & $\frac94$ & $\frac94$ & $\frac94$ & $\frac94$ & $0$ & $0$ & $0$ & $-\frac94$ & $-\frac94$ & $-18$ \\
$\langle T_{26}^{38} \rangle$ & $\frac94$ & $\frac94$ & $\frac94$ & $\frac94$ & $0$ & $0$ & $0$ & $-\frac94$ & $-\frac94$ & $-18$ \\
$\langle T_{27}^{38} \rangle$ & $\frac{15}{16}$ & $\frac{15}{16}$ & $\frac{15}{16}$ & $\frac{15}{16}$ & $0$ & $0$ & $0$ & $-\frac{27}{16}$ & $-\frac{27}{16}$ & $-\frac{15}{2}$ \\
$\langle T_{28}^{38} \rangle$ & $\frac{15}{16}$ & $\frac{15}{16}$ & $\frac{15}{16}$ & $\frac{15}{16}$ & $0$ & $0$ & $0$ & $-\frac{27}{16}$ & $-\frac{27}{16}$ & $-\frac{15}{2}$ \\
$\langle T_{29}^{38} \rangle$ & $-\frac{15}{8}$ & $-\frac{15}{8}$ & $-\frac{15}{8}$ & $-\frac{15}{8}$ & $-\frac32$ & $-\frac32$ & $-\frac32$ & $-\frac{27}{8}$ & $-\frac{27}{8}$ & $-\frac{15}{2}$ \\
$\langle T_{30}^{38} \rangle$ & $-\frac{15}{8}$ & $-\frac{15}{8}$ & $-\frac{15}{8}$ & $-\frac{15}{8}$ & $-\frac32$ & $-\frac32$ & $-\frac32$ & $-\frac{27}{8}$ & $-\frac{27}{8}$ & $-\frac{15}{2}$ \\
$\langle T_{31}^{38} \rangle$ & $-\frac{15}{8}$ & $-\frac{15}{8}$ & $-\frac{15}{8}$ & $-\frac{15}{8}$ & $0$ & $0$ & $0$ & $-\frac{15}{8}$ & $-\frac{15}{8}$ & $-\frac{15}{2}$ \\
$\langle T_{32}^{38} \rangle$ & $-\frac{15}{8}$ & $-\frac{15}{8}$ & $-\frac{15}{8}$ & $-\frac{15}{8}$ & $0$ & $0$ & $0$ & $-\frac{15}{8}$ & $-\frac{15}{8}$ & $-\frac{15}{2}$ \\
$\langle T_{33}^{38} \rangle$ & $-\frac{15}{8}$ & $-\frac{15}{8}$ & $-\frac{15}{8}$ & $-\frac{15}{8}$ & $0$ & $0$ & $0$ & $-\frac{15}{8}$ & $-\frac{15}{8}$ & $-\frac{15}{2}$ \\
$\langle T_{34}^{38} \rangle$ & $-\frac{15}{8}$ & $-\frac{15}{8}$ & $-\frac{15}{8}$ & $-\frac{15}{8}$ & $0$ & $0$ & $0$ & $-\frac{15}{8}$ & $-\frac{15}{8}$ & $-\frac{15}{2}$ \\
$\langle T_{46}^{38} \rangle$ & $\frac{27}{4}$ & $\frac{27}{4}$ & $\frac{27}{4}$ & $\frac{27}{4}$ & $0$ & $0$ & $0$ & $-\frac{27}{4}$ & $-\frac{27}{4}$ & $-54$ \\
$\langle T_{47}^{38} \rangle$ & $\frac{45}{16}$ & $\frac{45}{16}$ & $\frac{45}{16}$ & $\frac{45}{16}$ & $0$ & $0$ & $0$ & $-\frac{81}{16}$ & $-\frac{81}{16}$ & $-\frac{45}{2}$ \\
$\langle T_{48}^{38} \rangle$ & $\frac{45}{16}$ & $\frac{45}{16}$ & $\frac{45}{16}$ & $\frac{45}{16}$ & $0$ & $0$ & $0$ & $-\frac{81}{16}$ & $-\frac{81}{16}$ & $-\frac{45}{2}$ \\
$\langle T_{49}^{38} \rangle$ & $-\frac{45}{8}$ & $-\frac{45}{8}$ & $-\frac{45}{8}$ & $-\frac{45}{8}$ & $0$ & $0$ & $0$ & $-\frac{45}{8}$ & $-\frac{45}{8}$ & $-\frac{45}{2}$ \\
$\langle T_{50}^{38} \rangle$ & $-\frac{45}{8}$ & $-\frac{45}{8}$ & $-\frac{45}{8}$ & $-\frac{45}{8}$ & $0$ & $0$ & $0$ & $-\frac{45}{8}$ & $-\frac{45}{8}$ & $-\frac{45}{2}$ \\
$\langle T_{52}^{38} \rangle$ & $\frac{45}{8}$ & $\frac{45}{8}$ & $\frac{45}{8}$ & $\frac{45}{8}$ & $0$ & $0$ & $0$ & $-\frac{45}{8}$ & $-\frac{45}{8}$ & $-45$ \\
$\langle T_{53}^{38} \rangle$ & $\frac{45}{8}$ & $\frac{45}{8}$ & $\frac{45}{8}$ & $\frac{45}{8}$ & $0$ & $0$ & $0$ & $-\frac{45}{8}$ & $-\frac{45}{8}$ & $-45$ \\
$\langle T_{54}^{38} \rangle$ & $\frac{45}{8}$ & $\frac{45}{8}$ & $\frac{45}{8}$ & $\frac{45}{8}$ & $0$ & $0$ & $0$ & $-\frac{45}{8}$ & $-\frac{45}{8}$ & $-45$ \\
$\langle T_{94}^{38} \rangle$ & $\frac{225}{16}$ & $\frac{225}{16}$ & $\frac{225}{16}$ & $\frac{225}{16}$ & $0$ & $0$ & $0$ & $-\frac{225}{16}$ & $-\frac{225}{16}$ & $-\frac{225}{2}$ \\
$\langle T_{95}^{38} \rangle$ & $\frac{225}{16}$ & $\frac{225}{16}$ & $\frac{225}{16}$ & $\frac{225}{16}$ & $0$ & $0$ & $0$ & $-\frac{225}{16}$ & $-\frac{225}{16}$ & $-\frac{225}{2}$ \\
$\langle T_{96}^{38} \rangle$ & $\frac{225}{16}$ & $\frac{225}{16}$ & $\frac{225}{16}$ & $\frac{225}{16}$ & $0$ & $0$ & $0$ & $-\frac{405}{16}$ & $-\frac{405}{16}$ & $-\frac{225}{2}$ \\
$\langle T_{97}^{38} \rangle$ & $\frac{225}{16}$ & $\frac{225}{16}$ & $\frac{225}{16}$ & $\frac{225}{16}$ & $0$ & $0$ & $0$ & $-\frac{405}{16}$ & $-\frac{405}{16}$ & $-\frac{225}{2}$ \\
$\langle T_{98}^{38} \rangle$ & $\frac{135}{8}$ & $\frac{135}{8}$ & $\frac{135}{8}$ & $\frac{135}{8}$ & $0$ & $0$ & $0$ & $-\frac{135}{8}$ & $-\frac{135}{8}$ & $-135$ \\
$\langle T_{99}^{38} \rangle$ & $\frac{135}{8}$ & $\frac{135}{8}$ & $\frac{135}{8}$ & $\frac{135}{8}$ & $0$ & $0$ & $0$ & $-\frac{135}{8}$ & $-\frac{135}{8}$ & $-135$ \\
$\langle T_{100}^{38} \rangle$ & $-\frac{225}{8}$ & $-\frac{225}{8}$ & $-\frac{225}{8}$ & $-\frac{225}{8}$ & $0$ & $0$ & $0$ & $-\frac{225}{8}$ & $-\frac{225}{8}$ & $-\frac{225}{2}$ \\
$\langle T_{101}^{38} \rangle$ & $-\frac{225}{8}$ & $-\frac{225}{8}$ & $-\frac{225}{8}$ & $-\frac{225}{8}$ & $0$ & $0$ & $0$ & $-\frac{225}{8}$ & $-\frac{225}{8}$ & $-\frac{225}{2}$ \\
$\langle T_{120}^{38} \rangle$ & $\frac{675}{16}$ & $\frac{675}{16}$ & $\frac{675}{16}$ & $\frac{675}{16}$ & $0$ & $0$ & $0$ & $-\frac{675}{16}$ & $-\frac{675}{16}$ & $-\frac{675}{2}$ \\
$\langle T_{132}^{38} \rangle$ & $\frac{675}{16}$ & $\frac{675}{16}$ & $\frac{675}{16}$ & $\frac{675}{16}$ & $0$ & $0$ & $0$ & $-\frac{675}{16}$ & $-\frac{675}{16}$ & $-\frac{675}{2}$ \\
$\langle T_{133}^{38} \rangle$ & $\frac{675}{16}$ & $\frac{675}{16}$ & $\frac{675}{16}$ & $\frac{675}{16}$ & $0$ & $0$ & $0$ & $-\frac{675}{16}$ & $-\frac{675}{16}$ & $-\frac{675}{2}$ \\
$\langle T_{166}^{38} \rangle$ & $\frac{3375}{16}$ & $\frac{3375}{16}$ & $\frac{3375}{16}$ & $\frac{3375}{16}$ & $0$ & $0$ & $0$ & $-\frac{3375}{16}$ & $-\frac{3375}{16}$ & $-\frac{3375}{2}$ \\
\end{tabular}
\end{ruledtabular}
\end{table*}
 
\begin{table*}
\caption{\label{t:mm2733TO}Nontrivial matrix elements of the operators involved in the magnetic moments of decuplet baryons: flavor $\mathbf{27}$ representation. The entries correspond to $\sqrt{2} \langle T_i^{33} \rangle$.}
\begin{ruledtabular}
\begin{tabular}{lcccccccc}
& $\displaystyle \Delta^+p$ & $\displaystyle \Delta^0n$ & $\displaystyle {\Sigma^*}^0\Lambda$ & $\displaystyle {\Sigma^*}^0\Sigma^0$ & $\displaystyle {\Sigma^*}^+\Sigma^+$ & $\displaystyle {\Sigma^*}^-\Sigma^-$ & $\displaystyle {\Xi^*}^0\Xi^0$ & $\displaystyle {\Xi^*}^-\Xi^-$ \\[2mm]
\hline
$\langle T_{2}^{33} \rangle$ & $\frac29$ & $\frac29$ & $\frac{1}{3 \sqrt{3}}$ & $0$ & $\frac19$ & $-\frac19$ & $\frac19$ & $-\frac19$ \\
$\langle T_{3}^{33} \rangle$ & $0$ & $0$ & $0$ & $0$ & $0$ & $0$ & $0$ & $0$ \\
$\langle T_{9}^{33} \rangle$ & $\frac23$ & $\frac23$ & $0$ & $0$ & $0$ & $0$ & $-\frac13$ & $\frac13$ \\
$\langle T_{10}^{33} \rangle$ & $0$ & $0$ & $0$ & $0$ & $-\frac23$ & $\frac23$ & $-\frac13$ & $\frac13$ \\
$\langle T_{21}^{33} \rangle$ & $1$ & $1$ & $\frac{\sqrt{3}}{2}$ & $0$ & $\frac12$ & $-\frac12$ & $\frac12$ & $-\frac12$ \\
$\langle T_{22}^{33} \rangle$ & $-1$ & $-1$ & $-\frac{\sqrt{3}}{2}$ & $0$ & $-\frac12$ & $\frac12$ & $-\frac12$ & $\frac12$ \\
$\langle T_{23}^{33} \rangle$ & $0$ & $0$ & $0$ & $0$ & $0$ & $0$ & $0$ & $0$ \\
$\langle T_{25}^{33} \rangle$ & $2$ & $2$ & $0$ & $0$ & $0$ & $0$ & $1$ & $-1$ \\
$\langle T_{26}^{33} \rangle$ & $0$ & $0$ & $0$ & $0$ & $0$ & $0$ & $-1$ & $1$ \\
$\langle T_{27}^{33} \rangle$ & $\frac12$ & $\frac12$ & $\frac{1}{\sqrt{3}}$ & $0$ & $\frac23$ & $-\frac23$ & $\frac{13}{12}$ & $-\frac{13}{12}$ \\
$\langle T_{28}^{33} \rangle$ & $0$ & $0$ & $\frac{1}{2 \sqrt{3}}$ & $0$ & $\frac56$ & $-\frac56$ & $\frac{5}{12}$ & $-\frac{5}{12}$ \\
$\langle T_{31}^{33} \rangle$ & $1$ & $1$ & $-\frac{1}{2 \sqrt{3}}$ & $0$ & $\frac16$ & $-\frac16$ & $-\frac23$ & $\frac23$ \\
$\langle T_{32}^{33} \rangle$ & $0$ & $0$ & $-\frac{1}{2 \sqrt{3}}$ & $0$ & $\frac76$ & $-\frac76$ & $\frac13$ & $-\frac13$ \\
$\langle T_{33}^{33} \rangle$ & $-1$ & $-1$ & $\frac{1}{2 \sqrt{3}}$ & $0$ & $-\frac16$ & $\frac16$ & $\frac23$ & $-\frac23$ \\
$\langle T_{34}^{33} \rangle$ & $0$ & $0$ & $\frac{1}{2 \sqrt{3}}$ & $0$ & $-\frac76$ & $\frac76$ & $-\frac13$ & $\frac13$ \\
$\langle T_{45}^{33} \rangle$ & $0$ & $0$ & $0$ & $0$ & $0$ & $0$ & $0$ & $0$ \\
$\langle T_{52}^{33} \rangle$ & $0$ & $0$ & $0$ & $0$ & $0$ & $0$ & $-1$ & $1$ \\
$\langle T_{53}^{33} \rangle$ & $3$ & $3$ & $0$ & $0$ & $0$ & $0$ & $2$ & $-2$ \\
$\langle T_{54}^{33} \rangle$ & $0$ & $0$ & $0$ & $0$ & $1$ & $-1$ & $-2$ & $2$ \\
$\langle T_{55}^{33} \rangle$ & $0$ & $0$ & $-\frac{\sqrt{3}}{2}$ & $0$ & $-\frac52$ & $\frac52$ & $-\frac54$ & $\frac54$ \\
$\langle T_{56}^{33} \rangle$ & $-\frac32$ & $-\frac32$ & $-\sqrt{3}$ & $0$ & $-2$ & $2$ & $-\frac{13}{4}$ & $\frac{13}{4}$ \\
$\langle T_{57}^{33} \rangle$ & $0$ & $0$ & $-\frac{\sqrt{3}}{2}$ & $0$ & $0$ & $0$ & $0$ & $0$ \\
$\langle T_{59}^{33} \rangle$ & $0$ & $0$ & $-\frac{3 \sqrt{3}}{2}$ & $0$ & $\frac92$ & $-\frac92$ & $3$ & $-3$ \\
$\langle T_{65}^{33} \rangle$ & $-3$ & $-3$ & $-\frac{3 \sqrt{3}}{4}$ & $0$ & $\frac34$ & $-\frac34$ & $\frac34$ & $-\frac34$ \\
$\langle T_{66}^{33} \rangle$ & $0$ & $0$ & $-\frac{3 \sqrt{3}}{4}$ & $0$ & $-\frac94$ & $\frac94$ & $-\frac94$ & $\frac94$ \\
$\langle T_{67}^{33} \rangle$ & $-6$ & $-6$ & $\frac{\sqrt{3}}{2}$ & $0$ & $\frac12$ & $-\frac12$ & $-2$ & $2$ \\
$\langle T_{68}^{33} \rangle$ & $0$ & $0$ & $0$ & $0$ & $1$ & $-1$ & $\frac72$ & $-\frac72$ \\
$\langle T_{69}^{33} \rangle$ & $0$ & $0$ & $-\frac{\sqrt{3}}{4}$ & $0$ & $\frac74$ & $-\frac74$ & $\frac12$ & $-\frac12$ \\
$\langle T_{70}^{33} \rangle$ & $0$ & $0$ & $\frac{3 \sqrt{3}}{8}$ & $0$ & $\frac{39}{8}$ & $-\frac{39}{8}$ & $\frac{15}{4}$ & $-\frac{15}{4}$ \\
$\langle T_{94}^{33} \rangle$ & $\frac{13}{2}$ & $\frac{13}{2}$ & $\frac{\sqrt{3}}{2}$ & $0$ & $\frac12$ & $-\frac12$ & $\frac{17}{4}$ & $-\frac{17}{4}$ \\
$\langle T_{95}^{33} \rangle$ & $0$ & $0$ & $0$ & $0$ & $1$ & $-1$ & $-\frac{11}{4}$ & $\frac{11}{4}$ \\
$\langle T_{103}^{33} \rangle$ & $0$ & $0$ & $0$ & $0$ & $-9$ & $9$ & $-\frac92$ & $\frac92$ \\
$\langle T_{104}^{33} \rangle$ & $0$ & $0$ & $0$ & $0$ & $-3$ & $3$ & $-\frac{21}{2}$ & $\frac{21}{2}$ \\
$\langle T_{120}^{33} \rangle$ & $0$ & $0$ & $0$ & $0$ & $\frac92$ & $-\frac92$ & $-9$ & $9$ \\
$\langle T_{121}^{33} \rangle$ & $0$ & $0$ & $0$ & $0$ & $0$ & $0$ & $\frac{27}{2}$ & $-\frac{27}{2}$ \\
$\langle T_{122}^{33} \rangle$ & $0$ & $0$ & $-\frac{9 \sqrt{3}}{4}$ & $0$ & $-\frac{45}{4}$ & $\frac{45}{4}$ & $-\frac{45}{8}$ & $\frac{45}{8}$ \\
$\langle T_{123}^{33} \rangle$ & $-\frac{27}{4}$ & $-\frac{27}{4}$ & $-\frac{9 \sqrt{3}}{2}$ & $0$ & $-9$ & $9$ & $-\frac{117}{8}$ & $\frac{117}{8}$ \\
$\langle T_{134}^{33} \rangle$ & $-27$ & $-27$ & $0$ & $0$ & $0$ & $0$ & $-\frac{27}{4}$ & $\frac{27}{4}$ \\
$\langle T_{167}^{33} \rangle$ & $0$ & $0$ & $\frac{9 \sqrt{3}}{8}$ & $0$ & $-\frac{81}{8}$ & $\frac{81}{8}$ & $-\frac{351}{8}$ & $\frac{351}{8}$ \\
\end{tabular}
\end{ruledtabular}
\end{table*}

\begin{table*}
\caption{\label{t:mm2738TO}Nontrivial matrix elements of the operators involved in the magnetic moments of decuplet baryons: flavor $\mathbf{27}$ representation. The entries correspond to $\sqrt{6} \langle T_i^{38} \rangle$.}
\begin{ruledtabular}
\begin{tabular}{lcccccccc}
& $\displaystyle \Delta^+p$ & $\displaystyle \Delta^0n$ & $\displaystyle {\Sigma^*}^0\Lambda$ & $\displaystyle {\Sigma^*}^0\Sigma^0$ & $\displaystyle {\Sigma^*}^+\Sigma^+$ & $\displaystyle {\Sigma^*}^-\Sigma^-$ & $\displaystyle {\Xi^*}^0\Xi^0$ & $\displaystyle {\Xi^*}^-\Xi^-$ \\[2mm]
\hline
$\langle T_{2}^{38} \rangle$ & $0$ & $0$ & $0$ & $\frac13$ & $\frac13$ & $\frac13$ & $\frac13$ & $\frac13$ \\
$\langle T_{3}^{38} \rangle$ & $0$ & $0$ & $0$ & $1$ & $1$ & $1$ & $1$ & $1$ \\
$\langle T_{9}^{38} \rangle$ & $0$ & $0$ & $0$ & $0$ & $0$ & $0$ & $1$ & $1$ \\
$\langle T_{10}^{38} \rangle$ & $0$ & $0$ & $0$ & $0$ & $0$ & $0$ & $1$ & $1$ \\
$\langle T_{21}^{38} \rangle$ & $0$ & $0$ & $0$ & $\frac32$ & $\frac32$ & $\frac32$ & $\frac32$ & $\frac32$ \\
$\langle T_{22}^{38} \rangle$ & $0$ & $0$ & $0$ & $\frac32$ & $\frac32$ & $\frac32$ & $\frac32$ & $\frac32$ \\
$\langle T_{23}^{38} \rangle$ & $0$ & $0$ & $0$ & $\frac92$ & $\frac92$ & $\frac92$ & $\frac92$ & $\frac92$ \\
$\langle T_{25}^{38} \rangle$ & $0$ & $0$ & $0$ & $0$ & $0$ & $0$ & $3$ & $3$ \\
$\langle T_{26}^{38} \rangle$ & $0$ & $0$ & $0$ & $0$ & $0$ & $0$ & $3$ & $3$ \\
$\langle T_{27}^{38} \rangle$ & $0$ & $0$ & $0$ & $2$ & $2$ & $2$ & $\frac{13}{4}$ & $\frac{13}{4}$ \\
$\langle T_{28}^{38} \rangle$ & $0$ & $0$ & $0$ & $2$ & $2$ & $2$ & $\frac{13}{4}$ & $\frac{13}{4}$ \\
$\langle T_{31}^{38} \rangle$ & $0$ & $0$ & $0$ & $-\frac12$ & $-\frac12$ & $-\frac12$ & $2$ & $2$ \\
$\langle T_{32}^{38} \rangle$ & $0$ & $0$ & $0$ & $-\frac12$ & $-\frac12$ & $-\frac12$ & $2$ & $2$ \\
$\langle T_{33}^{38} \rangle$ & $0$ & $0$ & $0$ & $-\frac12$ & $-\frac12$ & $-\frac12$ & $2$ & $2$ \\
$\langle T_{34}^{38} \rangle$ & $0$ & $0$ & $0$ & $-\frac12$ & $-\frac12$ & $-\frac12$ & $2$ & $2$ \\
$\langle T_{45}^{38} \rangle$ & $0$ & $0$ & $0$ & $-\frac{27}{2}$ & $-\frac{27}{2}$ & $-\frac{27}{2}$ & $-\frac{27}{2}$ & $-\frac{27}{2}$ \\
$\langle T_{52}^{38} \rangle$ & $0$ & $0$ & $0$ & $0$ & $0$ & $0$ & $6$ & $6$ \\
$\langle T_{53}^{38} \rangle$ & $0$ & $0$ & $0$ & $0$ & $0$ & $0$ & $6$ & $6$ \\
$\langle T_{54}^{38} \rangle$ & $0$ & $0$ & $0$ & $0$ & $0$ & $0$ & $6$ & $6$ \\
$\langle T_{55}^{38} \rangle$ & $0$ & $0$ & $0$ & $-6$ & $-6$ & $-6$ & $-\frac{39}{4}$ & $-\frac{39}{4}$ \\
$\langle T_{56}^{38} \rangle$ & $0$ & $0$ & $0$ & $-6$ & $-6$ & $-6$ & $-\frac{39}{4}$ & $-\frac{39}{4}$ \\
$\langle T_{57}^{38} \rangle$ & $0$ & $0$ & $0$ & $0$ & $0$ & $0$ & $0$ & $0$ \\
$\langle T_{59}^{38} \rangle$ & $0$ & $0$ & $0$ & $\frac{45}{2}$ & $\frac{45}{2}$ & $\frac{45}{2}$ & $27$ & $27$ \\
$\langle T_{65}^{38} \rangle$ & $0$ & $0$ & $0$ & $-\frac94$ & $-\frac94$ & $-\frac94$ & $-\frac94$ & $-\frac94$ \\
$\langle T_{66}^{38} \rangle$ & $0$ & $0$ & $0$ & $-\frac94$ & $-\frac94$ & $-\frac94$ & $-\frac94$ & $-\frac94$ \\
$\langle T_{67}^{38} \rangle$ & $0$ & $0$ & $0$ & $\frac32$ & $\frac32$ & $\frac32$ & $-6$ & $-6$ \\
$\langle T_{68}^{38} \rangle$ & $0$ & $0$ & $0$ & $\frac32$ & $\frac32$ & $\frac32$ & $-6$ & $-6$ \\
$\langle T_{69}^{38} \rangle$ & $0$ & $0$ & $0$ & $\frac34$ & $\frac34$ & $\frac34$ & $-3$ & $-3$ \\
$\langle T_{70}^{38} \rangle$ & $0$ & $0$ & $0$ & $\frac{171}{8}$ & $\frac{171}{8}$ & $\frac{171}{8}$ & $\frac{99}{4}$ & $\frac{99}{4}$ \\
$\langle T_{94}^{38} \rangle$ & $0$ & $0$ & $0$ & $\frac32$ & $\frac32$ & $\frac32$ & $\frac{51}{4}$ & $\frac{51}{4}$ \\
$\langle T_{95}^{38} \rangle$ & $0$ & $0$ & $0$ & $\frac32$ & $\frac32$ & $\frac32$ & $\frac{51}{4}$ & $\frac{51}{4}$ \\
$\langle T_{103}^{38} \rangle$ & $0$ & $0$ & $0$ & $0$ & $0$ & $0$ & $-\frac{27}{2}$ & $-\frac{27}{2}$ \\
$\langle T_{104}^{38} \rangle$ & $0$ & $0$ & $0$ & $-\frac92$ & $-\frac92$ & $-\frac92$ & $18$ & $18$ \\
$\langle T_{120}^{38} \rangle$ & $0$ & $0$ & $0$ & $0$ & $0$ & $0$ & $27$ & $27$ \\
$\langle T_{121}^{38} \rangle$ & $0$ & $0$ & $0$ & $0$ & $0$ & $0$ & $-\frac{81}{2}$ & $-\frac{81}{2}$ \\
$\langle T_{122}^{38} \rangle$ & $0$ & $0$ & $0$ & $-27$ & $-27$ & $-27$ & $-\frac{351}{8}$ & $-\frac{351}{8}$ \\
$\langle T_{123}^{38} \rangle$ & $0$ & $0$ & $0$ & $-27$ & $-27$ & $-27$ & $-\frac{351}{8}$ & $-\frac{351}{8}$ \\
$\langle T_{134}^{38} \rangle$ & $0$ & $0$ & $0$ & $0$ & $0$ & $0$ & $-\frac{81}{4}$ & $-\frac{81}{4}$ \\
$\langle T_{167}^{38} \rangle$ & $0$ & $0$ & $0$ & $-\frac{189}{8}$ & $-\frac{189}{8}$ & $-\frac{189}{8}$ & $\frac{621}{8}$ & $\frac{621}{8}$ \\
\end{tabular}
\end{ruledtabular}
\end{table*}

Collecting all partial results, order $\mathcal{O}(m_q \ln m_q)$ corrections to baryon magnetic moments from diagrams \ref{fig:mmloop2}(a)-\ref{fig:mmloop2}(d), for the usual examples, read
\begin{eqnarray}
\delta \mu_{\Sigma^-}^{\mathrm{(loop\, 2ad)}} & = & \left[ \left( - \frac{1}{12} a_1^2 - \frac{13}{108} a_1b_2 - \frac{5}{81} a_1b_3 + \frac{1}{108} a_1c_3 - \frac{1}{36} b_2^2 - \frac{1}{36} b_2b_3 - \frac{1}{54} b_2c_3 - \frac{1}{81} b_3^2 + \frac{1}{162} b_3c_3 - \frac{1}{432} c_3^2 \right) m_1 \right. \nonumber \\
& & \mbox{} + \left( - \frac{13}{72} a_1^2 - \frac{7}{54} a_1b_2 - \frac{37}{324} a_1b_3 - \frac{1}{108} a_1c_3 - \frac{7}{216} b_2^2 - \frac{7}{162} b_2b_3 - \frac{37}{1944} b_3^2 - \frac{1}{432} c_3^2 \right) m_2 \nonumber \\
& & \mbox{} + \left( \frac{7}{324} a_1^2 - \frac{1}{36} a_1b_2 - \frac{2}{81} a_1b_3 + \frac{19}{324} a_1c_3 - \frac{1}{108} b_2^2 - \frac{1}{108} b_2b_3 - \frac{1}{243} b_3^2 + \frac{19}{1296} c_3^2 \right) m_3 \nonumber \\
& & \mbox{} + \left. \left( \frac{1}{54} a_1^2 - \frac{1}{54} a_1b_2 + \frac{1}{162} a_1b_3 + \frac{1}{108} a_1c_3 - \frac{1}{108} b_2c_3 + \frac{1}{324} b_3c_3 \right) m_4 \right] I_2(m_\pi,0,\mu) \nonumber \\
& & \mbox{} + \left[ \left( - \frac{11}{144} a_1^2 - \frac{31}{216} a_1b_2 - \frac{89}{648} a_1b_3 + \frac{7}{54} a_1c_3 - \frac{1}{48} b_2^2 - \frac{5}{216} b_2b_3 - \frac{1}{27} b_2c_3 - \frac{35}{1296} b_3^2 + \frac{1}{81} b_3c_3 + \frac{5}{216} c_3^2 \right) m_1 \right. \nonumber \\
& & \mbox{} + \left( - \frac{7}{48} a_1^2 - \frac{17}{216} a_1b_2 - \frac{103}{648} a_1b_3 + \frac{5}{54} a_1c_3 - \frac{7}{432} b_2^2 - \frac{17}{648} b_2b_3 - \frac{103}{3888} b_3^2 + \frac{5}{216} c_3^2 \right) m_2 \nonumber \\
& & \mbox{} + \left( \frac{575}{1296} a_1^2 - \frac{5}{216} a_1b_2 - \frac{35}{648} a_1b_3 + \frac{85}{162} a_1c_3 - \frac{1}{144} b_2^2 - \frac{5}{648} b_2b_3 - \frac{35}{3888} b_3^2 + \frac{85}{648} c_3^2 \right) m_3 \nonumber \\
& & \mbox{} + \left. \left( \frac{1}{27} a_1^2 - \frac{1}{27} a_1b_2 + \frac{1}{81} a_1b_3 + \frac{1}{54} a_1c_3 - \frac{1}{54} b_2c_3 + \frac{1}{162} b_3c_3 \right) m_4 \right] I_2(m_K,0,\mu) \nonumber \\
& & \mbox{} + \left[ \left( - \frac{1}{27} a_1b_3 + \frac{1}{18} a_1c_3 - \frac{1}{162} b_3^2 + \frac{1}{72} c_3^2 \right) m_1 + \left( - \frac{1}{27} a_1b_3 + \frac{1}{18} a_1c_3 - \frac{1}{162} b_3^2 + \frac{1}{72} c_3^2 \right) m_2 \right. \nonumber \\
& & \mbox{} + \left. \left( \frac{5}{27} a_1^2 - \frac{1}{81} a_1b_3 + \frac{11}{54} a_1c_3 - \frac{1}{486} b_3^2 + \frac{11}{216} c_3^2 \right) m_3 \right] I_2(m_\eta,0,\mu),
\end{eqnarray}
and
\begin{eqnarray}
\delta \mu_{{\Sigma^*}^-}^{\mathrm{(loop\, 2ad)}} & = & \left[ \left( - \frac18 a_1^2 - \frac{11}{36} a_1b_2 - \frac{55}{108} a_1b_3 + \frac{1}{36} a_1c_3 - \frac{19}{72} b_2^2 - \frac{95}{108} b_2b_3 + \frac19 b_2c_3 - \frac{475}{648} b_3^2 + \frac{5}{27} b_3c_3 - \frac{1}{48} c_3^2 \right) m_1 \right. \nonumber \\
& & \mbox{} + \left( - \frac{11}{24} a_1^2 - \frac{19}{36} a_1b_2 - \frac{95}{108} a_1b_3 - \frac{7}{36} a_1c_3 - \frac{19}{72} b_2^2 - \frac{95}{108} b_2b_3 - \frac{475}{648} b_3^2 - \frac{7}{144} c_3^2 \right) m_2 \nonumber \\
& & \mbox{} + \left( - \frac{161}{216} a_1^2 - \frac{95}{108} a_1b_2 - \frac{475}{324} a_1b_3 - \frac{11}{36} a_1c_3 - \frac{95}{216} b_2^2 - \frac{475}{324} b_2b_3 - \frac{2375}{1944} b_3^2 - \frac{11}{144} c_3^2 \right) m_3 \nonumber \\
& & \mbox{} + \left. \left( \frac19 a_1^2 + \frac19 a_1b_2 + \frac{5}{27} a_1b_3 + \frac{1}{18} a_1c_3 + \frac{1}{18} b_2c_3 + \frac{5}{54} b_3c_3 \right) m_4 \right] I_2(m_\pi,0,\mu) \nonumber \\
& & \mbox{} + \left[ \left( - \frac{13}{48} a_1^2 - \frac{35}{72} a_1b_2 - \frac{175}{216} a_1b_3 - \frac{1}{36} a_1c_3 - \frac{43}{144} b_2^2 - \frac{215}{216} b_2b_3 + \frac{1}{18} b_2c_3 - \frac{1075}{1296} b_3^2 + \frac{5}{54} b_3c_3 - \frac{1}{48} c_3^2 \right) m_1 \right. \nonumber \\
& & \mbox{} + \left( - \frac{7}{16} a_1^2 - \frac{43}{72} a_1b_2 - \frac{215}{216} a_1b_3 - \frac{5}{36} a_1c_3 - \frac{43}{144} b_2^2 - \frac{215}{216} b_2b_3 - \frac{1075}{1296} b_3^2 - \frac{5}{144} c_3^2 \right) m_2 \nonumber \\
& & \mbox{} + \left( - \frac{323}{432} a_1^2 - \frac{215}{216} a_1b_2 - \frac{1075}{648} a_1b_3 - \frac14 a_1c_3 - \frac{215}{432} b_2^2 - \frac{1075}{648} b_2b_3 - \frac{5375}{3888} b_3^2 - \frac{1}{16} c_3^2 \right) m_3 \nonumber \\
& & \mbox{} + \left. \left( \frac{1}{18} a_1^2 + \frac{1}{18} a_1b_2 + \frac{5}{54} a_1b_3 + \frac{1}{36} a_1c_3 + \frac{1}{36} b_2c_3 + \frac{5}{108} b_3c_3 \right) m_4 \right] I_2(m_K,0,\mu) \nonumber \\
& & \mbox{} + \left[ \left( - \frac{1}{12} a_1^2 - \frac{1}{12} a_1c_3 - \frac{1}{48} c_3^2 \right) m_1 + \left( - \frac{1}{12} a_1^2 - \frac{1}{12} a_1c_3 - \frac{1}{48} c_3^2 \right) m_2 \right. \nonumber \\
& & \mbox{} + \left. \left( - \frac{7}{36} a_1^2 - \frac{7}{36} a_1c_3 - \frac{7}{144} c_3^2 \right) m_3 \right] I_2(m_\eta,0,\mu),
\end{eqnarray}
All 27 allowed magnetic moments are listed in Appendix \ref{sec:Loop2ad}, Eqs.~(\ref{eq:mmnloop2ad}) to (\ref{eq:mmxsmxmloop2ad}).

The use of relations (\ref{eq:su3inv}) and (\ref{eq:rel1inv}) yields the magnetic moments expressed in terms of the $SU(3)$ invariants $\mu_D$, $\mu_F$, $\mu_C$, $\mu_T$, $D$, $F$, $\mathcal{C}$, and $\mathcal{H}$, namely,
\begin{eqnarray}
\delta \mu_{\Sigma^-}^{\mathrm{(loop\, 2ad)}} & = & \left[ \left( \frac29 D^2 + \frac23 D F + \frac83 F^2 + \frac19 \mathcal{C}^2 \right) \mu_ D + \left( - D^2 - 7 F^2 - \frac13 \mathcal{C}^2 \right) \mu_ F + \frac{5}{54} \mathcal{C}^2 \mu_C + \frac19 (D - F) \mathcal{C} \mu_T \right] I_2(m_\pi,0,\mu) \nonumber \\
& & \mbox{} + \left[ \left( \frac56 D^2 + D F + \frac56 F^2 + \frac59 \mathcal{C}^2 \right) \mu_ D + \left( - \frac72 D^2 - D F - \frac72 F^2 - \frac53 \mathcal{C}^2 \right) \mu_ F \right. \nonumber \\
& & \mbox{} + \left. \frac{20}{27} \mathcal{C}^2 \mu_C + \frac29 (D - F) \mathcal{C} \mu_T \right] I_2(m_K,0,\mu) \nonumber \\
& & \mbox{} + \left[ \left( \frac49 D^2 + \frac16 \mathcal{C}^2 \right) \mu_ D + \left( - \frac43 D^2 - \frac12 \mathcal{C}^2 \right) \mu_ F + \frac{5}{18} \mathcal{C}^2 \mu_C \right] I_2(m_\eta,0,\mu),
\end{eqnarray}
and
\begin{eqnarray}
\delta \mu_{{\Sigma^*}^-}^{\mathrm{(loop\, 2ad)}} & = & \left[ \frac{7}{36} \mathcal{C}^2 \mu_ D + \frac{1}{12} \mathcal{C}^2 \mu_ F + \left( - \frac{5}{12} \mathcal{C}^2 - \frac{19}{81} \mathcal{H}^2 \right) \mu_C - \frac{2}{27} \mathcal{C} \mathcal{H} \mu_T \right] I_2(m_\pi,0,\mu) \nonumber \\
& & \mbox{} + \left[ \frac{1}{18} \mathcal{C}^2 \mu_ D + \frac16 \mathcal{C}^2 \mu_ F + \left( - \frac13 \mathcal{C}^2 - \frac{43}{162} \mathcal{H}^2 \right) \mu_C - \frac{1}{27} \mathcal{C} \mathcal{H} \mu_T \right] I_2(m_K,0,\mu) \nonumber \\
& & \mbox{} + \left[ - \frac{1}{12} \mathcal{C}^2 \mu_ D + \frac14 \mathcal{C}^2 \mu_ F - \frac14 \mathcal{C}^2 \mu_C \right] I_2(m_\eta,0,\mu).
\end{eqnarray}

Equations (\ref{eq:mmnloop2adch}) to (\ref{eq:mmxsmxmloop2adch}) of Appendix \ref{sec:Loop2ad} are the counterparts of (\ref{eq:mmnloop2ad}) to (\ref{eq:mmxsmxmloop2ad}), respectively.

\subsection{\label{sec:mqlnmqe}Diagrams \ref{fig:mmloop2}(e)}

Corrections to magnetic moments from the diagram \ref{fig:mmloop2}(e) are straightforwardly evaluated as \cite{rfm09,rfm14}
\begin{equation}
\delta M_{\textrm{loop 2e}}^k = - \frac12 \left[T^a,\left[T^b,M^k \right] \right] \Pi^{ab}, \label{eq:corrloop3}
\end{equation}
where $\Pi^{ab}$ is the symmetric tensor already displayed in Eq.~(\ref{eq:pisym}), except for the fact that the corresponding loop integral is now $I_3(m,\mu)$. Retaining only the nonanalytic pieces of that integral, it turns out that
\begin{equation}
I_3(m,\mu) = - I_2(m,0,\mu),
\end{equation}
where $I_2(m,0,\mu)$ is given in Eq.~(\ref{eq:fprime}).

Explicit results for the case study are thus
\begin{equation}
\delta \mu_{\Sigma^-}^{\mathrm{(loop\, 2e)}} = \left[ \frac13 m_1 + \frac16 m_2 + \frac19 m_3 \right] I_2(m_\pi,0,\mu) + \left[ - \frac{1}{12} m_1 + \frac{1}{12} m_2 - \frac{1}{36} m_3 \right] I_2(m_K,0,\mu),
\end{equation}
and
\begin{equation}
\delta \mu_{{\Sigma^*}^-}^{\mathrm{(loop\, 2e)}} = \left[ \frac12 m_1 + \frac12 m_2 + \frac56 m_3 \right] I_2(m_\pi,0,\mu) + \left[ \frac14 m_1 + \frac14 m_2 + \frac{5}{12} m_3 \right] I_2(m_K,0,\mu),
\end{equation}
or equivalently, in terms of the $SU(3)$ invariants
\begin{equation}
\delta \mu_{\Sigma^-}^{\mathrm{(loop\, 2e)}} = \mu_F I_2(m_\pi,0,\mu) - \frac12 (\mu_D - \mu_F ) I_2(m_K,0,\mu),
\end{equation}
and
\begin{equation}
\delta \mu_{{\Sigma^*}^-}^{\mathrm{(loop\, 2e)}} = \mu_C I_2(m_\pi,0,\mu) + \frac12 \mu_C I_2(m_K,0,\mu).
\end{equation}
All allowed expressions are listed in Appendix \ref{app:Loop2}, Eqs.~(\ref{eq:mmnloop2e}) to (\ref{eq:mmxsmxmloop2e}), and their corresponding expressions in terms of the $SU(3)$ invariants are listed in Eqs.~(\ref{eq:mmnloop2ech}) to (\ref{eq:mmxsmxmloop2ech}).

\subsubsection{\label{sec:comL2}Comparison with heavy chiral perturbation theory results}

In HBCHPT, the corrections to magnetic moments from the Feynman diagrams displayed in Fig.~\ref{fig:mmloop2} can be organized as \cite{jen92}
\begin{equation}
\delta \mu_i^{\mathrm{(loop\, 2)}} = \sum_{P=\pi,K,\eta} -\frac12 (\overline{\gamma}_i^{(P)}-2\overline{\lambda}_i^{(P)}\alpha_i) \left[ -\frac{1}{16\pi^2 f^2}m_P^2 \ln\frac{m_P^2}{\mu^2} \right], \label{eq:l2ch}
\end{equation}
where the coefficients $\overline{\gamma}_i^{(P)}$, $\overline{\lambda}_i^{(P)}$, and $\alpha_i$ are listed in that reference.

The comparison between the expressions extracted from Eq.~(\ref{eq:l2ch}) fully agree with the ones found here for octet baryons listed in Appendix \ref{app:Loop2}, taking into account a missing factor of $-5/2$ in the contribution from the graph \ref{fig:mmloop2}(b) and the additional corrections noted in the erratum to \cite{jen92}.

\section{\label{sec:sb}Explicit $SU(3)$ symmetry breaking}

As it has already been discussed in Ref.~\cite{rfm14}, in the conventional chiral momentum counting scheme, tree diagrams involving higher order vertices will also contribute to the magnetic moments along with the one-loop contributions of orders $\mathcal{O}(m_q^{1/2})$ and $\mathcal{O}(m_q \ln m_q)$. These higher order contributions are needed as counterterms for the divergent parts of the loops integrals. The leading $SU(3)$ breaking effects to the magnetic moments thus will also have contributions from the effective Lagrangian of order $p^4$, which yield contributions linear in the quark mass \cite{krause}. In the combined formalism, a convenient way of accounting for terms of order $\mathcal{O}(m_q)$ springs from the fact that flavor $SU(3)$ SB transforms as a flavor octet. Neglecting isospin breaking and including first order $SU(3)$ SB. $M^{kc}$ thus has pieces transforming according to all $SU(3)$ representations contained in the tensor product $(1,\mathbf{8}\otimes \mathbf{8})=(1,\mathbf{1}) \oplus (1,\mathbf{8}_S) \oplus (1,\mathbf{8}_A) \oplus (1,\mathbf{10}+\overline{\mathbf{10}}) \oplus (1,\mathbf{27})$, namely,
\begin{equation}
\delta M_{\mathrm{SB}}^{kc} = \delta M_{\mathrm{SB},\mathbf{\mathbf{1}}}^{kc} + \delta M_{\mathrm{SB},\mathbf{\mathbf{8}}}^{kc} + \delta M_{\mathrm{SB},\mathbf{\mathbf{10}+\overline{\mathbf{10}}}}^{kc} + \delta M_{\mathrm{SB},\mathbf{\mathbf{27}}}^{kc}. \label{eq:akcsb}
\end{equation}
Following the detailed analysis presented in Ref.~\cite{rfm14}, explicit SB to the baryon magnetic operator can be cast into the form
\begin{eqnarray}
\delta M_{\mathrm{SB}}^{kc} & = & \left[ m_1^{1,\mathbf{1}} \delta^{c8}J^k + m_3^{1,\mathbf{1}} \frac{1}{N_c^2} \delta^{c8} \{J^2,J^k\} \right] \nonumber \\
& & \mbox{} + \left[ n_1^{1,\mathbf{8}} d^{ce8} G^{ke} + n_2^{1,\mathbf{8}} \frac{1}{N_c} d^{ce8} \mathcal{D}_2^{ke} + n_3^{1,\mathbf{8}} \frac{1}{N_c^2} d^{ce8} \mathcal{D}_3^{ke} + \bar{n}_3^{1,\mathbf{8}} \frac{1}{N_c^2} d^{ce8} \mathcal{O}_3^{ke} \right] \nonumber \\
& & \mbox{} + \left[ m_2^{1,\mathbf{10}+\overline{\mathbf{10}}} \frac{1}{N_c} \left( \{G^{kc},T^8\}-\{G^{k8},T^c\} \right)
 + m_3^{1,\mathbf{10}+\overline{\mathbf{10}}} \frac{1}{N_c^2} \left( \{G^{kc},\{J^r,G^{r8}\}\}-\{G^{k8},\{J^r,G^{rc}\}\} \right) \right] \nonumber \\
& & \mbox{} + \left[ m_2^{1,\mathbf{27}} \frac{1}{N_c} \left( \{G^{kc},T^8\}+\{G^{k8},T^c\} \right) + m_3^{1,\mathbf{27}} \frac{1}{N_c^2} \{J^k,\{T^c,T^8\}\} \right. \nonumber \\
& & \mbox{\hglue0.5truecm} \left. + \, \bar{m}_3^{1,\mathbf{27}} \frac{1}{N_c^2} \left( \{G^{kc},\{J^r,G^{r8}\}\}+\{G^{k8},\{J^r,G^{rc}\}\} \right) \right]. \label{eq:sb}
\end{eqnarray}
where the superscripts attached to the eleven unknown coefficients $m_i^{1,\mathbf{rep}}$ and $n_j^{1,\mathbf{rep}}$ indicate the spin-flavor representation $\mathbf{rep}$ they fall in. Although the series has been truncated at the $3$-body level, higher-order terms can be obtained by anticommuting the operators retained with $J^2$.

Equation (\ref{eq:sb}) is the one to be used in the numerical analysis. By using the appropriate matrix elements listed in Tables \ref{t:mm8O}-\ref{t:mm8TO}, the explicit SB contributions to magnetic moments in the usual examples read
\begin{equation}
\sqrt{3} \delta \mu_{\Sigma^-}^{\mathrm{SB}} = \frac12 m_1^{1,\mathbf{1}} + \frac{1}{12} m_3^{1,\mathbf{1}} - \frac12 n_1^{1,\mathbf{8}} - \frac16 n_2^{1,\mathbf{8}} - \frac16 n_3^{1,\mathbf{8}} + \frac13 m_2^{1,\mathbf{10}+\overline{\mathbf{10}}} - \frac13 m_2^{1,\mathbf{27}} - \frac19 \bar{c}_3^{1,\mathbf{27}},
\end{equation}
and
\begin{equation}
\sqrt{3} \delta \mu_{{\Sigma^*}^-}^{\mathrm{SB}} = \frac32 m_1^{1,\mathbf{1}} + \frac54 m_3^{1,\mathbf{1}} - \frac12 n_1^{1,\mathbf{8}} - \frac12 n_2^{1,\mathbf{8}} - \frac56 n_3^{1,\mathbf{8}}.
\end{equation}
The complete list of expressions is given in Appendix \ref{app:SB}.

\section{\label{sec:num}Numerical analysis}

A number of different fits to the experimental data can now be performed. These fits, however, are not intended to be definitive; instead, they can be useful in testing the working assumptions. The theoretical formulas are not as accurate enough as the experimental measurements so a theoretical error has to be included to get a meaningful $\chi^2$. Thus, the dominant error in all the fits is theoretical.

On the experimental bent, the Review of Particle Physics \cite{part} lists values for only ten magnetic moments: Seven out of the eight octet baryons ($\mu_{\Sigma^0}$ remains unknown), $\mu_{\Omega^-}$, and the transition moments $\mu_{\Sigma^0\Lambda}$ and $\mu_{\Delta^+p}$. The latter can be obtained from the $\Delta\to N \gamma$ helicity amplitudes $A_{\frac12}$ and $A_\frac32$. A consistent extraction of $\mu_\Delta^{++}$ can be used \cite{lopez}, together with two more pieces of information, namely, $\mu_{{\Sigma^*}^0\Lambda}$ and $\mu_{{\Sigma^*}^+\Sigma^+}$, which can be extracted from Refs.~\cite{clas1} and \cite{clas2}, respectively. Additional inputs are the physical masses of the $\pi$, $K$, and $\eta$ pseudoscalar mesons, the average decuplet-octet mass difference $\Delta=0.231$ GeV, which follows from the average baryon decuplet and octet and masses, $M_T=1.382$ GeV and $M_B=1.151$ GeV, respectively. Similarly, the pion decay constant is set to $f=93$ MeV and the scale of dimensional regularization used is $\mu = 1$ GeV.

The standard $\chi^2$ function to be minimized is written as
\begin{equation}
\chi^2 = \sum_{i=1}^N \left[ \frac{\mu_i^\mathrm{exp} - \mu_i^\mathrm{th}}{\Delta \mu_i^\mathrm{exp}} \right]^2, \label{eq:stchi2}
\end{equation}
where $\mu_i^\mathrm{exp}$ and $\Delta \mu_i^\mathrm{exp}$ are the available measured magnetic moments and their corresponding uncertainties, respectively, and $\mu_i^\mathrm{th}$ are their theoretical counterparts, which are constituted by the sum of tree-level values $\mu_i^{(0)}$, one-loop corrections $\delta \mu^{(\mathrm{loop}\, n)}$, and explicit SB corrections $\delta \mu^\mathrm{SB}$, i.e.,
\begin{equation}
\mu_i^\mathrm{th} = \mu_i^{(0)} + \delta \mu^{\mathrm{(loop\, 1)}} + \delta \mu^{\mathrm{(loop\, 2ad)}} + \mu^{\mathrm{(loop 2e)}}
+ \delta \mu^\mathrm{SB}.
\end{equation}

The free parameters in the theory are the operator coefficients $a_1$, $b_2$, $b_3$, and $c_3$ from the baryon axial current operator $A^{kc}$ (\ref{eq:akc}). Four additional parameters $m_k$ are introduced in the definition of the baryon magnetic moment operator $M^k$ (\ref{eq:mQ}). There are eleven additional parameters coming from explicit $SU(3)$ SB. In total, there are 19 free parameters to be determined and only $N=13$ pieces of experimental information.

The simplest possibility is an $SU(3)$ symmetric fit neglecting all $SU(3)$ breaking effects, which involves only the four parameters $m_i$. Keeping in mind that in most hadronic quantities $SU(3)$ breaking is around 20\%–30\% and that the theoretical errors are of order $\epsilon/N_c$, where $\epsilon$ is a measure of $SU(3)$ breaking, then a fair estimate of the theoretical error to be added in quadrature to the experimental ones is $\pm 0.30 \, \mu_N$ [recall that baryon magnetic moments are order $\mathcal{O}(N_c)$ at leading order in $N_c$]. The results are listed in the column labeled Fit 1 in Table \ref{t:bestfitp}. In this case, $\chi^2 = 12.22$ for nine degrees of freedom, but this particular value only reflects the choice of theoretical error. Adding smaller theoretical errors lowers the errors in the parameters at the expense of increasing $\chi^2$ and, except for $m_3$, the central values of the remaining coefficients change a little. The closeness of $\chi^2/\mathrm{dof}$ to one might be interpreted as a sign that $SU(3)$ SB is indeed around 30\%.

To proceed further, in order to gain predictive power, a few assumptions on the unknown parameters should be made. First, the values of the operator coefficients $a_1$, $b_2$, $b_3$, and $c_3$ can be borrowed from the recent analysis of the baryon axial current presented in Ref.~\cite{rfm21}, namely,
\begin{equation}
a_1=1.20 \pm 0.07, \quad b_2=-1.60 \pm 0.18, \quad b_3=1.25 \pm 0.07, \quad c_3=0.46 \pm 0.09,
\end{equation}
which are extracted from Table II of Ref.~\cite{rfm21}, labeled as Fit B.

The relevant parameters $m_k$ should be determined in full, so a few restrictions can be imposed on the parameters from explicit SB. The simplest one is to keep terms up to relative order $1/N_c$, so the relevant parameters become $m_1^{1,\mathbf{1}}$, $n_1^{1,\mathbf{8}}$, $n_2^{1,\mathbf{8}}$, $m_2^{1,\mathbf{10}+\overline{\mathbf{10}}}$, and $m_2^{1,\mathbf{27}}$.

In order to get a consistent least-squares fit, a theoretical uncertainty of $\pm 1/N_c^2 = \pm 0.11$ will be added in quadrature to the experimental errors to account for the omitted terms mentioned above. Without further ado, the fit yields the best-fit parameters listed in Table \ref{t:bestfitp} under the label Fit 2. In this case, $\chi^2 = 14.55/4$ dof and although it exceeds expectations, the best-fit parameters are fairly order 1 (except for $m_1$) and yield reasonable predictions, as it can be verified in the predicted magnetic moments listed in Table \ref{t:numbers}. Explicit SB and one-loop corrections to tree-level values roughly represent 30\%-40\%, which are in accordance with first-order SB.

\begin{table*}
\caption{\label{t:bestfitp}Best-fit parameters from least-squares fits: Fit 1 is an $SU(3)$ fit; Fit 2 includes one-loop and partial SB corrections (see the text); Fit 3 constitutes the so-called prior fit. The resulting values of the corresponding $SU(3)$ couplings $\mu_D$, $\mu_F$, $\mu_C$, and $\mu_T$ are also shown.}
\begin{ruledtabular}
\begin{tabular}{lrrr}
Parameter & Fit 1 & Fit 2 & Fit 3 \\
\hline
$m_1$ & $ 5.07 \pm 0.42$ & $ 7.86 \pm 0.09$ & $ 7.86 \pm 0.09$ \\
$m_2$ & $ 0.73 \pm 1.28$ & $-0.01 \pm 0.18$ & $ 0.01 \pm 0.19$ \\
$m_3$ & $-0.41 \pm 0.82$ & $-1.01 \pm 0.13$ & $-1.01 \pm 0.13$ \\
$m_4$ & $ 4.05 \pm 1.27$ & $ 1.67 \pm 0.23$ & $ 1.67 \pm 0.24$ \\
$m_1^{1,\mathbf{1}}$ & & $ 0.16 \pm 0.20$ & $0.16 \pm 0.20$ \\
$m_3^{1,\mathbf{1}}$ & &     & $0.021 \pm 0.100$ \\
$n_1^{1,\mathbf{8}}$ & & $-0.71 \pm 0.38$ & $-0.69 \pm 0.38$ \\
$n_2^{1,\mathbf{8}}$ & & $-2.61 \pm 0.89$ & $-2.65 \pm 0.90$ \\
$n_3^{1,\mathbf{8}}$ & &     & $0.010 \pm 0.100$ \\
$\bar{n}_3^{1,\mathbf{8}}$ & &     & $0.006 \pm 0.100$ \\
$m_2^{1,\mathbf{10}+\overline{\mathbf{10}}}$ & & $-2.35 \pm 0.23$ & $-2.35 \pm 0.23411$ \\ 
$m_3^{1,\mathbf{10}+\overline{\mathbf{10}}}$ & &     & $0.011 \pm 0.100$ \\
$m_2^{1,\mathbf{27}}$ & & $ 0.71 \pm 0.33$ & $0.68 \pm 0.35$ \\
$m_3^{1,\mathbf{27}}$ & &     & $0.025 \pm 0.100$ \\
$\bar{m}_3^{1,\mathbf{27}}$ & &     & $0.017 \pm 0.100$ \\
$\chi^2$ & $12.22$ & $14.56$ & $14.38$ \\
\hline
$\mu_D$ & $ 2.47 \pm 0.23$ & $ 3.76 \pm 0.05$ & $3.76 \pm 0.02$ \\
$\mu_F$ & $ 1.77 \pm 0.15$ & $ 2.51 \pm 0.03$ & $2.30 \pm 0.03$ \\
$\mu_C$ & $ 2.56 \pm 0.21$ & $ 3.08 \pm 0.08$ & $2.50 \pm 0.06$ \\
$\mu_T$ & $-14.18 \pm 0.95$ & $-17.38 \pm 0.33$ & $-17.39 \pm 0.24$ \\
$\mu_D/\mu_F$ & $1.40 \pm 0.13$ & $ 1.50 \pm 0.02$ & $1.62 \pm 0.02$ \\
\end{tabular}
\end{ruledtabular}
\end{table*}

In general, predictions are consistent with data and with other determinations. For instance, in the context of the $1/N_c$ expansion alone \cite{lebed}, there is an overall agreement. In the context of heavy baryon chiral perturbation theory \cite{m97} and relativistic baryon chiral perturbation theory \cite{k00}, there is a reasonable agreement with calculations for octet baryons to third order. These references, however, present more refined calculations to fourth order. At the level of precision presented in this work, no comparison is possible yet. Theoretical expressions need be improved, for instance, by lifting the $\Delta=0$ assumption in graphs \ref{fig:mmloop2}(a)-\ref{fig:mmloop2}(d). This could improve the determinations of $\mu_C$ and $\mu_T$ to a reasonable extent. Actually, the analysis of Ref.~\cite{rfm14} where partial terms containing a nonzero $\Delta$ in loop diagrams \ref{fig:mmloop2}(a)-\ref{fig:mmloop2}(d) seems to point in the right direction.

\begin{table*}
\caption{\label{t:numbers}Predicted baryon magnetic moments using the best-fit parameters from Fit 2.}
\begin{ruledtabular}
\begin{tabular}{lcrrrrrr}
& $\displaystyle \mu^{\mathrm{exp}}$ & $\displaystyle \mu^{\mathrm{th}}$ & $\displaystyle \mu^{(0)}$ & $\displaystyle \delta \mu^\mathrm{SB}$ & $\displaystyle \delta \mu^{\mathrm{(loop\, 1)}}$ & $\displaystyle \delta \mu^{\mathrm{(loop\, 2ad)}}$ & $\displaystyle \delta \mu^{\mathrm{(loop 2e)}}$ \\
\hline
$n$ & $-1.9130 \pm 0.000$ & $-2.079$ & $-2.507$ & $ 0.818$ & $ 0.804$ & $-0.861$ & $-0.334$ \\
$p$ & $ 2.7928 \pm 0.000$ & $ 2.852$ & $ 3.760$ & $-0.266$ & $-2.064$ & $ 0.616$ & $ 0.807$ \\
$\Sigma^-$ & $-1.160 \pm 0.025$ & $-1.108$ & $-1.253$ & $-0.085$ & $ 0.487$ & $-0.275$ & $ 0.017$ \\
$\Sigma^0$ & & $ 0.702$ & $ 1.253$ & $ 0.116$ & $-1.531$ & $ 0.390$ & $ 0.474$ \\
$\Sigma^+$ & $ 2.458 \pm 0.010$ & $ 2.512$ & $ 3.760$ & $ 0.317$ & $-3.550$ & $ 1.055$ & $ 0.930$ \\
$\Xi^-$ & $-0.6507 \pm 0.0025$ & $-0.602$ & $-1.253$ & $ 0.637$ & $ 1.059$ & $-0.449$ & $-0.596$ \\
$\Xi^0$ & $-1.250 \pm 0.014$ & $-1.279$ & $-2.507$ & $-0.587$ & $ 3.263$ & $-0.661$ & $-0.788$ \\
$\Lambda$ & $-0.613 \pm 0.004$ & $-0.487$ & $-1.253$ & $-0.021$ & $ 1.531$ & $-0.765$ & $ 0.021$ \\
$\Sigma^0\Lambda$ & $ 1.61 \pm 0.08$ & $ 1.239$ & $ 2.171$ & $-0.119$ & $-1.464$ & $ 0.255$ & $ 0.395$ \\
$\Delta^{++}$ & $ 6.14 \pm 0.51$\footnote{Value reported in Ref.~\cite{lopez}}& $ 5.695$ & $ 6.170$ & $ 0.007$ & $-3.273$ & $ 1.366$ & $ 1.426$ \\
$\Delta^+$ & & $ 2.821$ & $ 3.085$ & $ 0.554$ & $-2.278$ & $ 0.596$ & $ 0.864$ \\
$\Delta^0$ & & $-0.156$ & $ 0.000$ & $ 1.101$ & $-1.283$ & $-0.277$ & $ 0.302$ \\
$\Delta^-$ & & $-3.082$ & $-3.085$ & $ 1.649$ & $-0.288$ & $-1.098$ & $-0.260$ \\
${\Sigma^*}^+$ & & $ 2.044$ & $ 3.085$ & $-0.818$ & $-0.995$ & $ 0.210$ & $ 0.562$ \\
${\Sigma^*}^0$ & & $-0.361$ & $ 0.000$ & $ 0.142$ & $ 0.000$ & $-0.503$ & $ 0.000$ \\
${\Sigma^*}^-$ & & $-2.766$ & $-3.085$ & $ 1.101$ & $ 0.995$ & $-1.216$ & $-0.562$ \\
${\Xi^*}^0$ & & $-0.518$ & $ 0.000$ & $-0.818$ & $ 1.283$ & $-0.681$ & $-0.302$ \\
${\Xi^*}^-$ & & $-2.475$ & $-3.085$ & $ 0.554$ & $ 2.278$ & $-1.358$ & $-0.864$ \\
$\Omega^-$ & $-2.02 \pm 0.05$ & $-2.053$ & $-3.085$ & $ 0.007$ & $ 3.560$ & $-1.370$ & $-1.166$ \\
$\Delta^+ p$ & $ 3.51 \pm 0.09$ & $ 3.381$ & $ 4.097$ & $-0.638$ & $-3.071$ & $ 2.247$ & $ 0.746$ \\
$\Delta^0 n$ & & $ 3.381$ & $ 4.097$ & $-0.638$ & $-3.071$ & $ 2.247$ & $ 0.746$ \\
${\Sigma^*}^0\Lambda$ & $ 2.73 \pm 0.25$\footnote{Value extracted from Ref.~\cite{clas1}}& $ 2.885$ & $ 3.548$ & $-0.168$ & $-3.089$ & $ 2.071$ & $ 0.522$ \\
${\Sigma^*}^0\Sigma^0$ & & $ 1.284$ & $ 2.049$ & $ 0.097$ & $-3.048$ & $ 1.413$ & $ 0.774$ \\
${\Sigma^*}^+\Sigma^+$ & $ 3.17 \pm 0.36$\footnote{Value extracted from Ref.~\cite{clas2}} & $ 3.456$ & $ 4.097$ & $ 0.833$ & $-5.327$ & $ 2.705$ & $ 1.147$ \\
${\Sigma^*}^-\Sigma^-$ & & $-0.888$ & $ 0.000$ & $-0.640$ & $-0.769$ & $ 0.121$ & $ 0.401$ \\
${\Xi^*}^0\Xi^0$ & & $ 3.064$ & $ 4.097$ & $ 0.444$ & $-5.327$ & $ 2.702$ & $ 1.147$ \\
${\Xi^*}^-\Xi^-$ & & $-0.892$ & $ 0.000$ & $-0.640$ & $-0.769$ & $ 0.116$ & $ 0.401$ \\
\end{tabular}
\end{ruledtabular}
\end{table*}

An alternative approach to get at least an estimate of the size of the omitted free parameters of Fit 2 above can be achieved following the lines of the fitting procedure implemented in Ref.~\cite{severt}. The approach, adapted to the present analysis, consists in using the prior fit \cite{sch} to extend the standard $\chi^2$ of Eq.~(\ref{eq:stchi2}) to
\begin{equation}
\chi^2_\mathrm{prior} = \chi^2 + \left[ \frac{m_3^{1,\mathbf{1}}}{\Delta m_3^{1,\mathbf{1}}} \right]^2 +
\left[ \frac{n_3^{1,\mathbf{8}}}{\Delta n_3^{1,\mathbf{8}}} \right]^2 +
\left[ \frac{\bar{n}_3^{1,\mathbf{8}}}{\Delta \bar{n}_3^{1,\mathbf{8}}} \right]^2 +
\left[ \frac{m_3^{1,\mathbf{10}+\overline{\mathbf{10}}}}{\Delta m_3^{1,\mathbf{10}+\overline{\mathbf{10}}}} \right]^2 +
\left[ \frac{m_3^{1,\mathbf{27}}}{\Delta m_3^{1,\mathbf{27}}} \right]^2 +
\left[ \frac{\bar{m}_3^{1,\mathbf{27}}}{\Delta \bar{m}_3^{1,\mathbf{27}}} \right]^2,
\end{equation}
where $m_3^{1,\mathbf{rep}}$ and $n_3^{1,\mathbf{rep}}$ are the unknown coefficients that come along 3-body operators from explicit SB weighted by their respective errors. While the extra terms added to $\chi^2$ guarantees that these six parameters get values around zero (approximately Gaussian distributed \cite{severt}), the remaining nine parameters are the ones actually fitted to the experimental data. For definiteness, the nominal theoretical errors $\Delta m_3^{1,\mathbf{rep}} = \Delta n_3^{1,\mathbf{rep}} = 0.100$ have been used and the corresponding best-fit parameters are listed in Table \ref{t:bestfitp} under the label Fit 3. It is convenient to point out that nominal errors of $\pm 0.200$ and $\pm 0.050$ produce $\chi_\mathrm{prior}^2=13.97$ and $\chi_\mathrm{prior}^2=14.51$, respectively. In all cases, the six parameters referred to above are small compared to the ones retained in the standard fit, which suggest that the assumption of neglecting them in the analysis is justified.

\section{\label{sec:con}Concluding remarks}

Baryon magnetic moments to orders $\mathcal{O}(m_q^{1/2})$ and $\mathcal{O}(m_q \ln m_q)$ are evaluated in the present paper in the context of chiral perturbation theory in the large-$N_c$ limit. All the operator structures that appear for $N_c=3$ are accounted for in the analysis. Regrettably, the expressions obtained are rather long; however, including them in full is necessary to make the paper self-contained.

The approach presented here is twofold. On the one hand, previous analyses \cite{rfm09,rfm14} get improved with the addition of new terms not considered before, and, on the other hand, the complete structures presented allow one to carry out a full comparison with the conventional chiral perturbation theory results by using the relations between the operator coefficients $a_1$, $b_2$, $b_3$, and $c_3$ and the $SU(3)$ invariants $\mu_D$, $\mu_F$, $\mu_C$, and $\mu_T$.

The main conclusion obtained is that theoretical expressions of baryon magnetic moments agree in both theories at the physical value $N_c=3$ for $N_f=3$ flavors of light quarks.

A preliminary numerical analysis via a least-squares fit is also conducted to explore the free parameters in the theory. Although a stable fit is observed, the best-fit parameters are not entirely satisfactory with the assumptions made. The main issue is the lack of experimental data to perform a detailed determination of all the free parameters. In order to improve the theoretical expressions, also the effects of a nonzero decuplet-octet baryon mass difference in the diagrams of order $\mathcal{O}(m_q \ln m_q)$ are needed. The calculation of these contributions, however, involves a non-negligible effort which can be attempted elsewhere. The approach discussed here will constitute useful guidance for this enterprise. Of course, new and/or improved measurements of baryon magnetic moments will be welcome in the future.

\begin{acknowledgments}
The authors are grateful to Consejo Nacional de Ciencia y Tecnolog{\'\i}a (Mexico) for partial support.
\end{acknowledgments}

\appendix

\section{\label{app:rloop1}Reduction of baryon operators emerging from Fig.~\ref{fig:mmloop1}}

\subsection{Flavor $\mathbf{8}$ spin-independent operators}

\begin{equation}
\epsilon^{ijk} f^{abc} G^{ia} G^{jb} = - \frac12 (N_c+N_f)G^{kc} + \frac12 \mathcal{D}_2^{kc},
\end{equation}

\begin{equation}
\epsilon^{ijk} f^{abc} (G^{ia} \mathcal{D}_2^{jb} + \mathcal{D}_2^{ia} G^{jb}) = - N_f G^{kc} - \mathcal{O}_3^{kc},
\end{equation}

\begin{equation}
\epsilon^{ijk} f^{abc} \mathcal{D}_2^{ia} \mathcal{D}_2^{jb} = - \frac12 N_f \mathcal{D}_2^{kc},
\end{equation}

\begin{equation}
\epsilon^{ijk} f^{abc} (G^{ia} \mathcal{D}_3^{jb} + \mathcal{D}_3^{ia} G^{jb}) = - 2(N_c+N_f) G^{kc} - (N_f-2) \mathcal{D}_2^{kc} - (N_c+N_f) \mathcal{O}_3^{kc},
\end{equation}

\begin{equation}
\epsilon^{ijk} f^{abc} (G^{ia} \mathcal{O}_3^{jb} + \mathcal{O}_3^{ia} G^{jb}) = \frac32 N_f \mathcal{D}_2^{kc} - \frac12 (N_c+N_f) \mathcal{D}_3^{kc} - \frac12 (N_c+N_f) \mathcal{O}_3^{kc} + \mathcal{D}_4^{kc},
\end{equation}

\begin{equation}
\epsilon^{ijk} f^{abc} (\mathcal{D}_2^{ia} \mathcal{D}_3^{jb} + \mathcal{D}_3^{ia} \mathcal{D}_2^{jb}) = - N_f \mathcal{D}_3^{kc},
\end{equation}

\begin{equation}
\epsilon^{ijk} f^{abc} (\mathcal{D}_2^{ia} \mathcal{O}_3^{jb} + \mathcal{O}_3^{ia} \mathcal{D}_2^{jb}) = - N_f \mathcal{O}_3^{kc} - \mathcal{O}_5^{kc},
\end{equation}

\begin{equation}
\epsilon^{ijk} f^{abc} \mathcal{D}_3^{ia} \mathcal{D}_3^{jb} = - (N_c+N_f) \mathcal{D}_3^{kc} - (N_f-2) \mathcal{D}_4^{kc},
\end{equation}

\begin{equation}
\epsilon^{ijk} f^{abc} (\mathcal{D}_3^{ia} \mathcal{O}_3^{jb} + \mathcal{O}_3^{ia} \mathcal{D}_3^{jb}) = - 2(N_c+N_f) \mathcal{O}_3^{kc} - (N_c+N_f) \mathcal{O}_5^{kc},
\end{equation}

\begin{equation}
\epsilon^{ijk} f^{abc} \mathcal{O}_3^{ia} \mathcal{O}_3^{jb} = \frac32 N_f \mathcal{D}_2^{kc} - \frac34 (N_c+N_f) \mathcal{D}_3^{kc} + \frac14 (5N_f+6) \mathcal{D}_4^{kc} - \frac14 (N_c+N_f) \mathcal{D}_5^{kc} + \frac12 \mathcal{D}_6^{kc}.
\end{equation}

\subsection{Flavor $\mathbf{8}$ spin-dependent operators}

\begin{equation}
\epsilon^{ijk} f^{abc} G^{ia} J^2 G^{jb} = - \frac12 (N_c+N_f)G^{kc} + \frac12 (N_f+1) \mathcal{D}_2^{kc} - \frac18 (N_c+N_f) \mathcal{D}_3^{kc} - \frac14 (N_c+N_f) \mathcal{O}_3^{kc} + \frac14 \mathcal{D}_4^{kc},
\end{equation}

\begin{equation}
\epsilon^{ijk} f^{abc} (G^{ia} J^2 \mathcal{D}_2^{jb} + \mathcal{D}_2^{ia} J^2 G^{jb}) = - \frac14 N_f \mathcal{D}_3^{kc} - (N_f+1) \mathcal{O}_3^{kc} - \frac12 \mathcal{O}_5^{kc},
\end{equation}

\begin{equation}
\epsilon^{ijk} f^{abc} \mathcal{D}_2^{ia} J^2 \mathcal{D}_2^{jb} = - \frac14 N_f \mathcal{D}_4^{kc},
\end{equation}

\begin{equation}
\epsilon^{ijk} f^{abc} (G^{ia} J^2 \mathcal{D}_3^{jb} + \mathcal{D}_3^{ia} J^2 G^{jb}) = - \frac12 (N_c+N_f) \mathcal{D}_3^{kc} - 3(N_c+N_f) \mathcal{O}_3^{kc} - \frac12 (N_f-2) \mathcal{D}_4^{kc} - \frac12 (N_c+N_f) \mathcal{O}_5^{kc},
\end{equation}

\begin{eqnarray}
\epsilon^{ijk} f^{abc} (G^{ia} J^2 \mathcal{O}_3^{jb} + \mathcal{O}_3^{ia} J^2 G^{jb}) & = & 3 N_f \mathcal{D}_2^{kc} - \frac32 (N_c+N_f) \mathcal{D}_3^{kc} - \frac12 (N_c+N_f) \mathcal{O}_3^{kc} + \frac14 (7N_f+12) \mathcal{D}_4^{kc} \nonumber \\
& & \mbox{} - \frac14 (N_c+N_f) \mathcal{D}_5^{kc} - \frac14 (N_c+N_f) \mathcal{O}_5^{kc} + \frac12 \mathcal{D}_6^{kc},
\end{eqnarray}

\begin{equation}
\epsilon^{ijk} f^{abc} (\mathcal{D}_2^{ia} J^2 \mathcal{D}_3^{jb} + \mathcal{D}_3^{ia} J^2 \mathcal{D}_2^{jb}) = - \frac12 N_f \mathcal{D}_5^{kc},
\end{equation}

\begin{equation}
\epsilon^{ijk} f^{abc} (\mathcal{D}_2^{ia} J^2 \mathcal{O}_3^{jb} + \mathcal{O}_3^{ia} J^2 \mathcal{D}_2^{jb}) = - (N_f+1) \mathcal{O}_5^{kc} - \frac12 \mathcal{O}_7^{kc},
\end{equation}

\begin{equation}
\epsilon^{ijk} f^{abc} \mathcal{D}_3^{ia} J^2 \mathcal{D}_3^{jb} = - \frac12 (N_c+N_f) \mathcal{D}_5^{kc} - \frac12 (N_f-2) \mathcal{D}_6^{kc},
\end{equation}

\begin{equation}
\epsilon^{ijk} f^{abc} (\mathcal{D}_3^{ia} J^2 \mathcal{O}_3^{jb} + \mathcal{O}_3^{ia} J^2 \mathcal{D}_3^{jb}) = - 3(N_c+N_f) \mathcal{O}_5^{kc} - \frac12 (N_c+N_f) \mathcal{O}_7^{kc},
\end{equation}

\begin{eqnarray}
\epsilon^{ijk} f^{abc} \mathcal{O}_3^{ia} J^2 \mathcal{O}_3^{jb} & = & 3 N_f \mathcal{D}_2^{kc} - \frac32 (N_c+N_f) \mathcal{D}_3^{kc} + \frac14 (19N_f+12) \mathcal{D}_4^{kc} - \frac{13}{8} (N_c+N_f) \mathcal{D}_5^{kc} + \frac18 (9N_f+26) \mathcal{D}_6^{kc} \nonumber \\
& & \mbox{} - \frac18 (N_c+N_f) \mathcal{D}_7^{kc} + \frac14 \mathcal{D}_8^{kc}.
\end{eqnarray}

\subsection{Flavor $\mathbf{10}+\overline{\mathbf{10}}$ spin-independent operators}

\begin{equation}
\epsilon^{ijk}(f^{aec} d^{be8} - f^{bec} d^{ae8} - f^{abe} d^{ec8}) G^{ia} G^{jb} = - \frac12 \{G^{kc},T^8\} + \frac12 \{G^{k8},T^c\} - \frac{1}{N_f} if^{c8e} [J^2,G^{ke}],
\end{equation}

\begin{equation}
\epsilon^{ijk}(f^{aec} d^{be8} - f^{bec} d^{ae8} - f^{abe} d^{ec8}) (G^{ia} \mathcal{D}_2^{jb} + \mathcal{D}_2^{ia} G^{jb}) = - \frac{N_c+N_f}{N_f} if^{c8e} [J^2,G^{ke}] - \{G^{kc},\{J^r,G^{r8}\}\} + \{G^{k8},\{J^r,G^{rc}\}\},
\end{equation}

\begin{equation}
\epsilon^{ijk}(f^{aec} d^{be8} - f^{bec} d^{ae8} - f^{abe} d^{ec8}) \mathcal{D}_2^{ia} \mathcal{D}_2^{jb} = 0,
\end{equation}

\begin{eqnarray}
& & \epsilon^{ijk}(f^{aec} d^{be8} - f^{bec} d^{ae8} - f^{abe} d^{ec8}) (G^{ia} \mathcal{D}_3^{jb} + \mathcal{D}_3^{ia} G^{jb}) \nonumber \\
& & \mbox{\hglue0.2truecm} = - 2 \{G^{kc},T^8\} + 2 \{G^{k8},T^c\} - \frac{4}{N_f} if^{c8e} [J^2,G^{ke}] - \{\mathcal{D}_2^{kc},\{J^r,G^{r8}\}\} + \{\mathcal{D}_2^{k8},\{J^r,G^{rc}\}\} - \{J^2,\{G^{kc},T^8\}\} \nonumber \\
& & \mbox{\hglue0.6truecm} + \{J^2,\{G^{k8},T^c\}\} - \frac{2}{N_f} if^{c8e} \{J^2,[J^2,G^{ke}]\},
\end{eqnarray}

\begin{eqnarray}
& & \epsilon^{ijk}(f^{aec} d^{be8} - f^{bec} d^{ae8} - f^{abe} d^{ec8}) (G^{ia} \mathcal{O}_3^{jb} + \mathcal{O}_3^{ia} G^{jb}) \nonumber \\
& & \mbox{\hglue0.2truecm} = \frac12 \{\mathcal{D}_2^{kc},\{J^r,G^{r8}\}\} - \frac12 \{\mathcal{D}_2^{k8},\{J^r,G^{rc}\}\} - \frac12 \{J^2,\{G^{kc},T^8\}\} + \frac12 \{J^2,\{G^{k8},T^c\}\} - \frac{1}{N_f} if^{c8e} \{J^2,[J^2,G^{ke}]\}, \nonumber \\
\end{eqnarray}

\begin{equation}
\epsilon^{ijk}(f^{aec} d^{be8} - f^{bec} d^{ae8} - f^{abe} d^{ec8}) (\mathcal{D}_2^{ia} \mathcal{D}_3^{jb} + \mathcal{D}_3^{ia} \mathcal{D}_2^{jb}) = 0,
\end{equation}

\begin{eqnarray}
& & \epsilon^{ijk}(f^{aec} d^{be8} - f^{bec} d^{ae8} - f^{abe} d^{ec8}) (\mathcal{D}_2^{ia} \mathcal{O}_3^{jb} + \mathcal{O}_3^{ia} \mathcal{D}_2^{jb}) \nonumber \\
& & \mbox{\hglue0.2truecm} = - \frac{N_c+N_f}{N_f} if^{c8e} \{J^2,[J^2,G^{ke}]\} - \{J^2,\{G^{kc},\{J^r,G^{r8}\}\}\} + \{J^2,\{G^{k8},\{J^r,G^{rc}\}\}\}, \nonumber \\
\end{eqnarray}

\begin{equation}
\epsilon^{ijk}(f^{aec} d^{be8} - f^{bec} d^{ae8} - f^{abe} d^{ec8}) \mathcal{D}_3^{ia} \mathcal{D}_3^{jb} = 2 \{\mathcal{D}_2^{kc},\{J^r,G^{r8}\}\} - 2 \{\mathcal{D}_2^{k8},\{J^r,G^{rc}\}\},
\end{equation}

\begin{eqnarray}
& & \epsilon^{ijk}(f^{aec} d^{be8} - f^{bec} d^{ae8} - f^{abe} d^{ec8}) (\mathcal{D}_3^{ia} \mathcal{O}_3^{jb} + \mathcal{O}_3^{ia} \mathcal{D}_3^{jb}) \nonumber \\
& & \mbox{\hglue0.2truecm} = - 2 \{\mathcal{D}_2^{kc},\{J^r,G^{r8}\}\} + 2 \{\mathcal{D}_2^{k8},\{J^r,G^{rc}\}\} - 2 \{J^2,\{G^{kc},T^8\}\} + 2 \{J^2,\{G^{k8},T^c\}\} - \frac{4}{N_f} if^{c8e} \{J^2,[J^2,G^{ke}]\} \nonumber \\
& & \mbox{\hglue0.6truecm} - \{J^2,\{\mathcal{D}_2^{kc},\{J^r,G^{r8}\}\}\} + \{J^2,\{\mathcal{D}_2^{k8},\{J^r,G^{rc}\}\}\} - \{J^2,\{J^2,\{G^{kc},T^8\}\}\} + \{J^2,\{J^2,\{G^{k8},T^c\}\}\} \nonumber \\
& & \mbox{\hglue0.6truecm} - \frac{2}{N_f} if^{c8e} \{J^2,\{J^2,[J^2,G^{ke}]\}\},
\end{eqnarray}

\begin{eqnarray}
& & \epsilon^{ijk}(f^{aec} d^{be8} - f^{bec} d^{ae8} - f^{abe} d^{ec8}) \mathcal{O}_3^{ia} \mathcal{O}_3^{jb} \nonumber \\
& & \mbox{\hglue0.2truecm} = \frac32 \{\mathcal{D}_2^{kc},\{J^r,G^{r8}\}\} - \frac32 \{\mathcal{D}_2^{k8},\{J^r,G^{rc}\}\} + \frac12 \{J^2,\{\mathcal{D}_2^{kc},\{J^r,G^{r8}\}\}\} - \frac12 \{J^2,\{\mathcal{D}_2^{k8},\{J^r,G^{rc}\}\}\}.
\end{eqnarray}

\subsection{Flavor $\mathbf{10}+\overline{\mathbf{10}}$ spin-dependent operators}

\begin{eqnarray}
& & \epsilon^{ijk}(f^{aec} d^{be8} - f^{bec} d^{ae8} - f^{abe} d^{ec8}) G^{ia} J^2 G^{jb} \nonumber \\
& & \mbox{\hglue0.2truecm} = - \frac12 \{G^{kc},T^8\} + \frac12 \{G^{k8},T^c\} - \frac{1}{N_f} if^{c8e} [J^2,G^{ke}] - \frac14 \{J^2,\{G^{kc},T^8\}\} + \frac14 \{J^2,\{G^{k8},T^c\}\} \nonumber \\
& & \mbox{\hglue0.6truecm} - \frac{1}{2N_f} if^{c8e} \{J^2,[J^2,G^{ke}]\},
\end{eqnarray}

\begin{eqnarray}
& & \epsilon^{ijk}(f^{aec} d^{be8} - f^{bec} d^{ae8} - f^{abe} d^{ec8}) (G^{ia} J^2 \mathcal{D}_2^{jb} + \mathcal{D}_2^{ia} J^2 G^{jb}) \nonumber \\
& & \mbox{\hglue0.2truecm} = - \frac{N_c+N_f}{N_f} if^{c8e} [J^2,G^{ke}] - \{G^{kc},\{J^r,G^{r8}\}\} + \{G^{k8},\{J^r,G^{rc}\}\} - \frac{N_c+N_f}{2N_f} if^{c8e} \{J^2,[J^2,G^{ke}]\} \nonumber \\
& & \mbox{\hglue0.6truecm} - \frac12 \{J^2,\{G^{kc},\{J^r,G^{r8}\}\}\} + \frac12 \{J^2,\{G^{k8},\{J^r,G^{rc}\}\}\},
\end{eqnarray}

\begin{equation}
\epsilon^{ijk}(f^{aec} d^{be8} - f^{bec} d^{ae8} - f^{abe} d^{ec8}) \mathcal{D}_2^{ia} J^2 \mathcal{D}_2^{jb} = 0,
\end{equation}

\begin{eqnarray}
& & \epsilon^{ijk}(f^{aec} d^{be8} - f^{bec} d^{ae8} - f^{abe} d^{ec8}) (G^{ia} J^2 \mathcal{D}_3^{jb} + \mathcal{D}_3^{ia} J^2 G^{jb}) \nonumber \\
& & \mbox{\hglue0.2truecm} = - 2 \{\mathcal{D}_2^{kc},\{J^r,G^{r8}\}\} + 2 \{\mathcal{D}_2^{k8},\{J^r,G^{rc}\}\} - 3 \{J^2,\{G^{kc},T^8\}\} + 3 \{J^2,\{G^{k8},T^c\}\} - \frac{6}{N_f} if^{c8e} \{J^2,[J^2,G^{ke}]\} \nonumber \\
& & \mbox{\hglue0.6truecm} - \frac12 \{J^2,\{\mathcal{D}_2^{kc},\{J^r,G^{r8}\}\}\} + \frac12 \{J^2,\{\mathcal{D}_2^{k8},\{J^r,G^{rc}\}\}\} - \frac12 \{J^2,\{J^2,\{G^{kc},T^8\}\}\} + \frac12 \{J^2,\{J^2,\{G^{k8},T^c\}\}\} \nonumber \\
& & \mbox{\hglue0.6truecm} - \frac{1}{N_f} if^{c8e} \{J^2,\{J^2,[J^2,G^{ke}]\}\},
\end{eqnarray}

\begin{eqnarray}
& & \epsilon^{ijk}(f^{aec} d^{be8} - f^{bec} d^{ae8} - f^{abe} d^{ec8}) (G^{ia} J^2 \mathcal{O}_3^{jb} + \mathcal{O}_3^{ia} J^2 G^{jb}) \nonumber \\
& & \mbox{\hglue0.2truecm} = \frac52 \{\mathcal{D}_2^{kc},\{J^r,G^{r8}\}\} - \frac52 \{\mathcal{D}_2^{k8},\{J^r,G^{rc}\}\} - \frac12 \{J^2,\{G^{kc},T^8\}\} + \frac12 \{J^2,\{G^{k8},T^c\}\} - \frac{1}{N_f} if^{c8e} \{J^2,[J^2,G^{ke}]\} \nonumber \\
& & \mbox{\hglue0.6truecm} + \frac14 \{J^2,\{\mathcal{D}_2^{kc},\{J^r,G^{r8}\}\}\} - \frac14 \{J^2,\{\mathcal{D}_2^{k8},\{J^r,G^{rc}\}\}\} - \frac14 \{J^2,\{J^2,\{G^{kc},T^8\}\}\} + \frac14 \{J^2,\{J^2,\{G^{k8},T^c\}\}\} \nonumber \\
& & \mbox{\hglue0.6truecm} - \frac{1}{2N_f} if^{c8e} \{J^2,\{J^2,[J^2,G^{ke}]\}\},
\end{eqnarray}

\begin{equation}
\epsilon^{ijk}(f^{aec} d^{be8} - f^{bec} d^{ae8} - f^{abe} d^{ec8}) (\mathcal{D}_2^{ia} J^2 \mathcal{D}_3^{jb} + \mathcal{D}_3^{ia} J^2 \mathcal{D}_2^{jb}) = 0,
\end{equation}

\begin{eqnarray}
& & \epsilon^{ijk}(f^{aec} d^{be8} - f^{bec} d^{ae8} - f^{abe} d^{ec8}) (\mathcal{D}_2^{ia} J^2 \mathcal{O}_3^{jb} + \mathcal{O}_3^{ia} J^2 \mathcal{D}_2^{jb}) \nonumber \\
& & \mbox{\hglue0.2truecm} = - \frac{N_c+N_f}{N_f} if^{c8e} \{J^2,[J^2,G^{ke}]\} - \{J^2,\{G^{kc},\{J^r,G^{r8}\}\}\} + \{J^2,\{G^{k8},\{J^r,G^{rc}\}\}\} \nonumber \\
& & \mbox{\hglue0.6truecm} - \frac{N_c+N_f}{2N_f} if^{c8e} \{J^2,\{J^2,[J^2,G^{ke}]\}\} - \frac12 \{J^2,\{J^2,\{G^{kc},\{J^r,G^{r8}\}\}\}\} + \frac12 \{J^2,\{J^2,\{G^{k8},\{J^r,G^{rc}\}\}\}\}, \nonumber \\
\end{eqnarray}

\begin{equation}
\epsilon^{ijk}(f^{aec} d^{be8} - f^{bec} d^{ae8} - f^{abe} d^{ec8}) \mathcal{D}_3^{ia} J^2 \mathcal{D}_3^{jb} = \{J^2,\{\mathcal{D}_2^{kc},\{J^r,G^{r8}\}\}\} - \{J^2,\{\mathcal{D}_2^{k8},\{J^r,G^{rc}\}\}\},
\end{equation}

\begin{eqnarray}
& & \epsilon^{ijk}(f^{aec} d^{be8} - f^{bec} d^{ae8} - f^{abe} d^{ec8}) (\mathcal{D}_3^{ia} J^2 \mathcal{O}_3^{jb} + \mathcal{O}_3^{ia} J^2 \mathcal{D}_3^{jb}) \nonumber \\
& & \mbox{\hglue0.2truecm} = - 3 \{J^2,\{\mathcal{D}_2^{kc},\{J^r,G^{r8}\}\}\} + 3 \{J^2,\{\mathcal{D}_2^{k8},\{J^r,G^{rc}\}\}\} - 3 \{J^2,\{J^2,\{G^{kc},T^8\}\}\} + 3 \{J^2,\{J^2,\{G^{k8},T^c\}\}\} \nonumber \\
& & \mbox{\hglue0.6truecm} - \frac{6}{N_f} if^{c8e} \{J^2,\{J^2,[J^2,G^{ke}]\}\} - \frac12 \{J^2,\{J^2,\{\mathcal{D}_2^{kc},\{J^r,G^{r8}\}\}\}\} + \frac12 \{J^2,\{J^2,\{\mathcal{D}_2^{k8},\{J^r,G^{rc}\}\}\}\} \nonumber \\
& & \mbox{\hglue0.6truecm} - \frac12 \{J^2,\{J^2,\{J^2,\{G^{kc},T^8\}\}\}\} + \frac12 \{J^2,\{J^2,\{J^2,\{G^{k8},T^c\}\}\}\} - \frac{1}{N_f} if^{c8e} \{J^2,\{J^2,\{J^2,[J^2,G^{ke}]\}\}\},
\end{eqnarray}

\begin{eqnarray}
& & \epsilon^{ijk}(f^{aec} d^{be8} - f^{bec} d^{ae8} - f^{abe} d^{ec8}) \mathcal{O}_3^{ia} J^2 \mathcal{O}_3^{jb} \nonumber \\
& & \mbox{\hglue0.2truecm} = 3 \{\mathcal{D}_2^{kc},\{J^r,G^{r8}\}\} - 3 \{\mathcal{D}_2^{k8},\{J^r,G^{rc}\}\} + \frac{13}{4} \{J^2,\{\mathcal{D}_2^{kc},\{J^r,G^{r8}\}\}\} - \frac{13}{4} \{J^2,\{\mathcal{D}_2^{k8},\{J^r,G^{rc}\}\}\}
 \nonumber \\
& & \mbox{\hglue0.6truecm} + \frac14\{J^2,\{J^2,\{\mathcal{D}_2^{kc},\{J^r,G^{r8}\}\}\}\} - \frac14\{J^2,\{J^2,\{\mathcal{D}_2^{k8},\{J^r,G^{rc}\}\}\}\}.
\end{eqnarray}

\section{\label{app:Loop1}Complete expressions from order $\mathcal{O}(m_q^{1/2})$ corrections}

Order $\mathcal{O}(m_q^{1/2})$ corrections to baryon magnetic moments coming from Fig.~\ref{fig:mmloop1}, \textit{including all} the terms allowed for $N_f=N_c=3$, are given, for octet baryons, by
\begin{eqnarray}
\delta \mu_n^{\mathrm{(loop\, 1)}} & = & \left[ \frac{25}{36} a_1^2 + \frac{5}{18} a_1b_2 + \frac{1}{36} b_2^2 + \frac{25}{54} a_1b_3 + \frac{5}{54} b_2b_3 + \frac{25}{324} b_3^2 \right] I_1(m_\pi,0,\mu) \nonumber \\
& & \mbox{} + \left[ - \frac{1}{36} a_1^2 + \frac{1}{18} a_1b_2 - \frac{1}{36} b_2^2 - \frac{1}{54} a_1b_3 + \frac{1}{54} b_2b_3 - \frac{1}{324} b_3^2 \right] I_1(m_K,0,\mu) \nonumber \\
& & \mbox{} + \left[ \frac29 a_1^2 + \frac29 a_1c_3 + \frac{1}{18} c_3^2 \right] I_1(m_\pi,\Delta,\mu) + \left[ \frac19 a_1^2 + \frac19 a_1c_3 + \frac{1}{36} c_3^2 \right] I_1(m_K,\Delta,\mu), \label{eq:mun}
\end{eqnarray}

\begin{eqnarray}
\delta \mu_p^{\mathrm{(loop\, 1)}} & = & \left[ -\frac{25}{36} a_1^2 - \frac{5}{18} a_1b_2 - \frac{1}{36} b_2^2 - \frac{25}{54} a_1b_3 - \frac{5}{54} b_2b_3 - \frac{25}{324} b_3^2 \right] I_1(m_\pi,0,\mu) \nonumber \\
& & \mbox{} + \left[ -\frac{7}{18} a_1^2 - \frac29 a_1b_2 - \frac{1}{18} b_2^2 - \frac{7}{27} a_1b_3 - \frac{2}{27} b_2b_3 - \frac{7}{162} b_3^2 \right] I_1(m_K,0,\mu) \nonumber \\
& & \mbox{} + \left[ -\frac29 a_1^2 - \frac29 a_1c_3 - \frac{1}{18} c_3^2 \right] I_1(m_\pi,\Delta,\mu) + \left[ \frac{1}{18} a_1^2 + \frac{1}{18} a_1c_3 + \frac{1}{72} c_3^2 \right] I_1(m_K,\Delta,\mu),
\end{eqnarray}

\begin{eqnarray}
\delta \mu_{\Sigma^-}^{\mathrm{(loop\, 1)}} & = & \left[ \frac{7}{18} a_1^2 + \frac29 a_1b_2 + \frac{1}{18} b_2^2 + \frac{7}{27} a_1b_3 + \frac{2}{27} b_2b_3 + \frac{7}{162} b_3^2 \right] I_1(m_\pi,0,\mu) \nonumber \\
& & \mbox{} + \left[ \frac{1}{36} a_1^2 - \frac{1}{18} a_1b_2 + \frac{1}{36} b_2^2 + \frac{1}{54} a_1b_3 - \frac{1}{54} b_2b_3 + \frac{1}{324} b_3^2 \right] I_1(m_K,0,\mu) \nonumber \\
& & \mbox{} + \left[ -\frac{1}{18} a_1^2 - \frac{1}{18} a_1c_3 - \frac{1}{72} c_3^2 \right] I_1(m_\pi,\Delta,\mu) + \left[ -\frac19 a_1^2 - \frac19 a_1c_3 - \frac{1}{36} c_3^2 \right] I_1(m_K,\Delta,\mu),
\end{eqnarray}

\begin{equation}
\delta \mu_{\Sigma^0}^{\mathrm{(loop\, 1)}} = \left[ -\frac13 a_1^2 - \frac16 a_1b_2 - \frac29 a_1b_3 - \frac{1}{18} b_2b_3 - \frac{1}{27} b_3^2 \right] I_1(m_K,0,\mu) + \left[ -\frac16 a_1^2 - \frac16 a_1c_3 - \frac{1}{24} c_3^2 \right] I_1(m_K,\Delta,\mu),
\end{equation}

\begin{eqnarray}
\delta \mu_{\Sigma^+}^{\mathrm{(loop\, 1)}} & = & \left[ -\frac{7}{18} a_1^2 - \frac29 a_1b_2 - \frac{1}{18} b_2^2 - \frac{7}{27} a_1b_3 - \frac{2}{27} b_2b_3 - \frac{7}{162} b_3^2 \right] I_1(m_\pi,0,\mu) \nonumber \\
& & \mbox{} + \left[ -\frac{25}{36} a_1^2 - \frac{5}{18} a_1b_2 - \frac{1}{36} b_2^2 - \frac{25}{54} a_1b_3 - \frac{5}{54} b_2b_3 - \frac{25}{324} b_3^2 \right] I_1(m_K,0,\mu) \nonumber \\
& & \mbox{} + \left[ \frac{1}{18} a_1^2 + \frac{1}{18} a_1c_3 + \frac{1}{72} c_3^2 \right] I_1(m_\pi,\Delta,\mu) + \left[ -\frac29 a_1^2 - \frac29 a_1c_3 - \frac{1}{18} c_3^2 \right] I_1(m_K,\Delta,\mu),
\end{eqnarray}

\begin{eqnarray}
\delta \mu_{\Xi^-}^{\mathrm{(loop\, 1)}} & = & \left[ \frac{1}{36} a_1^2 - \frac{1}{18} a_1b_2 + \frac{1}{36} b_2^2 + \frac{1}{54} a_1b_3 - \frac{1}{54} b_2b_3 + \frac{1}{324} b_3^2 \right] I_1(m_\pi,0,\mu) \nonumber \\
& & \mbox{} + \left[ \frac{7}{18} a_1^2 + \frac29 a_1b_2 + \frac{1}{18} b_2^2 + \frac{7}{27} a_1b_3 + \frac{2}{27} b_2b_3 + \frac{7}{162} b_3^2 \right] I_1(m_K,0,\mu) \nonumber \\
& & \mbox{} + \left[ -\frac19 a_1^2 - \frac19 a_1c_3 - \frac{1}{36} c_3^2 \right] I_1(m_\pi,\Delta,\mu) + \left[ -\frac{1}{18} a_1^2 - \frac{1}{18} a_1c_3 - \frac{1}{72} c_3^2 \right] I_1(m_K,\Delta,\mu),
\end{eqnarray}

\begin{eqnarray}
\delta \mu_{\Xi^0}^{\mathrm{(loop\, 1)}} & = & \left[ - \frac{1}{36} a_1^2 + \frac{1}{18} a_1b_2 - \frac{1}{36} b_2^2 - \frac{1}{54} a_1b_3 + \frac{1}{54} b_2b_3 - \frac{1}{324} b_3^2 \right] I_1(m_\pi,0,\mu) \nonumber \\
& & \mbox{} + \left[ \frac{25}{36} a_1^2 + \frac{5}{18} a_1b_2 + \frac{1}{36} b_2^2 + \frac{25}{54} a_1b_3 + \frac{5}{54} b_2b_3 + \frac{25}{324} b_3^2 \right] I_1(m_K,0,\mu) \nonumber \\ 
& & \mbox{} + \left[ \frac19 a_1^2 + \frac19 a_1c_3 + \frac{1}{36} c_3^2 \right] I_1(m_\pi,\Delta,\mu) + \left[ \frac29 a_1^2 + \frac29 a_1c_3 + \frac{1}{18} c_3^2 \right] I_1(m_K,\Delta,\mu),
\end{eqnarray}

\begin{equation}
\delta \mu_{\Lambda}^{\mathrm{(loop\, 1)}} = \left[ \frac13 a_1^2 + \frac16 a_1b_2 + \frac29 a_1b_3 + \frac{1}{18} b_2b_3 + \frac{1}{27} b_3^2 \right] I_1(m_K,0,\mu) + \left[ \frac16 a_1^2 + \frac16 a_1c_3 + \frac{1}{24} c_3^2 \right] I_1(m_K,\Delta,\mu),
\end{equation}
for decuplet baryons, by
\begin{eqnarray}
\delta \mu_{\Delta^{++}}^{\mathrm{(loop\, 1)}} & = & \left[ -\frac14 a_1^2 - \frac12 a_1b_2 - \frac14 b_2^2 - \frac56 a_1b_3 - \frac56 b_2b_3 - \frac{25}{36} b_3^2 \right] I_1(m_\pi,0,\mu) \nonumber \\
& & \mbox{} + \left[ -\frac14 a_1^2 - \frac12 a_1b_2 - \frac14 b_2^2 - \frac56 a_1b_3 - \frac56 b_2b_3 - \frac{25}{36} b_3^2 \right] I_1(m_K,0,\mu) \nonumber \\
& & \mbox{} + \left[ -\frac12 a_1^2 - \frac12 a_1c_3 - \frac18 c_3^2 \right] I_1(m_\pi,-\Delta,\mu) + \left[ -\frac12 a_1^2 - \frac12 a_1c_3 - \frac18 c_3^2 \right] I_1(m_K,-\Delta,\mu),
\end{eqnarray}

\begin{eqnarray}
\delta \mu_{\Delta^+}^{\mathrm{(loop\, 1)}} & = & \left[ -\frac{1}{12} a_1^2 - \frac16 a_1b_2 - \frac{1}{12} b_2^2 - \frac{5}{18} a_1b_3 - \frac{5}{18} b_2b_3 - \frac{25}{108} b_3^2 \right] I_1(m_\pi,0,\mu) \nonumber \\
& & \mbox{} + \left[ -\frac16 a_1^2 - \frac13 a_1b_2 - \frac16 b_2^2 - \frac59 a_1b_3 - \frac59 b_2b_3 - \frac{25}{54} b_3^2 \right] I_1(m_K,0,\mu) \nonumber \\
& & \mbox{} + \left[ -\frac16 a_1^2 - \frac16 a_1c_3 - \frac{1}{24} c_3^2 \right] I_1(m_\pi,-\Delta,\mu) + \left[ -\frac13 a_1^2 - \frac13 a_1c_3 - \frac{1}{12} c_3^2 \right] I_1(m_K,-\Delta,\mu),
\end{eqnarray}

\begin{eqnarray}
\delta \mu_{\Delta^0}^{\mathrm{(loop\, 1)}} & = & \left[ \frac{1}{12} a_1^2 + \frac16 a_1b_2 + \frac{1}{12} b_2^2 + \frac{5}{18} a_1b_3 + \frac{5}{18} b_2b_3 + \frac{25}{108} b_3^2 \right] I_1(m_\pi,0,\mu) \nonumber \\
& & \mbox{} + \left[ -\frac{1}{12} a_1^2 - \frac16 a_1b_2 - \frac{1}{12} b_2^2 - \frac{5}{18} a_1b_3 - \frac{5}{18} b_2b_3 - \frac{25}{108} b_3^2 \right] I_1(m_K,0,\mu) \nonumber \\
& & \mbox{} + \left[ \frac16 a_1^2 + \frac16 a_1c_3 + \frac{1}{24} c_3^2 \right] I_1(m_\pi,-\Delta,\mu) + \left[ -\frac16 a_1^2 - \frac16 a_1c_3 - \frac{1}{24} c_3^2 \right] I_1(m_K,-\Delta,\mu),
\end{eqnarray}

\begin{equation}
\delta \mu_{\Delta^-}^{\mathrm{(loop\, 1)}} = \left[ \frac14 a_1^2 + \frac12 a_1b_2 + \frac14 b_2^2 + \frac56 a_1b_3 + \frac56 b_2b_3 + \frac{25}{36} b_3^2 \right] I_1(m_\pi,0,\mu) + \left[ \frac12 a_1^2 + \frac12 a_1c_3 + \frac18 c_3^2 \right] I_1(m_\pi,-\Delta,\mu),
\end{equation}

\begin{eqnarray}
\delta \mu_{{\Sigma^*}^+}^{\mathrm{(loop\, 1)}} & = & \left[ -\frac16 a_1^2 - \frac13 a_1b_2 - \frac16 b_2^2 - \frac59 a_1b_3 - \frac59 b_2b_3 - \frac{25}{54} b_3^2 \right] I_1(m_\pi,0,\mu) \nonumber \\
& & \mbox{} + \left[ -\frac{1}{12} a_1^2 - \frac16 a_1b_2 - \frac{1}{12} b_2^2 - \frac{5}{18} a_1b_3 - \frac{5}{18} b_2b_3 - \frac{25}{108} b_3^2 \right] I_1(m_K,0,\mu) \nonumber \\
& & \mbox{} + \left[ -\frac13 a_1^2 - \frac13 a_1c_3 - \frac{1}{12} c_3^2 \right] I_1(m_\pi,-\Delta,\mu) + \left[ -\frac16 a_1^2 - \frac16 a_1c_3 - \frac{1}{24} c_3^2 \right] I_1(m_K,-\Delta,\mu),
\end{eqnarray}

\begin{equation}
\delta \mu_{{\Sigma^*}^0}^{\mathrm{(loop\, 1)}} = 0,
\end{equation}

\begin{eqnarray}
\delta \mu_{{\Sigma^*}^-}^{\mathrm{(loop\, 1)}} & = & \left[ \frac16 a_1^2 + \frac13 a_1b_2 + \frac16 b_2^2 + \frac59 a_1b_3 + \frac59 b_2b_3 + \frac{25}{54} b_3^2 \right] I_1(m_\pi,0,\mu) \nonumber \\
& & \mbox{} + \left[ \frac{1}{12} a_1^2 + \frac16 a_1b_2 + \frac{1}{12} b_2^2 + \frac{5}{18} a_1b_3 + \frac{5}{18} b_2b_3 + \frac{25}{108} b_3^2 \right] I_1(m_K,0,\mu) \nonumber \\
& & \mbox{} + \left[ \frac13 a_1^2 + \frac13 a_1c_3 + \frac{1}{12} c_3^2 \right] I_1(m_\pi,-\Delta,\mu) + \left[ \frac16 a_1^2 + \frac16 a_1c_3 + \frac{1}{24} c_3^2 \right] I_1(m_K,-\Delta,\mu),
\end{eqnarray}

\begin{eqnarray}
\delta \mu_{{\Xi^*}^0}^{\mathrm{(loop\, 1)}} & = & \left[ - \frac{1}{12} a_1^2 - \frac16 a_1b_2 - \frac{1}{12} b_2^2 - \frac{5}{18} a_1b_3 - \frac{5}{18} b_2b_3 - \frac{25}{108} b_3^2 \right] I_1(m_\pi,0,\mu) \nonumber \\
& & \mbox{} + \left[ \frac{1}{12} a_1^2 + \frac16 a_1b_2 + \frac{1}{12} b_2^2 + \frac{5}{18} a_1b_3 + \frac{5}{18} b_2b_3 + \frac{25}{108} b_3^2 \right] I_1(m_K,0,\mu) \nonumber \\
& & \mbox{} + \left[ -\frac16 a_1^2 - \frac16 a_1c_3 - \frac{1}{24} c_3^2 \right] I_1(m_\pi,-\Delta,\mu) + \left[ \frac16 a_1^2 + \frac16 a_1c_3 + \frac{1}{24} c_3^2 \right] I_1(m_K,-\Delta,\mu),
\end{eqnarray}

\begin{eqnarray}
\delta \mu_{{\Xi^*}^-}^{\mathrm{(loop\, 1)}} & = & \left[ \frac{1}{12} a_1^2 + \frac16 a_1b_2 + \frac{1}{12} b_2^2 + \frac{5}{18} a_1b_3 + \frac{5}{18} b_2b_3 + \frac{25}{108} b_3^2 \right] I_1(m_\pi,0,\mu) \nonumber \\
& & \mbox{} + \left[ \frac16 a_1^2 + \frac13 a_1b_2 + \frac16 b_2^2 + \frac59 a_1b_3 + \frac59 b_2b_3 + \frac{25}{54} b_3^2 \right] I_1(m_K,0,\mu) \nonumber \\
& & \mbox{} + \left[ \frac16 a_1^2 + \frac16 a_1c_3 + \frac{1}{24} c_3^2 \right] I_1(m_\pi,-\Delta,\mu) + \left[ \frac13 a_1^2 + \frac13 a_1c_3 + \frac{1}{12} c_3^2 \right] I_1(m_K,-\Delta,\mu),
\end{eqnarray}

\begin{equation}
\delta \mu_{\Omega^-}^{\mathrm{(loop\, 1)}} = \left[ \frac14 a_1^2 + \frac12 a_1b_2 + \frac14 b_2^2 + \frac56 a_1b_3 + \frac56 b_2b_3 + \frac{25}{36} b_3^2 \right] I_1(m_K,0,\mu) + \left[ \frac12 a_1^2 + \frac12 a_1c_3 + \frac18 c_3^2 \right] I_1(m_K,-\Delta,\mu),
\end{equation}
and for octet-octet and decuplet-octet transitions, by
\begin{eqnarray}
\sqrt{3} \delta \mu_{\Sigma^0\Lambda}^{\mathrm{(loop\, 1)}} & = & \left[ -\frac23 a_1^2 - \frac13 a_1b_2 - \frac49 a_1b_3 - \frac19 b_2b_3 - \frac{2}{27} b_3^2 \right] I_1(m_\pi,0,\mu) \nonumber \\
& & \mbox{} + \left[ -\frac13 a_1^2 - \frac16 a_1b_2 - \frac29 a_1b_3 - \frac{1}{18} b_2b_3 - \frac{1}{27} b_3^2 \right] I_1(m_K,0,\mu) \nonumber \\
& & \mbox{} + \left[ -\frac13 a_1^2 - \frac13 a_1c_3 - \frac{1}{12} c_3^2 \right] I_1(m_\pi,\Delta,\mu) + \left[ -\frac16 a_1^2 - \frac16 a_1c_3 - \frac{1}{24} c_3^2 \right] I_1(m_K,\Delta,\mu),
\end{eqnarray}

\begin{eqnarray}
\sqrt{2} \delta \mu_{\Delta^+p}^{\mathrm{(loop\, 1)}} & = & \left[ - \frac{5}{18} a_1^2 - \frac{1}{18} a_1b_2 - \frac{5}{54} a_1b_3 - \frac{5}{36} a_1c_3 - \frac{1}{36} b_2c_3 - \frac{5}{108} b_3c_3 \right] I_1(m_\pi,0,\mu) \nonumber \\
& & \mbox{} + \left[ -\frac{1}{18} a_1^2 + \frac{1}{18} a_1b_2 - \frac{1}{54} a_1b_3 - \frac{1}{36} a_1c_3 + \frac{1}{36} b_2c_3
 - \frac{1}{108} b_3c_3 \right] I_1(m_K,0,\mu) \nonumber \\
& & \mbox{} + \left[ -\frac{25}{18} a_1^2 - \frac{25}{18} a_1b_2 - \frac{125}{54} a_1b_3 - \frac{25}{36} a_1c_3 - \frac{25}{36} b_2c_3 - \frac{125}{108} b_3c_3 \right] I_1(m_\pi,\Delta,\mu) \nonumber \\
& & \mbox{} + \left[ -\frac{5}{18} a_1^2 - \frac{5}{18} a_1b_2 - \frac{25}{54} a_1b_3 - \frac{5}{36} a_1c_3 - \frac{5}{36} b_2c_3 - \frac{25}{108} b_3c_3 \right] I_1(m_K,\Delta,\mu),
\end{eqnarray}

\begin{eqnarray}
\sqrt{2} \delta \mu_{\Delta^0n}^{\mathrm{(loop\, 1)}} & = & \left[ -\frac{5}{18} a_1^2 - \frac{1}{18} a_1b_2 - \frac{5}{54} a_1b_3 - \frac{5}{36} a_1c_3 - \frac{1}{36} b_2c_3 - \frac{5}{108} b_3c_3 \right] I_1(m_\pi,0,\mu) \nonumber \\
& & \mbox{} + \left[ -\frac{1}{18} a_1^2 + \frac{1}{18} a_1b_2 - \frac{1}{54} a_1b_3 - \frac{1}{36} a_1c_3 + \frac{1}{36} b_2c_3 - \frac{1}{108} b_3c_3 \right] I_1(m_K,0,\mu) \nonumber \\
& & \mbox{} + \left[ -\frac{25}{18} a_1^2 - \frac{25}{18} a_1b_2 - \frac{125}{54} a_1b_3 - \frac{25}{36} a_1c_3 - \frac{25}{36} b_2c_3 - \frac{125}{108} b_3c_3 \right] I_1(m_\pi,\Delta,\mu) \nonumber \\
& & \mbox{} + \left[ -\frac{5}{18} a_1^2 - \frac{5}{18} a_1b_2 - \frac{25}{54} a_1b_3 - \frac{5}{36} a_1c_3 - \frac{5}{36} b_2c_3 - \frac{25}{108} b_3c_3 \right] I_1(m_K,\Delta,\mu),
\end{eqnarray}

\begin{eqnarray}
\sqrt{6} \delta \mu_{{\Sigma^*}^0\Lambda}^{\mathrm{(loop\, 1)}} & = & \left[ -\frac13 a_1^2 - \frac19 a_1b_3 - \frac16 a_1c_3 - \frac{1}{18} b_3c_3 \right] I_1(m_\pi,0,\mu) \nonumber \\
& & \mbox{} + \left[ -\frac16 a_1^2 - \frac{1}{18} a_1b_3 - \frac{1}{12} a_1c_3 - \frac{1}{36} b_3c_3 \right] I_1(m_K,0,\mu) \nonumber \\
& & \mbox{} + \left[ -\frac53 a_1^2 - \frac53 a_1b_2 - \frac{25}{9} a_1b_3 - \frac56 a_1c_3 - \frac56 b_2c_3 - \frac{25}{18} b_3c_3 \right] I_1(m_\pi,\Delta,\mu) \nonumber \\
& & \mbox{} + \left[ -\frac56 a_1^2 - \frac56 a_1b_2 - \frac{25}{18} a_1b_3 - \frac{5}{12} a_1c_3 - \frac{5}{12} b_2c_3 - \frac{25}{36} b_3c_3 \right] I_1(m_K,\Delta,\mu),
\end{eqnarray}

\begin{eqnarray}
\sqrt{2} \delta \mu_{{\Sigma^*}^0\Sigma^0}^{\mathrm{(loop\, 1)}} & = & \left[ -\frac16 a_1^2 - \frac{1}{18} a_1b_3 - \frac{1}{12} a_1c_3 - \frac{1}{36} b_3c_3 \right] I_1(m_K,0,\mu) \nonumber \\
& & \mbox{} + \left[ -\frac56 a_1^2 - \frac56 a_1b_2 - \frac{25}{18} a_1b_3 - \frac{5}{12} a_1c_3 - \frac{5}{12} b_2c_3 - \frac{25}{36} b_3c_3 \right] I_1(m_K,\Delta,\mu),
\end{eqnarray}

\begin{eqnarray}
\sqrt{2} \delta \mu_{{\Sigma^*}^+\Sigma^+}^{\mathrm{(loop\, 1)}} & = & \left[ - \frac{1}{18} a_1^2 + \frac{1}{18} a_1b_2 - \frac{1}{54} a_1b_3 - \frac{1}{36} a_1c_3 + \frac{1}{36} b_2c_3 - \frac{1}{108} b_3c_3 \right] I_1(m_\pi,0,\mu) \nonumber \\
& & \mbox{} + \left[ -\frac{5}{18} a_1^2 - \frac{1}{18} a_1b_2 - \frac{5}{54} a_1b_3 - \frac{5}{36} a_1c_3 - \frac{1}{36} b_2c_3 - \frac{5}{108} b_3c_3 \right] I_1(m_K,0,\mu) \nonumber \\
& & \mbox{} + \left[ -\frac{5}{18} a_1^2 - \frac{5}{18} a_1b_2 - \frac{25}{54} a_1b_3 - \frac{5}{36} a_1c_3 - \frac{5}{36} b_2c_3 - \frac{25}{108} b_3c_3 \right] I_1(m_\pi,\Delta,\mu) \nonumber \\
& & \mbox{} + \left[ -\frac{25}{18} a_1^2 - \frac{25}{18} a_1b_2 - \frac{125}{54} a_1b_3 - \frac{25}{36} a_1c_3 - \frac{25}{36} b_2c_3 - \frac{125}{108} b_3c_3 \right] I_1(m_K,\Delta,\mu),
\end{eqnarray}

\begin{eqnarray}
\sqrt{2} \delta \mu_{{\Sigma^*}^-\Sigma^-}^{\mathrm{(loop\, 1)}} & = & \left[ \frac{1}{18} a_1^2 - \frac{1}{18} a_1b_2 + \frac{1}{54} a_1b_3 + \frac{1}{36} a_1c_3 - \frac{1}{36} b_2c_3 + \frac{1}{108} b_3c_3 \right] I_1(m_\pi,0,\mu) \nonumber \\
& & \mbox{} + \left[ -\frac{1}{18} a_1^2 + \frac{1}{18} a_1b_2 - \frac{1}{54} a_1b_3 - \frac{1}{36} a_1c_3 + \frac{1}{36} b_2c_3 - \frac{1}{108} b_3c_3 \right] I_1(m_K,0,\mu) \nonumber \\
& & \mbox{} + \left[ \frac{5}{18} a_1^2 + \frac{5}{18} a_1b_2 + \frac{25}{54} a_1b_3 + \frac{5}{36} a_1c_3 + \frac{5}{36} b_2c_3 + \frac{25}{108} b_3c_3 \right] I_1(m_\pi,\Delta,\mu) \nonumber \\
& & \mbox{} + \left[ -\frac{5}{18} a_1^2 - \frac{5}{18} a_1b_2 - \frac{25}{54} a_1b_3 - \frac{5}{36} a_1c_3 - \frac{5}{36} b_2c_3 - \frac{25}{108} b_3c_3 \right] I_1(m_K,\Delta,\mu),
\end{eqnarray}

\begin{eqnarray}
\sqrt{2} \delta \mu_{{\Xi^*}^0\Xi^0}^{\mathrm{(loop\, 1)}} & = & \left[ -\frac{1}{18} a_1^2 + \frac{1}{18} a_1b_2 - \frac{1}{54} a_1b_3 - \frac{1}{36} a_1c_3 + \frac{1}{36} b_2c_3 - \frac{1}{108} b_3c_3 \right] I_1(m_\pi,0,\mu) \nonumber \\
& & \mbox{} + \left[ -\frac{5}{18} a_1^2 - \frac{1}{18} a_1b_2 - \frac{5}{54} a_1b_3 - \frac{5}{36} a_1c_3 - \frac{1}{36} b_2c_3 - \frac{5}{108} b_3c_3 \right] I_1(m_K,0,\mu) \nonumber \\
& & \mbox{} + \left[ -\frac{5}{18} a_1^2 - \frac{5}{18} a_1b_2 - \frac{25}{54} a_1b_3 - \frac{5}{36} a_1c_3 - \frac{5}{36} b_2c_3 - \frac{25}{108} b_3c_3 \right] I_1(m_\pi,\Delta,\mu) \nonumber \\
& & \mbox{} + \left[ -\frac{25}{18} a_1^2 - \frac{25}{18} a_1b_2 - \frac{125}{54} a_1b_3 - \frac{25}{36} a_1c_3 - \frac{25}{36} b_2c_3 - \frac{125}{108} b_3c_3 \right] I_1(m_K,\Delta,\mu),
\end{eqnarray}

\begin{eqnarray}
\sqrt{2} \delta \mu_{{\Xi^*}^-\Xi^-}^{\mathrm{(loop\, 1)}} & = & \left[ \frac{1}{18} a_1^2 - \frac{1}{18} a_1b_2 + \frac{1}{54} a_1b_3 + \frac{1}{36} a_1c_3 - \frac{1}{36} b_2c_3 + \frac{1}{108} b_3c_3 \right] I_1(m_\pi,0,\mu) \nonumber \\
& & \mbox{} + \left[ -\frac{1}{18} a_1^2 + \frac{1}{18} a_1b_2 - \frac{1}{54} a_1b_3 - \frac{1}{36} a_1c_3 + \frac{1}{36} b_2c_3 - \frac{1}{108} b_3c_3 \right] I_1(m_K,0,\mu) \nonumber \\
& & \mbox{} + \left[ \frac{5}{18} a_1^2 + \frac{5}{18} a_1b_2 + \frac{25}{54} a_1b_3 + \frac{5}{36} a_1c_3 + \frac{5}{36} b_2c_3 + \frac{25}{108} b_3c_3 \right] I_1(m_\pi,\Delta,\mu) \nonumber \\
& & \mbox{} + \left[ -\frac{5}{18} a_1^2 - \frac{5}{18} a_1b_2 - \frac{25}{54} a_1b_3 - \frac{5}{36} a_1c_3 - \frac{5}{36} b_2c_3 - \frac{25}{108} b_3c_3 \right] I_1(m_K,\Delta,\mu). \label{eq:muxixi}
\end{eqnarray}

Using the inverse relations (\ref{eq:rel1inv}), $\delta \mu_i^{\mathrm{(loop\, 1)}}$ expressions can be rewritten, for octet baryons, as
\begin{equation}
\delta \mu_n^{\mathrm{(loop\, 1)}} = (D+F)^2 I_1(m_\pi,0,\mu) - (D-F)^2 I_1(m_K,0,\mu) + \frac29 \mathcal{C}^2 I_1(m_\pi,\Delta,\mu) + \frac19 \mathcal{C}^2 I_1(m_K,\Delta,\mu) \label{eq:munch},
\end{equation}

\begin{equation}
\delta \mu_p^{\mathrm{(loop\, 1)}} = - (D+F)^2 I_1(m_\pi,0,\mu) - \frac23 (D^2+3F^2) I_1(m_K,0,\mu) - \frac29 \mathcal{C}^2 I_1(m_\pi,\Delta,\mu) + \frac{1}{18} \mathcal{C}^2 I_1(m_K,\Delta,\mu),
\end{equation}

\begin{equation}
\delta \mu_{\Sigma^-}^{\mathrm{(loop\, 1)}} = \frac23(D^2+3F^2) I_1(m_\pi,0,\mu) + (D-F)^2 I_1(m_K,0,\mu) - \frac{1}{18} \mathcal{C}^2 I_1(m_\pi,\Delta,\mu) - \frac19 \mathcal{C}^2 I_1(m_K,\Delta,\mu),
\end{equation}

\begin{equation}
\delta \mu_{\Sigma^0}^{\mathrm{(loop\, 1)}} = - 2 DF I_1(m_K,0,\mu) - \frac16 \mathcal{C}^2 I_1(m_K,\Delta,\mu),
\end{equation}

\begin{equation}
\delta \mu_{\Sigma^+}^{\mathrm{(loop\, 1)}} = - \frac23 (D^2+3F^2) I_1(m_\pi,0,\mu) - (D+F)^2 I_1(m_K,0,\mu) + \frac{1}{18} \mathcal{C}^2 I_1(m_\pi,\Delta,\mu) - \frac29 \mathcal{C}^2 I_1(m_K,\Delta,\mu),
\end{equation}

\begin{equation}
\delta \mu_{\Xi^-}^{\mathrm{(loop\, 1)}} = (D-F)^2 I_1(m_\pi,0,\mu) + \frac23 (D^2+3F^2) I_1(m_K,0,\mu) - \frac19 \mathcal{C}^2 I_1(m_\pi,\Delta,\mu) - \frac{1}{18} \mathcal{C}^2 I_1(m_K,\Delta,\mu),
\end{equation}

\begin{equation}
\delta \mu_{\Xi^0}^{\mathrm{(loop\, 1)}} = - (D-F)^2 I_1(m_\pi,0,\mu) + (D+F)^2 I_1(m_K,0,\mu) + \frac19 \mathcal{C}^2 I_1(m_\pi,\Delta,\mu) + \frac29 \mathcal{C}^2 I_1(m_K,\Delta,\mu),
\end{equation}

\begin{equation}
\delta \mu_{\Lambda}^{\mathrm{(loop\, 1)}} = 2 DF I_1(m_K,0,\mu) + \frac16 \mathcal{C}^2 I_1(m_K,\Delta,\mu), \label{eq:mul}
\end{equation}
for decuplet baryons, as
\begin{equation}
\delta \mu_{\Delta^{++}}^{\mathrm{(loop\, 1)}} = - \frac19 \mathcal{H}^2 I_1(m_\pi,0,\mu) - \frac19 \mathcal{H}^2 I_1(m_K,0,\mu) - \frac12 \mathcal{C}^2 I_1(m_\pi,-\Delta,\mu) - \frac12 \mathcal{C}^2 I_1(m_K,-\Delta,\mu),
\end{equation}

\begin{equation}
\delta \mu_{\Delta^+}^{\mathrm{(loop\, 1)}} = - \frac{1}{27} \mathcal{H}^2 I_1(m_\pi,0,\mu) - \frac{2}{27} \mathcal{H}^2 I_1(m_K,0,\mu) - \frac16 \mathcal{C}^2 I_1(m_\pi,-\Delta,\mu) - \frac13 \mathcal{C}^2 I_1(m_K,-\Delta,\mu),
\end{equation}

\begin{equation}
\delta \mu_{\Delta^0}^{\mathrm{(loop\, 1)}} = \frac{1}{27} \mathcal{H}^2 I_1(m_\pi,0,\mu) - \frac{1}{27} \mathcal{H}^2 I_1(m_K,0,\mu) + \frac16 \mathcal{C}^2 I_1(m_\pi,-\Delta,\mu) - \frac16 \mathcal{C}^2 I_1(m_K,-\Delta,\mu),
\end{equation}

\begin{equation}
\delta \mu_{\Delta^-}^{\mathrm{(loop\, 1)}} = \frac19 \mathcal{H}^2 I_1(m_\pi,0,\mu) + \frac12 \mathcal{C}^2 I_1(m_\pi,-\Delta,\mu),
\end{equation}

\begin{equation}
\delta \mu_{{\Sigma^*}^+}^{\mathrm{(loop\, 1)}} = - \frac{2}{27} \mathcal{H}^2 I_1(m_\pi,0,\mu) - \frac{1}{27}\mathcal{H}^2 I_1(m_K,0,\mu) - \frac13 \mathcal{C}^2 I_1(m_\pi,-\Delta,\mu) - \frac16 \mathcal{C}^2 I_1(m_K,-\Delta,\mu),
\end{equation}

\begin{equation}
\delta \mu_{{\Sigma^*}^0}^{\mathrm{(loop\, 1)}} = 0,
\end{equation}

\begin{equation}
\delta \mu_{{\Sigma^*}^-}^{\mathrm{(loop\, 1)}} = \frac{2}{27} \mathcal{H}^2 I_1(m_\pi,0,\mu) + \frac{1}{27} \mathcal{H}^2 I_1(m_K,0,\mu) + \frac13 \mathcal{C}^2 I_1(m_\pi,-\Delta,\mu) + \frac16 \mathcal{C}^2 I_1(m_K,-\Delta,\mu),
\end{equation}

\begin{equation}
\delta \mu_{{\Xi^*}^0}^{\mathrm{(loop\, 1)}} = - \frac{1}{27} \mathcal{H}^2 I_1(m_\pi,0,\mu) + \frac{1}{27} \mathcal{H}^2 I_1(m_K,0,\mu) - \frac16 \mathcal{C}^2 I_1(m_\pi,-\Delta,\mu) + \frac16 \mathcal{C}^2 I_1(m_K,-\Delta,\mu),
\end{equation}

\begin{equation}
\delta \mu_{{\Xi^*}^-}^{\mathrm{(loop\, 1)}} = \frac{1}{27} \mathcal{H}^2 I_1(m_\pi,0,\mu) + \frac{2}{27} \mathcal{H}^2 I_1(m_K,0,\mu) + \frac16 \mathcal{C}^2 I_1(m_\pi,-\Delta,\mu) + \frac13 \mathcal{C}^2 I_1(m_K,-\Delta,\mu),
\end{equation}

\begin{equation}
\delta \mu_{\Omega^-}^{\mathrm{(loop\, 1)}} = 
\frac19 \mathcal{H}^2 I_1(m_K,0,\mu) + \frac12 \mathcal{C}^2 I_1(m_K,-\Delta,\mu),
\end{equation}
and for octet-octet and decuplet-octet transitions, as
\begin{equation}
\sqrt{3} \delta \mu_{\Sigma^0\Lambda}^{\mathrm{(loop\, 1)}} = - 4 DF I_1(m_\pi,0,\mu) - 2 DF I_1(m_K,0,\mu) - \frac13 \mathcal{C}^2 I_1(m_\pi,\Delta,\mu) - \frac16 \mathcal{C}^2 I_1(m_K,\Delta,\mu), \label{eq:musl}
\end{equation}

\begin{equation}
\sqrt{2} \delta \mu_{\Delta^+p}^{\mathrm{(loop\, 1)}} = \frac13 \mathcal{C} (D+F) I_1(m_\pi,0,\mu) + \frac13 \mathcal{C} (D-F) I_1(m_K,0,\mu) - \frac{25}{27} \mathcal{C} \mathcal{H} I_1(m_\pi,\Delta,\mu) - \frac{5}{27} \mathcal{C} \mathcal{H} I_1(m_K,\Delta,\mu),
\end{equation}

\begin{equation}
\sqrt{2} \delta \mu_{\Delta^0n}^{\mathrm{(loop\, 1)}} = \frac13 \mathcal{C} (D+F) I_1(m_\pi,0,\mu) + \frac13 \mathcal{C} (D-F) I_1(m_K,0,\mu) - \frac{25}{27} \mathcal{C} \mathcal{H} I_1(m_\pi,\Delta,\mu) - \frac{5}{27} \mathcal{C} \mathcal{H} I_1(m_K,\Delta,\mu),
\end{equation}

\begin{equation}
\sqrt{6} \delta \mu_{{\Sigma^*}^0\Lambda}^{\mathrm{(loop\, 1)}} = \frac23 \mathcal{C} D I_1(m_\pi,0,\mu) + \frac13 \mathcal{C} D I_1(m_K,0,\mu) - \frac{10}{9} \mathcal{C} \mathcal{H} I_1(m_\pi,\Delta,\mu) - \frac59 \mathcal{C} \mathcal{H} I_1(m_K,\Delta,\mu),
\end{equation}

\begin{equation}
\sqrt{2} \delta \mu_{{\Sigma^*}^0\Sigma^0}^{\mathrm{(loop\, 1)}} = \frac13 \mathcal{C} D I_1(m_K,0,\mu) - \frac59 \mathcal{C} \mathcal{H} I_1(m_K,\Delta,\mu),
\end{equation}

\begin{equation}
\sqrt{2} \delta \mu_{{\Sigma^*}^+\Sigma^+}^{\mathrm{(loop\, 1)}} = \frac13 \mathcal{C} (D-F) I_1(m_\pi,0,\mu) + \frac13 \mathcal{C} (D+F) I_1(m_K,0,\mu) - \frac{5}{27} \mathcal{C} \mathcal{H} I_1(m_\pi,\Delta,\mu) - \frac{25}{27} \mathcal{C} \mathcal{H} I_1(m_K,\Delta,\mu),
\end{equation}

\begin{equation}
\sqrt{2} \delta \mu_{{\Sigma^*}^-\Sigma^-}^{\mathrm{(loop\, 1)}} = - \frac13 \mathcal{C} (D-F) I_1(m_\pi,0,\mu) + \frac13 \mathcal{C} (D-F) I_1(m_K,0,\mu) + \frac{5}{27} \mathcal{C} \mathcal{H} I_1(m_\pi,\Delta,\mu) - \frac{5}{27} \mathcal{C} \mathcal{H} I_1(m_K,\Delta,\mu),
\end{equation}

\begin{equation}
\sqrt{2} \delta \mu_{{\Xi^*}^0\Xi^0}^{\mathrm{(loop\, 1)}} = \frac13 \mathcal{C} (D-F) I_1(m_\pi,0,\mu) + \frac13 \mathcal{C} (D+F) I_1(m_K,0,\mu) - \frac{5}{27} \mathcal{C} \mathcal{H} I_1(m_\pi,\Delta,\mu) - \frac{25}{27} \mathcal{C} \mathcal{H} I_1(m_K,\Delta,\mu),
\end{equation}

\begin{equation}
\sqrt{2} \delta \mu_{{\Xi^*}^-\Xi^-}^{\mathrm{(loop\, 1)}} = - \frac13 \mathcal{C} (D-F) I_1(m_\pi,0,\mu) + \frac13 \mathcal{C} (D-F) I_1(m_K,0,\mu) + \frac{5}{27} \mathcal{C} \mathcal{H} I_1(m_\pi,\Delta,\mu) - \frac{5}{27} \mathcal{C} \mathcal{H} I_1(m_K,\Delta,\mu). \label{eq:muxixich}
\end{equation}

\section{\label{app:rloop2}Reduction of baryon operators emerging from Fig.~\ref{fig:mmloop2}}

\subsection{Flavor $\mathbf{1}$ operators}

\begin{equation}
[G^{ia},[G^{ia},G^{kc}]] = \frac{3N_f^2-4}{4N_f} G^{kc},
\end{equation}

\begin{equation}
[G^{ia},[G^{ia},\mathcal{D}_2^{kc}]] = - (N_c+N_f) G^{kc} + \frac{7N_f^2+4N_f-4}{4N_f} \mathcal{D}_2^{kc},
\end{equation}

\begin{equation}
[\mathcal{D}_2^{ia},[G^{ia},G^{kc}]] + [G^{ia},[\mathcal{D}_2^{ia},G^{kc}]] = \frac{(N_c+N_f)(N_f-2)}{N_f} G^{kc} + \frac12 (N_f+2) \mathcal{D}_2^{kc},
\end{equation}

\begin{equation}
[G^{ia},[G^{ia},\mathcal{D}_3^{kc}]] = - [N_c(N_c+2N_f)+4] G^{kc} - 4 (N_c+N_f) \mathcal{D}_2^{kc} + \frac{11N_f^2+12N_f-4}{4N_f} \mathcal{D}_3^{kc},
\end{equation}

\begin{equation}
[\mathcal{D}_3^{ia},[G^{ia},G^{kc}]] + [G^{ia},[\mathcal{D}_3^{ia},G^{kc}]] = 2 (N_f-2) G^{kc} + (N_c+N_f) \mathcal{D}_2^{kc} + \frac{N_f^2+2N_f-4}{2N_f} \mathcal{D}_3^{kc} + \frac{(N_f+4)(N_f-2)}{N_f} \mathcal{O}_3^{kc},
\end{equation}

\begin{equation}
[G^{ia},[G^{ia},\mathcal{O}_3^{kc}]] = - [N_c(N_c+2N_f)-N_f] G^{kc} + (N_c+N_f) \mathcal{D}_2^{kc} + \frac{11N_f^2+12N_f-4}{4N_f} \mathcal{O}_3^{kc},
\end{equation}

\begin{equation}
 [G^{ia},[\mathcal{O}_3^{ia},G^{kc}]] + [\mathcal{O}_3^{ia},[G^{ia},G^{kc}]] = - \frac32 (N_c+N_f) \mathcal{D}_2^{kc} + \frac12 (N_f+1) \mathcal{D}_3^{kc} + N_f \mathcal{O}_3^{kc},
\end{equation}

\begin{equation}
[\mathcal{D}_2^{ia},[G^{ia},\mathcal{D}_2^{kc}]] + [G^{ia},[\mathcal{D}_2^{ia},\mathcal{D}_2^{kc}]] = -2 N_f G^{kc}
+ \frac{2(N_c+N_f)(N_f-1)}{N_f} \mathcal{D}_2^{kc} + \frac12 N_f \mathcal{D}_3^{kc} - 2 \mathcal{O}_3^{kc},
\end{equation}

\begin{equation}
[\mathcal{D}_2^{ia},[\mathcal{D}_2^{ia},G^{kc}]] = \frac{N_c(N_c+2N_f)(N_f-2)-2 N_f^2}{2N_f} G^{kc} + \frac14 (N_f+2) \mathcal{D}_3^{kc}
+ \frac12 (N_f+4) \mathcal{O}_3^{kc},
\end{equation}

\begin{eqnarray}
& & [\mathcal{D}_2^{ia},[G^{ia},\mathcal{D}_3^{kc}]] + [G^{ia},[\mathcal{D}_2^{ia},\mathcal{D}_3^{kc}]] \nonumber \\
& & \mbox{\hglue0.2truecm} = - 4 (N_c+N_f) G^{kc} - 2 (N_f-2) \mathcal{D}_2^{kc} + \frac{(N_c+N_f)(3N_f-2)}{N_f} \mathcal{D}_3^{kc} - 2 (N_c+N_f) \mathcal{O}_3^{kc} + (N_f-2) \mathcal{D}_4^{kc},
\end{eqnarray}

\begin{equation}
[\mathcal{D}_2^{ia},[G^{ia},\mathcal{O}_3^{kc}]] + [G^{ia},[\mathcal{D}_2^{ia},\mathcal{O}_3^{kc}]] = 3 N_f \mathcal{D}_2^{kc}
- (N_c+N_f) \mathcal{D}_3^{kc} + \frac{2(N_c+N_f)(N_f-1)}{N_f} \mathcal{O}_3^{kc} + 2 \mathcal{D}_4^{kc},
\end{equation}

\begin{eqnarray}
& & [\mathcal{D}_3^{ia},[G^{ia},\mathcal{D}_2^{kc}]] + [G^{ia},[\mathcal{D}_3^{ia},\mathcal{D}_2^{kc}]] \nonumber \\
& & \mbox{\hglue0.2truecm} = -4 (N_c+N_f) G^{kc} + [N_c(N_c+2N_f)+2N_f] \mathcal{D}_2^{kc} + (N_c+N_f) \mathcal{D}_3^{kc} - 2 (N_c+N_f) \mathcal{O}_3^{kc} + \frac{N_f^2-4}{N_f} \mathcal{D}_4^{kc},
\end{eqnarray}

\begin{equation}
[G^{ia},[\mathcal{O}_3^{ia},\mathcal{D}_2^{kc}]] + [\mathcal{O}_3^{ia},[G^{ia},\mathcal{D}_2^{kc}]] = - \frac32 [N_c(N_c+2N_f)-4N_f] \mathcal{D}_2^{kc} - \frac52 (N_c+N_f) \mathcal{D}_3^{kc} - (N_c+N_f) \mathcal{O}_3^{kc} + 3 (N_f+2) \mathcal{D}_4^{kc},
\end{equation}

\begin{equation}
[\mathcal{D}_2^{ia},[\mathcal{D}_2^{ia},\mathcal{D}_2^{kc}]] = \frac{N_c(N_c+2N_f)(N_f-2)-2N_f^2}{2N_f} \mathcal{D}_2^{kc} + \frac12 (N_f+2) \mathcal{D}_4^{kc},
\end{equation}

\begin{eqnarray}
& & [\mathcal{D}_2^{ia},[\mathcal{D}_3^{ia},G^{kc}]] + [\mathcal{D}_3^{ia},[\mathcal{D}_2^{ia},G^{kc}]] \nonumber \\
& & \mbox{\hglue0.2truecm} = - 4 (N_c+N_f) G^{kc} - 2 (N_f-2) \mathcal{D}_2^{kc} + \frac{(N_c+N_f)(3N_f-2)}{N_f} \mathcal{D}_3^{kc} + \frac{2(N_c+N_f)(5N_f-4)}{N_f} \mathcal{O}_3^{kc} + (N_f-2) \mathcal{D}_4^{kc}, \nonumber \\
\end{eqnarray}

\begin{equation}
[\mathcal{D}_2^{ia},[\mathcal{O}_3^{ia},G^{kc}]] + [\mathcal{O}_3^{ia},[\mathcal{D}_2^{ia},G^{kc}]] = 3N_f \mathcal{D}_2^{kc} - (N_c+N_f) \mathcal{D}_3^{kc} - (N_c+N_f) \mathcal{O}_3^{kc} + 2 \mathcal{D}_4^{kc},
\end{equation}

\begin{eqnarray}
& & [\mathcal{D}_3^{ia},[G^{ia},\mathcal{D}_3^{kc}]] + [G^{ia},[\mathcal{D}_3^{ia},\mathcal{D}_3^{kc}]] \nonumber \\
& & \mbox{\hglue0.2truecm} = - 4 [N_c(N_c+2N_f)+2N_f] G^{kc} + 4 (N_c+N_f) \mathcal{D}_2^{kc} + 2 [N_c(N_c+2N_f)+2N_f-2] \mathcal{D}_3^{kc} \nonumber \\
& & \mbox{\hglue0.6truecm} - 2 [N_c(N_c+2N_f)-2N_f+8] \mathcal{O}_3^{kc} - 2 (N_c+N_f) \mathcal{D}_4^{kc} + \frac{N_f^2+2N_f-4}{N_f} \mathcal{D}_5^{kc},
\end{eqnarray}

\begin{eqnarray}
& & [\mathcal{D}_3^{ia},[G^{ia},\mathcal{O}_3^{kc}]] + [G^{ia},[\mathcal{D}_3^{ia},\mathcal{O}_3^{kc}]] \nonumber \\
& & \mbox{\hglue0.2truecm} = - [N_c(N_c+2N_f)-N_f] \mathcal{D}_3^{kc} + [N_c(N_c+2N_f)+2N_f] \mathcal{O}_3^{kc} + 2 (N_c+N_f) \mathcal{D}_4^{kc} + \frac{(N_f+4)(N_f-2)}{N_f} \mathcal{O}_5^{kc}, \nonumber \\
\end{eqnarray}

\begin{eqnarray}
& & [G^{ia},[\mathcal{O}_3^{ia},\mathcal{D}_3^{kc}]] + [\mathcal{O}_3^{ia},[G^{ia},\mathcal{D}_3^{kc}]] \nonumber \\
& & \mbox{\hglue0.2truecm} = - 24 (N_c+N_f) \mathcal{D}_2^{kc} - [4N_c(N_c+2N_f)-13N_f] \mathcal{D}_3^{kc} - [N_c(N_c+2N_f)+4] \mathcal{O}_3^{kc} - 9 (N_c+N_f) \mathcal{D}_4^{kc} \nonumber \\
& & \mbox{\hglue0.6truecm} + (5N_f+11) \mathcal{D}_5^{kc},
\end{eqnarray}

\begin{eqnarray}
& & [G^{ia},[\mathcal{O}_3^{ia},\mathcal{O}_3^{kc}]] + [\mathcal{O}_3^{ia},[G^{ia},\mathcal{O}_3^{kc}]] \nonumber \\
& & \mbox{\hglue0.2truecm} = - 3 N_c (N_c+2 N_f) G^{kc} + 3 (N_c+N_f) \mathcal{D}_2^{kc} - \frac12 [N_c(N_c+2N_f)-3N_f] \mathcal{D}_3^{kc} - \frac12 [9N_c(N_c+2N_f)-34N_f-12] \mathcal{O}_3^{kc} \nonumber \\
& & \mbox{\hglue0.6truecm} + (N_c+N_f) \mathcal{D}_4^{kc} + 5 (N_f+2) \mathcal{O}_5^{kc},
\end{eqnarray}

\begin{equation}
[\mathcal{D}_2^{ia},[\mathcal{D}_2^{ia},\mathcal{D}_3^{kc}]] = \frac{N_c(N_c+2N_f)(N_f-2)-2N_f^2}{2N_f} \mathcal{D}_3^{kc}
+ \frac12 (N_f+2) \mathcal{D}_5^{kc},
\end{equation}

\begin{equation}
[\mathcal{D}_2^{ia},[\mathcal{D}_2^{ia},\mathcal{O}_3^{kc}]] = \frac{N_c(N_c+2N_f)(N_f-2) -2N_f^2}{2N_f} \mathcal{O}_3^{kc}
+ \frac12 (N_f+4) \mathcal{O}_5^{kc},
\end{equation}

\begin{equation}
[\mathcal{D}_2^{ia},[\mathcal{D}_3^{ia},\mathcal{D}_2^{kc}]] + [\mathcal{D}_3^{ia},[\mathcal{D}_2^{ia},\mathcal{D}_2^{kc}]] = -2 N_f \mathcal{D}_3^{kc} + \frac{4(N_c+N_f)(N_f-1)}{N_f} \mathcal{D}_4^{kc} + N_f \mathcal{D}_5^{kc},
\end{equation}

\begin{equation}
[\mathcal{D}_2^{ia},[\mathcal{O}_3^{ia},\mathcal{D}_2^{kc}]] + [\mathcal{O}_3^{ia},[\mathcal{D}_2^{ia},\mathcal{D}_2^{kc}]] = - 2 N_f \mathcal{O}_3^{kc} - 2 \mathcal{O}_5^{kc},
\end{equation}

\begin{eqnarray}
[\mathcal{D}_3^{ia},[\mathcal{D}_3^{ia},G^{kc}]] & = & - 2 [N_c(N_c+2N_f)+2N_f] G^{kc} + 2 (N_c+N_f) \mathcal{D}_2^{kc} + [N_c(N_c+2N_f)+2N_f-2] \mathcal{D}_3^{kc} \nonumber \\
& & \mbox{} + \frac{3N_cN_f(N_c+2N_f)+8N_f^2-8 N_f+8}{N_f} \mathcal{O}_3^{kc} - (N_c+N_f) \mathcal{D}_4^{kc} + \frac{N_f^2+2N_f-4}{2N_f} \mathcal{D}_5^{kc} \nonumber \\
& & \mbox{} + \frac{(N_f+10)(N_f-2)}{N_f} \mathcal{O}_5^{kc},
\end{eqnarray}

\begin{equation}
[\mathcal{D}_3^{ia},[\mathcal{O}_3^{ia},G^{kc}]] + [\mathcal{O}_3^{ia},[\mathcal{D}_3^{ia},G^{kc}]] = - [N_c(N_c+2N_f)-N_f] \mathcal{D}_3^{kc} - [N_c(N_c+2N_f)+4] \mathcal{O}_3^{kc} + 2 (N_c+N_f) \mathcal{D}_4^{kc},
\end{equation}

\begin{eqnarray}
[\mathcal{O}_3^{ia},[\mathcal{O}_3^{ia},G^{kc}]] & = & \frac32 N_c (N_c+2 N_f) G^{kc} - 3 (N_c+N_f) \mathcal{D}_2^{kc} + \frac14 N_c (N_c+2 N_f) \mathcal{D}_3^{kc} \nonumber \\
& & \mbox{} + \frac14 [5N_c(N_c+2N_f)-30N_f-12] \mathcal{O}_3^{kc} - \frac74 (N_c+N_f) \mathcal{D}_4^{kc} + \frac14 (N_f+3) \mathcal{D}_5^{kc} \nonumber \\
& & \mbox{} + \frac12 (N_f-4) \mathcal{O}_5^{kc},
\end{eqnarray}

\begin{eqnarray}
& & [\mathcal{D}_2^{ia},[\mathcal{D}_3^{ia},\mathcal{D}_3^{kc}]] + [\mathcal{D}_3^{ia},[\mathcal{D}_2^{ia},\mathcal{D}_3^{kc}]] \nonumber \\
& & \mbox{\hglue0.2truecm} = - 4 (N_c+N_f) \mathcal{D}_3^{kc} - 4 (N_f-2) \mathcal{D}_4^{kc} + \frac{2(N_c+N_f)(3N_f-2)}{N_f} \mathcal{D}_5^{kc} + 2 (N_f-2) \mathcal{D}_6^{kc},
\end{eqnarray}

\begin{equation}
[\mathcal{D}_2^{ia},[\mathcal{D}_3^{ia},\mathcal{O}_3^{kc}]] + [\mathcal{D}_3^{ia},[\mathcal{D}_2^{ia},\mathcal{O}_3^{kc}]] = - 4 (N_c+N_f) \mathcal{O}_3^{kc} + \frac{2(N_c+N_f)(5N_f-4)}{N_f} \mathcal{O}_5^{kc},
\end{equation}

\begin{equation}
[\mathcal{D}_2^{ia},[\mathcal{O}_3^{ia},\mathcal{D}_3^{kc}]] + [\mathcal{O}_3^{ia},[\mathcal{D}_2^{ia},\mathcal{D}_3^{kc}]] = - 4 (N_c+N_f) \mathcal{O}_3^{kc} - 2 (N_c+N_f) \mathcal{O}_5^{kc},
\end{equation}

\begin{equation}
[\mathcal{D}_2^{ia},[\mathcal{O}_3^{ia},\mathcal{O}_3^{kc}]] + [\mathcal{O}_3^{ia},[\mathcal{D}_2^{ia},\mathcal{O}_3^{kc}]] = 6 N_f \mathcal{D}_2^{kc} - 3 (N_c+N_f) \mathcal{D}_3^{kc} + (5N_f+6) \mathcal{D}_4^{kc} - (N_c+N_f) \mathcal{D}_5^{kc} + 2 \mathcal{D}_6^{kc},
\end{equation}

\begin{equation}
[\mathcal{D}_3^{ia},[\mathcal{D}_3^{ia},\mathcal{D}_2^{kc}]] = - 2 (N_c+N_f) \mathcal{D}_3^{kc} + [N_c(N_c+2N_f)+2N_f] \mathcal{D}_4^{kc} + (N_c+N_f) \mathcal{D}_5^{kc} + \frac{N_f^2-4}{N_f} \mathcal{D}_6^{kc},
\end{equation}

\begin{equation}
[\mathcal{D}_3^{ia},[\mathcal{O}_3^{ia},\mathcal{D}_2^{kc}]] + [\mathcal{O}_3^{ia},[\mathcal{D}_3^{ia},\mathcal{D}_2^{kc}]] = - 4 (N_c+N_f) \mathcal{O}_3^{kc} - 2 (N_c+N_f) \mathcal{O}_5^{kc},
\end{equation}

\begin{eqnarray}
& & [\mathcal{O}_3^{ia},[\mathcal{O}_3^{ia},\mathcal{D}_2^{kc}]] \nonumber \\
& & \mbox{\hglue0.2truecm} = - \frac32 [N_c(N_c+2N_f)-4N_f] \mathcal{D}_2^{kc} - 3 (N_c+N_f) \mathcal{D}_3^{kc} - \frac14 [5N_c(N_c+2N_f)-38N_f-24] \mathcal{D}_4^{kc} - \frac74 (N_c+N_f) \mathcal{D}_5^{kc} \nonumber \\
& & \mbox{\hglue0.6truecm} + \frac12 (3N_f+10) \mathcal{D}_6^{kc},
\end{eqnarray}

\begin{eqnarray}
& & [\mathcal{D}_3^{ia},[\mathcal{D}_3^{ia},\mathcal{D}_3^{kc}]] \nonumber \\
& & \mbox{\hglue0.2truecm} = - 2 [N_c(N_c+2N_f)+2N_f] \mathcal{D}_3^{kc} + 4 (N_c+N_f) \mathcal{D}_4^{kc} + 2 [N_c(N_c+2N_f)+2N_f-2] \mathcal{D}_5^{kc} - 2 (N_c+N_f) \mathcal{D}_6^{kc} \nonumber \\
& & \mbox{\hglue0.6truecm} + \frac{N_f^2+2N_f-4}{N_f} \mathcal{D}_7^{kc},
\end{eqnarray}

\begin{eqnarray}
& & [\mathcal{D}_3^{ia},[\mathcal{D}_3^{ia},\mathcal{O}_3^{kc}]] \nonumber \\
& & \mbox{\hglue0.2truecm} = -2 [N_c(N_c+2N_f)+2N_f] \mathcal{O}_3^{kc} + \frac{3N_cN_f(N_c+2N_f)+8N_f^2-8 N_f+8}{N_f} \mathcal{O}_5^{kc} + \frac{(N_f+10)(N_f-2)}{N_f} \mathcal{O}_7^{kc},
\end{eqnarray}

\begin{equation}
[\mathcal{D}_3^{ia},[\mathcal{O}_3^{ia},\mathcal{D}_3^{kc}]] + [\mathcal{O}_3^{ia},[\mathcal{D}_3^{ia},\mathcal{D}_3^{kc}]] = - 4 [N_c(N_c+2N_f)+2N_f] \mathcal{O}_3^{kc} - 2 [N_c(N_c+2N_f)-2N_f+8] \mathcal{O}_5^{kc},
\end{equation}

\begin{equation}
[\mathcal{D}_3^{ia},[\mathcal{O}_3^{ia},\mathcal{O}_3^{kc}]] + [\mathcal{O}_3^{ia},[\mathcal{D}_3^{ia},\mathcal{O}_3^{kc}]] = - 3 N_c (N_c+2 N_f) \mathcal{D}_3^{kc} + 6 (N_c+N_f) \mathcal{D}_4^{kc} - [N_c(N_c+2N_f)-3N_f] \mathcal{D}_5^{kc} + 2 (N_c+N_f) \mathcal{D}_6^{kc},
\end{equation}

\begin{eqnarray}
& & [\mathcal{O}_3^{ia},[\mathcal{O}_3^{ia},\mathcal{D}_3^{kc}]] \nonumber \\
& & \mbox{\hglue0.2truecm} = - 24 (N_c+N_f) \mathcal{D}_2^{kc} - \frac32 [3N_c(N_c+2N_f)-8N_f] \mathcal{D}_3^{kc} - 32 (N_c+N_f) \mathcal{D}_4^{kc} - [3 N_c(N_c+2N_f)-19N_f-12] \mathcal{D}_5^{kc} \nonumber \\
& & \mbox{\hglue0.6truecm} - \frac{11}{2} (N_c+N_f) \mathcal{D}_6^{kc} + \frac12 (5N_f+17) \mathcal{D}_7^{kc},
\end{eqnarray}

\begin{equation}
[\mathcal{O}_3^{ia},[\mathcal{O}_3^{ia},\mathcal{O}_3^{kc}]] = - \frac32 N_c (N_c+2 N_f) \mathcal{O}_3^{kc} - \frac14 [9N_c(N_c+2N_f)-34N_f-12] \mathcal{O}_5^{kc} + \frac52 (N_f+2) \mathcal{O}_7^{kc}.
\end{equation}

\subsection{Flavor $\mathbf{8}$ operators}

\begin{equation}
d^{ab8} [G^{ia},[G^{ib},G^{kc}]] = \frac{3N_f^2-16}{8N_f} d^{c8e} G^{ke} + \frac{N_f^2-4}{2N_f^2} \delta^{c8} J^k,
\end{equation}

\begin{eqnarray}
d^{ab8} [G^{ia},[G^{ib},\mathcal{D}_2^{kc}]] & = & - \frac12 (N_c+N_f) d^{c8e} G^{ke} + \frac18 (3N_f+4) d^{c8e} \mathcal{D}_2^{ke}
- \frac12 \{G^{kc},T^8\} + \frac{N_f^2+N_f-4}{2N_f} \{G^{k8},T^c\} \nonumber \\
& & \mbox{} - \frac{1}{N_f} i f^{c8e} [J^2,G^{ke}],
\end{eqnarray}

\begin{eqnarray}
& & d^{ab8}([\mathcal{D}_2^{ia},[G^{ib},G^{kc}]] + [G^{ia},[\mathcal{D}_2^{ib},G^{kc}]]) \nonumber \\
& & \mbox{\hglue0.2truecm} = \frac{(N_c+N_f)(N_f-4)}{2N_f} d^{c8e} G^{ke} + \frac{(N_c+N_f)(N_f-2)}{N_f^2} \delta^{c8} J^k + \frac14 (N_f+2) d^{c8e} \mathcal{D}_2^{ke} + \frac{N_f-4}{2N_f} \{G^{kc},T^8\} + \frac12 \{G^{k8},T^c\} \nonumber \\
& & \mbox{\hglue0.6truecm} - \frac{N_f^2+2N_f-4}{4N_f} i f^{c8e} [J^2,G^{ke}],
\end{eqnarray}

\begin{eqnarray}
& & d^{ab8} [G^{ia},[G^{ib},\mathcal{D}_3^{kc}]] \nonumber \\
& & \mbox{\hglue0.2truecm} = - 4 d^{c8e} G^{ke} - \frac{2[N_c(N_c+2N_f)-N_f+2]}{N_f} \delta^{c8} J^k - 2 (N_c+N_f) d^{c8e} \mathcal{D}_2^{ke} - (N_c+N_f) \{G^{kc},T^8\} \nonumber \\
& & \mbox{\hglue0.6truecm} - \frac12 (N_c+N_f) i f^{c8e} [J^2,G^{ke}] + \frac18 (3N_f+8) d^{c8e} \mathcal{D}_3^{ke} - \frac{2}{N_f} d^{c8e} \mathcal{O}_3^{ke} + \frac{2}{N_f} \{G^{kc},\{J^r,G^{r8}\}\} \nonumber \\
& & \mbox{\hglue0.6truecm} + \frac{N_f^2+2N_f-6}{N_f} \{G^{k8},\{J^r,G^{rc}\}\} - \{J^k,\{T^c,T^8\}\} + (N_f+2) \{J^k,\{G^{rc},G^{r8}\}\} + \frac{N_f+2}{N_f} \delta^{c8} \{J^2,J^k\}, \nonumber \\
\end{eqnarray}

\begin{eqnarray}
& & d^{ab8}([\mathcal{D}_3^{ia},[G^{ib},G^{kc}]] + [G^{ia},[\mathcal{D}_3^{ib},G^{kc}]]) \nonumber \\
& & \mbox{\hglue0.2truecm} = (N_f-4) d^{c8e} G^{ke} + \frac{N_c(N_c+2N_f)+4N_f-8}{2N_f} \delta^{c8} J^k + \frac12 (N_c+N_f) d^{c8e} \mathcal{D}_2^{ke} - (N_c+N_f) i f^{c8e} [J^2,G^{ke}] \nonumber \\
& & \mbox{\hglue0.6truecm} + \frac{(N_f+4)(N_f-2)}{4N_f} d^{c8e} \mathcal{D}_3^{ke} + \frac{N_f^2+2N_f-20}{2N_f} d^{c8e} \mathcal{O}_3^{ke} + \frac{N_f-6}{N_f} \{G^{kc},\{J^r,G^{r8}\}\} + \frac{N_f+2}{N_f} \{G^{k8},\{J^r,G^{rc}\}\} \nonumber \\
& & \mbox{\hglue0.6truecm} + \frac14 \{J^k,\{T^c,T^8\}\} - \{J^k,\{G^{rc},G^{r8}\}\} + \frac{N_f-4}{N_f^2} \delta^{c8} \{J^2,J^k\},
\end{eqnarray}

\begin{eqnarray}
& & d^{ab8} [G^{ia},[G^{ib},\mathcal{O}_3^{kc}]] \nonumber \\
& & \mbox{\hglue0.2truecm} = \frac12 N_f d^{c8e} G^{ke} + \frac{N_c(N_c+2 N_f)}{2N_f} \delta^{c8} J^k + \frac12 (N_c+N_f) d^{c8e} \mathcal{D}_2^{ke} - (N_c+N_f) \{G^{kc},T^8\} \nonumber \\
& & \mbox{\hglue0.6truecm} + \frac34 (N_c+N_f) i f^{c8e} [J^2,G^{ke}] - \frac{1}{N_f} d^{c8e} \mathcal{D}_3^{ke} + \frac{3N_f^2+8N_f-8}{8N_f} d^{c8e} \mathcal{O}_3^{ke} + \frac{N_f^2+2N_f-1}{N_f} \{G^{kc},\{J^r,G^{r8}\}\} \nonumber \\
& & \mbox{\hglue0.6truecm} - \frac{N_f^2+2N_f-2}{2N_f} \{G^{k8},\{J^r,G^{rc}\}\} + \frac14 \{J^k,\{T^c,T^8\}\} - \frac{N_f^2+2N_f-4}{2N_f} \{J^k,\{G^{rc},G^{r8}\}\} - \frac{2}{N_f^2} \delta^{c8} \{J^2,J^k\}, \nonumber \\
\end{eqnarray}

\begin{eqnarray}
& & d^{ab8}([G^{ia},[\mathcal{O}_3^{ib},G^{kc}]] + [\mathcal{O}_3^{ia},[G^{ib},G^{kc}]]) \nonumber \\
& & \mbox{\hglue0.2truecm} = - \frac{3N_c(N_c+2N_f)}{4N_f} \delta^{c8} J^k - \frac34 (N_c+N_f) d^{c8e} \mathcal{D}_2^{ke} + \frac14 (N_c+N_f) i f^{c8e} [J^2,G^{ke}] + \frac{N_f^2+N_f-4}{4N_f} d^{c8e} \mathcal{D}_3^{ke} \nonumber \\
& & \mbox{\hglue0.6truecm} + \frac{N_f^2-2}{2N_f} d^{c8e} \mathcal{O}_3^{ke} + \frac{1}{N_f} \{G^{kc},\{J^r,G^{r8}\}\} - \frac{1}{N_f} \{G^{k8},\{J^r,G^{rc}\}\} - \frac38 \{J^k,\{T^c,T^8\}\} \nonumber \\
& & \mbox{\hglue0.6truecm} + \frac{N_f+4}{2N_f} \{J^k,\{G^{rc},G^{r8}\}\} + \frac{2N_f^2+N_f-4}{2N_f^2} \delta^{c8} \{J^2,J^k\},
\end{eqnarray}

\begin{eqnarray}
& & d^{ab8} ([\mathcal{D}_2^{ia},[G^{ib},\mathcal{D}_2^{kc}]] + [G^{ia},[\mathcal{D}_2^{ib},\mathcal{D}_2^{kc}]]) \nonumber \\
& & \mbox{\hglue0.2truecm} = - N_f d^{c8e} G^{ke} + \frac{(N_c+N_f)(N_f-2)}{N_f} \{G^{k8},T^c\} - \frac{N_c+N_f}{N_f} i f^{c8e} [J^2,G^{ke}] + \frac14 N_f d^{c8e} \mathcal{D}_3^{ke} - d^{c8e} \mathcal{O}_3^{ke} \nonumber \\
& & \mbox{\hglue0.6truecm} - \{G^{kc},\{J^r,G^{r8}\}\} + \{G^{k8},\{J^r,G^{rc}\}\} + \frac{N_f-2}{2N_f} \{J^k,\{T^c,T^8\}\},
\end{eqnarray}

\begin{eqnarray}
& & d^{ab8} [\mathcal{D}_2^{ia},[\mathcal{D}_2^{ib},G^{kc}]] \nonumber \\
& & \mbox{\hglue0.2truecm} = - \frac12 N_f d^{c8e} G^{ke} + \frac{(N_c+N_f)(N_f-4)}{2N_f} \{G^{kc},T^8\} - \frac{(N_c+N_f)(N_f-4)}{4N_f} i f^{c8e} [J^2,G^{ke}] + \frac18 N_f d^{c8e} \mathcal{D}_3^{ke} \nonumber \\
& & \mbox{\hglue0.6truecm} + \frac14 (N_f+2) d^{c8e} \mathcal{O}_3^{ke} + \frac32 \{G^{kc},\{J^r,G^{r8}\}\} - \frac12 \{G^{k8},\{J^r,G^{rc}\}\},
\end{eqnarray}

\begin{eqnarray}
& & d^{ab8} ([\mathcal{D}_2^{ia},[G^{ib},\mathcal{D}_3^{kc}]] + [G^{ia},[\mathcal{D}_2^{ib},\mathcal{D}_3^{kc}]]) \nonumber \\
& & \mbox{\hglue0.2truecm} = - 2 (N_c+N_f) d^{c8e} G^{ke} - (N_f-2) d^{c8e} \mathcal{D}_2^{ke} - 2 \{G^{kc},T^8\} + 2 \{G^{k8},T^c\} - 2 (N_f-1) i f^{c8e} [J^2,G^{ke}] \nonumber \\
& & \mbox{\hglue0.6truecm} + \frac12 (N_c+N_f) d^{c8e} \mathcal{D}_3^{ke} - \frac{2 (N_c+N_f)}{N_f} d^{c8e} \mathcal{O}_3^{ke}
- \frac{(N_c+N_f)(N_f-2)}{N_f} \{G^{kc},\{J^r,G^{r8}\}\} \nonumber \\
& & \mbox{\hglue0.6truecm} + \frac{3(N_c+N_f)(N_f-2)}{N_f} \{G^{k8},\{J^r,G^{rc}\}\} + \frac12 (N_f-2) d^{c8e} \mathcal{D}_4^{ke} - 2 \{\mathcal{D}_2^{kc},\{J^r,G^{r8}\}\} \nonumber \\
& & \mbox{\hglue0.6truecm} + \frac{4(N_f-1)}{N_f} \{\mathcal{D}_2^{k8},\{J^r,G^{rc}\}\} - \{J^2,\{G^{kc},T^8\}\} + \{J^2,\{G^{k8},T^c\}\}
- i f^{c8e} \{J^2,[J^2,G^{ke}]\},
\end{eqnarray}

\begin{eqnarray}
& & d^{ab8}([\mathcal{D}_2^{ia},[G^{ib},\mathcal{O}_3^{kc}]] + [G^{ia},[\mathcal{D}_2^{ib},\mathcal{O}_3^{kc}]]) \nonumber \\
& & \mbox{\hglue0.2truecm} = \frac32 N_f d^{c8e} \mathcal{D}_2^{ke} + \frac12 (N_f-2) i f^{c8e} [J^2,G^{ke}] - \frac{N_c+N_f}{N_f} d^{c8e} \mathcal{D}_3^{ke} + \frac{(N_c+N_f)(N_f-2)}{2N_f} d^{c8e} \mathcal{O}_3^{ke} \nonumber \\
& & \mbox{\hglue0.6truecm} + \frac{(N_c+N_f)(N_f-2)}{2N_f} \{G^{kc},\{J^r,G^{r8}\}\} - \frac{(N_c+N_f)(N_f-2)}{2N_f} \{G^{k8},\{J^r,G^{rc}\}\} \nonumber \\
& & \mbox{\hglue0.6truecm} - \frac{(N_c+N_f)(N_f-2)}{N_f} \{J^k,\{G^{rc},G^{r8}\}\} + \frac{(N_c+N_f)(N_f-2)}{N_f^2} \delta^{c8} \{J^2,J^k\} + d^{c8e} \mathcal{D}_4^{ke} + \{\mathcal{D}_2^{kc},\{J^r,G^{r8}\}\} \nonumber \\
& & \mbox{\hglue0.6truecm} - \frac{2 (N_f-1)}{N_f} \{\mathcal{D}_2^{k8},\{J^r,G^{rc}\}\} + \frac{N_f-2}{N_f} \{J^2,\{G^{kc},T^8\}\}
- \frac{N_f^2-4}{4N_f} i f^{c8e} \{J^2,[J^2,G^{ke}]\},
\end{eqnarray}

\begin{eqnarray}
& & d^{ab8}([\mathcal{D}_3^{ia},[G^{ib},\mathcal{D}_2^{kc}]] + [G^{ia},[\mathcal{D}_3^{ib},\mathcal{D}_2^{kc}]]) \nonumber \\
& & \mbox{\hglue0.2truecm} = - 2 (N_c+N_f) d^{c8e} G^{ke} - (N_f-2) d^{c8e} \mathcal{D}_2^{ke} - 2 \{G^{kc},T^8\} + 2 (N_f-1) \{G^{k8},T^c\} - \frac{4}{N_f} i f^{c8e} [J^2,G^{ke}] \nonumber \\
& & \mbox{\hglue0.6truecm} + \frac12 (N_c+N_f) d^{c8e} \mathcal{D}_3^{ke} - (N_c+N_f) d^{c8e} \mathcal{O}_3^{ke}
+ \frac12 (N_c+N_f) \{J^k,\{T^c,T^8\}\} + \frac12 (N_f-2) d^{c8e} \mathcal{D}_4^{ke} \nonumber \\
& & \mbox{\hglue0.6truecm} - (N_f+2) \{\mathcal{D}_2^{kc},\{J^r,G^{r8}\}\} + 2 \{\mathcal{D}_2^{k8},\{J^r,G^{rc}\}\} - \{J^2,\{G^{kc},T^8\}\} + \frac{N_f^2+3N_f-8}{N_f} \{J^2,\{G^{k8},T^c\}\} \nonumber \\
& & \mbox{\hglue0.6truecm} - \frac{2}{N_f} i f^{c8e} \{J^2,[J^2,G^{ke}]\},
\end{eqnarray}

\begin{eqnarray}
& & d^{ab8} ([G^{ia},[\mathcal{O}_3^{ib},\mathcal{D}_2^{kc}]] + [\mathcal{O}_3^{ia},[G^{ib},\mathcal{D}_2^{kc}]]) \nonumber \\
& & \mbox{\hglue0.2truecm} = 3 N_f d^{c8e} \mathcal{D}_2^{ke} - \frac54 (N_c+N_f) d^{c8e} \mathcal{D}_3^{ke} - \frac12 (N_c+N_f) d^{c8e} \mathcal{O}_3^{ke} - \frac34 (N_c+N_f) \{J^k,\{T^c,T^8\}\} + \frac12 (N_f+5) d^{c8e} \mathcal{D}_4^{ke} \nonumber \\
& & \mbox{\hglue0.6truecm} + \frac{N_f^2+6N_f+4}{2N_f} \{\mathcal{D}_2^{kc},\{J^r,G^{r8}\}\} - 2 \{\mathcal{D}_2^{k8},\{J^r,G^{rc}\}\}
- \frac12 \{J^2,\{G^{kc},T^8\}\} + \frac{N_f^2+N_f-4}{2N_f} \{J^2,\{G^{k8},T^c\}\} \nonumber \\
& & \mbox{\hglue0.6truecm} - \frac{1}{N_f} i f^{c8e} \{J^2,[J^2,G^{ke}]\},
\end{eqnarray}

\begin{equation}
d^{ab8} [\mathcal{D}_2^{ia},[\mathcal{D}_2^{ib},\mathcal{D}_2^{kc}]] = - \frac12 N_f d^{c8e} \mathcal{D}_2^{ke} + \frac{(N_c+N_f)(N_f-4)}{4N_f} \{J^k,\{T^c,T^8\}\} + \frac14 N_f d^{c8e} \mathcal{D}_4^{ke} + \{\mathcal{D}_2^{kc},\{J^r,G^{r8}\}\},
\end{equation}

\begin{eqnarray}
& & d^{ab8} ([\mathcal{D}_2^{ia},[\mathcal{D}_3^{ib},G^{kc}]] + [\mathcal{D}_3^{ia},[\mathcal{D}_2^{ib},G^{kc}]]) \nonumber \\
& & \mbox{\hglue0.2truecm} = - 2 (N_c+N_f) d^{c8e} G^{ke} - (N_f-2) d^{c8e} \mathcal{D}_2^{ke} - 2 \{G^{kc},T^8\} + 2 \{G^{k8},T^c\}
+ \frac{N_f^2-2N_f-4}{N_f} i f^{c8e} [J^2,G^{ke}] \nonumber \\
& & \mbox{\hglue0.6truecm} + \frac12 (N_c+N_f) d^{c8e} \mathcal{D}_3^{ke} + \frac{2(N_c+N_f)(N_f-1)}{N_f} d^{c8e} \mathcal{O}_3^{ke}
+ \frac{3 (N_c+N_f)(N_f-2)}{N_f} \{G^{kc},\{J^r,G^{r8}\}\} \nonumber \\
& & \mbox{\hglue0.6truecm} - \frac{(N_c+N_f)(N_f-2)}{N_f} \{G^{k8},\{J^r,G^{rc}\}\} + \frac12 (N_f-2) d^{c8e} \mathcal{D}_4^{ke}
- \frac{2 (N_f-2)}{N_f} \{\mathcal{D}_2^{k8},\{J^r,G^{rc}\}\} \nonumber \\
& & \mbox{\hglue0.6truecm} + \frac{5N_f-8}{N_f} \{J^2,\{G^{kc},T^8\}\} - \{J^2,\{G^{k8},T^c\}\} - \frac{N_f^2+2N_f-12}{2N_f} i f^{c8e} \{J^2,[J^2,G^{ke}]\},
\end{eqnarray}

\begin{eqnarray}
& & d^{ab8} ([\mathcal{D}_2^{ia},[\mathcal{O}_3^{ib},G^{kc}]] + [\mathcal{O}_3^{ia},[\mathcal{D}_2^{ib},G^{kc}]]) \nonumber \\
& & \mbox{\hglue0.2truecm} = \frac32 N_f d^{c8e} \mathcal{D}_2^{ke} - \frac12 (N_f-2) i f^{c8e} [J^2,G^{ke}] - \frac{N_c+N_f}{N_f} d^{c8e} \mathcal{D}_3^{ke} - \frac{N_c+N_f}{N_f} d^{c8e} \mathcal{O}_3^{ke} \nonumber \\
& & \mbox{\hglue0.6truecm} - \frac{(N_c+N_f)(N_f-2)}{2N_f} \{G^{kc},\{J^r,G^{r8}\}\} + \frac{(N_c+N_f)(N_f-2)}{2N_f} \{G^{k8},\{J^r,G^{rc}\}\} \nonumber \\
& & \mbox{\hglue0.6truecm} - \frac{(N_c+N_f)(N_f-2)}{N_f} \{J^k,\{G^{rc},G^{r8}\}\} + \frac{(N_c+N_f)(N_f-2)}{N_f^2} \delta^{c8} \{J^2,J^k\} + d^{c8e} \mathcal{D}_4^{ke} + \frac12 \{\mathcal{D}_2^{kc},\{J^r,G^{r8}\}\} \nonumber \\
& & \mbox{\hglue0.6truecm} - \frac12 \{\mathcal{D}_2^{k8},\{J^r,G^{rc}\}\} - \frac12 \{J^2,\{G^{kc},T^8\}\} + \frac12 \{J^2,\{G^{k8},T^c\}\} - \frac12 i f^{c8e} \{J^2,[J^2,G^{ke}]\},
\end{eqnarray}

\begin{eqnarray}
& & d^{ab8} ([\mathcal{D}_3^{ia},[G^{ib},\mathcal{D}_3^{kc}]] + [G^{ia},[\mathcal{D}_3^{ib},\mathcal{D}_3^{kc}]]) \nonumber \\
& & \mbox{\hglue0.2truecm} = - 4 N_f d^{c8e} G^{ke} + \frac{2N_c(N_c+2N_f)}{N_f} \delta^{c8} J^k + 2 (N_c+N_f) d^{c8e} \mathcal{D}_2^{ke}
- 4 (N_c+N_f) \{G^{kc},T^8\} \nonumber \\
& & \mbox{\hglue0.6truecm} - 2 (N_c+N_f) i f^{c8e} [J^2,G^{ke}] - 2 d^{c8e} \mathcal{D}_3^{ke} - \frac{4(3N_f+2)}{N_f} d^{c8e} \mathcal{O}_3^{ke} + \frac{2(N_f^2-2N_f+4)}{N_f} \{G^{kc},\{J^r,G^{r8}\}\} \nonumber \\
& & \mbox{\hglue0.6truecm} + \frac{2(3N_f^2-2N_f-4)}{N_f} \{G^{k8},\{J^r,G^{rc}\}\} + \{J^k,\{T^c,T^8\}\} - 4 (N_f-1) \{J^k,\{G^{rc},G^{r8}\}\} \nonumber \\
& & \mbox{\hglue0.6truecm} - \frac{N_c(N_c+2N_f)+4}{N_f} \delta^{c8} \{J^2,J^k\} - (N_c+N_f) d^{c8e} \mathcal{D}_4^{ke} + 6 (N_c+N_f) \{\mathcal{D}_2^{k8},\{J^r,G^{rc}\}\} \nonumber \\
& & \mbox{\hglue0.6truecm} - 2 (N_c+N_f) \{J^2,\{G^{kc},T^8\}\} - (N_c+N_f) i f^{c8e} \{J^2,[J^2,G^{ke}]\} + \frac12 (N_f+2) d^{c8e} \mathcal{D}_5^{ke} - \frac{4}{N_f} d^{c8e} \mathcal{O}_5^{ke} \nonumber \\
& & \mbox{\hglue0.6truecm} + \frac{4}{N_f} \{J^2,\{G^{kc},\{J^r,G^{r8}\}\}\} + \frac{2(N_f^2+4N_f-10)}{N_f} \{J^2,\{G^{k8},\{J^r,G^{rc}\}\}\} - \frac12 \{J^2,\{J^k,\{T^c,T^8\}\}\} \nonumber \\
& & \mbox{\hglue0.6truecm} + 2 (N_f-1) \{J^2,\{J^k,\{G^{rc},G^{r8}\}\}\} - 2 (N_f+1) \{J^k,\{\{J^m,G^{mc}\},\{J^r,G^{r8}\}\}\} + \frac{2}{N_f} \delta^{c8} \{J^2,\{J^2,J^k\}\}, \nonumber \\
\end{eqnarray}

\begin{eqnarray}
& & d^{ab8} ([\mathcal{D}_3^{ia},[G^{ib},\mathcal{O}_3^{kc}]] + [G^{ia},[\mathcal{D}_3^{ib},\mathcal{O}_3^{kc}]]) \nonumber \\
& & \mbox{\hglue0.2truecm} = \frac12 N_f d^{c8e} \mathcal{D}_3^{ke} + N_f d^{c8e} \mathcal{O}_3^{ke} + \frac{N_c(N_c+2N_f)}{N_f} \delta^{c8} \{J^2,J^k\} + (N_c+N_f) d^{c8e} \mathcal{D}_4^{ke} - 3 (N_c+N_f) \{\mathcal{D}_2^{k8},\{J^r,G^{rc}\}\} \nonumber \\
& & \mbox{\hglue0.6truecm} + (N_c+N_f) \{J^2,\{G^{kc},T^8\}\} - \frac12 (N_c+N_f) i f^{c8e} \{J^2,[J^2,G^{ke}]\} - \frac{2}{N_f} d^{c8e} \mathcal{D}_5^{ke} + \frac{N_f^2+2N_f-16}{2N_f} d^{c8e} \mathcal{O}_5^{ke} \nonumber \\
& & \mbox{\hglue0.6truecm} + \frac{N_f-8}{N_f} \{J^2,\{G^{kc},\{J^r,G^{r8}\}\}\} - \frac{N_f^2+N_f-8}{N_f} \{J^2,\{G^{k8},\{J^r,G^{rc}\}\}\} + \frac12 \{J^2,\{J^k,\{T^c,T^8\}\}\} \nonumber \\
& & \mbox{\hglue0.6truecm} - \frac{N_f^2+2N_f-4}{N_f} \{J^2,\{J^k,\{G^{rc},G^{r8}\}\}\} + (N_f+1) \{J^k,\{\{J^m,G^{mc}\},\{J^r,G^{r8}\}\}\} - \frac{4}{N_f^2} \delta^{c8} \{J^2,\{J^2,J^k\}\}, \nonumber \\
\end{eqnarray}

\begin{eqnarray}
& & d^{ab8} ([G^{ia},[\mathcal{O}_3^{ib},\mathcal{D}_3^{kc}]] + [\mathcal{O}_3^{ia},[G^{ib},\mathcal{D}_3^{kc}]]) \nonumber \\
& & \mbox{\hglue0.2truecm} = - \frac{12N_c(N_c+2 N_f)}{N_f} \delta^{c8} J^k - 12 (N_c+N_f) d^{c8e} \mathcal{D}_2^{ke} + \frac12 (5N_f-8) d^{c8e} \mathcal{D}_3^{ke} - 4 d^{c8e} \mathcal{O}_3^{ke} - 6 \{J^k,\{T^c,T^8\}\} \nonumber \\
& & \mbox{\hglue0.6truecm} + 8 (N_f+1) \{J^k,\{G^{rc},G^{r8}\}\} - \frac{9N_c(N_c+2N_f)-32N_f+16}{2N_f} \delta^{c8} \{J^2,J^k\}
- \frac92 (N_c+N_f) d^{c8e} \mathcal{D}_4^{ke} \nonumber \\
& & \mbox{\hglue0.6truecm} - 7 (N_c+N_f) \{\mathcal{D}_2^{k8},\{J^r,G^{rc}\}\} - (N_c+N_f) \{J^2,\{G^{kc},T^8\}\}
- \frac12 (N_c+N_f) i f^{c8e} \{J^2,[J^2,G^{ke}]\} \nonumber \\
& & \mbox{\hglue0.6truecm} + \frac12 (N_f+5) d^{c8e} \mathcal{D}_5^{ke} - \frac{2}{N_f} d^{c8e} \mathcal{O}_5^{ke} + \frac{2}{N_f} \{J^2,\{G^{kc},\{J^r,G^{r8}\}\}\} + \frac{N_f^2+2N_f-6}{N_f} \{J^2,\{G^{k8},\{J^r,G^{rc}\}\}\} \nonumber \\
& & \mbox{\hglue0.6truecm} - \frac94 \{J^2,\{J^k,\{T^c,T^8\}\}\} + (N_f+7) \{J^2,\{J^k,\{G^{rc},G^{r8}\}\}\} + \frac{N_f^2+4N_f+2}{N_f} \{J^k,\{\{J^m,G^{mc}\},\{J^r,G^{r8}\}\}\} \nonumber \\
& & \mbox{\hglue0.6truecm} + \frac{2N_f+5}{N_f} \delta^{c8} \{J^2,\{J^2,J^k\}\},
\end{eqnarray}

\begin{eqnarray}
& & d^{ab8} ([G^{ia},[\mathcal{O}_3^{ib},\mathcal{O}_3^{kc}]] + [\mathcal{O}_3^{ia},[G^{ib},\mathcal{O}_3^{kc}]]) \nonumber \\
& & \mbox{\hglue0.2truecm} = \frac{3N_c(N_c+2N_f)}{2N_f} \delta^{c8} J^k + \frac32 (N_c+N_f) d^{c8e} \mathcal{D}_2^{ke} - 3 (N_c+N_f) \{G^{kc},T^8\} + 4 (N_c+N_f) i f^{c8e} [J^2,G^{ke}] \nonumber \\
& & \mbox{\hglue0.6truecm} + \frac{(N_f+4)(N_f-2)}{4N_f} d^{c8e} \mathcal{D}_3^{ke} + 3 (N_f+1) d^{c8e} \mathcal{O}_3^{ke} + \frac12 (11N_f+6) \{G^{kc},\{J^r,G^{r8}\}\} \nonumber \\
& & \mbox{\hglue0.6truecm} - \frac12 (5N_f+6) \{G^{k8},\{J^r,G^{rc}\}\} + \frac34 \{J^k,\{T^c,T^8\}\} - \frac{2N_f^2+N_f-4}{N_f} \{J^k,\{G^{rc},G^{r8}\}\} \nonumber \\
& & \mbox{\hglue0.6truecm} + \frac{N_cN_f(N_c+2N_f)-2N_f^2+2N_f-8}{2N_f^2} \delta^{c8} \{J^2,J^k\} + \frac12 (N_c+N_f) d^{c8e} \mathcal{D}_4^{ke} + \frac72 (N_c+N_f) \{\mathcal{D}_2^{k8},\{J^r,G^{rc}\}\} \nonumber \\
& & \mbox{\hglue0.6truecm} - \frac92 (N_c+N_f) \{J^2,\{G^{kc},T^8\}\} + \frac32 (N_c+N_f) i f^{c8e} \{J^2,[J^2,G^{ke}]\} - \frac{1}{N_f} d^{c8e} \mathcal{D}_5^{ke} + \frac12 (N_f+4) d^{c8e} \mathcal{O}_5^{ke} \nonumber \\
& & \mbox{\hglue0.6truecm} + 2 (N_f+4) \{J^2,\{G^{kc},\{J^r,G^{r8}\}\}\} - \frac12 (N_f+4) \{J^2,\{G^{k8},\{J^r,G^{rc}\}\}\} + \frac14 \{J^2,\{J^k,\{T^c,T^8\}\}\} \nonumber \\
& & \mbox{\hglue0.6truecm} - \frac{N_f^2+4N_f-8}{2N_f} \{J^2,\{J^k,\{G^{rc},G^{r8}\}\}\} - \frac{N_f^2+4N_f+2}{2N_f} \{J^k,\{\{J^m,G^{mc}\},\{J^r,G^{r8}\}\}\} \nonumber \\
& & \mbox{\hglue0.6truecm} - \frac{2}{N_f^2} \delta^{c8} \{J^2,\{J^2,J^k\}\},
\end{eqnarray}

\begin{eqnarray}
d^{ab8} [\mathcal{D}_2^{ia},[\mathcal{D}_2^{ib},\mathcal{D}_3^{kc}]] & = & - \frac12 N_f d^{c8e} \mathcal{D}_3^{ke}
+ \frac{(N_c+N_f)(N_f-4)}{N_f} \{\mathcal{D}_2^{k8},\{J^r,G^{rc}\}\} + \frac14 N_f d^{c8e} \mathcal{D}_5^{ke} \nonumber \\
& & \mbox{} + \{J^k,\{\{J^m,G^{mc}\},\{J^r,G^{r8}\}\}\},
\end{eqnarray}

\begin{eqnarray}
& & d^{ab8} [\mathcal{D}_2^{ia},[\mathcal{D}_2^{ib},\mathcal{O}_3^{kc}]] \nonumber \\
& & \mbox{\hglue0.2truecm} = - \frac12 N_f d^{c8e} \mathcal{O}_3^{ke} - \frac{(N_c+N_f)(N_f-4)}{2N_f} \{\mathcal{D}_2^{k8},\{J^r,G^{rc}\}\} + \frac{(N_c+N_f)(N_f-4)}{2N_f} \{J^2,\{G^{kc},T^8\}\} \nonumber \\
& & \mbox{\hglue0.6truecm} - \frac{(N_c+N_f)(N_f-4)}{4N_f} i f^{c8e} \{J^2,[J^2,G^{ke}]\} + \frac14 (N_f+2) d^{c8e} \mathcal{O}_5^{ke}
+ \frac32 \{J^2,\{G^{kc},\{J^r,G^{r8}\}\}\} \nonumber \\
& & \mbox{\hglue0.6truecm} - \frac12 \{J^2,\{G^{k8},\{J^r,G^{rc}\}\}\} - \frac12 \{J^k,\{\{J^m,G^{mc}\},\{J^r,G^{r8}\}\}\},
\end{eqnarray}

\begin{eqnarray}
& & d^{ab8} ([\mathcal{D}_2^{ia},[\mathcal{D}_3^{ib},\mathcal{D}_2^{kc}]] + [\mathcal{D}_3^{ia},[\mathcal{D}_2^{ib},\mathcal{D}_2^{kc}]]) \nonumber \\
& & \mbox{\hglue0.2truecm} = - N_f d^{c8e} \mathcal{D}_3^{ke} + \frac{2(N_c+N_f)(N_f-2)}{N_f} \{\mathcal{D}_2^{kc},\{J^r,G^{r8}\}\} + \frac12 N_f d^{c8e} \mathcal{D}_5^{ke} + \frac{N_f-2}{N_f} \{J^2,\{J^k,\{T^c,T^8\}\}\},
\end{eqnarray}

\begin{eqnarray}
& & d^{ab8} ([\mathcal{D}_2^{ia},[\mathcal{O}_3^{ib},\mathcal{D}_2^{kc}]] + [\mathcal{O}_3^{ia},[\mathcal{D}_2^{ib},\mathcal{D}_2^{kc}]]) \nonumber \\
& & \mbox{\hglue0.2truecm} = - N_f d^{c8e} \mathcal{O}_3^{ke} - \frac{(N_c+N_f)(N_f-2)}{N_f} \{\mathcal{D}_2^{kc},\{J^r,G^{r8}\}\} + \frac{(N_c+N_f)(N_f-2)}{N_f} \{J^2,\{G^{k8},T^c\}\} \nonumber \\
& & \mbox{\hglue0.6truecm} - \frac{N_c+N_f}{N_f} i f^{c8e} \{J^2,[J^2,G^{ke}]\} - d^{c8e} \mathcal{O}_5^{ke} - \{J^2,\{G^{kc},\{J^r,G^{r8}\}\}\} + \{J^2,\{G^{k8},\{J^r,G^{rc}\}\}\},
\end{eqnarray}

\begin{eqnarray}
& & d^{ab8} [\mathcal{D}_3^{ia},[\mathcal{D}_3^{ib},G^{kc}]] \nonumber \\
& & \mbox{\hglue0.2truecm} = - 2 N_f d^{c8e} G^{ke} + \frac{N_c(N_c+2 N_f)}{N_f} \delta^{c8} J^k + (N_c+N_f) d^{c8e} \mathcal{D}_2^{ke} - 2 (N_c+N_f) \{G^{kc},T^8\} - d^{c8e} \mathcal{D}_3^{ke} \nonumber \\
& & \mbox{\hglue0.6truecm} + \frac{N_f^2-2N_f+8}{N_f} d^{c8e} \mathcal{O}_3^{ke} + \frac{3N_f^2-6N_f+8}{N_f} \{G^{kc},\{J^r,G^{r8}\}\} + \frac{(N_f+4)(N_f-2)}{N_f} \{G^{k8},\{J^r,G^{rc}\}\} \nonumber \\
& & \mbox{\hglue0.6truecm} + \frac12 \{J^k,\{T^c,T^8\}\} - 2 (N_f-1) \{J^k,\{G^{rc},G^{r8}\}\} - \frac{N_c(N_c+2N_f)+4}{2N_f} \delta^{c8} \{J^2,J^k\} - \frac12 (N_c+N_f) d^{c8e} \mathcal{D}_4^{ke} \nonumber \\
& & \mbox{\hglue0.6truecm} - (N_c+N_f) \{\mathcal{D}_2^{k8},\{J^r,G^{rc}\}\} + 3 (N_c+N_f) \{J^2,\{G^{kc},T^8\}\} - (N_c+N_f) i f^{c8e} \{J^2,[J^2,G^{ke}]\} \nonumber \\
& & \mbox{\hglue0.6truecm} + \frac14 (N_f+2) d^{c8e} \mathcal{D}_5^{ke} + \frac{N_f^2+4N_f-24}{2N_f} d^{c8e} \mathcal{O}_5^{ke} + \frac{2(3N_f-14)}{N_f} \{J^2,\{G^{kc},\{J^r,G^{r8}\}\}\} \nonumber \\
& & \mbox{\hglue0.6truecm} - \frac{N_f^2+2N_f-12}{N_f} \{J^2,\{G^{k8},\{J^r,G^{rc}\}\}\} - \frac14 \{J^2,\{J^k,\{T^c,T^8\}\}\} + (N_f-1) \{J^2,\{J^k,\{G^{rc},G^{r8}\}\}\} \nonumber \\
& & \mbox{\hglue0.6truecm} - \frac{N_f-4}{N_f} \{J^k,\{\{J^m,G^{mc}\},\{J^r,G^{r8}\}\}\} + \frac{1}{N_f} \delta^{c8} \{J^2,\{J^2,J^k\}\},
\end{eqnarray}

\begin{eqnarray}
& & d^{ab8} ([\mathcal{D}_3^{ia},[\mathcal{O}_3^{ib},G^{kc}]] + [\mathcal{O}_3^{ia},[\mathcal{D}_3^{ib},G^{kc}]]) \nonumber \\
& & \mbox{\hglue0.2truecm} = \frac12 N_f d^{c8e} \mathcal{D}_3^{ke} - 4 d^{c8e} \mathcal{O}_3^{ke} + \frac{N_c (N_c+2 N_f)}{N_f} \delta^{c8} \{J^2,J^k\} + (N_c+N_f) d^{c8e} \mathcal{D}_4^{ke} - (N_c+N_f) \{\mathcal{D}_2^{k8},\{J^r,G^{rc}\}\} \nonumber \\
& & \mbox{\hglue0.6truecm} - (N_c+N_f) \{J^2,\{G^{kc},T^8\}\} - \frac12 (N_c+N_f) i f^{c8e} \{J^2,[J^2,G^{ke}]\}
- \frac{2}{N_f} d^{c8e} \mathcal{D}_5^{ke} - \frac{2}{N_f} d^{c8e} \mathcal{O}_5^{ke} \nonumber \\
& & \mbox{\hglue0.6truecm} + \frac{2}{N_f} \{J^2,\{G^{kc},\{J^r,G^{r8}\}\}\} + \frac{N_f^2+2N_f-6}{N_f} \{J^2,\{G^{k8},\{J^r,G^{rc}\}\}\}
+ \frac12 \{J^2,\{J^k,\{T^c,T^8\}\}\} \nonumber \\
& & \mbox{\hglue0.6truecm} - \frac{N_f^2+2N_f-4}{N_f} \{J^2,\{J^k,\{G^{rc},G^{r8}\}\}\} + \frac{2}{N_f} \{J^k,\{\{J^m,G^{mc}\},\{J^r,G^{r8}\}\}\} - \frac{4}{N_f^2} \delta^{c8} \{J^2,\{J^2,J^k\}\},
\end{eqnarray}

\begin{eqnarray}
& & d^{ab8} [\mathcal{O}_3^{ia},[\mathcal{O}_3^{ib},G^{kc}]] \nonumber \\
& & \mbox{\hglue0.2truecm} = - \frac{3N_c(N_c+2 N_f)}{2N_f} \delta^{c8} J^k - \frac32 (N_c+N_f) d^{c8e} \mathcal{D}_2^{ke} + \frac32 (N_c+N_f) \{G^{kc},T^8\} - 2 (N_c+N_f) i f^{c8e} [J^2,G^{ke}] \nonumber \\
& & \mbox{\hglue0.6truecm} + \frac14 (N_f-2) d^{c8e} \mathcal{D}_3^{ke} - \frac12 (2N_f+3) d^{c8e} \mathcal{O}_3^{ke} - \frac14 (11N_f+6) \{G^{kc},\{J^r,G^{r8}\}\} + \frac14 (5N_f+6) \{G^{k8},\{J^r,G^{rc}\}\} \nonumber \\
& & \mbox{\hglue0.6truecm} - \frac34 \{J^k,\{T^c,T^8\}\} + (N_f+1) \{J^k,\{G^{rc},G^{r8}\}\} - \frac{7N_c(N_c+2N_f)-16N_f+8}{8N_f} \delta^{c8} \{J^2,J^k\} \nonumber \\
& & \mbox{\hglue0.6truecm} - \frac78 (N_c+N_f) d^{c8e} \mathcal{D}_4^{ke} - \frac34 (N_c+N_f) \{\mathcal{D}_2^{k8},\{J^r,G^{rc}\}\} + \frac54 (N_c+N_f) \{J^2,\{G^{kc},T^8\}\} \nonumber \\
& & \mbox{\hglue0.6truecm} + \frac14 (N_c+N_f) i f^{c8e} \{J^2,[J^2,G^{ke}]\} + \frac18 (N_f+3) d^{c8e} \mathcal{D}_5^{ke} + \frac14 N_f d^{c8e} \mathcal{O}_5^{ke} - 2 \{J^2,\{G^{kc},\{J^r,G^{r8}\}\}\} \nonumber \\
& & \mbox{\hglue0.6truecm} - \frac14 N_f \{J^2,\{G^{k8},\{J^r,G^{rc}\}\}\} - \frac{7}{16} \{J^2,\{J^k,\{T^c,T^8\}\}\} + \frac14 (N_f+5) \{J^2,\{J^k,\{G^{rc},G^{r8}\}\}\} \nonumber \\
& & \mbox{\hglue0.6truecm} + \frac34 \{J^k,\{\{J^m,G^{mc}\},\{J^r,G^{r8}\}\}\} + \frac{2N_f+3}{4N_f} \delta^{c8} \{J^2,\{J^2,J^k\}\},
\end{eqnarray}

\begin{eqnarray}
& & d^{ab8} ([\mathcal{D}_2^{ia},[\mathcal{D}_3^{ib},\mathcal{D}_3^{kc}]] + [\mathcal{D}_3^{ia},[\mathcal{D}_2^{ib},\mathcal{D}_3^{kc}]]) \nonumber \\
& & \mbox{\hglue0.2truecm} = - 2 (N_c+N_f) d^{c8e} \mathcal{D}_3^{ke} - 2 (N_f-2) d^{c8e} \mathcal{D}_4^{ke} + 4 \{\mathcal{D}_2^{kc},\{J^r,G^{r8}\}\} - 4 \{\mathcal{D}_2^{k8},\{J^r,G^{rc}\}\} + (N_c+N_f) d^{c8e} \mathcal{D}_5^{ke} \nonumber \\
& & \mbox{\hglue0.6truecm} + \frac{2 (N_c+N_f)(N_f-2)}{N_f} \{J^k,\{\{J^m,G^{mc}\},\{J^r,G^{r8}\}\}\} + (N_f-2) d^{c8e} \mathcal{D}_6^{ke} - 2 \{J^2,\{\mathcal{D}_2^{kc},\{J^r,G^{r8}\}\}\} \nonumber \\
& & \mbox{\hglue0.6truecm} + \frac{2(3N_f-4)}{N_f} \{J^2,\{\mathcal{D}_2^{k8},\{J^r,G^{rc}\}\}\},
\end{eqnarray}

\begin{eqnarray}
& & d^{ab8} ([\mathcal{D}_2^{ia},[\mathcal{D}_3^{ib},\mathcal{O}_3^{kc}]] + [\mathcal{D}_3^{ia},[\mathcal{D}_2^{ib},\mathcal{O}_3^{kc}]]) \nonumber \\
& & \mbox{\hglue0.2truecm} = - 2 (N_c+N_f) d^{c8e} \mathcal{O}_3^{ke} - 2 \{\mathcal{D}_2^{kc},\{J^r,G^{r8}\}\} + 2 \{\mathcal{D}_2^{k8},\{J^r,G^{rc}\}\} - 2 \{J^2,\{G^{kc},T^8\}\} + 2 \{J^2,\{G^{k8},T^c\}\} \nonumber \\
& & \mbox{\hglue0.6truecm} + \frac{N_f^2-2N_f-4}{N_f} i f^{c8e} \{J^2,[J^2,G^{ke}]\} + \frac{2(N_c+N_f)(N_f-1)}{N_f} d^{c8e} \mathcal{O}_5^{ke} \nonumber \\
& & \mbox{\hglue0.6truecm} + \frac{3 (N_c+N_f)(N_f-2)}{N_f} \{J^2,\{G^{kc},\{J^r,G^{r8}\}\}\} - \frac{(N_c+N_f)(N_f-2)}{N_f} \{J^2,\{G^{k8},\{J^r,G^{rc}\}\}\} \nonumber \\
& & \mbox{\hglue0.6truecm} - \frac{(N_c+N_f)(N_f-2)}{N_f} \{J^k,\{\{J^m,G^{mc}\},\{J^r,G^{r8}\}\}\} + \{J^2,\{\mathcal{D}_2^{kc},\{J^r,G^{r8}\}\}\} \nonumber \\
& & \mbox{\hglue0.6truecm} - \frac{5N_f-8}{N_f} \{J^2,\{\mathcal{D}_2^{k8},\{J^r,G^{rc}\}\}\} + \frac{5N_f-8}{N_f} \{J^2,\{J^2,\{G^{kc},T^8\}\}\} - \{J^2,\{J^2,\{G^{k8},T^c\}\}\} \nonumber \\
& & \mbox{\hglue0.6truecm} - \frac{N_f^2+2N_f-12}{2N_f} i f^{c8e} \{J^2,\{J^2,[J^2,G^{ke}]\}\},
\end{eqnarray}

\begin{eqnarray}
& & d^{ab8} ([\mathcal{D}_2^{ia},[\mathcal{O}_3^{ib},\mathcal{D}_3^{kc}]] + [\mathcal{O}_3^{ia},[\mathcal{D}_2^{ib},\mathcal{D}_3^{kc}]]) \nonumber \\
& & \mbox{\hglue0.2truecm} = - 2 (N_c+N_f) d^{c8e} \mathcal{O}_3^{ke} - 2 \{\mathcal{D}_2^{kc},\{J^r,G^{r8}\}\} + 2 \{\mathcal{D}_2^{k8},\{J^r,G^{rc}\}\} - 2 \{J^2,\{G^{kc},T^8\}\} + 2 \{J^2,\{G^{k8},T^c\}\} \nonumber \\
& & \mbox{\hglue0.6truecm} - 2 (N_f-1) i f^{c8e} \{J^2,[J^2,G^{ke}]\} - \frac{2 (N_c+N_f)}{N_f} d^{c8e} \mathcal{O}_5^{ke} - \frac{(N_c+N_f)(N_f-2)}{N_f} \{J^2,\{G^{kc},\{J^r,G^{r8}\}\}\} \nonumber \\
& & \mbox{\hglue0.6truecm} + \frac{3(N_c+N_f)(N_f-2)}{N_f} \{J^2,\{G^{k8},\{J^r,G^{rc}\}\}\} - \frac{(N_c+N_f)(N_f-2)}{N_f} \{J^k,\{\{J^m,G^{mc}\},\{J^r,G^{r8}\}\}\} \nonumber \\
& & \mbox{\hglue0.6truecm} - \{J^2,\{\mathcal{D}_2^{kc},\{J^r,G^{r8}\}\}\} + \{J^2,\{\mathcal{D}_2^{k8},\{J^r,G^{rc}\}\}\} - \{J^2,\{J^2,\{G^{kc},T^8\}\}\} + \{J^2,\{J^2,\{G^{k8},T^c\}\}\} \nonumber \\
& & \mbox{\hglue0.6truecm} - i f^{c8e} \{J^2,\{J^2,[J^2,G^{ke}]\}\},
\end{eqnarray}

\begin{eqnarray}
& & d^{ab8} ([\mathcal{D}_2^{ia},[\mathcal{O}_3^{ib},\mathcal{O}_3^{kc}]] + [\mathcal{O}_3^{ia},[\mathcal{D}_2^{ib},\mathcal{O}_3^{kc}]]) \nonumber \\
& & \mbox{\hglue0.2truecm} = 3 N_f d^{c8e} \mathcal{D}_2^{ke} - \frac{(N_f+4) (N_c+N_f)}{2N_f} d^{c8e} \mathcal{D}_3^{ke}
- \frac{2 (N_c+N_f)(N_f-2)}{N_f} \{J^k,\{G^{rc},G^{r8}\}\} \nonumber \\
& & \mbox{\hglue0.6truecm} + \frac{2(N_c+N_f)(N_f-2)}{N_f^2} \delta^{c8} \{J^2,J^k\} + \frac12 (5N_f+6) d^{c8e} \mathcal{D}_4^{ke} + 3 \{\mathcal{D}_2^{kc},\{J^r,G^{r8}\}\} - 3 \{\mathcal{D}_2^{k8},\{J^r,G^{rc}\}\} \nonumber \\
& & \mbox{\hglue0.6truecm} - \frac{N_c+N_f}{N_f} d^{c8e} \mathcal{D}_5^{ke} - \frac{2(N_c+N_f)(N_f-2)}{N_f} \{J^2,\{J^k,\{G^{rc},G^{r8}\}\}\} \nonumber \\
& & \mbox{\hglue0.6truecm} + \frac{(N_c+N_f)(N_f-2)}{2N_f} \{J^k,\{\{J^m,G^{mc}\},\{J^r,G^{r8}\}\}\} + \frac{(N_c+N_f)(N_f-2)}{N_f^2} \delta^{c8} \{J^2,\{J^2,J^k\}\} + d^{c8e} \mathcal{D}_6^{ke} \nonumber \\
& & \mbox{\hglue0.6truecm} + \{J^2,\{\mathcal{D}_2^{kc},\{J^r,G^{r8}\}\}\} - \{J^2,\{\mathcal{D}_2^{k8},\{J^r,G^{rc}\}\}\},
\end{eqnarray}

\begin{eqnarray}
& & d^{ab8} [\mathcal{D}_3^{ia},[\mathcal{D}_3^{ib},\mathcal{D}_2^{kc}]] \nonumber \\
& & \mbox{\hglue0.2truecm} = - (N_c+N_f) d^{c8e} \mathcal{D}_3^{ke} - (N_f-2) d^{c8e} \mathcal{D}_4^{ke} + 2 (N_f-1) \{\mathcal{D}_2^{kc},\{J^r,G^{r8}\}\} - 2 \{\mathcal{D}_2^{k8},\{J^r,G^{rc}\}\} \nonumber \\
& & \mbox{\hglue0.6truecm} + \frac12 (N_c+N_f) d^{c8e} \mathcal{D}_5^{ke} + \frac12 (N_c+N_f) \{J^2,\{J^k,\{T^c,T^8\}\}\}
+ \frac12 (N_f-2) d^{c8e} \mathcal{D}_6^{ke} \nonumber \\
& & \mbox{\hglue0.6truecm} + \frac{N_f-8}{N_f} \{J^2,\{\mathcal{D}_2^{kc},\{J^r,G^{r8}\}\}\} + \{J^2,\{\mathcal{D}_2^{k8},\{J^r,G^{rc}\}\}\},
\end{eqnarray}

\begin{eqnarray}
& & d^{ab8} ([\mathcal{D}_3^{ia},[\mathcal{O}_3^{ib},\mathcal{D}_2^{kc}]] + [\mathcal{O}_3^{ia},[\mathcal{D}_3^{ib},\mathcal{D}_2^{kc}]]) \nonumber \\
& & \mbox{\hglue0.2truecm} = - 2 (N_c+N_f) d^{c8e} \mathcal{O}_3^{ke} - 2 (N_f-1) \{\mathcal{D}_2^{kc},\{J^r,G^{r8}\}\} + 2 \{\mathcal{D}_2^{k8},\{J^r,G^{rc}\}\} - 2 \{J^2,\{G^{kc},T^8\}\} \nonumber \\
& & \mbox{\hglue0.6truecm} + 2 (N_f-1) \{J^2,\{G^{k8},T^c\}\} - \frac{4}{N_f} i f^{c8e} \{J^2,[J^2,G^{ke}]\}
- (N_c+N_f) d^{c8e} \mathcal{O}_5^{ke} \nonumber \\
& & \mbox{\hglue0.6truecm} - \frac{N_f^2+3N_f-8}{N_f} \{J^2,\{\mathcal{D}_2^{kc},\{J^r,G^{r8}\}\}\} + \{J^2,\{\mathcal{D}_2^{k8},\{J^r,G^{rc}\}\}\} - \{J^2,\{J^2,\{G^{kc},T^8\}\}\} \nonumber \\
& & \mbox{\hglue0.6truecm} + \frac{N_f^2+3N_f-8}{N_f} \{J^2,\{J^2,\{G^{k8},T^c\}\}\} - \frac{2}{N_f} i f^{c8e} \{J^2,\{J^2,[J^2,G^{ke}]\}\},
\end{eqnarray}

\begin{eqnarray}
& & d^{ab8} [\mathcal{O}_3^{ia},[\mathcal{O}_3^{ib},\mathcal{D}_2^{kc}]] \nonumber \\
& & \mbox{\hglue0.2truecm} = 3 N_f d^{c8e} \mathcal{D}_2^{ke} - \frac32 (N_c+N_f) d^{c8e} \mathcal{D}_3^{ke} - \frac34 (N_c+N_f) \{J^k,\{T^c,T^8\}\} + \frac14 (13N_f+12) d^{c8e} \mathcal{D}_4^{ke} \nonumber \\
& & \mbox{\hglue0.6truecm} + \frac32 (N_f+2) \{\mathcal{D}_2^{kc},\{J^r,G^{r8}\}\} - 3 \{\mathcal{D}_2^{k8},\{J^r,G^{rc}\}\}
- \frac78 (N_c+N_f) d^{c8e} \mathcal{D}_5^{ke} \nonumber \\
& & \mbox{\hglue0.6truecm} - \frac58 (N_c+N_f) \{J^2,\{J^k,\{T^c,T^8\}\}\} + \frac14 (N_f+7) d^{c8e} \mathcal{D}_6^{ke} + \frac14 (2N_f+13) \{J^2,\{\mathcal{D}_2^{kc},\{J^r,G^{r8}\}\}\} \nonumber \\
& & \mbox{\hglue0.6truecm} - \frac74 \{J^2,\{\mathcal{D}_2^{k8},\{J^r,G^{rc}\}\}\},
\end{eqnarray}

\begin{eqnarray}
& & d^{ab8} [\mathcal{D}_3^{ia},[\mathcal{D}_3^{ib},\mathcal{D}_3^{kc}]] \nonumber \\
& & \mbox{\hglue0.2truecm} = - 2 N_f d^{c8e} \mathcal{D}_3^{ke} + \frac{2N_c(N_c+2 N_f)}{N_f} \delta^{c8} \{J^2,J^k\} + 2 (N_c+N_f) d^{c8e} \mathcal{D}_4^{ke} - 4 (N_c+N_f) \{\mathcal{D}_2^{k8},\{J^r,G^{rc}\}\} \nonumber \\
& & \mbox{\hglue0.6truecm} - 2 d^{c8e} \mathcal{D}_5^{ke} + \{J^2,\{J^k,\{T^c,T^8\}\}\} - 4 (N_f-1) \{J^2,\{J^k,\{G^{rc},G^{r8}\}\}\} \nonumber \\
& & \mbox{\hglue0.6truecm} + 4 (N_f-1) \{J^k,\{\{J^m,G^{mc}\},\{J^r,G^{r8}\}\}\} - \frac{N_c(N_c+2N_f)+4}{N_f} \delta^{c8} \{J^2,\{J^2,J^k\}\} - (N_c+N_f) d^{c8e} \mathcal{D}_6^{ke} \nonumber \\
& & \mbox{\hglue0.6truecm} + 4 (N_c+N_f) \{J^2,\{\mathcal{D}_2^{k8},\{J^r,G^{rc}\}\}\} + \frac12 (N_f+2) d^{c8e} \mathcal{D}_7^{ke} - \frac12 \{J^2,\{J^2,\{J^k,\{T^c,T^8\}\}\}\} \nonumber \\
& & \mbox{\hglue0.6truecm} + 2 (N_f-1) \{J^2,\{J^2,\{J^k,\{G^{rc},G^{r8}\}\}\}\} - \frac{N_f^2-2N_f+8}{N_f} \{J^2,\{J^k,\{\{J^m,G^{mc}\},\{J^r,G^{r8}\}\}\}\} \nonumber \\
& & \mbox{\hglue0.6truecm} + \frac{2}{N_f} \delta^{c8} \{J^2,\{J^2,\{J^2,J^k\}\}\},
\end{eqnarray}

\begin{eqnarray}
& & d^{ab8} [\mathcal{D}_3^{ia},[\mathcal{D}_3^{ib},\mathcal{O}_3^{kc}]] \nonumber \\
& & \mbox{\hglue0.2truecm} = - 2 N_f d^{c8e} \mathcal{O}_3^{ke} + 2 (N_c+N_f) \{\mathcal{D}_2^{k8},\{J^r,G^{rc}\}\} - 2 (N_c+N_f) \{J^2,\{G^{kc},T^8\}\} + \frac{N_f^2-2N_f+8}{N_f} d^{c8e} \mathcal{O}_5^{ke} \nonumber \\
& & \mbox{\hglue0.6truecm} + \frac{3N_f^2-6N_f+8}{N_f} \{J^2,\{G^{kc},\{J^r,G^{r8}\}\}\} + \frac{(N_f+4)(N_f-2)}{N_f} \{J^2,\{G^{k8},\{J^r,G^{rc}\}\}\} \nonumber \\
& & \mbox{\hglue0.6truecm} - 2 (N_f-1) \{J^k,\{\{J^m,G^{mc}\},\{J^r,G^{r8}\}\}\} - 3 (N_c+N_f) \{J^2,\{\mathcal{D}_2^{k8},\{J^r,G^{rc}\}\}\} \nonumber \\
& & \mbox{\hglue0.6truecm} + 3 (N_c+N_f) \{J^2,\{J^2,\{G^{kc},T^8\}\}\} - (N_c+N_f) i f^{c8e} \{J^2,\{J^2,[J^2,G^{ke}]\}\}
+ \frac{N_f^2+4N_f-24}{2N_f} d^{c8e} \mathcal{O}_7^{ke} \nonumber \\
& & \mbox{\hglue0.6truecm} + \frac{2(3N_f-14)}{N_f} \{J^2,\{J^2,\{G^{kc},\{J^r,G^{r8}\}\}\}\} - \frac{N_f^2+2N_f-12}{N_f} \{J^2,\{J^2,\{G^{k8},\{J^r,G^{rc}\}\}\}\} \nonumber \\
& & \mbox{\hglue0.6truecm} + \frac{N_f^2-4N_f+16}{2N_f} \{J^2,\{J^k,\{\{J^m,G^{mc}\},\{J^r,G^{r8}\}\}\}\},
\end{eqnarray}

\begin{eqnarray}
& & d^{ab8} ([\mathcal{D}_3^{ia},[\mathcal{O}_3^{ib},\mathcal{D}_3^{kc}]] + [\mathcal{O}_3^{ia},[\mathcal{D}_3^{ib},\mathcal{D}_3^{kc}]]) \nonumber \\
& & \mbox{\hglue0.2truecm} = - 4 N_f d^{c8e} \mathcal{O}_3^{ke} + 4 (N_c+N_f) \{\mathcal{D}_2^{k8},\{J^r,G^{rc}\}\} - 4 (N_c+N_f) \{J^2,\{G^{kc},T^8\}\} - 2 (N_c+N_f) i f^{c8e} \{J^2,[J^2,G^{ke}]\} \nonumber \\
& & \mbox{\hglue0.6truecm} - \frac{4(3N_f+2)}{N_f} d^{c8e} \mathcal{O}_5^{ke} + \frac{2(N_f^2-2N_f+4)}{N_f} \{J^2,\{G^{kc},\{J^r,G^{r8}\}\}\} + \frac{2(3N_f^2-2N_f-4)}{N_f} \{J^2,\{G^{k8},\{J^r,G^{rc}\}\}\} \nonumber \\
& & \mbox{\hglue0.6truecm} - 4 (N_f-1) \{J^k,\{\{J^m,G^{mc}\},\{J^r,G^{r8}\}\}\} + 2 (N_c+N_f) \{J^2,\{\mathcal{D}_2^{k8},\{J^r,G^{rc}\}\}\} \nonumber \\
& & \mbox{\hglue0.6truecm} - 2 (N_c+N_f) \{J^2,\{J^2,\{G^{kc},T^8\}\}\} - (N_c+N_f) i f^{c8e} \{J^2,\{J^2,[J^2,G^{ke}]\}\} - \frac{4}{N_f} d^{c8e} \mathcal{O}_7^{ke} \nonumber \\
& & \mbox{\hglue0.6truecm} + \frac{4}{N_f} \{J^2,\{J^2,\{G^{kc},\{J^r,G^{r8}\}\}\}\} + \frac{2(N_f^2+4N_f-10)}{N_f} \{J^2,\{J^2,\{G^{k8},\{J^r,G^{rc}\}\}\}\} \nonumber \\
& & \mbox{\hglue0.6truecm} - \frac{N_f^2+4N_f-8}{N_f} \{J^2,\{J^k,\{\{J^m,G^{mc}\},\{J^r,G^{r8}\}\}\}\},
\end{eqnarray}

\begin{eqnarray}
& & d^{ab8} ([\mathcal{D}_3^{ia},[\mathcal{O}_3^{ib},\mathcal{O}_3^{kc}]] + [\mathcal{O}_3^{ia},[\mathcal{D}_3^{ib},\mathcal{O}_3^{kc}]]) \nonumber \\
& & \mbox{\hglue0.2truecm} = \frac{3N_c(N_c+2N_f)}{N_f} \delta^{c8} \{J^2,J^k\} + 3 (N_c+N_f) d^{c8e} \mathcal{D}_4^{ke} - 6 (N_c+N_f) \{\mathcal{D}_2^{k8},\{J^r,G^{rc}\}\} + \frac{(N_f+4)(N_f-2)}{2N_f} d^{c8e} \mathcal{D}_5^{ke} \nonumber \\
& & \mbox{\hglue0.6truecm} + \frac32 \{J^2,\{J^k,\{T^c,T^8\}\}\} - \frac{2(2N_f^2+N_f-4)}{N_f} \{J^2,\{J^k,\{G^{rc},G^{r8}\}\}\}
+ 3 N_f \{J^k,\{\{J^m,G^{mc}\},\{J^r,G^{r8}\}\}\} \nonumber \\
& & \mbox{\hglue0.6truecm} + \frac{N_cN_f(N_c+2N_f)-2N_f^2+2N_f-8}{N_f^2} \delta^{c8} \{J^2,\{J^2,J^k\}\}
+ (N_c+N_f) d^{c8e} \mathcal{D}_6^{ke} \nonumber \\
& & \mbox{\hglue0.6truecm} - 2 (N_c+N_f) \{J^2,\{\mathcal{D}_2^{k8},\{J^r,G^{rc}\}\}\} - \frac{2}{N_f} d^{c8e} \mathcal{D}_7^{ke} + \frac12 \{J^2,\{J^2,\{J^k,\{T^c,T^8\}\}\}\} \nonumber \\
& & \mbox{\hglue0.6truecm} - \frac{N_f^2+4N_f-8}{N_f} \{J^2,\{J^2,\{J^k,\{G^{rc},G^{r8}\}\}\}\} + \frac{N_f^2+4N_f-4}{2N_f} \{J^2,\{J^k,\{\{J^m,G^{mc}\},\{J^r,G^{r8}\}\}\}\} \nonumber \\
& & \mbox{\hglue0.6truecm} - \frac{4}{N_f^2} \delta^{c8} \{J^2,\{J^2,\{J^2,J^k\}\}\},
\end{eqnarray}

\begin{eqnarray}
& & d^{ab8} [\mathcal{O}_3^{ia},[\mathcal{O}_3^{ib},\mathcal{D}_3^{kc}]] \nonumber \\
& & \mbox{\hglue0.2truecm} = - \frac{12N_c(N_c+2 N_f)}{N_f} \delta^{c8} J^k - 12 (N_c+N_f) d^{c8e} \mathcal{D}_2^{ke} + 2 (N_f-2) d^{c8e} \mathcal{D}_3^{ke} - 6 \{J^k,\{T^c,T^8\}\} \nonumber \\
& & \mbox{\hglue0.6truecm} + 8 (N_f+1) \{J^k,\{G^{rc},G^{r8}\}\} - \frac{8[2N_c(N_c+2N_f)-2N_f+1]}{N_f} \delta^{c8} \{J^2,J^k\} - 16 (N_c+N_f) d^{c8e} \mathcal{D}_4^{ke} \nonumber \\
& & \mbox{\hglue0.6truecm} - 9 (N_c+N_f) \{\mathcal{D}_2^{k8},\{J^r,G^{rc}\}\} + 3 N_f d^{c8e} \mathcal{D}_5^{ke}
 - 8 \{J^2,\{J^k,\{T^c,T^8\}\}\} + 8 (N_f+2) \{J^2,\{J^k,\{G^{rc},G^{r8}\}\}\} \nonumber \\
& & \mbox{\hglue0.6truecm} + \frac12 (5N_f+8) \{J^k,\{\{J^m,G^{mc}\},\{J^r,G^{r8}\}\}\} - \frac{11N_c(N_c+2N_f)-64N_f}{4N_f} \delta^{c8} \{J^2,\{J^2,J^k\}\} \nonumber \\
& & \mbox{\hglue0.6truecm} - \frac{11}{4} (N_c+N_f) d^{c8e} \mathcal{D}_6^{ke} - 6 (N_c+N_f) \{J^2,\{\mathcal{D}_2^{k8},\{J^r,G^{rc}\}\}\} + \frac14 (N_f+7) d^{c8e} \mathcal{D}_7^{ke} \nonumber \\
& & \mbox{\hglue0.6truecm} - \frac{11}{8} \{J^2,\{J^2,\{J^k,\{T^c,T^8\}\}\}\} + \frac12 (N_f+9) \{J^2,\{J^2,\{J^k,\{G^{rc},G^{r8}\}\}\}\} \nonumber \\
& & \mbox{\hglue0.6truecm} + \frac34 (N_f+6) \{J^2,\{J^k,\{\{J^m,G^{mc}\},\{J^r,G^{r8}\}\}\}\} + \frac{2N_f+7}{2N_f} \delta^{c8} \{J^2,\{J^2,\{J^2,J^k\}\}\},
\end{eqnarray}

\begin{eqnarray}
& & d^{ab8} [\mathcal{O}_3^{ia},[\mathcal{O}_3^{ib},\mathcal{O}_3^{kc}]] \nonumber \\
& & \mbox{\hglue0.2truecm} = \frac32 (N_c+N_f) \{\mathcal{D}_2^{k8},\{J^r,G^{rc}\}\} - \frac32 (N_c+N_f) \{J^2,\{G^{kc},T^8\}\} + 2 (N_c+N_f) i f^{c8e} \{J^2,[J^2,G^{ke}]\} \nonumber \\
& & \mbox{\hglue0.6truecm} + \frac32 (N_f+1) d^{c8e} \mathcal{O}_5^{ke}
+ \frac14 (11N_f+6) \{J^2,\{G^{kc},\{J^r,G^{r8}\}\}\} - \frac14 (5N_f+6) \{J^2,\{G^{k8},\{J^r,G^{rc}\}\}\} \nonumber \\
& & \mbox{\hglue0.6truecm} - \frac34 N_f \{J^k,\{\{J^m,G^{mc}\},\{J^r,G^{r8}\}\}\} + \frac94 (N_c+N_f) \{J^2,\{\mathcal{D}_2^{k8},\{J^r,G^{rc}\}\}\} - \frac94 (N_c+N_f) \{J^2,\{J^2,\{G^{kc},T^8\}\}\} \nonumber \\
& & \mbox{\hglue0.6truecm} + \frac34 (N_c+N_f) i f^{c8e} \{J^2,\{J^2,[J^2,G^{ke}]\}\} + \frac14 (N_f+4) d^{c8e} \mathcal{O}_7^{ke} + (N_f+4) \{J^2,\{J^2,\{G^{kc},\{J^r,G^{r8}\}\}\}\} \nonumber \\
& & \mbox{\hglue0.6truecm} - \frac14 (N_f+4) \{J^2,\{J^2,\{G^{k8},\{J^r,G^{rc}\}\}\}\} - \frac38 (N_f+4) \{J^2,\{J^k,\{\{J^m,G^{mc}\},\{J^r,G^{r8}\}\}\}\}.
\end{eqnarray}

\subsection{Flavor $\mathbf{27}$ operators}

\begin{equation}
[G^{i8},[G^{i8},G^{kc}]] = \frac14 f^{c8e} f^{8eg} G^{kg} + \frac12 d^{c8e} d^{8eg} G^{kg} + \frac{1}{N_f} \delta^{c8} G^{k8} + \frac{1}{2N_f} d^{c88} J^k,
\end{equation}

\begin{eqnarray}
[G^{i8},[G^{i8},\mathcal{D}_2^{kc}]] & = & \frac74 f^{c8e} f^{8eg} \mathcal{D}_2^{kg} + \frac12 d^{c8e} d^{8eg} \mathcal{D}_2^{kg} - \frac12 d^{ceg} d^{88e} \mathcal{D}_2^{kg} + \frac{1}{N_f} \delta^{c8} \mathcal{D}_2^{k8} + \frac12 d^{88e} \{G^{ke},T^c\} \nonumber \\
& & \mbox{} - \frac12 i f^{c8e} [G^{ke},\{J^r,G^{r8}\}] + \frac12 i f^{c8e} [G^{k8},\{J^r,G^{re}\}],
\end{eqnarray}

\begin{eqnarray}
& & [\mathcal{D}_2^{i8},[G^{i8},G^{kc}]] + [G^{i8},[\mathcal{D}_2^{i8},G^{kc}]] \nonumber \\
& & \mbox{\hglue0.2truecm} = \frac12 f^{c8e} f^{8eg} \mathcal{D}_2^{kg} + \frac{2}{N_f} \delta^{c8} \mathcal{D}_2^{k8} + d^{c8e} \{G^{ke},T^8\} + \frac12 i f^{c8e} [G^{ke},\{J^r,G^{r8}\}] + \frac12 i f^{c8e} [G^{k8},\{J^r,G^{re}\}],
\end{eqnarray}

\begin{eqnarray}
& & [G^{i8},[G^{i8},\mathcal{D}_3^{kc}]] \nonumber \\
& & \mbox{\hglue0.2truecm} = - \frac32 f^{c8e} f^{8eg} G^{kg} + \frac54 f^{c8e} f^{8eg} \mathcal{D}_3^{kg} + \frac32 d^{c8e} d^{8eg} \mathcal{D}_3^{kg} - d^{ceg} d^{88e} \mathcal{D}_3^{kg} + \frac{1}{N_f} \delta^{c8} \mathcal{D}_3^{k8} + \frac{1}{N_f} d^{c88} \{J^2,J^k\} \nonumber \\
& & \mbox{\hglue0.6truecm} - 2 \{G^{kc},\{G^{r8},G^{r8}\}\} + 2 \{G^{k8},\{G^{rc},G^{r8}\}\} - 3 d^{c8e} \{J^k,\{G^{re},G^{r8}\}\}
+ d^{88e} \{J^k,\{G^{rc},G^{re}\}\} \nonumber \\
& & \mbox{\hglue0.6truecm} + d^{c8e} \{G^{ke},\{J^r,G^{r8}\}\} + d^{88e} \{G^{ke},\{J^r,G^{rc}\}\} - \frac12 \epsilon^{kim} f^{c8e} \{T^e,\{J^i,G^{m8}\}\},
\end{eqnarray}

\begin{eqnarray}
& & [\mathcal{D}_3^{i8},[G^{i8},G^{kc}]] + [G^{i8},[\mathcal{D}_3^{i8},G^{kc}]] \nonumber \\
& & \mbox{\hglue0.2truecm} = \frac12 f^{c8e} f^{8eg} \mathcal{D}_3^{kg} + \frac{2}{N_f} \delta^{c8} \mathcal{D}_3^{k8} + d^{c8e} d^{8eg} \mathcal{O}_3^{kg} + 3 d^{c8e} \{G^{ke},\{J^r,G^{r8}\}\} - d^{c8e} \{G^{k8},\{J^r,G^{re}\}\},
\end{eqnarray}

\begin{eqnarray}
& & [G^{i8},[G^{i8},\mathcal{O}_3^{kc}]] \nonumber \\
& & \mbox{\hglue0.2truecm} = \frac34 f^{c8e} f^{8eg} G^{kg} + \frac{1}{N_f} \delta^{c8} \mathcal{D}_3^{k8} + \frac54 f^{c8e} f^{8eg} \mathcal{O}_3^{kg} + \frac32 d^{c8e} d^{8eg} \mathcal{O}_3^{kg} - d^{ceg} d^{88e} \mathcal{O}_3^{kg} + \frac{5}{N_f} \delta^{c8} \mathcal{O}_3^{k8} \nonumber \\
& & \mbox{\hglue0.6truecm} - \{G^{kc},\{G^{r8},G^{r8}\}\} - \{G^{k8},\{G^{rc},G^{r8}\}\} + \frac12 d^{c8e} \{J^k,\{G^{re},G^{r8}\}\} - \frac12 d^{88e} \{J^k,\{G^{rc},G^{re}\}\} \nonumber \\
& & \mbox{\hglue0.6truecm} - \frac12 d^{c8e} \{G^{ke},\{J^r,G^{r8}\}\} + d^{c8e} \{G^{k8},\{J^r,G^{re}\}\} + d^{88e} \{G^{kc},\{J^r,G^{re}\}\} \nonumber \\
& & \mbox{\hglue0.6truecm} - \frac12 d^{88e} \{G^{ke},\{J^r,G^{rc}\}\} + \frac34 \epsilon^{kim} f^{c8e} \{T^e,\{J^i,G^{m8}\}\},
\end{eqnarray}

\begin{eqnarray}
& & [G^{i8},[\mathcal{O}_3^{i8},G^{kc}]] + [\mathcal{O}_3^{i8},[G^{i8},G^{kc}]] \nonumber \\
& & \mbox{\hglue0.2truecm} = \frac12 d^{c8e} d^{8eg} \mathcal{D}_3^{kg} + \frac12 f^{c8e} f^{8eg} \mathcal{O}_3^{kg} + \frac12 d^{c8e} d^{8eg} \mathcal{O}_3^{kg} + \frac{2}{N_f} \delta^{c8} \mathcal{O}_3^{k8} + \frac{1}{N_f} d^{c88} \{J^2,J^k\} - d^{c8e} \{J^k,\{G^{re},G^{r8}\}\} \nonumber \\
& & \mbox{\hglue0.6truecm} - \frac12 d^{c8e} \{G^{ke},\{J^r,G^{r8}\}\} + \frac12 d^{c8e} \{G^{k8},\{J^r,G^{re}\}\},
\end{eqnarray}

\begin{eqnarray}
& & [\mathcal{D}_2^{i8},[G^{i8},\mathcal{D}_2^{kc}]] + [G^{i8},[\mathcal{D}_2^{i8},\mathcal{D}_2^{kc}]] \nonumber \\
& & \mbox{\hglue0.2truecm} = - f^{c8e} f^{8eg} G^{kg} + \frac12 f^{c8e} f^{8eg} \mathcal{D}_3^{kg} + \{G^{k8},\{T^c,T^8\}\} - \frac12 \epsilon^{kim} f^{c8e} \{T^e,\{J^i,G^{m8}\}\} + \frac12 \epsilon^{kim} f^{c8e} \{T^8,\{J^i,G^{me}\}\}, \nonumber \\
\end{eqnarray}

\begin{equation}
[\mathcal{D}_2^{i8},[\mathcal{D}_2^{i8},G^{kc}]] = - f^{c8e} f^{8eg} G^{kg} + \frac14 f^{c8e} f^{8eg} \mathcal{D}_3^{kg} + \frac12 f^{c8e} f^{8eg} \mathcal{O}_3^{kg} + \frac12 \{G^{kc},\{T^8,T^8\}\} - \frac12 \epsilon^{kim} f^{c8e} \{T^8,\{J^i,G^{me}\}\},
\end{equation}

\begin{eqnarray}
& & [\mathcal{D}_2^{i8},[G^{i8},\mathcal{D}_3^{kc}]] + [G^{i8},[\mathcal{D}_2^{i8},\mathcal{D}_3^{kc}]] \nonumber \\
& & \mbox{\hglue0.2truecm} = 5 i f^{c8e} [G^{k8},\{J^r,G^{re}\}] + d^{c8e} \{J^2,\{G^{ke},T^8\}\} - d^{c8e} \{\mathcal{D}_2^{k8},\{J^r,G^{re}\}\} + 3 \{\{J^r,G^{rc}\},\{G^{k8},T^8\}\} \nonumber \\
& & \mbox{\hglue0.6truecm} - \{\{J^r,G^{r8}\},\{G^{kc},T^8\}\} + i f^{c8e} \{J^k,[\{J^i,G^{ie}\},\{J^r,G^{r8}\}]\} - i f^{c8e} \{\{J^r,G^{re}\},[J^2,G^{k8}]\},
\end{eqnarray}

\begin{eqnarray}
& & [\mathcal{D}_2^{i8},[G^{i8},\mathcal{O}_3^{kc}]] + [G^{i8},[\mathcal{D}_2^{i8},\mathcal{O}_3^{kc}]] \nonumber \\
& & \mbox{\hglue0.2truecm} = \frac32 f^{c8e} f^{8eg} \mathcal{D}_2^{kg} - \frac12 i f^{c8e} [G^{k8},\{J^r,G^{re}\}] + \frac12 f^{c8e} f^{8eg} \mathcal{D}_4^{kg} + \frac{2}{N_f} \delta^{c8} \mathcal{D}_4^{k8} + \frac12 d^{c8e} \{J^2,\{G^{ke},T^8\}\} \nonumber \\
& & \mbox{\hglue0.6truecm} - 2 \{\mathcal{D}_2^{k8},\{G^{rc},G^{r8}\}\} + \frac12 d^{c8e} \{\mathcal{D}_2^{k8},\{J^r,G^{re}\}\} - \frac12 \{\{J^r,G^{rc}\},\{G^{k8},T^8\}\} + \frac12 \{\{J^r,G^{r8}\},\{G^{kc},T^8\}\} \nonumber \\
& & \mbox{\hglue0.6truecm} - \frac12 i f^{c8e} \{J^k,[\{J^i,G^{ie}\},\{J^r,G^{r8}\}]\} + \frac12 i f^{c8e} \{\{J^r,G^{re}\},[J^2,G^{k8}]\} + \frac12 i f^{c8e} \{J^2,[G^{ke},\{J^r,G^{r8}\}]\} \nonumber \\
& & \mbox{\hglue0.6truecm} + \frac12 i f^{c8e} \{J^2,[G^{k8},\{J^r,G^{re}\}]\},
\end{eqnarray}

\begin{eqnarray}
& & [\mathcal{D}_3^{i8},[G^{i8},\mathcal{D}_2^{kc}]] + [G^{i8},[\mathcal{D}_3^{i8},\mathcal{D}_2^{kc}]] \nonumber \\
& & \mbox{\hglue0.2truecm} = - 2 i f^{c8e} [G^{ke},\{J^r,G^{r8}\}] + d^{88e} \{J^2,\{G^{ke},T^c\}\} - d^{88e} \{\mathcal{D}_2^{kc},\{J^r,G^{re}\}\} + 2 \{\{J^r,G^{r8}\},\{G^{k8},T^c\}\} \nonumber \\
& & \mbox{\hglue0.6truecm} + i f^{c8e} \{J^k,[\{J^i,G^{ie}\},\{J^r,G^{r8}\}]\} - i f^{c8e} \{\{J^r,G^{re}\},[J^2,G^{k8}]\} + i f^{c8e} \{\{J^r,G^{r8}\},[J^2,G^{ke}]\},
\end{eqnarray}

\begin{eqnarray}
& & [G^{i8},[\mathcal{O}_3^{i8},\mathcal{D}_2^{kc}]] + [\mathcal{O}_3^{i8},[G^{i8},\mathcal{D}_2^{kc}]] \nonumber \\
& & \mbox{\hglue0.2truecm} = 6 f^{c8e} f^{8eg} \mathcal{D}_2^{kg} + \frac92 f^{c8e} f^{8eg} \mathcal{D}_4^{kg} + d^{c8e} d^{8eg} \mathcal{D}_4^{kg} - d^{ceg} d^{88e} \mathcal{D}_4^{kg} + \frac{2}{N_f} \delta^{c8} \mathcal{D}_4^{k8} + \frac12 d^{88e} \{J^2,\{G^{ke},T^c\}\} \nonumber \\
& & \mbox{\hglue0.6truecm} - 2 \{\mathcal{D}_2^{kc},\{G^{r8},G^{r8}\}\} + \frac12 d^{88e} \{\mathcal{D}_2^{kc},\{J^r,G^{re}\}\} - \frac32 i f^{c8e} \{J^k,[\{J^i,G^{ie}\},\{J^r,G^{r8}\}]\} \nonumber \\
& & \mbox{\hglue0.6truecm} + \frac12 i f^{c8e} \{\{J^r,G^{re}\},[J^2,G^{k8}]\} - \frac12 i f^{c8e} \{\{J^r,G^{r8}\},[J^2,G^{ke}]\} - i f^{c8e} \{J^2,[G^{ke},\{J^r,G^{r8}\}]\} \nonumber \\
& & \mbox{\hglue0.6truecm} + i f^{c8e} \{J^2,[G^{k8},\{J^r,G^{re}\}]\},
\end{eqnarray}

\begin{equation}
[\mathcal{D}_2^{i8},[\mathcal{D}_2^{i8},\mathcal{D}_2^{kc}]] = - f^{c8e} f^{8eg} \mathcal{D}_2^{kg} + \frac12 f^{c8e} f^{8eg} \mathcal{D}_4^{kg} + \frac12 \{\mathcal{D}_2^{kc},\{T^8,T^8\}\},
\end{equation}

\begin{eqnarray}
& & [\mathcal{D}_2^{i8},[\mathcal{D}_3^{i8},G^{kc}]] + [\mathcal{D}_3^{i8},[\mathcal{D}_2^{i8},G^{kc}]] \nonumber \\
& & \mbox{\hglue0.2truecm} = - 2 i f^{c8e} [G^{ke},\{J^r,G^{r8}\}] - i f^{c8e} [G^{k8},\{J^r,G^{re}\}] + d^{c8e} \{J^2,\{G^{ke},T^8\}\} - d^{c8e} \{\mathcal{D}_2^{k8},\{J^r,G^{re}\}\} \nonumber \\
& & \mbox{\hglue0.6truecm} - \{\{J^r,G^{rc}\},\{G^{k8},T^8\}\} + 3 \{\{J^r,G^{r8}\},\{G^{kc},T^8\}\} - i f^{c8e} \{\{J^r,G^{r8}\},[J^2,G^{ke}]\} \nonumber \\
& & \mbox{\hglue0.6truecm} + 2 i f^{c8e} \{J^2,[G^{ke},\{J^r,G^{r8}\}]\},
\end{eqnarray}

\begin{eqnarray}
& & [\mathcal{D}_2^{i8},[\mathcal{O}_3^{i8},G^{kc}]] + [\mathcal{O}_3^{i8},[\mathcal{D}_2^{i8},G^{kc}]] \nonumber \\
& & \mbox{\hglue0.2truecm} = \frac32 f^{c8e} f^{8eg} \mathcal{D}_2^{kg} + \frac12 i f^{c8e} [G^{k8},\{J^r,G^{re}\}] + \frac12 f^{c8e} f^{8eg} \mathcal{D}_4^{kg} + \frac{2}{N_f} \delta^{c8} \mathcal{D}_4^{k8} + \frac12 d^{c8e} \{J^2,\{G^{ke},T^8\}\} \nonumber \\
& & \mbox{\hglue0.6truecm} - 2 \{\mathcal{D}_2^{k8},\{G^{rc},G^{r8}\}\} + \frac12 d^{c8e} \{\mathcal{D}_2^{k8},\{J^r,G^{re}\}\} + \frac12 \{\{J^r,G^{rc}\},\{G^{k8},T^8\}\} - \frac12 \{\{J^r,G^{r8}\},\{G^{kc},T^8\}\} \nonumber \\
& & \mbox{\hglue0.6truecm} - \frac12 i f^{c8e} \{\{J^r,G^{r8}\},[J^2,G^{ke}]\} - \frac12 i f^{c8e} \{J^2,[G^{ke},\{J^r,G^{r8}\}]\}
+ \frac12 i f^{c8e} \{J^2,[G^{k8},\{J^r,G^{re}\}]\},
\end{eqnarray}

\begin{eqnarray}
& & [\mathcal{D}_3^{i8},[G^{i8},\mathcal{D}_3^{kc}]] + [G^{i8},[\mathcal{D}_3^{i8},\mathcal{D}_3^{kc}]] \nonumber \\
& & \mbox{\hglue0.2truecm} = 3 f^{c8e} f^{8eg} G^{kg} + \frac12 i \epsilon^{kim} f^{c8e} f^{8eg} \{J^i,G^{mg}\} + \frac52 f^{c8e} f^{8eg} \mathcal{D}_3^{kg} - \frac{N_c}{2} i d^{8eg} f^{c8e} \mathcal{D}_3^{kg} - \frac{N_c}{2} i d^{c8e} f^{8eg} \mathcal{D}_3^{kg} \nonumber \\
& & \mbox{\hglue0.6truecm} - 2 f^{c8e} f^{8eg} \mathcal{O}_3^{kg} + 6 d^{c8e} d^{8eg} \mathcal{O}_3^{kg} - 6 d^{ceg} d^{88e} \mathcal{O}_3^{kg} + 4 \{G^{kc},\{G^{r8},G^{r8}\}\} - 4 \{G^{k8},\{G^{rc},G^{r8}\}\} \nonumber \\
& & \mbox{\hglue0.6truecm} - 2 d^{c8e} \{J^k,\{G^{re},G^{r8}\}\} + 2 d^{88e} \{J^k,\{G^{rc},G^{re}\}\} + 8 d^{c8e} \{G^{ke},\{J^r,G^{r8}\}\} - 6 d^{c8e} \{G^{k8},\{J^r,G^{re}\}\} \nonumber \\
& & \mbox{\hglue0.6truecm} + 6 d^{88e} \{G^{kc},\{J^r,G^{re}\}\} - 8 d^{88e} \{G^{ke},\{J^r,G^{rc}\}\} + \frac14 (3N_f-8) \epsilon^{kim} f^{c8e} \{T^e,\{J^i,G^{m8}\}\} + i d^{8eg} f^{c8e} \mathcal{D}_4^{kg} \nonumber \\
& & \mbox{\hglue0.6truecm} + \frac{2}{N_f} i \epsilon^{kim} \delta^{c8} \{J^2,\{J^i,G^{m8}\}\} + i f^{c8e} \{\mathcal{D}_2^{ke},\{J^r,G^{r8}\}\} - 4 i \epsilon^{kim} \{\{J^i,G^{m8}\},\{G^{r8},G^{rc}\}\} \nonumber \\
& & \mbox{\hglue0.6truecm} + 2 i \epsilon^{kim} \{\{J^i,G^{mc}\},\{G^{r8},G^{r8}\}\} - 2 i \epsilon^{rim} \{G^{k8},\{J^r,\{G^{ic},G^{m8}\}\}\} + i \epsilon^{rim} d^{c8e} \{J^k,\{J^r,\{G^{i8},G^{me}\}\}\} \nonumber \\
& & \mbox{\hglue0.6truecm} + \frac34 i \epsilon^{kim} f^{cae} f^{8eb} \{\{J^i,G^{m8}\},\{T^a,T^b\}\} + \frac72 i f^{c8e} \{J^k,[\{J^i,G^{ie}\},\{J^r,G^{r8}\}]\} + \frac72 i f^{c8e} \{\{J^r,G^{re}\},[J^2,G^{k8}]\} \nonumber \\
& & \mbox{\hglue0.6truecm} - \frac72 i f^{c8e} \{\{J^r,G^{r8}\},[J^2,G^{ke}]\} - \frac72 i f^{c8e} \{J^2,[G^{ke},\{J^r,G^{r8}\}]\} + \frac72 i f^{c8e} \{J^2,[G^{k8},\{J^r,G^{re}\}]\} \nonumber \\
& & \mbox{\hglue0.6truecm} - d^{c8e} \{J^2,[G^{ke},\{J^r,G^{r8}\}]\} + d^{c8e} \{J^2,[G^{k8},\{J^r,G^{re}\}]\} + 2 [G^{k8},\{\{J^m,G^{m8}\},\{J^r,G^{rc}\}\}] \nonumber \\
& & \mbox{\hglue0.6truecm} - 2 \{\{J^m,G^{mc}\},[G^{k8},\{J^r,G^{r8}\}]\} - i \epsilon^{kim} f^{cea} f^{e8b} \{\{J^i,G^{m8}\},\{G^{ra},G^{rb}\}\} + f^{c8e} f^{8eg} \mathcal{D}_5^{kg} \nonumber \\
& & \mbox{\hglue0.6truecm} - 2 d^{c8e} \{J^2,\{J^k,\{G^{re},G^{r8}\}\}\} + 2 d^{88e} \{J^2,\{J^k,\{G^{rc},G^{re}\}\}\}
+ 8 d^{c8e} \{J^2,\{G^{ke},\{J^r,G^{r8}\}\}\} \nonumber \\
& & \mbox{\hglue0.6truecm} + 8 d^{88e} \{J^2,\{G^{ke},\{J^r,G^{rc}\}\}\} - \epsilon^{kim} f^{c8e} \{J^2,\{T^e,\{J^i,G^{m8}\}\}\} - 8 \{G^{kc},\{\{J^m,G^{m8}\},\{J^r,G^{r8}\}\}\} \nonumber \\
& & \mbox{\hglue0.6truecm} + 12 \{G^{k8},\{\{J^m,G^{m8}\},\{J^r,G^{rc}\}\}\} + 2 \{J^k,\{\{J^m,G^{mc}\},\{G^{r8},G^{r8}\}\}\} - 2 \{J^k,\{\{J^m,G^{m8}\},\{G^{r8},G^{rc}\}\}\} \nonumber \\
& & \mbox{\hglue0.6truecm} - 3 d^{c8e} \{\mathcal{D}_3^{ke},\{J^r,G^{r8}\}\} - 5 d^{88e} \{\mathcal{D}_3^{kc},\{J^r,G^{re}\}\}
- 2 \epsilon^{kim} f^{ab8} \{\{J^i,G^{m8}\},\{T^a,\{G^{rb},G^{rc}\}\}\} \nonumber \\
& & \mbox{\hglue0.6truecm} - 6 i \epsilon^{kil} [\{J^i,G^{l8}\},\{\{J^m,G^{m8}\},\{J^r,G^{rc}\}\}],
\end{eqnarray}

\begin{eqnarray}
& & [\mathcal{D}_3^{i8},[G^{i8},\mathcal{O}_3^{kc}]] + [G^{i8},[\mathcal{D}_3^{i8},\mathcal{O}_3^{kc}]] \nonumber \\
& & \mbox{\hglue0.2truecm} = - \frac{15}{4} f^{c8e} f^{8eg} G^{kg} - \frac12 i \epsilon^{kim} f^{c8e} f^{8eg} \{J^i,G^{mg}\} - \frac12 f^{c8e} f^{8eg} \mathcal{D}_3^{kg} + \frac{5}{32} N_c i d^{8eg} f^{c8e} \mathcal{D}_3^{kg} + \frac{5}{32} N_c i d^{c8e} f^{8eg} \mathcal{D}_3^{kg} \nonumber \\
& & \mbox{\hglue0.6truecm} + f^{c8e} f^{8eg} \mathcal{O}_3^{kg} - \frac52 d^{c8e} d^{8eg} \mathcal{O}_3^{kg} + \frac52 d^{ceg} d^{88e} \mathcal{O}_3^{kg} - 5 \{G^{kc},\{G^{r8},G^{r8}\}\} + 5 \{G^{k8},\{G^{rc},G^{r8}\}\} \nonumber \\
& & \mbox{\hglue0.6truecm} + \frac52 d^{c8e} \{J^k,\{G^{re},G^{r8}\}\} - \frac52 d^{88e} \{J^k,\{G^{rc},G^{re}\}\} - 5 d^{c8e} \{G^{ke},\{J^r,G^{r8}\}\} + \frac52 d^{c8e} \{G^{k8},\{J^r,G^{re}\}\} \nonumber \\
& & \mbox{\hglue0.6truecm} - \frac52 d^{88e} \{G^{kc},\{J^r,G^{re}\}\} + 5 d^{88e} \{G^{ke},\{J^r,G^{rc}\}\} - \frac18 (3N_f-2) \epsilon^{kim} f^{c8e} \{T^e,\{J^i,G^{m8}\}\} - \frac{5}{16} i d^{8eg} f^{c8e} \mathcal{D}_4^{kg} \nonumber \\
& & \mbox{\hglue0.6truecm} - \frac{5}{8N_f} i \epsilon^{kim} \delta^{c8} \{J^2,\{J^i,G^{m8}\}\} - i f^{c8e} \{\mathcal{D}_2^{ke},\{J^r,G^{r8}\}\} + \frac{21}{8} i \epsilon^{kim} \{\{J^i,G^{m8}\},\{G^{r8},G^{rc}\}\} \nonumber \\
& & \mbox{\hglue0.6truecm} - 2 i \epsilon^{kim} \{\{J^i,G^{mc}\},\{G^{r8},G^{r8}\}\} + 2 i \epsilon^{rim} \{G^{k8},\{J^r,\{G^{ic},G^{m8}\}\}\} - \frac{5}{16} i \epsilon^{rim} d^{c8e} \{J^k,\{J^r,\{G^{i8},G^{me}\}\}\} \nonumber \\
& & \mbox{\hglue0.6truecm} - \frac{15}{64} i \epsilon^{kim} f^{cae} f^{8eb} \{\{J^i,G^{m8}\},\{T^a,T^b\}\}
 - \frac{59}{16} i f^{c8e} \{J^k,[\{J^i,G^{ie}\},\{J^r,G^{r8}\}]\} - \frac{59}{16} i f^{c8e} \{\{J^r,G^{re}\},[J^2,G^{k8}]\} \nonumber \\
& & \mbox{\hglue0.6truecm} + \frac{59}{16} i f^{c8e} \{\{J^r,G^{r8}\},[J^2,G^{ke}]\} + \frac{59}{16} i f^{c8e} \{J^2,[G^{ke},\{J^r,G^{r8}\}]\} - \frac{59}{16} i f^{c8e} \{J^2,[G^{k8},\{J^r,G^{re}\}]\} \nonumber \\
& & \mbox{\hglue0.6truecm}
+ \frac{5}{16} d^{c8e} \{J^2,[G^{ke},\{J^r,G^{r8}\}]\}
- \frac{5}{16} d^{c8e} \{J^2,[G^{k8},\{J^r,G^{re}\}]\}
+ \frac{11}{32} [G^{kc},\{\{J^m,G^{m8}\},\{J^r,G^{r8}\}\}] \nonumber \\
& & \mbox{\hglue0.6truecm}
- \frac{21}{16} [G^{k8},\{\{J^m,G^{m8}\},\{J^r,G^{rc}\}\}]
+ \frac{21}{16} \{\{J^m,G^{mc}\},[G^{k8},\{J^r,G^{r8}\}]\} \nonumber \\
& & \mbox{\hglue0.6truecm}
+ \frac{5}{16} i \epsilon^{kim} f^{cea} f^{e8b} \{\{J^i,G^{m8}\},\{G^{ra},G^{rb}\}\}
+ \frac{2}{N_f} \delta^{c8} \mathcal{D}_5^{k8}
+ d^{c8e} d^{8eg} \mathcal{O}_5^{kg}
+ d^{c8e} \{J^2,\{J^k,\{G^{re},G^{r8}\}\}\} \nonumber \\
& & \mbox{\hglue0.6truecm}
- d^{88e} \{J^2,\{J^k,\{G^{rc},G^{re}\}\}\}
- 2 d^{c8e} \{J^2,\{G^{ke},\{J^r,G^{r8}\}\}\}
- d^{c8e} \{J^2,\{G^{k8},\{J^r,G^{re}\}\}\} \nonumber \\
& & \mbox{\hglue0.6truecm}
- 5 d^{88e} \{J^2,\{G^{ke},\{J^r,G^{rc}\}\}\}
+ \frac12 \epsilon^{kim} f^{c8e} \{J^2,\{T^e,\{J^i,G^{m8}\}\}\}
+ 5 \{G^{kc},\{\{J^m,G^{m8}\},\{J^r,G^{r8}\}\}\} \nonumber \\
& & \mbox{\hglue0.6truecm}
- 5 \{G^{k8},\{\{J^m,G^{m8}\},\{J^r,G^{rc}\}\}\}
- \{J^k,\{\{J^m,G^{mc}\},\{G^{r8},G^{r8}\}\}\}
- \{J^k,\{\{J^m,G^{m8}\},\{G^{r8},G^{rc}\}\}\} \nonumber \\
& & \mbox{\hglue0.6truecm} + 2 d^{c8e} \{\mathcal{D}_3^{ke},\{J^r,G^{r8}\}\}
+ 3 d^{88e} \{\mathcal{D}_3^{kc},\{J^r,G^{re}\}\}
+ \epsilon^{kim} f^{ab8} \{\{J^i,G^{m8}\},\{T^a,\{G^{rb},G^{rc}\}\}\} \nonumber \\
& & \mbox{\hglue0.6truecm}
+ 4 i \epsilon^{kil} [\{J^i,G^{l8}\},\{\{J^m,G^{m8}\},\{J^r,G^{rc}\}\}],
\end{eqnarray}

\begin{eqnarray}
& & [G^{i8},[\mathcal{O}_3^{i8},\mathcal{D}_3^{kc}]] + [\mathcal{O}_3^{i8},[G^{i8},\mathcal{D}_3^{kc}]] \nonumber \\
& & \mbox{\hglue0.2truecm} = - 3 f^{c8e} f^{8eg} G^{kg} + i \epsilon^{kim} f^{c8e} f^{8eg} \{J^i,G^{mg}\} + \frac54 f^{c8e} f^{8eg} \mathcal{D}_3^{kg} + 8 d^{c8e} d^{8eg} \mathcal{D}_3^{kg} - 4 d^{ceg} d^{88e} \mathcal{D}_3^{kg} \nonumber \\
& & \mbox{\hglue0.6truecm} + \frac{7}{16} N_c i d^{8eg} f^{c8e} \mathcal{D}_3^{kg} + \frac{7}{16} N_c i d^{c8e} f^{8eg} \mathcal{D}_3^{kg} - 2 f^{c8e} f^{8eg} \mathcal{O}_3^{kg} - 2 d^{c8e} d^{8eg} \mathcal{O}_3^{kg} + 2 d^{ceg} d^{88e} \mathcal{O}_3^{kg} \nonumber \\
& & \mbox{\hglue0.6truecm} + \frac{8}{N_f} d^{c88} \{J^2,J^k\} - 4 \{G^{kc},\{G^{r8},G^{r8}\}\} + 4 \{G^{k8},\{G^{rc},G^{r8}\}\} - 14 d^{c8e} \{J^k,\{G^{re},G^{r8}\}\} \nonumber \\
& & \mbox{\hglue0.6truecm} + 6 d^{88e} \{J^k,\{G^{rc},G^{re}\}\} - 4 d^{c8e} \{G^{ke},\{J^r,G^{r8}\}\} + 2 d^{c8e} \{G^{k8},\{J^r,G^{re}\}\} - 2 d^{88e} \{G^{kc},\{J^r,G^{re}\}\} \nonumber \\
& & \mbox{\hglue0.6truecm} + 4 d^{88e} \{G^{ke},\{J^r,G^{rc}\}\} - \frac18 (3N_f-4) \epsilon^{kim} f^{c8e} \{T^e,\{J^i,G^{m8}\}\}
- \frac78 i d^{8eg} f^{c8e} \mathcal{D}_4^{kg} \nonumber \\
& & \mbox{\hglue0.6truecm} - \frac{7}{4N_f} i \epsilon^{kim} \delta^{c8} \{J^2,\{J^i,G^{m8}\}\} + 2 i f^{c8e} \{\mathcal{D}_2^{ke},\{J^r,G^{r8}\}\} - \frac94 i \epsilon^{kim} \{\{J^i,G^{m8}\},\{G^{r8},G^{rc}\}\} \nonumber \\
& & \mbox{\hglue0.6truecm}
+ 4 i \epsilon^{kim} \{\{J^i,G^{mc}\},\{G^{r8},G^{r8}\}\}
- 4 i \epsilon^{rim} \{G^{k8},\{J^r,\{G^{ic},G^{m8}\}\}\}
- \frac78 i \epsilon^{rim} d^{c8e} \{J^k,\{J^r,\{G^{i8},G^{me}\}\}\} \nonumber \\
& & \mbox{\hglue0.6truecm} - \frac{21}{32} i \epsilon^{kim} f^{cae} f^{8eb} \{\{J^i,G^{m8}\},\{T^a,T^b\}\}
- \frac38 i f^{c8e} \{J^k,[\{J^i,G^{ie}\},\{J^r,G^{r8}\}]\}
- \frac38 i f^{c8e} \{\{J^r,G^{re}\},[J^2,G^{k8}]\} \nonumber \\
& & \mbox{\hglue0.6truecm}
+ \frac38 i f^{c8e} \{\{J^r,G^{r8}\},[J^2,G^{ke}]\} 
+ \frac38 i f^{c8e} \{J^2,[G^{ke},\{J^r,G^{r8}\}]\}
- \frac38 i f^{c8e} \{J^2,[G^{k8},\{J^r,G^{re}\}]\} \nonumber \\
& & \mbox{\hglue0.6truecm}
+ \frac78 d^{c8e} \{J^2,[G^{ke},\{J^r,G^{r8}\}]\}
- \frac78 d^{c8e} \{J^2,[G^{k8},\{J^r,G^{re}\}]\}
- \frac{23}{16} [G^{kc},\{\{J^m,G^{m8}\},\{J^r,G^{r8}\}\}] \nonumber \\
& & \mbox{\hglue0.6truecm}
+ \frac98 [G^{k8},\{\{J^m,G^{m8}\},\{J^r,G^{rc}\}\}]
- \frac98 \{\{J^m,G^{mc}\},[G^{k8},\{J^r,G^{r8}\}]\} \nonumber \\
& & \mbox{\hglue0.6truecm}
+ \frac78 i \epsilon^{kim} f^{cea} f^{e8b} \{\{J^i,G^{m8}\},\{G^{ra},G^{rb}\}\}
+ 2 f^{c8e} f^{8eg} \mathcal{D}_5^{kg}
+ 3 d^{c8e} d^{8eg} \mathcal{D}_5^{kg}
- 2 d^{ceg} d^{88e} \mathcal{D}_5^{kg} \nonumber \\
& & \mbox{\hglue0.6truecm}
+ \frac{2}{N_f} \delta^{c8} \mathcal{D}_5^{k8}
+ \frac{2}{N_f} d^{c88} \{J^2,\{J^2,J^k\}\}
- 4 \{J^2,\{G^{kc},\{G^{r8},G^{r8}\}\}\}
+ 4 \{J^2,\{G^{k8},\{G^{rc},G^{r8}\}\}\} \nonumber \\
& & \mbox{\hglue0.6truecm}
- 7 d^{c8e} \{J^2,\{J^k,\{G^{re},G^{r8}\}\}\}
+ d^{88e} \{J^2,\{J^k,\{G^{rc},G^{re}\}\}\}
- 2 d^{c8e} \{J^2,\{G^{ke},\{J^r,G^{r8}\}\}\} \nonumber \\
& & \mbox{\hglue0.6truecm}
- 2 d^{88e} \{J^2,\{G^{ke},\{J^r,G^{rc}\}\}\}
- \frac12 \epsilon^{kim} f^{c8e} \{J^2,\{T^e,\{J^i,G^{m8}\}\}\}
+ 4 \{G^{kc},\{\{J^m,G^{m8}\},\{J^r,G^{r8}\}\}\} \nonumber \\
& & \mbox{\hglue0.6truecm}
- 4 \{G^{k8},\{\{J^m,G^{m8}\},\{J^r,G^{rc}\}\}\}
- 5 \{J^k,\{\{J^m,G^{mc}\},\{G^{r8},G^{r8}\}\}\}
+ 3 \{J^k,\{\{J^m,G^{m8}\},\{G^{r8},G^{rc}\}\}\} \nonumber \\
& & \mbox{\hglue0.6truecm}
+ \frac52 d^{c8e} \{\mathcal{D}_3^{ke},\{J^r,G^{r8}\}\}
+ \frac52 d^{88e} \{\mathcal{D}_3^{kc},\{J^r,G^{re}\}\}
+ \epsilon^{kim} f^{ab8} \{\{J^i,G^{m8}\},\{T^a,\{G^{rb},G^{rc}\}\}\} \nonumber \\
& & \mbox{\hglue0.6truecm}
+ 3 i \epsilon^{kil} [\{J^i,G^{l8}\},\{\{J^m,G^{m8}\},\{J^r,G^{rc}\}\}],
\end{eqnarray}

\begin{eqnarray}
& & [G^{i8},[\mathcal{O}_3^{i8},\mathcal{O}_3^{kc}]] + [\mathcal{O}_3^{i8},[G^{i8},\mathcal{O}_3^{kc}]] \nonumber \\
& & \mbox{\hglue0.2truecm} = \frac98 f^{c8e} f^{8eg} G^{kg} - \frac14 i \epsilon^{kim} f^{c8e} f^{8eg} \{J^i,G^{mg}\} + f^{c8e} f^{8eg} \mathcal{D}_3^{kg} - \frac{23}{64} N_c i d^{8eg} f^{c8e} \mathcal{D}_3^{kg}
- \frac{23}{64} N_c i d^{c8e} f^{8eg} \mathcal{D}_3^{kg} \nonumber \\
& & \mbox{\hglue0.6truecm} + \frac{2}{N_f} \delta^{c8} \mathcal{D}_3^{k8} + 7 f^{c8e} f^{8eg} \mathcal{O}_3^{kg} + \frac{27}{4} d^{c8e} d^{8eg} \mathcal{O}_3^{kg} - \frac{19}{4} d^{ceg} d^{88e} \mathcal{O}_3^{kg} + \frac{24}{N_f} \delta^{c8} \mathcal{O}_3^{k8}
- \frac52 \{G^{kc},\{G^{r8},G^{r8}\}\} \nonumber \\
& & \mbox{\hglue0.6truecm} - \frac32 \{G^{k8},\{G^{rc},G^{r8}\}\} + \frac54 d^{c8e} \{J^k,\{G^{re},G^{r8}\}\} - \frac54 d^{88e} \{J^k,\{G^{rc},G^{re}\}\} - \frac92 d^{c8e} \{G^{ke},\{J^r,G^{r8}\}\} \nonumber \\
& & \mbox{\hglue0.6truecm} + \frac{21}{4} d^{c8e} \{G^{k8},\{J^r,G^{re}\}\} + \frac{19}{4} d^{88e} \{G^{kc},\{J^r,G^{re}\}\} - \frac72 d^{88e} \{G^{ke},\{J^r,G^{rc}\}\} \nonumber \\
& & \mbox{\hglue0.6truecm} + \frac{3(N_f+14)}{16} \epsilon^{kim} f^{c8e} \{T^e,\{J^i,G^{m8}\}\}
+ \frac{23}{32} i d^{8eg} f^{c8e} \mathcal{D}_4^{kg}
+ \frac{23}{16 N_f} i \epsilon^{kim} \delta^{c8} \{J^2,\{J^i,G^{m8}\}\} \nonumber \\
& & \mbox{\hglue0.6truecm}
- \frac12 i f^{c8e} \{\mathcal{D}_2^{ke},\{J^r,G^{r8}\}\}
- \frac{7}{16} i \epsilon^{kim} \{\{J^i,G^{m8}\},\{G^{r8},G^{rc}\}\}
- i \epsilon^{kim} \{\{J^i,G^{mc}\},\{G^{r8},G^{r8}\}\} \nonumber \\
& & \mbox{\hglue0.6truecm}
+ i \epsilon^{rim} \{G^{k8},\{J^r,\{G^{ic},G^{m8}\}\}\}
+ \frac{23}{32} i \epsilon^{rim} d^{c8e} \{J^k,\{J^r,\{G^{i8},G^{me}\}\}\}
+ \frac{69}{128} i \epsilon^{kim} f^{cae} f^{8eb} \{\{J^i,G^{m8}\},\{T^a,T^b\}\} \nonumber \\
& & \mbox{\hglue0.6truecm}
- \frac{7}{32} i f^{c8e} \{J^k,[\{J^i,G^{ie}\},\{J^r,G^{r8}\}]\}
- \frac{7}{32} i f^{c8e} \{\{J^r,G^{re}\},[J^2,G^{k8}]\}
+ \frac{7}{32} i f^{c8e} \{\{J^r,G^{r8}\},[J^2,G^{ke}]\} \nonumber \\
& & \mbox{\hglue0.6truecm}
+ \frac{7}{32} i f^{c8e} \{J^2,[G^{ke},\{J^r,G^{r8}\}]\}
- \frac{7}{32} i f^{c8e} \{J^2,[G^{k8},\{J^r,G^{re}\}]\} \nonumber \\
& & \mbox{\hglue0.6truecm}
- \frac{23}{32} d^{c8e} \{J^2,[G^{ke},\{J^r,G^{r8}\}]\}
+ \frac{23}{32} d^{c8e} \{J^2,[G^{k8},\{J^r,G^{re}\}]\}
+ \frac{39}{64} [G^{kc},\{\{J^m,G^{m8}\},\{J^r,G^{r8}\}\}] \nonumber \\
& & \mbox{\hglue0.6truecm}
+ \frac{7}{32} [G^{k8},\{\{J^m,G^{m8}\},\{J^r,G^{rc}\}\}]
- \frac{7}{32} \{\{J^m,G^{mc}\},[G^{k8},\{J^r,G^{r8}\}]\} \nonumber \\
& & \mbox{\hglue0.6truecm}
- \frac{23}{32} i \epsilon^{kim} f^{cea} f^{e8b} \{\{J^i,G^{m8}\},\{G^{ra},G^{rb}\}\}
+ \frac{1}{N_f} \delta^{c8} \mathcal{D}_5^{k8}
+ \frac52 f^{c8e} f^{8eg} \mathcal{O}_5^{kg}
+ \frac52 d^{c8e} d^{8eg} \mathcal{O}_5^{kg}
- 2 d^{ceg} d^{88e} \mathcal{O}_5^{kg} \nonumber \\
& & \mbox{\hglue0.6truecm}
+ \frac{10}{N_f} \delta^{c8} \mathcal{O}_5^{k8}
- 6 \{J^2,\{G^{kc},\{G^{r8},G^{r8}\}\}\}
- 2 \{J^2,\{G^{k8},\{G^{rc},G^{r8}\}\}\}
+ \frac12 d^{c8e} \{J^2,\{J^k,\{G^{re},G^{r8}\}\}\} \nonumber \\
& & \mbox{\hglue0.6truecm}
- \frac12 d^{88e} \{J^2,\{J^k,\{G^{rc},G^{re}\}\}\}
+ d^{c8e} \{J^2,\{G^{ke},\{J^r,G^{r8}\}\}\}
+ \frac52 d^{c8e} \{J^2,\{G^{k8},\{J^r,G^{re}\}\}\} \nonumber \\
& & \mbox{\hglue0.6truecm}
+ 2 d^{88e} \{J^2,\{G^{kc},\{J^r,G^{re}\}\}\}
+ \frac32 d^{88e} \{J^2,\{G^{ke},\{J^r,G^{rc}\}\}\}
+ \frac54 \epsilon^{kim} f^{c8e} \{J^2,\{T^e,\{J^i,G^{m8}\}\}\} \nonumber \\
& & \mbox{\hglue0.6truecm}
- \frac12 \{G^{kc},\{\{J^m,G^{m8}\},\{J^r,G^{r8}\}\}\}
+ \frac32 \{G^{k8},\{\{J^m,G^{m8}\},\{J^r,G^{rc}\}\}\}
+ \frac52 \{J^k,\{\{J^m,G^{mc}\},\{G^{r8},G^{r8}\}\}\} \nonumber \\
& & \mbox{\hglue0.6truecm}
- \frac12 \{J^k,\{\{J^m,G^{m8}\},\{G^{r8},G^{rc}\}\}\}
- \frac32 d^{c8e} \{\mathcal{D}_3^{ke},\{J^r,G^{r8}\}\}
- \frac32 d^{88e} \{\mathcal{D}_3^{kc},\{J^r,G^{re}\}\} \nonumber \\
& & \mbox{\hglue0.6truecm}
- \frac12 \epsilon^{kim} f^{ab8} \{\{J^i,G^{m8}\},\{T^a,\{G^{rb},G^{rc}\}\}\}
- 2 i \epsilon^{kil} [\{J^i,G^{l8}\},\{\{J^m,G^{m8}\},\{J^r,G^{rc}\}\}],
\end{eqnarray}

\begin{equation}
[\mathcal{D}_2^{i8},[\mathcal{D}_2^{i8},\mathcal{D}_3^{kc}]] = - \frac12 f^{c8e} f^{8eg} \mathcal{D}_3^{kg} + \frac12 f^{c8e} f^{8eg} \mathcal{D}_5^{kg} + \{\mathcal{D}_2^{k8},\{T^8,\{J^r,G^{rc}\}\}\},
\end{equation}

\begin{eqnarray}
[\mathcal{D}_2^{i8},[\mathcal{D}_2^{i8},\mathcal{O}_3^{kc}]] & = & - \frac14 f^{c8e} f^{8eg} \mathcal{D}_3^{kg} - f^{c8e} f^{8eg} \mathcal{O}_3^{kg} + \frac12 f^{c8e} f^{8eg} \mathcal{O}_5^{kg} + \frac12 \{J^2,\{G^{kc},\{T^8,T^8\}\}\} \nonumber \\
& & \mbox{} - \frac12 \epsilon^{kim} f^{c8e} \{J^2,\{T^8,\{J^i,G^{me}\}\}\} - \frac12 \{\mathcal{D}_2^{k8},\{T^8,\{J^r,G^{rc}\}\}\},
\end{eqnarray}

\begin{equation}
[\mathcal{D}_2^{i8},[\mathcal{D}_3^{i8},\mathcal{D}_2^{kc}]] + [\mathcal{D}_3^{i8},[\mathcal{D}_2^{i8},\mathcal{D}_2^{kc}]] = - 2 f^{c8e} f^{8eg} \mathcal{D}_3^{kg} + f^{c8e} f^{8eg} \mathcal{D}_5^{kg} + 2 \{\mathcal{D}_2^{kc},\{T^8,\{J^r,G^{r8}\}\}\},
\end{equation}

\begin{eqnarray}
& & [\mathcal{D}_2^{i8},[\mathcal{O}_3^{i8},\mathcal{D}_2^{kc}]] + [\mathcal{O}_3^{i8},[\mathcal{D}_2^{i8},\mathcal{D}_2^{kc}]] \nonumber \\
& & \mbox{\hglue0.2truecm} = \frac12 f^{c8e} f^{8eg} \mathcal{D}_3^{kg} - f^{c8e} f^{8eg} \mathcal{O}_3^{kg} + \{J^2,\{G^{k8},\{T^c,T^8\}\}\} - \frac12 \epsilon^{kim} f^{c8e} \{J^2,\{T^e,\{J^i,G^{m8}\}\}\} \nonumber \\
& & \mbox{\hglue0.6truecm} + \frac12 \epsilon^{kim} f^{c8e} \{J^2,\{T^8,\{J^i,G^{me}\}\}\} - \{\mathcal{D}_2^{kc},\{T^8,\{J^r,G^{r8}\}\}\},
\end{eqnarray}

\begin{eqnarray}
& & [\mathcal{D}_3^{i8},[\mathcal{D}_3^{i8},G^{kc}]] \nonumber \\
& & \mbox{\hglue0.2truecm} = - \frac{15}{2} f^{c8e} f^{8eg} G^{kg} - \frac34 i \epsilon^{kim} f^{c8e} f^{8eg} \{J^i,G^{mg}\} - \frac74 f^{c8e} f^{8eg} \mathcal{D}_3^{kg} + \frac58 N_c i d^{8eg} f^{c8e} \mathcal{D}_3^{kg} + \frac58 N_c i d^{c8e} f^{8eg} \mathcal{D}_3^{kg} \nonumber \\
& & \mbox{\hglue0.6truecm} + \frac32 f^{c8e} f^{8eg} \mathcal{O}_3^{kg} - 5 d^{c8e} d^{8eg} \mathcal{O}_3^{kg} + 3 d^{ceg} d^{88e} \mathcal{O}_3^{kg} - 10 \{G^{kc},\{G^{r8},G^{r8}\}\} + 10 \{G^{k8},\{G^{rc},G^{r8}\}\} \nonumber \\
& & \mbox{\hglue0.6truecm} + 5 d^{c8e} \{J^k,\{G^{re},G^{r8}\}\} - 5 d^{88e} \{J^k,\{G^{rc},G^{re}\}\} - 10 d^{c8e} \{G^{ke},\{J^r,G^{r8}\}\} + 5 d^{c8e} \{G^{k8},\{J^r,G^{re}\}\} \nonumber \\
& & \mbox{\hglue0.6truecm} - 3 d^{88e} \{G^{kc},\{J^r,G^{re}\}\} + 8 d^{88e} \{G^{ke},\{J^r,G^{rc}\}\} - \frac52 \epsilon^{kim} f^{c8e} \{T^e,\{J^i,G^{m8}\}\}
- \frac54 i d^{8eg} f^{c8e} \mathcal{D}_4^{kg} \nonumber \\
& & \mbox{\hglue0.6truecm}
- \frac{5}{2N_f} i \epsilon^{kim} \delta^{c8} \{J^2,\{J^i,G^{m8}\}\}
- \frac32 i f^{c8e} \{\mathcal{D}_2^{ke},\{J^r,G^{r8}\}\}
+ \frac{11}{2} i \epsilon^{kim} \{\{J^i,G^{m8}\},\{G^{r8},G^{rc}\}\} \nonumber \\
& & \mbox{\hglue0.6truecm}
- 3 i \epsilon^{kim} \{\{J^i,G^{mc}\},\{G^{r8},G^{r8}\}\}
+ 3 i \epsilon^{rim} \{G^{k8},\{J^r,\{G^{ic},G^{m8}\}\}\} - \frac54 i \epsilon^{rim} d^{c8e} \{J^k,\{J^r,\{G^{i8},G^{me}\}\}\} \nonumber \\
& & \mbox{\hglue0.6truecm}
- \frac{15}{16} i \epsilon^{kim} f^{cae} f^{8eb} \{\{J^i,G^{m8}\},\{T^a,T^b\}\}
- \frac{13}{4} i f^{c8e} \{J^k,[\{J^i,G^{ie}\},\{J^r,G^{r8}\}]\} \nonumber \\
& & \mbox{\hglue0.6truecm}
- \frac{13}{4} i f^{c8e} \{\{J^r,G^{re}\},[J^2,G^{k8}]\}
+ \frac{13}{4} i f^{c8e} \{\{J^r,G^{r8}\},[J^2,G^{ke}]\}
+ \frac{13}{4} i f^{c8e} \{J^2,[G^{ke},\{J^r,G^{r8}\}]\} \nonumber \\
& & \mbox{\hglue0.6truecm}
- \frac{13}{4} i f^{c8e} \{J^2,[G^{k8},\{J^r,G^{re}\}]\}
+ \frac54 d^{c8e} \{J^2,[G^{ke},\{J^r,G^{r8}\}]\}
- \frac54 d^{c8e} \{J^2,[G^{k8},\{J^r,G^{re}\}]\} \nonumber \\
& & \mbox{\hglue0.6truecm}
+ \frac18 [G^{kc},\{\{J^m,G^{m8}\},\{J^r,G^{r8}\}\}]
- \frac{11}{4} [G^{k8},\{\{J^m,G^{m8}\},\{J^r,G^{rc}\}\}]
+ \frac{11}{4} \{\{J^m,G^{mc}\},[G^{k8},\{J^r,G^{r8}\}]\} \nonumber \\
& & \mbox{\hglue0.6truecm}
+ \frac54 i \epsilon^{kim} f^{cea} f^{e8b} \{\{J^i,G^{m8}\},\{G^{ra},G^{rb}\}\}
+ \frac12 f^{c8e} f^{8eg} \mathcal{D}_5^{kg}
+ d^{c8e} d^{8eg} \mathcal{O}_5^{kg} + 2 \{J^2,\{G^{kc},\{G^{r8},G^{r8}\}\}\} \nonumber \\
& & \mbox{\hglue0.6truecm}
- 2 \{J^2,\{G^{k8},\{G^{rc},G^{r8}\}\}\}
- d^{c8e} \{J^2,\{J^k,\{G^{re},G^{r8}\}\}\}
+ d^{88e} \{J^2,\{J^k,\{G^{rc},G^{re}\}\}\} \nonumber \\
& & \mbox{\hglue0.6truecm}
- d^{c8e} \{J^2,\{G^{k8},\{J^r,G^{re}\}\}\}
- 5 d^{88e} \{J^2,\{G^{ke},\{J^r,G^{rc}\}\}\}
+ \frac12 \epsilon^{kim} f^{c8e} \{J^2,\{T^e,\{J^i,G^{m8}\}\}\} \nonumber \\
& & \mbox{\hglue0.6truecm}
+ 8 \{G^{kc},\{\{J^m,G^{m8}\},\{J^r,G^{r8}\}\}\}
- 6 \{G^{k8},\{\{J^m,G^{m8}\},\{J^r,G^{rc}\}\}\}
+ d^{c8e} \{\mathcal{D}_3^{ke},\{J^r,G^{r8}\}\} \nonumber \\
& & \mbox{\hglue0.6truecm}
+ 2 d^{88e} \{\mathcal{D}_3^{kc},\{J^r,G^{re}\}\}
+ 4 i \epsilon^{kil} [\{J^i,G^{l8}\},\{\{J^m,G^{m8}\},\{J^r,G^{rc}\}\}],
\end{eqnarray}

\begin{eqnarray}
& & [\mathcal{D}_3^{i8},[\mathcal{O}_3^{i8},G^{kc}]] + [\mathcal{O}_3^{i8},[\mathcal{D}_3^{i8},G^{kc}]] \nonumber \\
& & \mbox{\hglue0.2truecm} = - 3 f^{c8e} f^{8eg} G^{kg} - \frac{3}{16} i \epsilon^{kim} f^{c8e} f^{8eg} \{J^i,G^{mg}\} - \frac14 f^{c8e} f^{8eg} \mathcal{D}_3^{kg} + \frac{17}{32} N_c i d^{8eg} f^{c8e} \mathcal{D}_3^{kg} + \frac{17}{32} N_c i d^{c8e} f^{8eg} \mathcal{D}_3^{kg} \nonumber \\
& & \mbox{\hglue0.6truecm} - \frac32 f^{c8e} f^{8eg} \mathcal{O}_3^{kg} - 2 d^{c8e} d^{8eg} \mathcal{O}_3^{kg} + 2 d^{ceg} d^{88e} \mathcal{O}_3^{kg} - 4 \{G^{kc},\{G^{r8},G^{r8}\}\} + 4 \{G^{k8},\{G^{rc},G^{r8}\}\} \nonumber \\
& & \mbox{\hglue0.6truecm} + 2 d^{c8e} \{J^k,\{G^{re},G^{r8}\}\} - 2 d^{88e} \{J^k,\{G^{rc},G^{re}\}\} - 4 d^{c8e} \{G^{ke},\{J^r,G^{r8}\}\} + 2 d^{c8e} \{G^{k8},\{J^r,G^{re}\}\} \nonumber \\
& & \mbox{\hglue0.6truecm} - 2 d^{88e} \{G^{kc},\{J^r,G^{re}\}\} + 4 d^{88e} \{G^{ke},\{J^r,G^{rc}\}\} - \epsilon^{kim} f^{c8e} \{T^e,\{J^i,G^{m8}\}\} - \frac{17}{16} i d^{8eg} f^{c8e} \mathcal{D}_4^{kg} \nonumber \\
& & \mbox{\hglue0.6truecm} - \frac{17}{8N_f} i \epsilon^{kim} \delta^{c8} \{J^2,\{J^i,G^{m8}\}\} - \frac38 i f^{c8e} \{\mathcal{D}_2^{ke},\{J^r,G^{r8}\}\} + \frac{23}{8} i \epsilon^{kim} \{\{J^i,G^{m8}\},\{G^{r8},G^{rc}\}\} \nonumber \\
& & \mbox{\hglue0.6truecm} - \frac34 i \epsilon^{kim} \{\{J^i,G^{mc}\},\{G^{r8},G^{r8}\}\} + \frac34 i \epsilon^{rim} \{G^{k8},\{J^r,\{G^{ic},G^{m8}\}\}\} - \frac{17}{16} i \epsilon^{rim} d^{c8e} \{J^k,\{J^r,\{G^{i8},G^{me}\}\}\} \nonumber \\
& & \mbox{\hglue0.6truecm}
- \frac{51}{64} i \epsilon^{kim} f^{cae} f^{8eb} \{\{J^i,G^{m8}\},\{T^a,T^b\}\}
- \frac{41}{16} i f^{c8e} \{J^k,[\{J^i,G^{ie}\},\{J^r,G^{r8}\}]\}
- \frac{41}{16} i f^{c8e} \{\{J^r,G^{re}\},[J^2,G^{k8}]\} \nonumber \\
& & \mbox{\hglue0.6truecm}
+ \frac{41}{16} i f^{c8e} \{\{J^r,G^{r8}\},[J^2,G^{ke}]\}
+ \frac{41}{16} i f^{c8e} \{J^2,[G^{ke},\{J^r,G^{r8}\}]\}
- \frac{41}{16} i f^{c8e} \{J^2,[G^{k8},\{J^r,G^{re}\}]\} \nonumber \\
& & \mbox{\hglue0.6truecm}
+ \frac{17}{16} d^{c8e} \{J^2,[G^{ke},\{J^r,G^{r8}\}]\}
- \frac{17}{16} d^{c8e} \{J^2,[G^{k8},\{J^r,G^{re}\}]\}
- \frac{11}{32} [G^{kc},\{\{J^m,G^{m8}\},\{J^r,G^{r8}\}\}] \nonumber \\
& & \mbox{\hglue0.6truecm}
- \frac{23}{16} [G^{k8},\{\{J^m,G^{m8}\},\{J^r,G^{rc}\}\}]
+ \frac{23}{16} \{\{J^m,G^{mc}\},[G^{k8},\{J^r,G^{r8}\}]\} \nonumber \\
& & \mbox{\hglue0.6truecm}
+ \frac{17}{16} i \epsilon^{kim} f^{cea} f^{e8b} \{\{J^i,G^{m8}\},\{G^{ra},G^{rb}\}\}
+ \frac{2}{N_f} \delta^{c8} \mathcal{D}_5^{k8}
- 2 \{J^2,\{G^{kc},\{G^{r8},G^{r8}\}\}\} \nonumber \\
& & \mbox{\hglue0.6truecm}
+ 2 \{J^2,\{G^{k8},\{G^{rc},G^{r8}\}\}\}
+ d^{c8e} \{J^2,\{J^k,\{G^{re},G^{r8}\}\}\}
- d^{88e} \{J^2,\{J^k,\{G^{rc},G^{re}\}\}\} \nonumber \\
& & \mbox{\hglue0.6truecm}
- 3 d^{c8e} \{J^2,\{G^{ke},\{J^r,G^{r8}\}\}\}
- 3 d^{88e} \{J^2,\{G^{ke},\{J^r,G^{rc}\}\}\}
- \frac12 \epsilon^{kim} f^{c8e} \{J^2,\{T^e,\{J^i,G^{m8}\}\}\} \nonumber \\
& & \mbox{\hglue0.6truecm}
+ 4 \{G^{kc},\{\{J^m,G^{m8}\},\{J^r,G^{r8}\}\}\}
- 4 \{G^{k8},\{\{J^m,G^{m8}\},\{J^r,G^{rc}\}\}\}
- 2 \{J^k,\{\{J^m,G^{m8}\},\{G^{r8},G^{rc}\}\}\} \nonumber \\
& & \mbox{\hglue0.6truecm}
+ 2 d^{c8e} \{\mathcal{D}_3^{ke},\{J^r,G^{r8}\}\}
+ 2 d^{88e} \{\mathcal{D}_3^{kc},\{J^r,G^{re}\}\}
+ 4 i \epsilon^{kil} [\{J^i,G^{l8}\},\{\{J^m,G^{m8}\},\{J^r,G^{rc}\}\}],
\end{eqnarray}

\begin{eqnarray}
& & [\mathcal{O}_3^{i8},[\mathcal{O}_3^{i8},G^{kc}]] \nonumber \\
& & \mbox{\hglue0.2truecm} = \frac{21}{8} f^{c8e} f^{8eg} G^{kg} + \frac{5}{32} i \epsilon^{kim} f^{c8e} f^{8eg} \{J^i,G^{mg}\} + \frac{9}{16} f^{c8e} f^{8eg} \mathcal{D}_3^{kg} + \frac12 d^{c8e} d^{8eg} \mathcal{D}_3^{kg} - \frac{11}{32} N_c i d^{8eg} f^{c8e} \mathcal{D}_3^{kg} \nonumber \\
& & \mbox{\hglue0.6truecm} - \frac{11}{32} N_c i d^{c8e} f^{8eg} \mathcal{D}_3^{kg} - \frac{1}{N_f} \delta^{c8} \mathcal{D}_3^{k8} - \frac{21}{8} f^{c8e} f^{8eg} \mathcal{O}_3^{kg} - \frac54 d^{c8e} d^{8eg} \mathcal{O}_3^{kg} + \frac14 d^{ceg} d^{88e} \mathcal{O}_3^{kg} - \frac{12}{N_f} \delta^{c8} \mathcal{O}_3^{k8} \nonumber \\
& & \mbox{\hglue0.6truecm} + \frac{1}{N_f} d^{c88} \{J^2,J^k\} + \frac{11}{2} \{G^{kc},\{G^{r8},G^{r8}\}\}
- \frac72 \{G^{k8},\{G^{rc},G^{r8}\}\} - \frac{15}{4} d^{c8e} \{J^k,\{G^{re},G^{r8}\}\} \nonumber \\
& & \mbox{\hglue0.6truecm} + \frac{11}{4} d^{88e} \{J^k,\{G^{rc},G^{re}\}\} + \frac{13}{2} d^{c8e} \{G^{ke},\{J^r,G^{r8}\}\} - \frac{19}{4} d^{c8e} \{G^{k8},\{J^r,G^{re}\}\} - \frac14 d^{88e} \{G^{kc},\{J^r,G^{re}\}\} \nonumber \\
& & \mbox{\hglue0.6truecm} - \frac52 d^{88e} \{G^{ke},\{J^r,G^{rc}\}\} - \frac58 \epsilon^{kim} f^{c8e} \{T^e,\{J^i,G^{m8}\}\}
+ \frac{11}{16} i d^{8eg} f^{c8e} \mathcal{D}_4^{kg}
+ \frac{11}{8N_f} i \epsilon^{kim} \delta^{c8} \{J^2,\{J^i,G^{m8}\}\} \nonumber \\
& & \mbox{\hglue0.6truecm}
+ \frac{5}{16} i f^{c8e} \{\mathcal{D}_2^{ke},\{J^r,G^{r8}\}\}
- 2 i \epsilon^{kim} \{\{J^i,G^{m8}\},\{G^{r8},G^{rc}\}\}
+ \frac58 i \epsilon^{kim} \{\{J^i,G^{mc}\},\{G^{r8},G^{r8}\}\} \nonumber \\
& & \mbox{\hglue0.6truecm}
- \frac58 i \epsilon^{rim} \{G^{k8},\{J^r,\{G^{ic},G^{m8}\}\}\}
+ \frac{11}{16} i \epsilon^{rim} d^{c8e} \{J^k,\{J^r,\{G^{i8},G^{me}\}\}\} \nonumber \\
& & \mbox{\hglue0.6truecm}
+ \frac{33}{64} i \epsilon^{kim} f^{cae} f^{8eb} \{\{J^i,G^{m8}\},\{T^a,T^b\}\}
+ \frac{43}{16} i f^{c8e} \{J^k,[\{J^i,G^{ie}\},\{J^r,G^{r8}\}]\} \nonumber \\
& & \mbox{\hglue0.6truecm}
+ \frac{43}{16} i f^{c8e} \{\{J^r,G^{re}\},[J^2,G^{k8}]\}
- \frac{43}{16} i f^{c8e} \{\{J^r,G^{r8}\},[J^2,G^{ke}]\}
- \frac{43}{16} i f^{c8e} \{J^2,[G^{ke},\{J^r,G^{r8}\}]\} \nonumber \\
& & \mbox{\hglue0.6truecm}
+ \frac{43}{16} i f^{c8e} \{J^2,[G^{k8},\{J^r,G^{re}\}]\}
- \frac{11}{16} d^{c8e} \{J^2,[G^{ke},\{J^r,G^{r8}\}]\}
+ \frac{11}{16} d^{c8e} \{J^2,[G^{k8},\{J^r,G^{re}\}]\} \nonumber \\
& & \mbox{\hglue0.6truecm}
+ \frac{3}{16} [G^{kc},\{\{J^m,G^{m8}\},\{J^r,G^{r8}\}\}]
+ [G^{k8},\{\{J^m,G^{m8}\},\{J^r,G^{rc}\}\}]
- \{\{J^m,G^{mc}\},[G^{k8},\{J^r,G^{r8}\}]\} \nonumber \\
& & \mbox{\hglue0.6truecm}
- \frac{11}{16} i \epsilon^{kim} f^{cea} f^{e8b} \{\{J^i,G^{m8}\},\{G^{ra},G^{rb}\}\}
+ \frac14 d^{c8e} d^{8eg} \mathcal{D}_5^{kg}
- \frac{1}{2N_f} \delta^{c8} \mathcal{D}_5^{k8}
+ \frac14 f^{c8e} f^{8eg} \mathcal{O}_5^{kg} \nonumber \\
& & \mbox{\hglue0.6truecm}
+ \frac14 d^{c8e} d^{8eg} \mathcal{O}_5^{kg}
+ \frac{1}{N_f} \delta^{c8} \mathcal{O}_5^{k8}
+ \frac{1}{2N_f} d^{c88} \{J^2,\{J^2,J^k\}\}
+ \frac52 \{J^2,\{G^{kc},\{G^{r8},G^{r8}\}\}\} \nonumber \\
& & \mbox{\hglue0.6truecm}
- \frac12 \{J^2,\{G^{k8},\{G^{rc},G^{r8}\}\}\}
- \frac54 d^{c8e} \{J^2,\{J^k,\{G^{re},G^{r8}\}\}\}
+ \frac14 d^{88e} \{J^2,\{J^k,\{G^{rc},G^{re}\}\}\} \nonumber \\
& & \mbox{\hglue0.6truecm}
+ \frac52 d^{c8e} \{J^2,\{G^{ke},\{J^r,G^{r8}\}\}\}
+ \frac14 d^{c8e} \{J^2,\{G^{k8},\{J^r,G^{re}\}\}\}
+ \frac{11}{4} d^{88e} \{J^2,\{G^{ke},\{J^r,G^{rc}\}\}\} \nonumber \\
& & \mbox{\hglue0.6truecm}
+ \frac18 \epsilon^{kim} f^{c8e} \{J^2,\{T^e,\{J^i,G^{m8}\}\}\}
- 4 \{G^{kc},\{\{J^m,G^{m8}\},\{J^r,G^{r8}\}\}\}
+ \frac72 \{G^{k8},\{\{J^m,G^{m8}\},\{J^r,G^{rc}\}\}\} \nonumber \\
& & \mbox{\hglue0.6truecm}
- \{J^k,\{\{J^m,G^{mc}\},\{G^{r8},G^{r8}\}\}\}
+ \{J^k,\{\{J^m,G^{m8}\},\{G^{r8},G^{rc}\}\}\}
- \frac54 d^{c8e} \{\mathcal{D}_3^{ke},\{J^r,G^{r8}\}\} \nonumber \\
& & \mbox{\hglue0.6truecm}
- \frac32 d^{88e} \{\mathcal{D}_3^{kc},\{J^r,G^{re}\}\}
- 3 i \epsilon^{kil} [\{J^i,G^{l8}\},\{\{J^m,G^{m8}\},\{J^r,G^{rc}\}\}],
\end{eqnarray}

\begin{eqnarray}
& & [\mathcal{D}_2^{i8},[\mathcal{D}_3^{i8},\mathcal{D}_3^{kc}]] + [\mathcal{D}_3^{i8},[\mathcal{D}_2^{i8},\mathcal{D}_3^{kc}]] \nonumber \\
& & \mbox{\hglue0.2truecm} = \frac{7}{44} i \epsilon^{kim} f^{c8e} f^{8eg} \{J^i,G^{mg}\} - \frac{3}{22} N_c i d^{8eg} f^{c8e} \mathcal{D}_3^{kg} - \frac{3}{22} N_c i d^{c8e} f^{8eg} \mathcal{D}_3^{kg} + \frac{3}{11} i d^{8eg} f^{c8e} \mathcal{D}_4^{kg} \nonumber \\
& & \mbox{\hglue0.6truecm}
+ \frac{6}{11 N_f} i \epsilon^{kim} \delta^{c8} \{J^2,\{J^i,G^{m8}\}\}
+ \frac{7}{22} i f^{c8e} \{\mathcal{D}_2^{ke},\{J^r,G^{r8}\}\}
- \frac{13}{11} i \epsilon^{kim} \{\{J^i,G^{m8}\},\{G^{r8},G^{rc}\}\} \nonumber \\
& & \mbox{\hglue0.6truecm}
+ \frac{7}{11} i \epsilon^{kim} \{\{J^i,G^{mc}\},\{G^{r8},G^{r8}\}\}
- \frac{7}{11} i \epsilon^{rim} \{G^{k8},\{J^r,\{G^{ic},G^{m8}\}\}\}
+ \frac{3}{11} i \epsilon^{rim} d^{c8e} \{J^k,\{J^r,\{G^{i8},G^{me}\}\}\} \nonumber \\
& & \mbox{\hglue0.6truecm}
+ \frac{9}{44} i \epsilon^{kim} f^{cae} f^{8eb} \{\{J^i,G^{m8}\},\{T^a,T^b\}\}
- \frac{23}{11} i f^{c8e} \{J^k,[\{J^i,G^{ie}\},\{J^r,G^{r8}\}]\}
- \frac{1}{11} i f^{c8e} \{\{J^r,G^{re}\},[J^2,G^{k8}]\} \nonumber \\
& & \mbox{\hglue0.6truecm}
+ \frac{1}{11} i f^{c8e} \{\{J^r,G^{r8}\},[J^2,G^{ke}]\}
+ \frac{1}{11} i f^{c8e} \{J^2,[G^{ke},\{J^r,G^{r8}\}]\}
+ \frac{3}{11} i f^{c8e} \{J^2,[G^{k8},\{J^r,G^{re}\}]\} \nonumber \\
& & \mbox{\hglue0.6truecm}
- \frac{3}{11} d^{c8e} \{J^2,[G^{ke},\{J^r,G^{r8}\}]\}
+ \frac{3}{11} d^{c8e} \{J^2,[G^{k8},\{J^r,G^{re}\}]\}
- \frac{1}{44} [G^{kc},\{\{J^m,G^{m8}\},\{J^r,G^{r8}\}\}] \nonumber \\
& & \mbox{\hglue0.6truecm}
+ \frac{13}{22} [G^{k8},\{\{J^m,G^{m8}\},\{J^r,G^{rc}\}\}]
- \frac{13}{22} \{\{J^m,G^{mc}\},[G^{k8},\{J^r,G^{r8}\}]\} \nonumber \\
& & \mbox{\hglue0.6truecm}
- \frac{3}{11} i \epsilon^{kim} f^{cea} f^{e8b} \{\{J^i,G^{m8}\},\{G^{ra},G^{rb}\}\}
- \frac{4}{11} d^{c8e} \{J^2,\{J^2,\{G^{ke},T^8\}\}\}
+ \frac{4}{11} d^{c8e} \{J^2,\{\mathcal{D}_2^{k8},\{J^r,G^{re}\}\}\} \nonumber \\
& & \mbox{\hglue0.6truecm}
+ \frac{4}{11} \{J^2,\{\{J^r,G^{rc}\},\{G^{k8},T^8\}\}\}
- \frac{4}{11} \{J^2,\{\{J^r,G^{r8}\},\{G^{kc},T^8\}\}\}
+ 2 i f^{c8e} \{J^2,\{J^k,[\{J^i,G^{ie}\},\{J^r,G^{r8}\}]\}\} \nonumber \\
& & \mbox{\hglue0.6truecm}
+ \frac{4}{11} i f^{c8e} \{J^2,\{\{J^r,G^{r8}\},[J^2,G^{ke}]\}\}
+ 4 \{\mathcal{D}_2^{k8},\{\{J^m,G^{mc}\},\{J^r,G^{r8}\}\}\} \nonumber \\
& & \mbox{\hglue0.6truecm}
- \frac{4}{11} i \epsilon^{kim} [\{T^8,\{J^r,G^{r8}\}\},\{J^2,\{J^i,G^{mc}\}\}],
\end{eqnarray}

\begin{eqnarray}
& & [\mathcal{D}_2^{i8},[\mathcal{D}_3^{i8},\mathcal{O}_3^{kc}]] + [\mathcal{D}_3^{i8},[\mathcal{D}_2^{i8},\mathcal{O}_3^{kc}]] \nonumber \\
& & \mbox{\hglue0.2truecm} = - \frac{7}{11} i \epsilon^{kim} f^{c8e} f^{8eg} \{J^i,G^{mg}\} + \frac{137}{352} N_c i d^{8eg} f^{c8e} \mathcal{D}_3^{kg} + \frac{137}{352} N_c i d^{c8e} f^{8eg} \mathcal{D}_3^{kg} - \frac{137}{176} i d^{8eg} f^{c8e} \mathcal{D}_4^{kg} \nonumber \\
& & \mbox{\hglue0.6truecm}
- \frac{137}{88 N_f} i \epsilon^{kim} \delta^{c8} \{J^2,\{J^i,G^{m8}\}\}
- \frac{14}{11} i f^{c8e} \{\mathcal{D}_2^{ke},\{J^r,G^{r8}\}\}
+ \frac{361}{88} i \epsilon^{kim} \{\{J^i,G^{m8}\},\{G^{r8},G^{rc}\}\} \nonumber \\
& & \mbox{\hglue0.6truecm}
- \frac{28}{11} i \epsilon^{kim} \{\{J^i,G^{mc}\},\{G^{r8},G^{r8}\}\}
+ \frac{28}{11} i \epsilon^{rim} \{G^{k8},\{J^r,\{G^{ic},G^{m8}\}\}\}
- \frac{137}{176} i \epsilon^{rim} d^{c8e} \{J^k,\{J^r,\{G^{i8},G^{me}\}\}\} \nonumber \\
& & \mbox{\hglue0.6truecm}
- \frac{411}{704} i \epsilon^{kim} f^{cae} f^{8eb} \{\{J^i,G^{m8}\},\{T^a,T^b\}\}
- \frac{79}{176} i f^{c8e} \{J^k,[\{J^i,G^{ie}\},\{J^r,G^{r8}\}]\} \nonumber \\
& & \mbox{\hglue0.6truecm}
- \frac{255}{176} i f^{c8e} \{\{J^r,G^{re}\},[J^2,G^{k8}]\}
+ \frac{255}{176} i f^{c8e} \{\{J^r,G^{r8}\},[J^2,G^{ke}]\}
- \frac{97}{176} i f^{c8e} \{J^2,[G^{ke},\{J^r,G^{r8}\}]\} \nonumber \\
& & \mbox{\hglue0.6truecm}
- \frac{511}{176} i f^{c8e} \{J^2,[G^{k8},\{J^r,G^{re}\}]\}
+ \frac{137}{176} d^{c8e} \{J^2,[G^{ke},\{J^r,G^{r8}\}]\}
- \frac{137}{176} d^{c8e} \{J^2,[G^{k8},\{J^r,G^{re}\}]\} \nonumber \\
& & \mbox{\hglue0.6truecm}
+ \frac{87}{352} [G^{kc},\{\{J^m,G^{m8}\},\{J^r,G^{r8}\}\}]
- \frac{361}{176} [G^{k8},\{\{J^m,G^{m8}\},\{J^r,G^{rc}\}\}] \nonumber \\
& & \mbox{\hglue0.6truecm}
+ \frac{361}{176} \{\{J^m,G^{mc}\},[G^{k8},\{J^r,G^{r8}\}]\}
+ \frac{137}{176} i \epsilon^{kim} f^{cea} f^{e8b} \{\{J^i,G^{m8}\},\{G^{ra},G^{rb}\}\} \nonumber \\
& & \mbox{\hglue0.6truecm}
+ \frac{16}{11} d^{c8e} \{J^2,\{J^2,\{G^{ke},T^8\}\}\}
- \frac{16}{11} d^{c8e} \{J^2,\{\mathcal{D}_2^{k8},\{J^r,G^{re}\}\}\}
- \frac{16}{11} \{J^2,\{\{J^r,G^{rc}\},\{G^{k8},T^8\}\}\} \nonumber \\
& & \mbox{\hglue0.6truecm}
+ \frac{38}{11} \{J^2,\{\{J^r,G^{r8}\},\{G^{kc},T^8\}\}\}
- i f^{c8e} \{J^2,\{J^k,[\{J^i,G^{ie}\},\{J^r,G^{r8}\}]\}\} \nonumber \\
& & \mbox{\hglue0.6truecm}
- \frac{16}{11} i f^{c8e} \{J^2,\{\{J^r,G^{r8}\},[J^2,G^{ke}]\}\}
+ 2 i f^{c8e} \{J^2,\{J^2,[G^{ke},\{J^r,G^{r8}\}]\}\} \nonumber \\
& & \mbox{\hglue0.6truecm}
- 2 \{\mathcal{D}_2^{k8},\{\{J^m,G^{mc}\},\{J^r,G^{r8}\}\}\}
+ \frac{5}{11} i \epsilon^{kim} [\{T^8,\{J^r,G^{r8}\}\},\{J^2,\{J^i,G^{mc}\}\}],
\end{eqnarray}

\begin{eqnarray}
& & [\mathcal{D}_2^{i8},[\mathcal{O}_3^{i8},\mathcal{D}_3^{kc}]] + [\mathcal{O}_3^{i8},[\mathcal{D}_2^{i8},\mathcal{D}_3^{kc}]] \nonumber \\
& & \mbox{\hglue0.2truecm} = \frac{63}{176} i \epsilon^{kim} f^{c8e} f^{8eg} \{J^i,G^{mg}\} + \frac{67}{176} N_c i d^{8eg} f^{c8e} \mathcal{D}_3^{kg} + \frac{67}{176} N_c i d^{c8e} f^{8eg} \mathcal{D}_3^{kg} - \frac{67}{88} i d^{8eg} f^{c8e} \mathcal{D}_4^{kg} \nonumber \\
& & \mbox{\hglue0.6truecm}
- \frac{67}{44 N_f} i \epsilon^{kim} \delta^{c8} \{J^2,\{J^i,G^{m8}\}\}
+ \frac{63}{88} i f^{c8e} \{\mathcal{D}_2^{ke},\{J^r,G^{r8}\}\}
+ \frac{1}{11} i \epsilon^{kim} \{\{J^i,G^{m8}\},\{G^{r8},G^{rc}\}\} \nonumber \\
& & \mbox{\hglue0.6truecm}
+ \frac{63}{44} i \epsilon^{kim} \{\{J^i,G^{mc}\},\{G^{r8},G^{r8}\}\}
- \frac{63}{44} i \epsilon^{rim} \{G^{k8},\{J^r,\{G^{ic},G^{m8}\}\}\}
- \frac{67}{88} i \epsilon^{rim} d^{c8e} \{J^k,\{J^r,\{G^{i8},G^{me}\}\}\} \nonumber \\
& & \mbox{\hglue0.6truecm}
- \frac{201}{352} i \epsilon^{kim} f^{cae} f^{8eb} \{\{J^i,G^{m8}\},\{T^a,T^b\}\}
+ \frac{101}{44} i f^{c8e} \{J^k,[\{J^i,G^{ie}\},\{J^r,G^{r8}\}]\} \nonumber \\
& & \mbox{\hglue0.6truecm}
+ \frac{57}{44} i f^{c8e} \{\{J^r,G^{re}\},[J^2,G^{k8}]\}
- \frac{57}{44} i f^{c8e} \{\{J^r,G^{r8}\},[J^2,G^{ke}]\}
- \frac{57}{44} i f^{c8e} \{J^2,[G^{ke},\{J^r,G^{r8}\}]\} \nonumber \\
& & \mbox{\hglue0.6truecm}
+ \frac{269}{44} i f^{c8e} \{J^2,[G^{k8},\{J^r,G^{re}\}]\}
+ \frac{67}{88} d^{c8e} \{J^2,[G^{ke},\{J^r,G^{r8}\}]\}
- \frac{67}{88} d^{c8e} \{J^2,[G^{k8},\{J^r,G^{re}\}]\} \nonumber \\
& & \mbox{\hglue0.6truecm}
- \frac{65}{88} [G^{kc},\{\{J^m,G^{m8}\},\{J^r,G^{r8}\}\}]
- \frac{1}{22} [G^{k8},\{\{J^m,G^{m8}\},\{J^r,G^{rc}\}\}]
+ \frac{1}{22} \{\{J^m,G^{mc}\},[G^{k8},\{J^r,G^{r8}\}]\} \nonumber \\
& & \mbox{\hglue0.6truecm}
+ \frac{67}{88} i \epsilon^{kim} f^{cea} f^{e8b} \{\{J^i,G^{m8}\},\{G^{ra},G^{rb}\}\}
+ \frac{13}{11} d^{c8e} \{J^2,\{J^2,\{G^{ke},T^8\}\}\}
- \frac{13}{11} d^{c8e} \{J^2,\{\mathcal{D}_2^{k8},\{J^r,G^{re}\}\}\} \nonumber \\
& & \mbox{\hglue0.6truecm}
+ \frac{31}{11} \{J^2,\{\{J^r,G^{rc}\},\{G^{k8},T^8\}\}\}
- \frac{9}{11} \{J^2,\{\{J^r,G^{r8}\},\{G^{kc},T^8\}\}\}
- i f^{c8e} \{J^2,\{\{J^r,G^{re}\},[J^2,G^{k8}]\}\} \nonumber \\
& & \mbox{\hglue0.6truecm}
- \frac{2}{11} i f^{c8e} \{J^2,\{\{J^r,G^{r8}\},[J^2,G^{ke}]\}\}
- 2 \{\mathcal{D}_2^{k8},\{\{J^m,G^{mc}\},\{J^r,G^{r8}\}\}\} \nonumber \\
& & \mbox{\hglue0.6truecm}
+ \frac{2}{11} i \epsilon^{kim} [\{T^8,\{J^r,G^{r8}\}\},\{J^2,\{J^i,G^{mc}\}\}],
\end{eqnarray}

\begin{eqnarray}
& & [\mathcal{D}_2^{i8},[\mathcal{O}_3^{i8},\mathcal{O}_3^{kc}]] + [\mathcal{O}_3^{i8},[\mathcal{D}_2^{i8},\mathcal{O}_3^{kc}]] \nonumber \\
& & \mbox{\hglue0.2truecm} = 3 f^{c8e} f^{8eg} \mathcal{D}_2^{kg} - \frac{23}{22} i \epsilon^{kim} f^{c8e} f^{8eg} \{J^i,G^{mg}\} - \frac{45}{176} N_c i d^{8eg} f^{c8e} \mathcal{D}_3^{kg} - \frac{45}{176} N_c i d^{c8e} f^{8eg} \mathcal{D}_3^{kg} + 4 f^{c8e} f^{8eg} \mathcal{D}_4^{kg} \nonumber \\
& & \mbox{\hglue0.6truecm}
+ \frac{45}{88} i d^{8eg} f^{c8e} \mathcal{D}_4^{kg}
+ \frac{4}{N_f} \delta^{c8} \mathcal{D}_4^{k8}
+ \frac{45}{44 N_f} i \epsilon^{kim} \delta^{c8} \{J^2,\{J^i,G^{m8}\}\}
- 4 \{\mathcal{D}_2^{k8},\{G^{rc},G^{r8}\}\}
+ 2 d^{c8e} \{\mathcal{D}_2^{k8},\{J^r,G^{re}\}\} \nonumber \\
& & \mbox{\hglue0.6truecm}
- \frac{23}{11} i f^{c8e} \{\mathcal{D}_2^{ke},\{J^r,G^{r8}\}\}
+ \frac{139}{44} i \epsilon^{kim} \{\{J^i,G^{m8}\},\{G^{r8},G^{rc}\}\}
- \frac{46}{11} i \epsilon^{kim} \{\{J^i,G^{mc}\},\{G^{r8},G^{r8}\}\} \nonumber \\
& & \mbox{\hglue0.6truecm}
+ \frac{46}{11} i \epsilon^{rim} \{G^{k8},\{J^r,\{G^{ic},G^{m8}\}\}\}
+ \frac{45}{88} i \epsilon^{rim} d^{c8e} \{J^k,\{J^r,\{G^{i8},G^{me}\}\}\} \nonumber \\
& & \mbox{\hglue0.6truecm}
+ \frac{135}{352} i \epsilon^{kim} f^{cae} f^{8eb} \{\{J^i,G^{m8}\},\{T^a,T^b\}\}
- \frac{123}{44} i f^{c8e} \{J^k,[\{J^i,G^{ie}\},\{J^r,G^{r8}\}]\} \nonumber \\
& & \mbox{\hglue0.6truecm}
- \frac{57}{44} i f^{c8e} \{\{J^r,G^{re}\},[J^2,G^{k8}]\}
+ \frac{57}{44} i f^{c8e} \{\{J^r,G^{r8}\},[J^2,G^{ke}]\}
+ \frac{57}{44} i f^{c8e} \{J^2,[G^{ke},\{J^r,G^{r8}\}]\} \nonumber \\
& & \mbox{\hglue0.6truecm}
- \frac{93}{44} i f^{c8e} \{J^2,[G^{k8},\{J^r,G^{re}\}]\}
- \frac{45}{88} d^{c8e} \{J^2,[G^{ke},\{J^r,G^{r8}\}]\}
+ \frac{45}{88} d^{c8e} \{J^2,[G^{k8},\{J^r,G^{re}\}]\} \nonumber \\
& & \mbox{\hglue0.6truecm}
+ \frac{229}{176} [G^{kc},\{\{J^m,G^{m8}\},\{J^r,G^{r8}\}\}]
- \frac{139}{88} [G^{k8},\{\{J^m,G^{m8}\},\{J^r,G^{rc}\}\}] \nonumber \\
& & \mbox{\hglue0.6truecm}
+ \frac{139}{88} \{\{J^m,G^{mc}\},[G^{k8},\{J^r,G^{r8}\}]\}
- \frac{45}{88} i \epsilon^{kim} f^{cea} f^{e8b} \{\{J^i,G^{m8}\},\{G^{ra},G^{rb}\}\}
+ \frac12 f^{c8e} f^{8eg} \mathcal{D}_6^{kg} \nonumber \\
& & \mbox{\hglue0.6truecm}
+ \frac{2}{N_f} \delta^{c8} \mathcal{D}_6^{k8}
+ \frac{9}{11} d^{c8e} \{J^2,\{J^2,\{G^{ke},T^8\}\}\}
- 4 \{J^2,\{\mathcal{D}_2^{k8},\{G^{rc},G^{r8}\}\}\}
+ \frac{2}{11} d^{c8e} \{J^2,\{\mathcal{D}_2^{k8},\{J^r,G^{re}\}\}\} \nonumber \\
& & \mbox{\hglue0.6truecm}
- \frac{9}{11} \{J^2,\{\{J^r,G^{rc}\},\{G^{k8},T^8\}\}\}
+ \frac{9}{11} \{J^2,\{\{J^r,G^{r8}\},\{G^{kc},T^8\}\}\}
+ \frac12 i f^{c8e} \{J^2,\{\{J^r,G^{re}\},[J^2,G^{k8}]\}\} \nonumber \\
& & \mbox{\hglue0.6truecm}
- \frac{29}{22} i f^{c8e} \{J^2,\{\{J^r,G^{r8}\},[J^2,G^{ke}]\}\}
- \frac12 i f^{c8e} \{J^2,\{J^2,[G^{ke},\{J^r,G^{r8}\}]\}\} \nonumber \\
& & \mbox{\hglue0.6truecm}
+ \frac12 i f^{c8e} \{J^2,\{J^2,[G^{k8},\{J^r,G^{re}\}]\}\}
+ \{\mathcal{D}_2^{k8},\{\{J^m,G^{mc}\},\{J^r,G^{r8}\}\}\} \nonumber \\
& & \mbox{\hglue0.6truecm}
+ \frac{9}{11} i \epsilon^{kim} [\{T^8,\{J^r,G^{r8}\}\},\{J^2,\{J^i,G^{mc}\}\}],
\end{eqnarray}

\begin{eqnarray}
& & [\mathcal{D}_3^{i8},[\mathcal{D}_3^{i8},\mathcal{D}_2^{kc}]] \nonumber \\
& & \mbox{\hglue0.6truecm} = - 3 f^{c8e} f^{8eg} G^{kg} - \frac14 i \epsilon^{kim} f^{c8e} f^{8eg} \{J^i,G^{mg}\} - f^{c8e} f^{8eg} \mathcal{D}_3^{kg} + \frac38 N_c i d^{8eg} f^{c8e} \mathcal{D}_3^{kg} + \frac38 N_c i d^{c8e} f^{8eg} \mathcal{D}_3^{kg} \nonumber \\
& & \mbox{\hglue0.6truecm} - 2 d^{c8e} d^{8eg} \mathcal{O}_3^{kg} + 2 d^{ceg} d^{88e} \mathcal{O}_3^{kg} - 4 \{G^{kc},\{G^{r8},G^{r8}\}\} + 4 \{G^{k8},\{G^{rc},G^{r8}\}\} + 2 d^{c8e} \{J^k,\{G^{re},G^{r8}\}\} \nonumber \\
& & \mbox{\hglue0.6truecm} - 2 d^{88e} \{J^k,\{G^{rc},G^{re}\}\} - 4 d^{c8e} \{G^{ke},\{J^r,G^{r8}\}\} + 2 d^{c8e} \{G^{k8},\{J^r,G^{re}\}\} - 2 d^{88e} \{G^{kc},\{J^r,G^{re}\}\} \nonumber \\
& & \mbox{\hglue0.6truecm} + 4 d^{88e} \{G^{ke},\{J^r,G^{rc}\}\} - \epsilon^{kim} f^{c8e} \{T^e,\{J^i,G^{m8}\}\}
- \frac34 i d^{8eg} f^{c8e} \mathcal{D}_4^{kg}
- \frac{3}{2N_f} i \epsilon^{kim} \delta^{c8} \{J^2,\{J^i,G^{m8}\}\} \nonumber \\
& & \mbox{\hglue0.6truecm}
- \frac12 i f^{c8e} \{\mathcal{D}_2^{ke},\{J^r,G^{r8}\}\}
+ \frac52 i \epsilon^{kim} \{\{J^i,G^{m8}\},\{G^{r8},G^{rc}\}\} - i \epsilon^{kim} \{\{J^i,G^{mc}\},\{G^{r8},G^{r8}\}\} \nonumber \\
& & \mbox{\hglue0.6truecm} + i \epsilon^{rim} \{G^{k8},\{J^r,\{G^{ic},G^{m8}\}\}\}
- \frac34 i \epsilon^{rim} d^{c8e} \{J^k,\{J^r,\{G^{i8},G^{me}\}\}\}
- \frac{9}{16} i \epsilon^{kim} f^{cae} f^{8eb} \{\{J^i,G^{m8}\},\{T^a,T^b\}\} \nonumber \\
& & \mbox{\hglue0.6truecm} - \frac{19}{4} i f^{c8e} \{J^k,[\{J^i,G^{ie}\},\{J^r,G^{r8}\}]\} - \frac{11}{4} i f^{c8e} \{\{J^r,G^{re}\},[J^2,G^{k8}]\}
+ \frac{11}{4} i f^{c8e} \{\{J^r,G^{r8}\},[J^2,G^{ke}]\} \nonumber \\
& & \mbox{\hglue0.6truecm}
+ \frac{11}{4} i f^{c8e} \{J^2,[G^{ke},\{J^r,G^{r8}\}]\}
- \frac{11}{4} i f^{c8e} \{J^2,[G^{k8},\{J^r,G^{re}\}]\}
+ \frac34 d^{c8e} \{J^2,[G^{ke},\{J^r,G^{r8}\}]\} \nonumber \\
& & \mbox{\hglue0.6truecm}
- \frac34 d^{c8e} \{J^2,[G^{k8},\{J^r,G^{re}\}]\}
- \frac18 [G^{kc},\{\{J^m,G^{m8}\},\{J^r,G^{r8}\}\}]
- \frac54 [G^{k8},\{\{J^m,G^{m8}\},\{J^r,G^{rc}\}\}] \nonumber \\
& & \mbox{\hglue0.6truecm}
+ \frac54 \{\{J^m,G^{mc}\},[G^{k8},\{J^r,G^{r8}\}]\}
+ \frac34 i \epsilon^{kim} f^{cea} f^{e8b} \{\{J^i,G^{m8}\},\{G^{ra},G^{rb}\}\}
- 4 d^{c8e} \{J^2,\{G^{ke},\{J^r,G^{r8}\}\}\} \nonumber \\
& & \mbox{\hglue0.6truecm}
- 4 d^{88e} \{J^2,\{G^{ke},\{J^r,G^{rc}\}\}\}
+ 4 \{G^{kc},\{\{J^m,G^{m8}\},\{J^r,G^{r8}\}\}\} 
- 4 \{G^{k8},\{\{J^m,G^{m8}\},\{J^r,G^{rc}\}\}\} \nonumber \\
& & \mbox{\hglue0.6truecm}
+ 2 d^{c8e} \{\mathcal{D}_3^{ke},\{J^r,G^{r8}\}\}
+ 2 d^{88e} \{\mathcal{D}_3^{kc},\{J^r,G^{re}\}\}
+ 4 i \epsilon^{kil} [\{J^i,G^{l8}\},\{\{J^m,G^{m8}\},\{J^r,G^{rc}\}\}] \nonumber \\
& & \mbox{\hglue0.6truecm}
+ i f^{c8e} \{J^2,\{J^k,[\{J^i,G^{ie}\},\{J^r,G^{r8}\}]\}\}
+ 2 \{\mathcal{D}_2^{kc},\{\{J^m,G^{m8}\},\{J^r,G^{r8}\}\}\},
\end{eqnarray}

\begin{eqnarray}
& & [\mathcal{D}_3^{i8},[\mathcal{O}_3^{i8},\mathcal{D}_2^{kc}]] + [\mathcal{O}_3^{i8},[\mathcal{D}_3^{i8},\mathcal{D}_2^{kc}]] \nonumber \\
& & \mbox{\hglue0.2truecm} = \frac{21}{2} f^{c8e} f^{8eg} G^{kg} + \frac78 i \epsilon^{kim} f^{c8e} f^{8eg} \{J^i,G^{mg}\} + \frac72 f^{c8e} f^{8eg} \mathcal{D}_3^{kg} - \frac{21}{16} N_c i d^{8eg} f^{c8e} \mathcal{D}_3^{kg} - \frac{21}{16} N_c i d^{c8e} f^{8eg} \mathcal{D}_3^{kg} \nonumber \\
& & \mbox{\hglue0.6truecm} + 7 d^{c8e} d^{8eg} \mathcal{O}_3^{kg} - 7 d^{ceg} d^{88e} \mathcal{O}_3^{kg} + 14 \{G^{kc},\{G^{r8},G^{r8}\}\} - 14 \{G^{k8},\{G^{rc},G^{r8}\}\} - 7 d^{c8e} \{J^k,\{G^{re},G^{r8}\}\} \nonumber \\
& & \mbox{\hglue0.6truecm} + 7 d^{88e} \{J^k,\{G^{rc},G^{re}\}\} + 14 d^{c8e} \{G^{ke},\{J^r,G^{r8}\}\} - 7 d^{c8e} \{G^{k8},\{J^r,G^{re}\}\} + 7 d^{88e} \{G^{kc},\{J^r,G^{re}\}\} \nonumber \\
& & \mbox{\hglue0.6truecm}
- 14 d^{88e} \{G^{ke},\{J^r,G^{rc}\}\}
+ \frac72 \epsilon^{kim} f^{c8e} \{T^e,\{J^i,G^{m8}\}\}
+ \frac{21}{8} i d^{8eg} f^{c8e} \mathcal{D}_4^{kg}
+ \frac{21}{4N_f} i \epsilon^{kim} \delta^{c8} \{J^2,\{J^i,G^{m8}\}\} \nonumber \\
& & \mbox{\hglue0.6truecm}
+ \frac74 i f^{c8e} \{\mathcal{D}_2^{ke},\{J^r,G^{r8}\}\}
- \frac{35}{4} i \epsilon^{kim} \{\{J^i,G^{m8}\},\{G^{r8},G^{rc}\}\}
+ \frac72 i \epsilon^{kim} \{\{J^i,G^{mc}\},\{G^{r8},G^{r8}\}\} \nonumber \\
& & \mbox{\hglue0.6truecm}
- \frac72 i \epsilon^{rim} \{G^{k8},\{J^r,\{G^{ic},G^{m8}\}\}\}
+ \frac{21}{8} i \epsilon^{rim} d^{c8e} \{J^k,\{J^r,\{G^{i8},G^{me}\}\}\} \nonumber \\
& & \mbox{\hglue0.6truecm}
+ \frac{63}{32} i \epsilon^{kim} f^{cae} f^{8eb} \{\{J^i,G^{m8}\},\{T^a,T^b\}\}
+ \frac{95}{8} i f^{c8e} \{J^k,[\{J^i,G^{ie}\},\{J^r,G^{r8}\}]\} \nonumber \\
& & \mbox{\hglue0.6truecm}
+ \frac{79}{8} i f^{c8e} \{\{J^r,G^{re}\},[J^2,G^{k8}]\}
- \frac{79}{8} i f^{c8e} \{\{J^r,G^{r8}\},[J^2,G^{ke}]\}
- \frac{95}{8} i f^{c8e} \{J^2,[G^{ke},\{J^r,G^{r8}\}]\} \nonumber \\
& & \mbox{\hglue0.6truecm}
+ \frac{79}{8} i f^{c8e} \{J^2,[G^{k8},\{J^r,G^{re}\}]\}
- \frac{21}{8} d^{c8e} \{J^2,[G^{ke},\{J^r,G^{r8}\}]\}
+ \frac{21}{8} d^{c8e} \{J^2,[G^{k8},\{J^r,G^{re}\}]\} \nonumber \\
& & \mbox{\hglue0.6truecm}
+ \frac{7}{16} [G^{kc},\{\{J^m,G^{m8}\},\{J^r,G^{r8}\}\}]
+ \frac{35}{8} [G^{k8},\{\{J^m,G^{m8}\},\{J^r,G^{rc}\}\}]
- \frac{35}{8} \{\{J^m,G^{mc}\},[G^{k8},\{J^r,G^{r8}\}]\} \nonumber \\
& & \mbox{\hglue0.6truecm}
- \frac{21}{8} i \epsilon^{kim} f^{cea} f^{e8b} \{\{J^i,G^{m8}\},\{G^{ra},G^{rb}\}\}
+ 14 d^{c8e} \{J^2,\{G^{ke},\{J^r,G^{r8}\}\}\}
+ 14 d^{88e} \{J^2,\{G^{ke},\{J^r,G^{rc}\}\}\} \nonumber \\
& & \mbox{\hglue0.6truecm}
- 14 \{G^{kc},\{\{J^m,G^{m8}\},\{J^r,G^{r8}\}\}\}
+ 14 \{G^{k8},\{\{J^m,G^{m8}\},\{J^r,G^{rc}\}\}\}
- 7 d^{c8e} \{\mathcal{D}_3^{ke},\{J^r,G^{r8}\}\} \nonumber \\
& & \mbox{\hglue0.6truecm}
- 7 d^{88e} \{\mathcal{D}_3^{kc},\{J^r,G^{re}\}\}
- 14 i \epsilon^{kil} [\{J^i,G^{l8}\},\{\{J^m,G^{m8}\},\{J^r,G^{rc}\}\}]
+ d^{88e} \{J^2,\{J^2,\{G^{ke},T^c\}\}\} \nonumber \\
& & \mbox{\hglue0.6truecm}
- d^{88e} \{J^2,\{\mathcal{D}_2^{kc},\{J^r,G^{re}\}\}\}
+ 2 \{J^2,\{\{J^r,G^{r8}\},\{G^{k8},T^c\}\}\}
- i f^{c8e} \{J^2,\{\{J^r,G^{re}\},[J^2,G^{k8}]\}\} \nonumber \\
& & \mbox{\hglue0.6truecm}
+ i f^{c8e} \{J^2,\{\{J^r,G^{r8}\},[J^2,G^{ke}]\}\}
- 2 \{\mathcal{D}_2^{kc},\{\{J^m,G^{m8}\},\{J^r,G^{r8}\}\}\},
\end{eqnarray}

\begin{eqnarray}
& & [\mathcal{O}_3^{i8},[\mathcal{O}_3^{i8},\mathcal{D}_2^{kc}]] \nonumber \\
& & \mbox{\hglue0.2truecm} = - \frac92 f^{c8e} f^{8eg} G^{kg}
+ 6 f^{c8e} f^{8eg} \mathcal{D}_2^{kg}
+ \frac{5}{16} i \epsilon^{kim} f^{c8e} f^{8eg} \{J^i,G^{mg}\}
- \frac32 f^{c8e} f^{8eg} \mathcal{D}_3^{kg}
+ \frac{9}{16} N_c i d^{8eg} f^{c8e} \mathcal{D}_3^{kg} \nonumber \\
& & \mbox{\hglue0.6truecm}
+ \frac{9}{16} N_c i d^{c8e} f^{8eg} \mathcal{D}_3^{kg}
- 3 d^{c8e} d^{8eg} \mathcal{O}_3^{kg}
+ 3 d^{ceg} d^{88e} \mathcal{O}_3^{kg}
- 6 \{G^{kc},\{G^{r8},G^{r8}\}\}
+ 6 \{G^{k8},\{G^{rc},G^{r8}\}\} \nonumber \\
& & \mbox{\hglue0.6truecm}
+ 3 d^{c8e} \{J^k,\{G^{re},G^{r8}\}\}
- 3 d^{88e} \{J^k,\{G^{rc},G^{re}\}\}
- 6 d^{c8e} \{G^{ke},\{J^r,G^{r8}\}\}
+ 3 d^{c8e} \{G^{k8},\{J^r,G^{re}\}\} \nonumber \\
& & \mbox{\hglue0.6truecm}
- 3 d^{88e} \{G^{kc},\{J^r,G^{re}\}\}
+ 6 d^{88e} \{G^{ke},\{J^r,G^{rc}\}\}
- \frac32 \epsilon^{kim} f^{c8e} \{T^e,\{J^i,G^{m8}\}\}
+ \frac{21}{2} f^{c8e} f^{8eg} \mathcal{D}_4^{kg}
+ d^{c8e} d^{8eg} \mathcal{D}_4^{kg} \nonumber \\
& & \mbox{\hglue0.6truecm}
- d^{ceg} d^{88e} \mathcal{D}_4^{kg}
- \frac98 i d^{8eg} f^{c8e} \mathcal{D}_4^{kg}
+ \frac{2}{N_f} \delta^{c8} \mathcal{D}_4^{k8}
- \frac{9}{4N_f} i \epsilon^{kim} \delta^{c8} \{J^2,\{J^i,G^{m8}\}\}
- 2 \{\mathcal{D}_2^{kc},\{G^{r8},G^{r8}\}\} \nonumber \\
& & \mbox{\hglue0.6truecm}
+ d^{88e} \{\mathcal{D}_2^{kc},\{J^r,G^{re}\}\}
+ \frac58 i f^{c8e} \{\mathcal{D}_2^{ke},\{J^r,G^{r8}\}\}
+ i \epsilon^{kim} \{\{J^i,G^{m8}\},\{G^{r8},G^{rc}\}\} \nonumber \\
& & \mbox{\hglue0.6truecm}
+ \frac54 i \epsilon^{kim} \{\{J^i,G^{mc}\},\{G^{r8},G^{r8}\}\}
- \frac54 i \epsilon^{rim} \{G^{k8},\{J^r,\{G^{ic},G^{m8}\}\}\}
- \frac98 i \epsilon^{rim} d^{c8e} \{J^k,\{J^r,\{G^{i8},G^{me}\}\}\} \nonumber \\
& & \mbox{\hglue0.6truecm}
- \frac{27}{32} i \epsilon^{kim} f^{cae} f^{8eb} \{\{J^i,G^{m8}\},\{T^a,T^b\}\}
- \frac{57}{8} i f^{c8e} \{J^k,[\{J^i,G^{ie}\},\{J^r,G^{r8}\}]\}
- \frac{33}{8} i f^{c8e} \{\{J^r,G^{re}\},[J^2,G^{k8}]\} \nonumber \\
& & \mbox{\hglue0.6truecm}
+ \frac{33}{8} i f^{c8e} \{\{J^r,G^{r8}\},[J^2,G^{ke}]\}
+ \frac{33}{8} i f^{c8e} \{J^2,[G^{ke},\{J^r,G^{r8}\}]\}
- \frac{33}{8} i f^{c8e} \{J^2,[G^{k8},\{J^r,G^{re}\}]\} \nonumber \\
& & \mbox{\hglue0.6truecm}
+ \frac98 d^{c8e} \{J^2,[G^{ke},\{J^r,G^{r8}\}]\}
- \frac98 d^{c8e} \{J^2,[G^{k8},\{J^r,G^{re}\}]\}
- \frac78 [G^{kc},\{\{J^m,G^{m8}\},\{J^r,G^{r8}\}\}] \nonumber \\
& & \mbox{\hglue0.6truecm}
- \frac12 [G^{k8},\{\{J^m,G^{m8}\},\{J^r,G^{rc}\}\}]
+ \frac12 \{\{J^m,G^{mc}\},[G^{k8},\{J^r,G^{r8}\}]\}
+ \frac98 i \epsilon^{kim} f^{cea} f^{e8b} \{\{J^i,G^{m8}\},\{G^{ra},G^{rb}\}\} \nonumber \\
& & \mbox{\hglue0.6truecm}
- 6 d^{c8e} \{J^2,\{G^{ke},\{J^r,G^{r8}\}\}\}
- 6 d^{88e} \{J^2,\{G^{ke},\{J^r,G^{rc}\}\}\}
+ 6 \{G^{kc},\{\{J^m,G^{m8}\},\{J^r,G^{r8}\}\}\} \nonumber \\
& & \mbox{\hglue0.6truecm}
- 6 \{G^{k8},\{\{J^m,G^{m8}\},\{J^r,G^{rc}\}\}\}
+ 3 d^{c8e} \{\mathcal{D}_3^{ke},\{J^r,G^{r8}\}\}
+ 3 d^{88e} \{\mathcal{D}_3^{kc},\{J^r,G^{re}\}\} \nonumber \\
& & \mbox{\hglue0.6truecm}
+ 6 i \epsilon^{kil} [\{J^i,G^{l8}\},\{\{J^m,G^{m8}\},\{J^r,G^{rc}\}\}]
+ \frac{11}{4} f^{c8e} f^{8eg} \mathcal{D}_6^{kg}
+ \frac12 d^{c8e} d^{8eg} \mathcal{D}_6^{kg}
- \frac12 d^{ceg} d^{88e} \mathcal{D}_6^{kg}
+ \frac{1}{N_f} \delta^{c8} \mathcal{D}_6^{k8} \nonumber \\
& & \mbox{\hglue0.6truecm}
- 2 \{J^2,\{\mathcal{D}_2^{kc},\{G^{r8},G^{r8}\}\}\}
+ \frac12 d^{88e} \{J^2,\{\mathcal{D}_2^{kc},\{J^r,G^{re}\}\}\}
- \frac54 i f^{c8e} \{J^2,\{J^k,[\{J^i,G^{ie}\},\{J^r,G^{r8}\}]\}\} \nonumber \\
& & \mbox{\hglue0.6truecm}
+ \frac12 i f^{c8e} \{J^2,\{\{J^r,G^{re}\},[J^2,G^{k8}]\}\}
- \frac12 i f^{c8e} \{J^2,\{\{J^r,G^{r8}\},[J^2,G^{ke}]\}\}
- \frac12 i f^{c8e} \{J^2,\{J^2,[G^{ke},\{J^r,G^{r8}\}]\}\} \nonumber \\
& & \mbox{\hglue0.6truecm}
+ \frac12 i f^{c8e} \{J^2,\{J^2,[G^{k8},\{J^r,G^{re}\}]\}\}
+ \frac12 \{\mathcal{D}_2^{kc},\{\{J^m,G^{m8}\},\{J^r,G^{r8}\}\}\},
\end{eqnarray}

\begin{eqnarray}
& & [\mathcal{D}_3^{i8},[\mathcal{D}_3^{i8},\mathcal{D}_3^{kc}]] \nonumber \\
& & \mbox{\hglue0.2truecm} = \frac{176N_c-2625}{48} f^{c8e} f^{8eg} G^{kg} - \frac{6248N_c+5155}{6336} i \epsilon^{kim} f^{c8e} f^{8eg} \{J^i,G^{mg}\} + \frac{176N_c-2985}{144} f^{c8e} f^{8eg} \mathcal{D}_3^{kg} \nonumber \\
& & \mbox{\hglue0.6truecm} - \frac{N_c(3960N_c-48923)}{6336} (i f^{c8e} d^{8eg} \mathcal{D}_3^{kg} + i d^{c8e} f^{8eg} \mathcal{D}_3^{kg}) + \frac{437}{144} f^{c8e} f^{8eg} \mathcal{O}_3^{kg} + \frac{176N_c-2625}{72} d^{c8e} d^{8eg} \mathcal{O}_3^{kg} \nonumber \\
& & \mbox{\hglue0.6truecm} - \frac{176N_c-2625}{72} d^{ceg} d^{88e} \mathcal{O}_3^{kg} + \frac{176N_c-2625}{36} \{G^{kc},\{G^{r8},G^{r8}\}\} - \frac{176N_c-2625}{36} \{G^{k8},\{G^{rc},G^{r8}\}\} \nonumber \\
& & \mbox{\hglue0.6truecm} - \frac{176N_c-2625}{72} d^{c8e} \{J^k,\{G^{re},G^{r8}\}\} + \frac{176N_c-2625}{72} d^{88e} \{J^k,\{G^{rc},G^{re}\}\} + \frac{176N_c-2625}{36} d^{c8e} \{G^{ke},\{J^r,G^{r8}\}\} \nonumber \\
& & \mbox{\hglue0.6truecm} - \frac{176N_c-2625}{72} d^{c8e} \{G^{k8},\{J^r,G^{re}\}\} + \frac{176N_c-2625}{72} d^{88e} \{G^{kc},\{J^r,G^{re}\}\} - \frac{176N_c-2625}{36} d^{88e} \{G^{ke},\{J^r,G^{rc}\}\} \nonumber \\
& & \mbox{\hglue0.6truecm} + \frac{704N_c+879N_f-14016}{576} \epsilon^{kim} f^{c8e} \{T^e,\{J^i,G^{m8}\}\} + \frac{3960N_c-48923}{3168} i f^{c8e} d^{8eg} \mathcal{D}_4^{kg} \nonumber \\
& & \mbox{\hglue0.6truecm} + \frac{1}{12} i \epsilon^{kim} f^{c8e} f^{8eg} \{J^2,\{J^i,G^{mg}\}\} + \frac{N_c(88N_c+176N_f+3960)-48923}{1584N_f} i \epsilon^{kim} \delta^{c8} \{J^2,\{J^i,G^{m8}\}\} \nonumber \\
& & \mbox{\hglue0.6truecm} - \frac{6248N_c+5155}{3168} i f^{c8e} \{\mathcal{D}_2^{ke},\{J^r,G^{r8}\}\} + \frac{1144N_c+27039}{792} i \epsilon^{kim} \{\{J^i,G^{m8}\},\{G^{r8},G^{rc}\}\} \nonumber \\
& & \mbox{\hglue0.6truecm} - \frac{6248N_c+5155}{1584} i \epsilon^{kim} \{\{J^i,G^{mc}\},\{G^{r8},G^{r8}\}\} + \frac{6248N_c+5155}{1584} i \epsilon^{rim} \{G^{k8},\{J^r,\{G^{ic},G^{m8}\}\}\} \nonumber \\
& & \mbox{\hglue0.6truecm} + \frac{3960N_c-48923}{3168} i \epsilon^{rim} d^{c8e} \{J^k,\{J^r,\{G^{i8},G^{me}\}\}\} + \frac{3960N_c-48923}{4224} i \epsilon^{kim} f^{cae} f^{8eb} \{\{J^i,G^{m8}\},\{T^a,T^b\}\} \nonumber \\
& & \mbox{\hglue0.6truecm} + \frac{1936N_c-66977}{3168} i f^{c8e} \{J^k,[\{J^i,G^{ie}\},\{J^r,G^{r8}\}]\} + \frac{1936N_c-66977}{3168} i f^{c8e} \{\{J^r,G^{re}\},[J^2,G^{k8}]\} \nonumber \\
& & \mbox{\hglue0.6truecm} - \frac{1936N_c-66977}{3168} i f^{c8e} \{\{J^r,G^{r8}\},[J^2,G^{ke}]\} - \frac{1936N_c-66977}{3168} i f^{c8e} \{J^2,[G^{ke},\{J^r,G^{r8}\}]\} \nonumber \\
& & \mbox{\hglue0.6truecm} + \frac{1936N_c-65665}{3168} i f^{c8e} \{J^2,[G^{k8},\{J^r,G^{re}\}]\} - \frac{3960N_c-48923}{3168} d^{c8e} \{J^2,[G^{ke},\{J^r,G^{r8}\}]\} \nonumber \\
& & \mbox{\hglue0.6truecm} + \frac{3960N_c-48923}{3168} d^{c8e} \{J^2,[G^{k8},\{J^r,G^{re}\}]\} + \frac{1276N_c-5471}{792} [G^{kc},\{\{J^m,G^{m8}\},\{J^r,G^{r8}\}\}] \nonumber \\
& & \mbox{\hglue0.6truecm} - \frac{1144N_c+27039}{1584} [G^{k8},\{\{J^m,G^{m8}\},\{J^r,G^{rc}\}\}] + \frac{1144N_c+27039}{1584} \{\{J^m,G^{mc}\},[G^{k8},\{J^r,G^{r8}\}]\} \nonumber \\
& & \mbox{\hglue0.6truecm} - \frac{3960N_c-48923}{3168} i \epsilon^{kim} f^{cea} f^{e8b} \{\{J^i,G^{m8}\},\{G^{ra},G^{rb}\}\} - \frac16 f^{c8e} f^{8eg} \mathcal{D}_5^{kg} - \frac{1}{36} N_c i f^{c8e} d^{8eg} \mathcal{D}_5^{kg} \nonumber \\
& & \mbox{\hglue0.6truecm} - \frac{1}{36} N_c i d^{c8e} f^{8eg} \mathcal{D}_5^{kg} + \frac23 d^{c8e} d^{8eg} \mathcal{O}_5^{kg} - \frac{14}{9} d^{ceg} d^{88e} \mathcal{O}_5^{kg} + \frac{341}{36} \{J^2,\{G^{kc},\{G^{r8},G^{r8}\}\}\} \nonumber \\
& & \mbox{\hglue0.6truecm} - \frac{341}{36} \{J^2,\{G^{k8},\{G^{rc},G^{r8}\}\}\} + \frac{10}{3} d^{c8e} \{J^2,\{J^k,\{G^{re},G^{r8}\}\}\} - \frac{10}{3} d^{88e} \{J^2,\{J^k,\{G^{rc},G^{re}\}\}\} \nonumber \\
& & \mbox{\hglue0.6truecm} + \frac{352N_c-5447}{72} d^{c8e} \{J^2,\{G^{ke},\{J^r,G^{r8}\}\}\} - \frac23 d^{c8e} \{J^2,\{G^{k8},\{J^r,G^{re}\}\}\} + \frac{14}{9} d^{88e} \{J^2,\{G^{kc},\{J^r,G^{re}\}\}\} \nonumber \\
& & \mbox{\hglue0.6truecm} + \frac{352N_c-5703}{72} d^{88e} \{J^2,\{G^{ke},\{J^r,G^{rc}\}\}\} + \frac13 \epsilon^{kim} f^{c8e} \{J^2,\{T^e,\{J^i,G^{m8}\}\}\} \nonumber \\
& & \mbox{\hglue0.6truecm} - \frac{176N_c-2625}{36} \{G^{kc},\{\{J^m,G^{m8}\},\{J^r,G^{r8}\}\}\} + \frac{176N_c-2625}{36} \{G^{k8},\{\{J^m,G^{m8}\},\{J^r,G^{rc}\}\}\} \nonumber \\
& & \mbox{\hglue0.6truecm} - \frac{581}{72} \{J^k,\{\{J^m,G^{mc}\},\{G^{r8},G^{r8}\}\}\} + \frac{581}{72} \{J^k,\{\{J^m,G^{m8}\},\{G^{r8},G^{rc}\}\}\} \nonumber \\
& & \mbox{\hglue0.6truecm}
 - \frac{352N_c-5255}{144} d^{c8e} \{\mathcal{D}_3^{ke},\{J^r,G^{r8}\}\} - \frac{352N_c-5831}{144} d^{88e} \{\mathcal{D}_3^{kc},\{J^r,G^{re}\}\} \nonumber \\
& & \mbox{\hglue0.6truecm} - \frac{293}{72} \epsilon^{kim} f^{ab8} \{\{J^i,G^{m8}\},\{T^a,\{G^{rb},G^{rc}\}\}\} + \frac{1}{36} (N_c+N_f) i \epsilon^{kim} d^{c8e} \{J^2,\{T^e,\{J^i,G^{m8}\}\}\} \nonumber \\
& & \mbox{\hglue0.6truecm} - \frac{352N_c-5543}{72} i \epsilon^{kil} [\{J^i,G^{l8}\},\{\{J^m,G^{m8}\},\{J^r,G^{rc}\}\}] + \frac{1}{18} i f^{c8e} d^{8eg} \mathcal{D}_6^{kg} - \frac{41}{99} d^{c8e} \{J^2,\{J^2,\{G^{ke},T^8\}\}\} \nonumber \\
& & \mbox{\hglue0.6truecm} + \frac{1}{9N_f} i \epsilon^{kim} \delta^{c8} \{J^2,\{J^2,\{J^i,G^{m8}\}\}\} + \frac{41}{99} d^{c8e} \{J^2,\{\mathcal{D}_2^{k8},\{J^r,G^{re}\}\}\} + \frac16 i f^{c8e} \{J^2,\{\mathcal{D}_2^{ke},\{J^r,G^{r8}\}\}\} \nonumber \\
& & \mbox{\hglue0.6truecm} + \frac{41}{99} \{J^2,\{\{J^r,G^{rc}\},\{G^{k8},T^8\}\}\} - \frac{41}{99} \{J^2,\{\{J^r,G^{r8}\},\{G^{kc},T^8\}\}\} - \frac{1}{36} i \epsilon^{kim} \{J^2,\{\{T^c,T^8\},\{J^i,G^{m8}\}\}\} \nonumber
\end{eqnarray}
\begin{eqnarray}
& & \mbox{\hglue0.6truecm} - \frac49 i \epsilon^{kim} \{J^2,\{\{G^{rc},G^{r8}\},\{J^i,G^{m8}\}\}\} + \frac13 i \epsilon^{kim} \{J^2,\{\{G^{r8},G^{r8}\},\{J^i,G^{mc}\}\}\} \nonumber \\
& & \mbox{\hglue0.6truecm} - \frac13 i \epsilon^{rim} \{J^2,\{G^{k8},\{J^r,\{G^{ic},G^{m8}\}\}\}\} + \frac{1}{18} i \epsilon^{rim} d^{c8e} \{J^2,\{J^k,\{J^r,\{G^{i8},G^{me}\}\}\}\} \nonumber \\
& & \mbox{\hglue0.6truecm} + \frac{1}{18} i \epsilon^{kim} f^{cae} f^{8eb} \{J^2,\{\{J^i,G^{m8}\},\{T^a,T^b\}\}\} - \frac56 i f^{c8e} \{J^2,\{J^k,[\{J^i,G^{ie}\},\{J^r,G^{r8}\}]\}\} \nonumber \\
& & \mbox{\hglue0.6truecm} - \frac56 i f^{c8e} \{J^2,\{\{J^r,G^{re}\},[J^2,G^{k8}]\}\} + \frac{247}{198} i f^{c8e} \{J^2,\{\{J^r,G^{r8}\},[J^2,G^{ke}]\}\} \nonumber \\
& & \mbox{\hglue0.6truecm} + \frac56 i f^{c8e} \{J^2,\{J^2,[G^{ke},\{J^r,G^{r8}\}]\}\} - \frac56 i f^{c8e} \{J^2,\{J^2,[G^{k8},\{J^r,G^{re}\}]\}\} \nonumber \\
& & \mbox{\hglue0.6truecm} - \frac{41}{99} i \epsilon^{kim} [\{T^8,\{J^r,G^{r8}\}\},\{J^2,\{J^i,G^{mc}\}\}] - \frac{1}{18} d^{c8e} \{J^2,\{J^2,[G^{ke},\{J^r,G^{r8}\}]\}\} \nonumber \\
& & \mbox{\hglue0.6truecm} + \frac{1}{18} d^{c8e} \{J^2,\{J^2,[G^{k8},\{J^r,G^{re}\}]\}\} - \frac{1}{18} \{J^2,[G^{kc},\{\{J^m,G^{m8}\},\{J^r,G^{r8}\}\}]\} \nonumber \\
& & \mbox{\hglue0.6truecm} + \frac29 \{J^2,[G^{k8},\{\{J^m,G^{m8}\},\{J^r,G^{rc}\}\}]\} - \frac29 \{J^2,\{\{J^m,G^{mc}\},[G^{k8},\{J^r,G^{r8}\}]\}\} + f^{c8e} f^{8eg} \mathcal{D}_7^{kg} \nonumber \\
& & \mbox{\hglue0.6truecm} - 2 d^{c8e} \{J^2,\{J^2,\{J^k,\{G^{re},G^{r8}\}\}\}\} + 2 d^{88e} \{J^2,\{J^2,\{J^k,\{G^{rc},G^{re}\}\}\}\} + \frac43 d^{c8e} \{J^2,\{J^2,\{G^{ke},\{J^r,G^{r8}\}\}\}\} \nonumber \\
& & \mbox{\hglue0.6truecm} + \frac43 d^{88e} \{J^2,\{J^2,\{G^{ke},\{J^r,G^{rc}\}\}\}\} - \frac43 \{J^2,\{G^{kc},\{\{J^m,G^{m8}\},\{J^r,G^{r8}\}\}\}\} \nonumber \\
& & \mbox{\hglue0.6truecm} + \frac43 \{J^2,\{G^{k8},\{\{J^m,G^{m8}\},\{J^r,G^{rc}\}\}\}\} + 2 \{J^2,\{J^k,\{\{J^m,G^{mc}\},\{G^{r8},G^{r8}\}\}\}\} \nonumber \\
& & \mbox{\hglue0.6truecm} - 2 \{J^2,\{J^k,\{\{J^m,G^{m8}\},\{G^{r8},G^{rc}\}\}\}\} + \frac13 d^{c8e} \{J^2,\{\mathcal{D}_3^{ke},\{J^r,G^{r8}\}\}\} - \frac53 d^{88e} \{J^2,\{\mathcal{D}_3^{kc},\{J^r,G^{re}\}\}\} \nonumber \\
& & \mbox{\hglue0.6truecm} + \frac49 i \epsilon^{kil} \{J^2,[\{J^i,G^{l8}\},\{\{J^m,G^{m8}\},\{J^r,G^{rc}\}\}]\} + 2 \{\mathcal{D}_3^{kc},\{\{J^m,G^{m8}\},\{J^r,G^{r8}\}\}\} \nonumber \\
& & \mbox{\hglue0.6truecm} - \frac{16}{9} i \epsilon^{kil} \{J^2,\{J^i,\{J^r,[G^{l8},\{G^{r8},\{J^m,G^{mc}\}\}]\}\}\},
\end{eqnarray}

\begin{eqnarray}
& & [\mathcal{D}_3^{i8},[\mathcal{D}_3^{i8},\mathcal{O}_3^{kc}]] \nonumber \\
& & \mbox{\hglue0.2truecm} = \frac{4976N_c-7845}{192} f^{c8e} f^{8eg} G^{kg} + \frac{159016N_c-399703}{25344} i \epsilon^{kim} f^{c8e} f^{8eg} \{J^i,G^{mg}\} + \frac{4976N_c-8637}{576} f^{c8e} f^{8eg} \mathcal{D}_3^{kg} \nonumber \\
& & \mbox{\hglue0.6truecm} - \frac{N_c(234432N_c-716249)}{25344} (i f^{c8e} d^{8eg} \mathcal{D}_3^{kg} + i d^{c8e} f^{8eg} \mathcal{D}_3^{kg}) - \frac{2431}{576} f^{c8e} f^{8eg} \mathcal{O}_3^{kg} + \frac{4976N_c-7845}{288} d^{c8e} d^{8eg} \mathcal{O}_3^{kg} \nonumber \\
& & \mbox{\hglue0.6truecm} - \frac{4976N_c-7845}{288} d^{ceg} d^{88e} \mathcal{O}_3^{kg} + \frac{4976N_c-7845}{144} \{G^{kc},\{G^{r8},G^{r8}\}\} - \frac{4976N_c-7845}{144} \{G^{k8},\{G^{rc},G^{r8}\}\} \nonumber \\
& & \mbox{\hglue0.6truecm} - \frac{4976N_c-7845}{288} d^{c8e} \{J^k,\{G^{re},G^{r8}\}\} + \frac{4976N_c-7845}{288} d^{88e} \{J^k,\{G^{rc},G^{re}\}\} \nonumber \\
& & \mbox{\hglue0.6truecm} + \frac{4976N_c-7845}{144} d^{c8e} \{G^{ke},\{J^r,G^{r8}\}\} - \frac{4976N_c-7845}{288} d^{c8e} \{G^{k8},\{J^r,G^{re}\}\} \nonumber \\
& & \mbox{\hglue0.6truecm} + \frac{4976N_c-7845}{288} d^{88e} \{G^{kc},\{J^r,G^{re}\}\} - \frac{4976N_c-7845}{144} d^{88e} \{G^{ke},\{J^r,G^{rc}\}\} \nonumber \\
& & \mbox{\hglue0.6truecm} + \frac{19904N_c+2643N_f-41952}{2304} \epsilon^{kim} f^{c8e} \{T^e,\{J^i,G^{m8}\}\} + \frac{234432N_c-716249}{12672} i f^{c8e} d^{8eg} \mathcal{D}_4^{kg} \nonumber \\
& & \mbox{\hglue0.6truecm} - \frac{47}{48} i \epsilon^{kim} f^{c8e} f^{8eg} \{J^2,\{J^i,G^{mg}\}\} + \frac{N_c(1408N_c+2816N_f+234432)-716249}{6336N_f} i \epsilon^{kim} \delta^{c8} \{J^2,\{J^i,G^{m8}\}\} \nonumber \\
& & \mbox{\hglue0.6truecm} + \frac{159016N_c-399703}{12672} i f^{c8e} \{\mathcal{D}_2^{ke},\{J^r,G^{r8}\}\} - \frac{49181N_c-139494}{792} i \epsilon^{kim} \{\{J^i,G^{m8}\},\{G^{r8},G^{rc}\}\} \nonumber \\
& & \mbox{\hglue0.6truecm} + \frac{159016N_c-399703}{6336} i \epsilon^{kim} \{\{J^i,G^{mc}\},\{G^{r8},G^{r8}\}\} - \frac{159016N_c-399703}{6336} i \epsilon^{rim} \{G^{k8},\{J^r,\{G^{ic},G^{m8}\}\}\} \nonumber \\
& & \mbox{\hglue0.6truecm} + \frac{234432N_c-716249}{12672} i \epsilon^{rim} d^{c8e} \{J^k,\{J^r,\{G^{i8},G^{me}\}\}\} + \frac{234432N_c-716249}{16896} i \epsilon^{kim} f^{cae} f^{8eb} \{\{J^i,G^{m8}\},\{T^a,T^b\}\} \nonumber \\
& & \mbox{\hglue0.6truecm} + \frac{236896N_c-382457}{12672} i f^{c8e} \{J^k,[\{J^i,G^{ie}\},\{J^r,G^{r8}\}]\} + \frac{236896N_c-382457}{12672} i f^{c8e} \{\{J^r,G^{re}\},[J^2,G^{k8}]\} \nonumber \\
& & \mbox{\hglue0.6truecm} - \frac{236896N_c-382457}{12672} i f^{c8e} \{\{J^r,G^{r8}\},[J^2,G^{ke}]\} - \frac{236896N_c-382457}{12672} i f^{c8e} \{J^2,[G^{ke},\{J^r,G^{r8}\}]\} \nonumber \\
& & \mbox{\hglue0.6truecm} + \frac{236896N_c-429721}{12672} i f^{c8e} \{J^2,[G^{k8},\{J^r,G^{re}\}]\} - \frac{234432N_c-716249}{12672} d^{c8e} \{J^2,[G^{ke},\{J^r,G^{r8}\}]\} \nonumber \\
& & \mbox{\hglue0.6truecm} + \frac{234432N_c-716249}{12672} d^{c8e} \{J^2,[G^{k8},\{J^r,G^{re}\}]\} + \frac{37708N_c-158273}{12672} [G^{kc},\{\{J^m,G^{m8}\},\{J^r,G^{r8}\}\}] \nonumber \\
& & \mbox{\hglue0.6truecm} + \frac{49181N_c-139494}{1584} [G^{k8},\{\{J^m,G^{m8}\},\{J^r,G^{rc}\}\}] - \frac{49181N_c-139494}{1584} \{\{J^m,G^{mc}\},[G^{k8},\{J^r,G^{r8}\}]\} \nonumber \\
& & \mbox{\hglue0.6truecm} - \frac{234432N_c-716249}{12672} i \epsilon^{kim} f^{cea} f^{e8b} \{\{J^i,G^{m8}\},\{G^{ra},G^{rb}\}\} - \frac{11}{12} f^{c8e} f^{8eg} \mathcal{D}_5^{kg} - \frac19 N_c i f^{c8e} d^{8eg} \mathcal{D}_5^{kg} \nonumber \\
& & \mbox{\hglue0.6truecm} - \frac19 N_c i d^{c8e} f^{8eg} \mathcal{D}_5^{kg} + \frac32 f^{c8e} f^{8eg} \mathcal{O}_5^{kg} - \frac{23}{6} d^{c8e} d^{8eg} \mathcal{O}_5^{kg} + \frac{59}{18} d^{ceg} d^{88e} \mathcal{O}_5^{kg} - \frac{223}{144} \{J^2,\{G^{kc},\{G^{r8},G^{r8}\}\}\} \nonumber \\
& & \mbox{\hglue0.6truecm} + \frac{223}{144} \{J^2,\{G^{k8},\{G^{rc},G^{r8}\}\}\} + \frac{11}{6} d^{c8e} \{J^2,\{J^k,\{G^{re},G^{r8}\}\}\} - \frac{11}{6} d^{88e} \{J^2,\{J^k,\{G^{rc},G^{re}\}\}\} \nonumber \\
& & \mbox{\hglue0.6truecm} + \frac{9952N_c-18779}{288} d^{c8e} \{J^2,\{G^{ke},\{J^r,G^{r8}\}\}\} + \frac{23}{6} d^{c8e} \{J^2,\{G^{k8},\{J^r,G^{re}\}\}\} - \frac{59}{18} d^{88e} \{J^2,\{G^{kc},\{J^r,G^{re}\}\}\} \nonumber \\
& & \mbox{\hglue0.6truecm} + \frac{9952N_c-14523}{288} d^{88e} \{J^2,\{G^{ke},\{J^r,G^{rc}\}\}\} - \frac{23}{12} \epsilon^{kim} f^{c8e} \{J^2,\{T^e,\{J^i,G^{m8}\}\}\} \nonumber \\
& & \mbox{\hglue0.6truecm} - \frac{4976N_c-7845}{144} \{G^{kc},\{\{J^m,G^{m8}\},\{J^r,G^{r8}\}\}\} + \frac{4976N_c-7845}{144} \{G^{k8},\{\{J^m,G^{m8}\},\{J^r,G^{rc}\}\}\} \nonumber \\
& & \mbox{\hglue0.6truecm} - \frac{305}{288} \{J^k,\{\{J^m,G^{mc}\},\{G^{r8},G^{r8}\}\}\} + \frac{305}{288} \{J^k,\{\{J^m,G^{m8}\},\{G^{r8},G^{rc}\}\}\} \nonumber \\
& & \mbox{\hglue0.6truecm} - \frac{9952N_c-17147}{576} d^{c8e} \{\mathcal{D}_3^{ke},\{J^r,G^{r8}\}\} - \frac{9952N_c-15995}{576} d^{88e} \{\mathcal{D}_3^{kc},\{J^r,G^{re}\}\} \nonumber \\
& & \mbox{\hglue0.6truecm} - \frac{881}{288} \epsilon^{kim} f^{ab8} \{\{J^i,G^{m8}\},\{T^a,\{G^{rb},G^{rc}\}\}\} + \frac19 (N_c+N_f) i \epsilon^{kim} d^{c8e} \{J^2,\{T^e,\{J^i,G^{m8}\}\}\} \nonumber \\
& & \mbox{\hglue0.6truecm} - \frac{9952N_c-16571}{288} i \epsilon^{kil} [\{J^i,G^{l8}\},\{\{J^m,G^{m8}\},\{J^r,G^{rc}\}\}] + \frac29 i f^{c8e} d^{8eg} \mathcal{D}_6^{kg} + \frac{1477}{396} d^{c8e} \{J^2,\{J^2,\{G^{ke},T^8\}\}\} \nonumber \\
& & \mbox{\hglue0.6truecm} + \frac{4}{9N_f} i \epsilon^{kim} \delta^{c8} \{J^2,\{J^2,\{J^i,G^{m8}\}\}\} - \frac{1477}{396} d^{c8e} \{J^2,\{\mathcal{D}_2^{k8},\{J^r,G^{re}\}\}\} - \frac{47}{24} i f^{c8e} \{J^2,\{\mathcal{D}_2^{ke},\{J^r,G^{r8}\}\}\} \nonumber
\end{eqnarray}
\begin{eqnarray}
& & \mbox{\hglue0.6truecm} - \frac{1477}{396} \{J^2,\{\{J^r,G^{rc}\},\{G^{k8},T^8\}\}\} + \frac{1477}{396} \{J^2,\{\{J^r,G^{r8}\},\{G^{kc},T^8\}\}\} - \frac19 i \epsilon^{kim} \{J^2,\{\{T^c,T^8\},\{J^i,G^{m8}\}\}\} \nonumber \\
& & \mbox{\hglue0.6truecm} + \frac{125}{36} i \epsilon^{kim} \{J^2,\{\{G^{rc},G^{r8}\},\{J^i,G^{m8}\}\}\} - \frac{47}{12} i \epsilon^{kim} \{J^2,\{\{G^{r8},G^{r8}\},\{J^i,G^{mc}\}\}\} \nonumber \\
& & \mbox{\hglue0.6truecm} + \frac{47}{12} i \epsilon^{rim} \{J^2,\{G^{k8},\{J^r,\{G^{ic},G^{m8}\}\}\}\} + \frac29 i \epsilon^{rim} d^{c8e} \{J^2,\{J^k,\{J^r,\{G^{i8},G^{me}\}\}\}\} \nonumber \\
& & \mbox{\hglue0.6truecm} + \frac29 i \epsilon^{kim} f^{cae} f^{8eb} \{J^2,\{\{J^i,G^{m8}\},\{T^a,T^b\}\}\} - \frac{25}{12} i f^{c8e} \{J^2,\{J^k,[\{J^i,G^{ie}\},\{J^r,G^{r8}\}]\}\} \nonumber \\
& & \mbox{\hglue0.6truecm} - \frac{25}{12} i f^{c8e} \{J^2,\{\{J^r,G^{re}\},[J^2,G^{k8}]\}\} - \frac{163}{99} i f^{c8e} \{J^2,\{\{J^r,G^{r8}\},[J^2,G^{ke}]\}\} + \frac{25}{12} i f^{c8e} \{J^2,\{J^2,[G^{ke},\{J^r,G^{r8}\}]\}\} \nonumber \\
& & \mbox{\hglue0.6truecm} - \frac{25}{12} i f^{c8e} \{J^2,\{J^2,[G^{k8},\{J^r,G^{re}\}]\}\} + \frac{1477}{396} i \epsilon^{kim} [\{T^8,\{J^r,G^{r8}\}\},\{J^2,\{J^i,G^{mc}\}\}] \nonumber \\
& & \mbox{\hglue0.6truecm} - \frac29 d^{c8e} \{J^2,\{J^2,[G^{ke},\{J^r,G^{r8}\}]\}\} + \frac29 d^{c8e} \{J^2,\{J^2,[G^{k8},\{J^r,G^{re}\}]\}\} + \frac{157}{144} \{J^2,[G^{kc},\{\{J^m,G^{m8}\},\{J^r,G^{r8}\}\}]\} \nonumber \\
& & \mbox{\hglue0.6truecm} - \frac{125}{72} \{J^2,[G^{k8},\{\{J^m,G^{m8}\},\{J^r,G^{rc}\}\}]\} + \frac{125}{72} \{J^2,\{\{J^m,G^{mc}\},[G^{k8},\{J^r,G^{r8}\}]\}\} + d^{c8e} d^{8eg} \mathcal{O}_7^{kg} \nonumber \\
& & \mbox{\hglue0.6truecm} + 2 \{J^2,\{J^2,\{G^{kc},\{G^{r8},G^{r8}\}\}\}\} - 2 \{J^2,\{J^2,\{G^{k8},\{G^{rc},G^{r8}\}\}\}\} + \frac73 d^{c8e} \{J^2,\{J^2,\{G^{ke},\{J^r,G^{r8}\}\}\}\} \nonumber \\
& & \mbox{\hglue0.6truecm} - d^{c8e} \{J^2,\{J^2,\{G^{k8},\{J^r,G^{re}\}\}\}\} - \frac83 d^{88e} \{J^2,\{J^2,\{G^{ke},\{J^r,G^{rc}\}\}\}\} + \frac12 \epsilon^{kim} f^{c8e} \{J^2,\{J^2,\{T^e,\{J^i,G^{m8}\}\}\}\} \nonumber \\
& & \mbox{\hglue0.6truecm} + \frac{17}{3} \{J^2,\{G^{kc},\{\{J^m,G^{m8}\},\{J^r,G^{r8}\}\}\}\} - \frac{11}{3} \{J^2,\{G^{k8},\{\{J^m,G^{m8}\},\{J^r,G^{rc}\}\}\}\} \nonumber \\
& & \mbox{\hglue0.6truecm} - \{J^2,\{J^k,\{\{J^m,G^{mc}\},\{G^{r8},G^{r8}\}\}\}\} + \{J^2,\{J^k,\{\{J^m,G^{m8}\},\{G^{r8},G^{rc}\}\}\}\} - \frac23 d^{c8e} \{J^2,\{\mathcal{D}_3^{ke},\{J^r,G^{r8}\}\}\} \nonumber \\
& & \mbox{\hglue0.6truecm} + \frac43 d^{88e} \{J^2,\{\mathcal{D}_3^{kc},\{J^r,G^{re}\}\}\} - \frac{11}{9} i \epsilon^{kil} \{J^2,[\{J^i,G^{l8}\},\{\{J^m,G^{m8}\},\{J^r,G^{rc}\}\}]\} \nonumber \\
& & \mbox{\hglue0.6truecm} - \{\mathcal{D}_3^{kc},\{\{J^m,G^{m8}\},\{J^r,G^{r8}\}\}\} + \frac{26}{9} i \epsilon^{kil} \{J^2,\{J^i,\{J^r,[G^{l8},\{G^{r8},\{J^m,G^{mc}\}\}]\}\}\},
\end{eqnarray}

\begin{eqnarray}
& & [\mathcal{D}_3^{i8},[\mathcal{O}_3^{i8},\mathcal{D}_3^{kc}]] + [\mathcal{O}_3^{i8},[\mathcal{D}_3^{i8},\mathcal{D}_3^{kc}]] \nonumber \\
& & \mbox{\hglue0.2truecm} = - \frac{1864N_c-129}{96} f^{c8e} f^{8eg} G^{kg} + \frac{10120N_c-146461}{12672} i \epsilon^{kim} f^{c8e} f^{8eg} \{J^i,G^{mg}\} - \frac{1864N_c-633}{288} f^{c8e} f^{8eg} \mathcal{D}_3^{kg} \nonumber \\
& & \mbox{\hglue0.6truecm} + \frac{N_c(9306N_c+36325)}{6336} (i f^{c8e} d^{8eg} \mathcal{D}_3^{kg} + i d^{c8e} f^{8eg} \mathcal{D}_3^{kg}) - \frac{1333}{288} f^{c8e} f^{8eg} \mathcal{O}_3^{kg} - \frac{1864N_c-129}{144} d^{c8e} d^{8eg} \mathcal{O}_3^{kg} \nonumber \\
& & \mbox{\hglue0.6truecm} + \frac{1864N_c-129}{144} d^{ceg} d^{88e} \mathcal{O}_3^{kg} - \frac{1864N_c-129}{72} \{G^{kc},\{G^{r8},G^{r8}\}\} + \frac{1864N_c-129}{72} \{G^{k8},\{G^{rc},G^{r8}\}\} \nonumber \\
& & \mbox{\hglue0.6truecm} + \frac{1864N_c-129}{144} d^{c8e} \{J^k,\{G^{re},G^{r8}\}\} - \frac{1864N_c-129}{144} d^{88e} \{J^k,\{G^{rc},G^{re}\}\} - \frac{1864N_c-129}{72} d^{c8e} \{G^{ke},\{J^r,G^{r8}\}\} \nonumber \\
& & \mbox{\hglue0.6truecm} + \frac{1864N_c-129}{144} d^{c8e} \{G^{k8},\{J^r,G^{re}\}\} - \frac{1864N_c-129}{144} d^{88e} \{G^{kc},\{J^r,G^{re}\}\} + \frac{1864N_c-129}{72} d^{88e} \{G^{ke},\{J^r,G^{rc}\}\} \nonumber \\
& & \mbox{\hglue0.6truecm} - \frac{7456N_c+1839N_f-7872}{1152} \epsilon^{kim} f^{c8e} \{T^e,\{J^i,G^{m8}\}\} - \frac{9306N_c+36325}{3168} i f^{c8e} d^{8eg} \mathcal{D}_4^{kg} \nonumber \\
& & \mbox{\hglue0.6truecm} + \frac{1}{24} i \epsilon^{kim} f^{c8e} f^{8eg} \{J^2,\{J^i,G^{mg}\}\} + \frac{N_c(110N_c+220N_f-9306)-36325}{1584N_f} i \epsilon^{kim} \delta^{c8} \{J^2,\{J^i,G^{m8}\}\} \nonumber \\
& & \mbox{\hglue0.6truecm} + \frac{10120N_c-146461}{6336} i f^{c8e} \{\mathcal{D}_2^{ke},\{J^r,G^{r8}\}\} + \frac{8492N_c+ 219111}{3168} i \epsilon^{kim} \{\{J^i,G^{m8}\},\{G^{r8},G^{rc}\}\} \nonumber \\
& & \mbox{\hglue0.6truecm} + \frac{10120N_c-146461}{3168} i \epsilon^{kim} \{\{J^i,G^{mc}\},\{G^{r8},G^{r8}\}\} - \frac{10120N_c-146461}{3168} i \epsilon^{rim} \{G^{k8},\{J^r,\{G^{ic},G^{m8}\}\}\} \nonumber \\
& & \mbox{\hglue0.6truecm} - \frac{9306N_c+36325}{3168} i \epsilon^{rim} d^{c8e} \{J^k,\{J^r,\{G^{i8},G^{me}\}\}\} - \frac{9306N_c+36325}{4224} i \epsilon^{kim} f^{cae} f^{8eb} \{\{J^i,G^{m8}\},\{T^a,T^b\}\} \nonumber \\
& & \mbox{\hglue0.6truecm} - \frac{83864N_c+75953}{6336} i f^{c8e} \{J^k,[\{J^i,G^{ie}\},\{J^r,G^{r8}\}]\} - \frac{83864N_c+75953}{6336} i f^{c8e} \{\{J^r,G^{re}\},[J^2,G^{k8}]\} \nonumber \\
& & \mbox{\hglue0.6truecm} + \frac{83864N_c+75953}{6336} i f^{c8e} \{\{J^r,G^{r8}\},[J^2,G^{ke}]\} + \frac{83864N_c+75953}{6336} i f^{c8e} \{J^2,[G^{ke},\{J^r,G^{r8}\}]\} \nonumber \\
& & \mbox{\hglue0.6truecm} - \frac{83864N_c+104209}{6336} i f^{c8e} \{J^2,[G^{k8},\{J^r,G^{re}\}]\} + \frac{9306N_c+36325}{3168} d^{c8e} \{J^2,[G^{ke},\{J^r,G^{r8}\}]\} \nonumber \\
& & \mbox{\hglue0.6truecm} - \frac{9306N_c+36325}{3168} d^{c8e} \{J^2,[G^{k8},\{J^r,G^{re}\}]\} - \frac{28732N_c-73811}{12672} [G^{kc},\{\{J^m,G^{m8}\},\{J^r,G^{r8}\}\}] \nonumber \\
& & \mbox{\hglue0.6truecm} - \frac{8492N_c+219111}{6336} [G^{k8},\{\{J^m,G^{m8}\},\{J^r,G^{rc}\}\}] + \frac{8492N_c+219111}{6336} \{\{J^m,G^{mc}\},[G^{k8},\{J^r,G^{r8}\}]\} \nonumber \\
& & \mbox{\hglue0.6truecm} + \frac{9306N_c+36325}{3168} i \epsilon^{kim} f^{cea} f^{e8b} \{\{J^i,G^{m8}\},\{G^{ra},G^{rb}\}\} + \frac76 f^{c8e} f^{8eg} \mathcal{D}_5^{kg} - \frac{5}{144} N_c i f^{c8e} d^{8eg} \mathcal{D}_5^{kg} \nonumber \\
& & \mbox{\hglue0.6truecm} - \frac{5}{144} N_c i d^{c8e} f^{8eg} \mathcal{D}_5^{kg} - 2 f^{c8e} f^{8eg} \mathcal{O}_5^{kg} + \frac73 d^{c8e} d^{8eg} \mathcal{O}_5^{kg} - \frac{22}{9} d^{ceg} d^{88e} \mathcal{O}_5^{kg} - \frac{853}{72} \{J^2,\{G^{kc},\{G^{r8},G^{r8}\}\}\} \nonumber \\
& & \mbox{\hglue0.6truecm} + \frac{853}{72} \{J^2,\{G^{k8},\{G^{rc},G^{r8}\}\}\} - \frac73 d^{c8e} \{J^2,\{J^k,\{G^{re},G^{r8}\}\}\} + \frac73 d^{88e} \{J^2,\{J^k,\{G^{rc},G^{re}\}\}\} \nonumber \\
& & \mbox{\hglue0.6truecm} - \frac{3728N_c-967}{144} d^{c8e} \{J^2,\{G^{ke},\{J^r,G^{r8}\}\}\} - \frac73 d^{c8e} \{J^2,\{G^{k8},\{J^r,G^{re}\}\}\} + \frac{22}{9} d^{88e} \{J^2,\{G^{kc},\{J^r,G^{re}\}\}\} \nonumber \\
& & \mbox{\hglue0.6truecm} - \frac{3728N_c-759}{144} d^{88e} \{J^2,\{G^{ke},\{J^r,G^{rc}\}\}\} + \frac{9N_f-46}{12} \epsilon^{kim} f^{c8e} \{J^2,\{T^e,\{J^i,G^{m8}\}\}\} \nonumber \\
& & \mbox{\hglue0.6truecm} + \frac{1864N_c-129}{72} \{G^{kc},\{\{J^m,G^{m8}\},\{J^r,G^{r8}\}\}\} - \frac{1864N_c-129}{72} \{G^{k8},\{\{J^m,G^{m8}\},\{J^r,G^{rc}\}\}\} \nonumber \\
& & \mbox{\hglue0.6truecm} + \frac{1189}{144} \{J^k,\{\{J^m,G^{mc}\},\{G^{r8},G^{r8}\}\}\} - \frac{1189}{144} \{J^k,\{\{J^m,G^{m8}\},\{G^{r8},G^{rc}\}\}\} + \frac{3728N_c-295}{288} d^{c8e} \{\mathcal{D}_3^{ke},\{J^r,G^{r8}\}\} \nonumber \\
& & \mbox{\hglue0.6truecm} + \frac{3728N_c-1447}{288} d^{88e} \{\mathcal{D}_3^{kc},\{J^r,G^{re}\}\} + \frac{613}{144} \epsilon^{kim} f^{ab8} \{\{J^i,G^{m8}\},\{T^a,\{G^{rb},G^{rc}\}\}\} \nonumber \\
& & \mbox{\hglue0.6truecm} + \frac{5}{144} (N_c+N_f) i \epsilon^{kim} d^{c8e} \{J^2,\{T^e,\{J^i,G^{m8}\}\}\} + \frac{3728N_c-871}{144} i \epsilon^{kil} [\{J^i,G^{l8}\},\{\{J^m,G^{m8}\},\{J^r,G^{rc}\}\}] \nonumber \\
& & \mbox{\hglue0.6truecm} + \frac{5}{72} i f^{c8e} d^{8eg} \mathcal{D}_6^{kg} + \frac{883}{198} d^{c8e} \{J^2,\{J^2,\{G^{ke},T^8\}\}\} + \frac{5}{36N_f} i \epsilon^{kim} \delta^{c8} \{J^2,\{J^2,\{J^i,G^{m8}\}\}\} \nonumber \\
& & \mbox{\hglue0.6truecm} - \frac{883}{198} d^{c8e} \{J^2,\{\mathcal{D}_2^{k8},\{J^r,G^{re}\}\}\} + \frac{1}{12} i f^{c8e} \{J^2,\{\mathcal{D}_2^{ke},\{J^r,G^{r8}\}\}\} - \frac{883}{198} \{J^2,\{\{J^r,G^{rc}\},\{G^{k8},T^8\}\}\} \nonumber
\end{eqnarray}
\begin{eqnarray}
& & \mbox{\hglue0.6truecm} + \frac{883}{198} \{J^2,\{\{J^r,G^{r8}\},\{G^{kc},T^8\}\}\} - \frac{5}{144} i \epsilon^{kim} \{J^2,\{\{T^c,T^8\},\{J^i,G^{m8}\}\}\} - \frac{11}{36} i \epsilon^{kim} \{J^2,\{\{G^{rc},G^{r8}\},\{J^i,G^{m8}\}\}\} \nonumber \\
& & \mbox{\hglue0.6truecm} + \frac16 i \epsilon^{kim} \{J^2,\{\{G^{r8},G^{r8}\},\{J^i,G^{mc}\}\}\} - \frac16 i \epsilon^{rim} \{J^2,\{G^{k8},\{J^r,\{G^{ic},G^{m8}\}\}\}\} \nonumber \\
& & \mbox{\hglue0.6truecm} + \frac{5}{72} i \epsilon^{rim} d^{c8e} \{J^2,\{J^k,\{J^r,\{G^{i8},G^{me}\}\}\}\} + \frac{5}{72} i \epsilon^{kim} f^{cae} f^{8eb} \{J^2,\{\{J^i,G^{m8}\},\{T^a,T^b\}\}\} \nonumber \\
& & \mbox{\hglue0.6truecm} + \frac43 i f^{c8e} \{J^2,\{J^k,[\{J^i,G^{ie}\},\{J^r,G^{r8}\}]\}\} + \frac43 i f^{c8e} \{J^2,\{\{J^r,G^{re}\},[J^2,G^{k8}]\}\} \nonumber \\
& & \mbox{\hglue0.6truecm} - \frac{1147}{198} i f^{c8e} \{J^2,\{\{J^r,G^{r8}\},[J^2,G^{ke}]\}\} - \frac43 i f^{c8e} \{J^2,\{J^2,[G^{ke},\{J^r,G^{r8}\}]\}\} + \frac43 i f^{c8e} \{J^2,\{J^2,[G^{k8},\{J^r,G^{re}\}]\}\} \nonumber \\
& & \mbox{\hglue0.6truecm} + \frac{883}{198} i \epsilon^{kim} [\{T^8,\{J^r,G^{r8}\}\},\{J^2,\{J^i,G^{mc}\}\}] - \frac{5}{72} d^{c8e} \{J^2,\{J^2,[G^{ke},\{J^r,G^{r8}\}]\}\} \nonumber \\
& & \mbox{\hglue0.6truecm} + \frac{5}{72} d^{c8e} \{J^2,\{J^2,[G^{k8},\{J^r,G^{re}\}]\}\} - \frac{1}{144} \{J^2,[G^{kc},\{\{J^m,G^{m8}\},\{J^r,G^{r8}\}\}]\} \nonumber \\
& & \mbox{\hglue0.6truecm} + \frac{11}{72} \{J^2,[G^{k8},\{\{J^m,G^{m8}\},\{J^r,G^{rc}\}\}]\} - \frac{11}{72} \{J^2,\{\{J^m,G^{mc}\},[G^{k8},\{J^r,G^{r8}\}]\}\} \nonumber \\
& & \mbox{\hglue0.6truecm} + \frac23 d^{c8e} \{J^2,\{J^2,\{G^{ke},\{J^r,G^{r8}\}\}\}\} + \frac23 d^{88e} \{J^2,\{J^2,\{G^{ke},\{J^r,G^{rc}\}\}\}\} - \epsilon^{kim} f^{c8e} \{J^2,\{J^2,\{T^e,\{J^i,G^{m8}\}\}\}\} \nonumber \\
& & \mbox{\hglue0.6truecm} - \frac23 \{J^2,\{G^{kc},\{\{J^m,G^{m8}\},\{J^r,G^{r8}\}\}\}\} + \frac{14}{3} \{J^2,\{G^{k8},\{\{J^m,G^{m8}\},\{J^r,G^{rc}\}\}\}\} \nonumber \\
& & \mbox{\hglue0.6truecm} - \frac13 d^{c8e} \{J^2,\{\mathcal{D}_3^{ke},\{J^r,G^{r8}\}\}\} - \frac13 d^{88e} \{J^2,\{\mathcal{D}_3^{kc},\{J^r,G^{re}\}\}\} - 2 \epsilon^{kim} f^{ab8} \{J^2,\{\{J^i,G^{m8}\},\{T^a,\{G^{rb},G^{rc}\}\}\}\} \nonumber \\
& & \mbox{\hglue0.6truecm} + \frac{14}{9} i \epsilon^{kil} \{J^2,[\{J^i,G^{l8}\},\{\{J^m,G^{m8}\},\{J^r,G^{rc}\}\}]\} - 2 \{\mathcal{D}_3^{kc},\{\{J^m,G^{m8}\},\{J^r,G^{r8}\}\}\} \nonumber \\
& & \mbox{\hglue0.6truecm} - \frac29 i \epsilon^{kil} \{J^2,\{J^i,\{J^r,[G^{l8},\{G^{r8},\{J^m,G^{mc}\}\}]\}\}\},
\end{eqnarray}

\begin{eqnarray}
& & [\mathcal{D}_3^{i8},[\mathcal{O}_3^{i8},\mathcal{O}_3^{kc}]] + [\mathcal{O}_3^{i8},[\mathcal{D}_3^{i8},\mathcal{O}_3^{kc}]] \nonumber \\
& & \mbox{\hglue0.2truecm} = \frac{1612N_c-1695}{48} f^{c8e} f^{8eg} G^{kg} + \frac{34826N_c-65975}{6336} i \epsilon^{kim} f^{c8e} f^{8eg} \{J^i,G^{mg}\} + \frac{1612N_c-1875}{144} f^{c8e} f^{8eg} \mathcal{D}_3^{kg} \nonumber \\
& & \mbox{\hglue0.6truecm} + \frac{N_c(2376N_c-184945)}{12672} (i f^{c8e} d^{8eg} \mathcal{D}_3^{kg} + i d^{c8e} f^{8eg} \mathcal{D}_3^{kg}) - \frac{13}{72} f^{c8e} f^{8eg} \mathcal{O}_3^{kg} + \frac{1612N_c-1695}{72} d^{c8e} d^{8eg} \mathcal{O}_3^{kg} \nonumber \\
& & \mbox{\hglue0.6truecm} - \frac{1612N_c-1695}{72} d^{ceg} d^{88e} \mathcal{O}_3^{kg} + \frac{1612N_c-1695}{36} \{G^{kc},\{G^{r8},G^{r8}\}\} - \frac{1612N_c-1695}{36} \{G^{k8},\{G^{rc},G^{r8}\}\} \nonumber \\
& & \mbox{\hglue0.6truecm} - \frac{1612N_c-1695}{72} d^{c8e} \{J^k,\{G^{re},G^{r8}\}\} + \frac{1612N_c-1695}{72} d^{88e} \{J^k,\{G^{rc},G^{re}\}\} + \frac{1612N_c-1695}{36} d^{c8e} \{G^{ke},\{J^r,G^{r8}\}\} \nonumber \\
& & \mbox{\hglue0.6truecm} - \frac{1612N_c-1695}{72} d^{c8e} \{G^{k8},\{J^r,G^{re}\}\} + \frac{1612N_c-1695}{72} d^{88e} \{G^{kc},\{J^r,G^{re}\}\} \nonumber \\
& & \mbox{\hglue0.6truecm} - \frac{1612N_c-1695}{36} d^{88e} \{G^{ke},\{J^r,G^{rc}\}\} + \frac{3224N_c+501N_f-5394}{288} \epsilon^{kim} f^{c8e} \{T^e,\{J^i,G^{m8}\}\} \nonumber \\
& & \mbox{\hglue0.6truecm} - \frac{2376N_c-184945}{6336} i f^{c8e} d^{8eg} \mathcal{D}_4^{kg} + \frac{1}{24} i \epsilon^{kim} f^{c8e} f^{8eg} \{J^2,\{J^i,G^{mg}\}\} \nonumber \\
& & \mbox{\hglue0.6truecm} - \frac{N_c(4004N_c+8008N_f+2376)-184945}{3168N_f} i \epsilon^{kim} \delta^{c8} \{J^2,\{J^i,G^{m8}\}\} + \frac{34826N_c-65975}{3168} i f^{c8e} \{\mathcal{D}_2^{ke},\{J^r,G^{r8}\}\} \nonumber \\
& & \mbox{\hglue0.6truecm} - \frac{67276N_c+52995}{3168} i \epsilon^{kim} \{\{J^i,G^{m8}\},\{G^{r8},G^{rc}\}\} + \frac{34826N_c-65975}{1584} i \epsilon^{kim} \{\{J^i,G^{mc}\},\{G^{r8},G^{r8}\}\} \nonumber \\
& & \mbox{\hglue0.6truecm} - \frac{34826N_c-65975}{1584} i \epsilon^{rim} \{G^{k8},\{J^r,\{G^{ic},G^{m8}\}\}\} - \frac{2376N_c-184945}{6336} i \epsilon^{rim} d^{c8e} \{J^k,\{J^r,\{G^{i8},G^{me}\}\}\} \nonumber \\
& & \mbox{\hglue0.6truecm} - \frac{2376N_c-184945}{8448} i \epsilon^{kim} f^{cae} f^{8eb} \{\{J^i,G^{m8}\},\{T^a,T^b\}\} + \frac{155848N_c-123989}{6336} i f^{c8e} \{J^k,[\{J^i,G^{ie}\},\{J^r,G^{r8}\}]\} \nonumber \\
& & \mbox{\hglue0.6truecm} + \frac{155848N_c-123989}{6336} i f^{c8e} \{\{J^r,G^{re}\},[J^2,G^{k8}]\} - \frac{155848N_c-123989}{6336} i f^{c8e} \{\{J^r,G^{r8}\},[J^2,G^{ke}]\} \nonumber \\
& & \mbox{\hglue0.6truecm} - \frac{155848N_c-123989}{6336} i f^{c8e} \{J^2,[G^{ke},\{J^r,G^{r8}\}]\} + \frac{155848N_c-109525}{6336} i f^{c8e} \{J^2,[G^{k8},\{J^r,G^{re}\}]\} \nonumber \\
& & \mbox{\hglue0.6truecm} + \frac{2376N_c-184945}{6336} d^{c8e} \{J^2,[G^{ke},\{J^r,G^{r8}\}]\} - \frac{2376N_c-184945}{6336} d^{c8e} \{J^2,[G^{k8},\{J^r,G^{re}\}]\} \nonumber \\
& & \mbox{\hglue0.6truecm} - \frac{72028N_c-316895}{12672} [G^{kc},\{\{J^m,G^{m8}\},\{J^r,G^{r8}\}\}] + \frac{67276N_c+52995}{6336} [G^{k8},\{\{J^m,G^{m8}\},\{J^r,G^{rc}\}\}] \nonumber \\
& & \mbox{\hglue0.6truecm} - \frac{67276N_c+52995}{6336} \{\{J^m,G^{mc}\},[G^{k8},\{J^r,G^{r8}\}]\} + \frac{2376N_c-184945}{6336} i \epsilon^{kim} f^{cea} f^{e8b} \{\{J^i,G^{m8}\},\{G^{ra},G^{rb}\}\} \nonumber \\
& & \mbox{\hglue0.6truecm} + \frac{5}{12} f^{c8e} f^{8eg} \mathcal{D}_5^{kg} + \frac{91}{144} N_c i f^{c8e} d^{8eg} \mathcal{D}_5^{kg} + \frac{91}{144} N_c i d^{c8e} f^{8eg} \mathcal{D}_5^{kg} + \frac{4}{N_f} \delta^{c8} \mathcal{D}_5^{k8} - \frac12 f^{c8e} f^{8eg} \mathcal{O}_5^{kg} - \frac53 d^{c8e} d^{8eg} \mathcal{O}_5^{kg} \nonumber \\
& & \mbox{\hglue0.6truecm} + \frac{44}{9} d^{ceg} d^{88e} \mathcal{O}_5^{kg} + \frac{107}{18} \{J^2,\{G^{kc},\{G^{r8},G^{r8}\}\}\} - \frac{107}{18} \{J^2,\{G^{k8},\{G^{rc},G^{r8}\}\}\} + \frac{17}{3} d^{c8e} \{J^2,\{J^k,\{G^{re},G^{r8}\}\}\} \nonumber \\
& & \mbox{\hglue0.6truecm} - \frac{17}{3} d^{88e} \{J^2,\{J^k,\{G^{rc},G^{re}\}\}\} + \frac{806N_c-991}{18} d^{c8e} \{J^2,\{G^{ke},\{J^r,G^{r8}\}\}\} + \frac53 d^{c8e} \{J^2,\{G^{k8},\{J^r,G^{re}\}\}\} \nonumber \\
& & \mbox{\hglue0.6truecm} - \frac{44}{9} d^{88e} \{J^2,\{G^{kc},\{J^r,G^{re}\}\}\} + \frac{806N_c-813}{18} d^{88e} \{J^2,\{G^{ke},\{J^r,G^{rc}\}\}\} \nonumber \\
& & \mbox{\hglue0.6truecm} + \frac{1}{24} (16-9N_f) \epsilon^{kim} f^{c8e} \{J^2,\{T^e,\{J^i,G^{m8}\}\}\} - \frac{1612N_c-1695}{36} \{G^{kc},\{\{J^m,G^{m8}\},\{J^r,G^{r8}\}\}\} \nonumber \\
& & \mbox{\hglue0.6truecm} + \frac{1612N_c-1695}{36} \{G^{k8},\{\{J^m,G^{m8}\},\{J^r,G^{rc}\}\}\} - \frac{311}{36} \{J^k,\{\{J^m,G^{mc}\},\{G^{r8},G^{r8}\}\}\} \nonumber \\
& & \mbox{\hglue0.6truecm} + \frac{167}{36} \{J^k,\{\{J^m,G^{m8}\},\{G^{r8},G^{rc}\}\}\} - \frac{806N_c-931}{36} d^{c8e} \{\mathcal{D}_3^{ke},\{J^r,G^{r8}\}\} - \frac{806N_c-1003}{36} d^{88e} \{\mathcal{D}_3^{kc},\{J^r,G^{re}\}\} \nonumber \\
& & \mbox{\hglue0.6truecm} - \frac{167}{36} \epsilon^{kim} f^{ab8} \{\{J^i,G^{m8}\},\{T^a,\{G^{rb},G^{rc}\}\}\} - \frac{91}{144} (N_c+N_f) i \epsilon^{kim} d^{c8e} \{J^2,\{T^e,\{J^i,G^{m8}\}\}\} \nonumber \\
& & \mbox{\hglue0.6truecm} - \frac{806N_c-931}{18} i \epsilon^{kil} [\{J^i,G^{l8}\},\{\{J^m,G^{m8}\},\{J^r,G^{rc}\}\}] - \frac{91}{72} i f^{c8e} d^{8eg} \mathcal{D}_6^{kg} - \frac{226}{99} d^{c8e} \{J^2,\{J^2,\{G^{ke},T^8\}\}\} \nonumber \\
& & \mbox{\hglue0.6truecm} - \frac{91}{36N_f} i \epsilon^{kim} \delta^{c8} \{J^2,\{J^2,\{J^i,G^{m8}\}\}\} + \frac{226}{99} d^{c8e} \{J^2,\{\mathcal{D}_2^{k8},\{J^r,G^{re}\}\}\} + \frac{1}{12} i f^{c8e} \{J^2,\{\mathcal{D}_2^{ke},\{J^r,G^{r8}\}\}\} \nonumber
\end{eqnarray}
\begin{eqnarray}
& & \mbox{\hglue0.6truecm} + \frac{226}{99} \{J^2,\{\{J^r,G^{rc}\},\{G^{k8},T^8\}\}\} - \frac{226}{99} \{J^2,\{\{J^r,G^{r8}\},\{G^{kc},T^8\}\}\} + \frac{91}{144} i \epsilon^{kim} \{J^2,\{\{T^c,T^8\},\{J^i,G^{m8}\}\}\} \nonumber \\
& & \mbox{\hglue0.6truecm} + \frac{85}{36} i \epsilon^{kim} \{J^2,\{\{G^{rc},G^{r8}\},\{J^i,G^{m8}\}\}\} + \frac16 i \epsilon^{kim} \{J^2,\{\{G^{r8},G^{r8}\},\{J^i,G^{mc}\}\}\} \nonumber \\
& & \mbox{\hglue0.6truecm} - \frac16 i \epsilon^{rim} \{J^2,\{G^{k8},\{J^r,\{G^{ic},G^{m8}\}\}\}\} - \frac{91}{72} i \epsilon^{rim} d^{c8e} \{J^2,\{J^k,\{J^r,\{G^{i8},G^{me}\}\}\}\} \nonumber \\
& & \mbox{\hglue0.6truecm} - \frac{91}{72} i \epsilon^{kim} f^{cae} f^{8eb} \{J^2,\{\{J^i,G^{m8}\},\{T^a,T^b\}\}\} - \frac{41}{12} i f^{c8e} \{J^2,\{J^k,[\{J^i,G^{ie}\},\{J^r,G^{r8}\}]\}\} \nonumber \\
& & \mbox{\hglue0.6truecm} - \frac{41}{12} i f^{c8e} \{J^2,\{\{J^r,G^{re}\},[J^2,G^{k8}]\}\} + \frac{2257}{396} i f^{c8e} \{J^2,\{\{J^r,G^{r8}\},[J^2,G^{ke}]\}\} + \frac{41}{12} i f^{c8e} \{J^2,\{J^2,[G^{ke},\{J^r,G^{r8}\}]\}\} \nonumber \\
& & \mbox{\hglue0.6truecm} - \frac{41}{12} i f^{c8e} \{J^2,\{J^2,[G^{k8},\{J^r,G^{re}\}]\}\} - \frac{226}{99} i \epsilon^{kim} [\{T^8,\{J^r,G^{r8}\}\},\{J^2,\{J^i,G^{mc}\}\}] \nonumber \\
& & \mbox{\hglue0.6truecm} + \frac{91}{72} d^{c8e} \{J^2,\{J^2,[G^{ke},\{J^r,G^{r8}\}]\}\} - \frac{91}{72} d^{c8e} \{J^2,\{J^2,[G^{k8},\{J^r,G^{re}\}]\}\} - \frac{97}{144} \{J^2,[G^{kc},\{\{J^m,G^{m8}\},\{J^r,G^{r8}\}\}]\} \nonumber \\
& & \mbox{\hglue0.6truecm} - \frac{85}{72} \{J^2,[G^{k8},\{\{J^m,G^{m8}\},\{J^r,G^{rc}\}\}]\} + \frac{85}{72} \{J^2,\{\{J^m,G^{mc}\},[G^{k8},\{J^r,G^{r8}\}]\}\} + \frac{2}{N_f} \delta^{c8} \mathcal{D}_7^{k8} \nonumber \\
& & \mbox{\hglue0.6truecm} - 2 \{J^2,\{J^2,\{G^{kc},\{G^{r8},G^{r8}\}\}\}\} + 2 \{J^2,\{J^2,\{G^{k8},\{G^{rc},G^{r8}\}\}\}\} + d^{c8e} \{J^2,\{J^2,\{J^k,\{G^{re},G^{r8}\}\}\}\} \nonumber \\
& & \mbox{\hglue0.6truecm} - d^{88e} \{J^2,\{J^2,\{J^k,\{G^{rc},G^{re}\}\}\}\} - \frac73 d^{c8e} \{J^2,\{J^2,\{G^{ke},\{J^r,G^{r8}\}\}\}\} - \frac73 d^{88e} \{J^2,\{J^2,\{G^{ke},\{J^r,G^{rc}\}\}\}\} \nonumber \\
& & \mbox{\hglue0.6truecm} + \frac{10}{3} \{J^2,\{G^{kc},\{\{J^m,G^{m8}\},\{J^r,G^{r8}\}\}\}\} - \frac{10}{3} \{J^2,\{G^{k8},\{\{J^m,G^{m8}\},\{J^r,G^{rc}\}\}\}\} \nonumber \\
& & \mbox{\hglue0.6truecm} - 4 \{J^2,\{J^k,\{\{J^m,G^{m8}\},\{G^{r8},G^{rc}\}\}\}\} + \frac53 d^{c8e} \{J^2,\{\mathcal{D}_3^{ke},\{J^r,G^{r8}\}\}\} + \frac53 d^{88e} \{J^2,\{\mathcal{D}_3^{kc},\{J^r,G^{re}\}\}\} \nonumber \\
& & \mbox{\hglue0.6truecm} + \epsilon^{kim} f^{ab8} \{J^2,\{\{J^i,G^{m8}\},\{T^a,\{G^{rb},G^{rc}\}\}\}\} - \frac{37}{9} i \epsilon^{kil} \{J^2,[\{J^i,G^{l8}\},\{\{J^m,G^{m8}\},\{J^r,G^{rc}\}\}]\} \nonumber \\
& & \mbox{\hglue0.6truecm} + \{\mathcal{D}_3^{kc},\{\{J^m,G^{m8}\},\{J^r,G^{r8}\}\}\} + \frac{58}{9} i \epsilon^{kil} \{J^2,\{J^i,\{J^r,[G^{l8},\{G^{r8},\{J^m,G^{mc}\}\}]\}\}\},
\end{eqnarray}

\begin{eqnarray}
& & [\mathcal{O}_3^{i8},[\mathcal{O}_3^{i8},\mathcal{D}_3^{kc}]] \nonumber \\
& & \mbox{\hglue0.2truecm} = \frac{422N_c-195}{48} f^{c8e} f^{8eg} G^{kg} - \frac{38489N_c-191759}{6336} i \epsilon^{kim} f^{c8e} f^{8eg} \{J^i,G^{mg}\} + \frac{422N_c-15}{144} f^{c8e} f^{8eg} \mathcal{D}_3^{kg} + 8 d^{c8e} d^{8eg} \mathcal{D}_3^{kg} \nonumber \\
& & \mbox{\hglue0.6truecm} - 4 d^{ceg} d^{88e} \mathcal{D}_3^{kg} - \frac{N_c(44352N_c+22915)}{25344} i f^{c8e} d^{8eg} \mathcal{D}_3^{kg} - \frac{N_c(44352N_c+22915)}{25344} i d^{c8e} f^{8eg} \mathcal{D}_3^{kg} - \frac{577}{144} f^{c8e} f^{8eg} \mathcal{O}_3^{kg} \nonumber \\
& & \mbox{\hglue0.6truecm} + \frac{422N_c-195}{72} d^{c8e} d^{8eg} \mathcal{O}_3^{kg} - \frac{422N_c-195}{72} d^{ceg} d^{88e} \mathcal{O}_3^{kg} + \frac{8}{N_f} d^{c88} \{J^2,J^k\} + \frac{422N_c-195}{36} \{G^{kc},\{G^{r8},G^{r8}\}\} \nonumber \\
& & \mbox{\hglue0.6truecm} - \frac{422N_c-195}{36} \{G^{k8},\{G^{rc},G^{r8}\}\} - \frac{422N_c+957}{72} d^{c8e} \{J^k,\{G^{re},G^{r8}\}\} + \frac{422N_c+381}{72} d^{88e} \{J^k,\{G^{rc},G^{re}\}\} \nonumber \\
& & \mbox{\hglue0.6truecm} + \frac{422N_c-195}{36} d^{c8e} \{G^{ke},\{J^r,G^{r8}\}\} - \frac{422N_c-195}{72} d^{c8e} \{G^{k8},\{J^r,G^{re}\}\} + \frac{422N_c-195}{72} d^{88e} \{G^{kc},\{J^r,G^{re}\}\} \nonumber \\
& & \mbox{\hglue0.6truecm} - \frac{422N_c-195}{36} d^{88e} \{G^{ke},\{J^r,G^{rc}\}\} + \frac{1688N_c-1515N_f+5280}{576} \epsilon^{kim} f^{c8e} \{T^e,\{J^i,G^{m8}\}\} \nonumber \\
& & \mbox{\hglue0.6truecm} + \frac{44352N_c+22915}{12672} i f^{c8e} d^{8eg} \mathcal{D}_4^{kg} + \frac{29}{24} i \epsilon^{kim} f^{c8e} f^{8eg} \{J^2,\{J^i,G^{mg}\}\} \nonumber \\
& & \mbox{\hglue0.6truecm} - \frac{N_c(2684N_c+5368N_f-44352)-22915}{6336N_f} i \epsilon^{kim} \delta^{c8} \{J^2,\{J^i,G^{m8}\}\} - \frac{38489N_c-191759}{3168} i f^{c8e} \{\mathcal{D}_2^{ke},\{J^r,G^{r8}\}\} \nonumber \\
& & \mbox{\hglue0.6truecm} + \frac{109604N_c-789951}{6336} i \epsilon^{kim} \{\{J^i,G^{m8}\},\{G^{r8},G^{rc}\}\} - \frac{38489N_c-191759}{1584} i \epsilon^{kim} \{\{J^i,G^{mc}\},\{G^{r8},G^{r8}\}\} \nonumber \\
& & \mbox{\hglue0.6truecm} + \frac{38489N_c-191759}{1584} i \epsilon^{rim} \{G^{k8},\{J^r,\{G^{ic},G^{m8}\}\}\} + \frac{44352N_c+22915}{12672} i \epsilon^{rim} d^{c8e} \{J^k,\{J^r,\{G^{i8},G^{me}\}\}\} \nonumber \\
& & \mbox{\hglue0.6truecm} + \frac{44352N_c+22915}{16896} i \epsilon^{kim} f^{cae} f^{8eb} \{\{J^i,G^{m8}\},\{T^a,T^b\}\} + \frac{39974N_c-48217}{6336} i f^{c8e} \{J^k,[\{J^i,G^{ie}\},\{J^r,G^{r8}\}]\} \nonumber \\
& & \mbox{\hglue0.6truecm} + \frac{39974N_c-48217}{6336} i f^{c8e} \{\{J^r,G^{re}\},[J^2,G^{k8}]\} - \frac{39974N_c-48217}{6336} i f^{c8e} \{\{J^r,G^{r8}\},[J^2,G^{ke}]\} \nonumber \\
& & \mbox{\hglue0.6truecm} - \frac{39974N_c-48217}{6336} i f^{c8e} \{J^2,[G^{ke},\{J^r,G^{r8}\}]\} + \frac{39974N_c-20921}{6336} i f^{c8e} \{J^2,[G^{k8},\{J^r,G^{re}\}]\} \nonumber \\
& & \mbox{\hglue0.6truecm} - \frac{44352N_c+22915}{12672} d^{c8e} \{J^2,[G^{ke},\{J^r,G^{r8}\}]\} + \frac{44352N_c+22915}{12672} d^{c8e} \{J^2,[G^{k8},\{J^r,G^{re}\}]\} \nonumber \\
& & \mbox{\hglue0.6truecm} + \frac{198308N_c-744121}{25344} [G^{kc},\{\{J^m,G^{m8}\},\{J^r,G^{r8}\}\}] - \frac{109604N_c-789951}{12672} [G^{k8},\{\{J^m,G^{m8}\},\{J^r,G^{rc}\}\}] \nonumber \\
& & \mbox{\hglue0.6truecm} + \frac{109604N_c-789951}{12672} \{\{J^m,G^{mc}\},[G^{k8},\{J^r,G^{r8}\}]\} - \frac{44352N_c+22915}{12672} i \epsilon^{kim} f^{cea} f^{e8b} \{\{J^i,G^{m8}\},\{G^{ra},G^{rb}\}\} \nonumber \\
& & \mbox{\hglue0.6truecm} + \frac{113}{24} f^{c8e} f^{8eg} \mathcal{D}_5^{kg} + 11 d^{c8e} d^{8eg} \mathcal{D}_5^{kg} - 6 d^{ceg} d^{88e} \mathcal{D}_5^{kg} + \frac{61}{288} N_c i f^{c8e} d^{8eg} \mathcal{D}_5^{kg} + \frac{61}{288} N_c i d^{c8e} f^{8eg} \mathcal{D}_5^{kg} + \frac{2}{N_f} \delta^{c8} \mathcal{D}_5^{k8} \nonumber \\
& & \mbox{\hglue0.6truecm} - \frac12 f^{c8e} f^{8eg} \mathcal{O}_5^{kg} - \frac13 d^{c8e} d^{8eg} \mathcal{O}_5^{kg} + \frac{11}{18} d^{ceg} d^{88e} \mathcal{O}_5^{kg} + \frac{10}{N_f} d^{c88} \{J^2,\{J^2,J^k\}\} - \frac{529}{36} \{J^2,\{G^{kc},\{G^{r8},G^{r8}\}\}\} \nonumber \\
& & \mbox{\hglue0.6truecm} + \frac{529}{36} \{J^2,\{G^{k8},\{G^{rc},G^{r8}\}\}\} - \frac{65}{3} d^{c8e} \{J^2,\{J^k,\{G^{re},G^{r8}\}\}\} + \frac{23}{3} d^{88e} \{J^2,\{J^k,\{G^{rc},G^{re}\}\}\} \nonumber \\
& & \mbox{\hglue0.6truecm} + \frac{844N_c+67}{72} d^{c8e} \{J^2,\{G^{ke},\{J^r,G^{r8}\}\}\} + \frac13 d^{c8e} \{J^2,\{G^{k8},\{J^r,G^{re}\}\}\} - \frac{11}{18} d^{88e} \{J^2,\{G^{kc},\{J^r,G^{re}\}\}\} \nonumber \\
& & \mbox{\hglue0.6truecm} + \frac{844N_c+183}{72} d^{88e} \{J^2,\{G^{ke},\{J^r,G^{rc}\}\}\} - \frac{9N_f-32}{24} \epsilon^{kim} f^{c8e} \{J^2,\{T^e,\{J^i,G^{m8}\}\}\} \nonumber \\
& & \mbox{\hglue0.6truecm} - \frac{422N_c-195}{36} \{G^{kc},\{\{J^m,G^{m8}\},\{J^r,G^{r8}\}\}\} + \frac{422N_c-195}{36} \{G^{k8},\{\{J^m,G^{m8}\},\{J^r,G^{rc}\}\}\} \nonumber \\
& & \mbox{\hglue0.6truecm} - \frac{71}{72} \{J^k,\{\{J^m,G^{mc}\},\{G^{r8},G^{r8}\}\}\} - \frac{73}{72} \{J^k,\{\{J^m,G^{m8}\},\{G^{r8},G^{rc}\}\}\} - \frac{844N_c-29}{144} d^{c8e} \{\mathcal{D}_3^{ke},\{J^r,G^{r8}\}\} \nonumber \\
& & \mbox{\hglue0.6truecm} - \frac{844N_c-173}{144} d^{88e} \{\mathcal{D}_3^{kc},\{J^r,G^{re}\}\} + \frac{505}{72} \epsilon^{kim} f^{ab8} \{\{J^i,G^{m8}\},\{T^a,\{G^{rb},G^{rc}\}\}\} \nonumber \\
& & \mbox{\hglue0.6truecm} - \frac{61}{288} (N_c+N_f) i \epsilon^{kim} d^{c8e} \{J^2,\{T^e,\{J^i,G^{m8}\}\}\} - \frac{844N_c+115}{72} i \epsilon^{kil} [\{J^i,G^{l8}\},\{\{J^m,G^{m8}\},\{J^r,G^{rc}\}\}] \nonumber \\
& & \mbox{\hglue0.6truecm} - \frac{61}{144} i f^{c8e} d^{8eg} \mathcal{D}_6^{kg} - \frac{853}{198} d^{c8e} \{J^2,\{J^2,\{G^{ke},T^8\}\}\} - \frac{61}{72N_f} i \epsilon^{kim} \delta^{c8} \{J^2,\{J^2,\{J^i,G^{m8}\}\}\} \nonumber
\end{eqnarray}
\begin{eqnarray}
& & \mbox{\hglue0.6truecm} + \frac{853}{198} d^{c8e} \{J^2,\{\mathcal{D}_2^{k8},\{J^r,G^{re}\}\}\} + \frac{29}{12} i f^{c8e} \{J^2,\{\mathcal{D}_2^{ke},\{J^r,G^{r8}\}\}\} + \frac{853}{198} \{J^2,\{\{J^r,G^{rc}\},\{G^{k8},T^8\}\}\} \nonumber \\
& & \mbox{\hglue0.6truecm} - \frac{853}{198} \{J^2,\{\{J^r,G^{r8}\},\{G^{kc},T^8\}\}\} + \frac{61}{288} i \epsilon^{kim} \{J^2,\{\{T^c,T^8\},\{J^i,G^{m8}\}\}\} - \frac{287}{72} i \epsilon^{kim} \{J^2,\{\{G^{rc},G^{r8}\},\{J^i,G^{m8}\}\}\} \nonumber \\
& & \mbox{\hglue0.6truecm} + \frac{29}{6} i \epsilon^{kim} \{J^2,\{\{G^{r8},G^{r8}\},\{J^i,G^{mc}\}\}\} - \frac{29}{6} i \epsilon^{rim} \{J^2,\{G^{k8},\{J^r,\{G^{ic},G^{m8}\}\}\}\} \nonumber \\
& & \mbox{\hglue0.6truecm} - \frac{61}{144} i \epsilon^{rim} d^{c8e} \{J^2,\{J^k,\{J^r,\{G^{i8},G^{me}\}\}\}\} - \frac{61}{144} i \epsilon^{kim} f^{cae} f^{8eb} \{J^2,\{\{J^i,G^{m8}\},\{T^a,T^b\}\}\} \nonumber \\
& & \mbox{\hglue0.6truecm} + \frac{11}{12} i f^{c8e} \{J^2,\{J^k,[\{J^i,G^{ie}\},\{J^r,G^{r8}\}]\}\} + \frac{11}{12} i f^{c8e} \{J^2,\{\{J^r,G^{re}\},[J^2,G^{k8}]\}\} \nonumber \\
& & \mbox{\hglue0.6truecm} + \frac{1343}{396} i f^{c8e} \{J^2,\{\{J^r,G^{r8}\},[J^2,G^{ke}]\}\} - \frac{11}{12} i f^{c8e} \{J^2,\{J^2,[G^{ke},\{J^r,G^{r8}\}]\}\} + \frac{11}{12} i f^{c8e} \{J^2,\{J^2,[G^{k8},\{J^r,G^{re}\}]\}\} \nonumber \\
& & \mbox{\hglue0.6truecm} - \frac{853}{198} i \epsilon^{kim} [\{T^8,\{J^r,G^{r8}\}\},\{J^2,\{J^i,G^{mc}\}\}] + \frac{61}{144} d^{c8e} \{J^2,\{J^2,[G^{ke},\{J^r,G^{r8}\}]\}\} \nonumber \\
& & \mbox{\hglue0.6truecm} - \frac{61}{144} d^{c8e} \{J^2,\{J^2,[G^{k8},\{J^r,G^{re}\}]\}\} - \frac{409}{288} \{J^2,[G^{kc},\{\{J^m,G^{m8}\},\{J^r,G^{r8}\}\}]\} \nonumber \\
& & \mbox{\hglue0.6truecm} + \frac{287}{144} \{J^2,[G^{k8},\{\{J^m,G^{m8}\},\{J^r,G^{rc}\}\}]\} - \frac{287}{144} \{J^2,\{\{J^m,G^{mc}\},[G^{k8},\{J^r,G^{r8}\}]\}\} + f^{c8e} f^{8eg} \mathcal{D}_7^{kg} \nonumber \\
& & \mbox{\hglue0.6truecm} + \frac32 d^{c8e} d^{8eg} \mathcal{D}_7^{kg} - d^{ceg} d^{88e} \mathcal{D}_7^{kg} + \frac{1}{N_f} \delta^{c8} \mathcal{D}_7^{k8} + \frac{1}{N_f} d^{c88} \{J^2,\{J^2,\{J^2,J^k\}\}\} - 2 \{J^2,\{J^2,\{G^{kc},\{G^{r8},G^{r8}\}\}\}\} \nonumber \\
& & \mbox{\hglue0.6truecm} + 2 \{J^2,\{J^2,\{G^{k8},\{G^{rc},G^{r8}\}\}\}\} - \frac92 d^{c8e} \{J^2,\{J^2,\{J^k,\{G^{re},G^{r8}\}\}\}\} + \frac12 d^{88e} \{J^2,\{J^2,\{J^k,\{G^{rc},G^{re}\}\}\}\} \nonumber \\
& & \mbox{\hglue0.6truecm} + \frac13 d^{c8e} \{J^2,\{J^2,\{G^{ke},\{J^r,G^{r8}\}\}\}\} + \frac13 d^{88e} \{J^2,\{J^2,\{G^{ke},\{J^r,G^{rc}\}\}\}\} + \frac23 \{J^2,\{G^{kc},\{\{J^m,G^{m8}\},\{J^r,G^{r8}\}\}\}\} \nonumber \\
& & \mbox{\hglue0.6truecm} - \frac23 \{J^2,\{G^{k8},\{\{J^m,G^{m8}\},\{J^r,G^{rc}\}\}\}\} - \frac92 \{J^2,\{J^k,\{\{J^m,G^{mc}\},\{G^{r8},G^{r8}\}\}\}\} \nonumber \\
& & \mbox{\hglue0.6truecm} + \frac52 \{J^2,\{J^k,\{\{J^m,G^{m8}\},\{G^{r8},G^{rc}\}\}\}\} + \frac{13}{12} d^{c8e} \{J^2,\{\mathcal{D}_3^{ke},\{J^r,G^{r8}\}\}\} + \frac{7}{12} d^{88e} \{J^2,\{\mathcal{D}_3^{kc},\{J^r,G^{re}\}\}\} \nonumber \\
& & \mbox{\hglue0.6truecm} + \epsilon^{kim} f^{ab8} \{J^2,\{\{J^i,G^{m8}\},\{T^a,\{G^{rb},G^{rc}\}\}\}\} - \frac89 i \epsilon^{kil} \{J^2,[\{J^i,G^{l8}\},\{\{J^m,G^{m8}\},\{J^r,G^{rc}\}\}]\} \nonumber \\
& & \mbox{\hglue0.6truecm} + \frac12 \{\mathcal{D}_3^{kc},\{\{J^m,G^{m8}\},\{J^r,G^{r8}\}\}\} + \frac59 i \epsilon^{kil} \{J^2,\{J^i,\{J^r,[G^{l8},\{G^{r8},\{J^m,G^{mc}\}\}]\}\}\},
\end{eqnarray}

\begin{eqnarray}
& & [\mathcal{O}_3^{i8},[\mathcal{O}_3^{i8},\mathcal{O}_3^{kc}]] \nonumber \\
& & \mbox{\hglue0.2truecm} = - \frac{17872N_c-34725}{768} f^{c8e} f^{8eg} G^{kg} - \frac{196856N_c-65447}{101376} i \epsilon^{kim} f^{c8e} f^{8eg} \{J^i,G^{mg}\} - \frac{17872N_c-36093}{2304} f^{c8e} f^{8eg} \mathcal{D}_3^{kg} \nonumber \\
& & \mbox{\hglue0.6truecm} + \frac{N_c(323928N_c-571207)}{101376} (i f^{c8e} d^{8eg} \mathcal{D}_3^{kg} + i d^{c8e} f^{8eg} \mathcal{D}_3^{kg}) + \frac{3431}{2304} f^{c8e} f^{8eg} \mathcal{O}_3^{kg} - \frac{17872N_c-34725}{1152} d^{c8e} d^{8eg} \mathcal{O}_3^{kg} \nonumber \\
& & \mbox{\hglue0.6truecm} + \frac{17872N_c-34725}{1152} d^{ceg} d^{88e} \mathcal{O}_3^{kg} - \frac{17872N_c-34725}{576} \{G^{kc},\{G^{r8},G^{r8}\}\} + \frac{17872N_c-34725}{576} \{G^{k8},\{G^{rc},G^{r8}\}\} \nonumber \\
& & \mbox{\hglue0.6truecm} + \frac{17872N_c-34725}{1152} d^{c8e} \{J^k,\{G^{re},G^{r8}\}\} - \frac{17872N_c-34725}{1152} d^{88e} \{J^k,\{G^{rc},G^{re}\}\} \nonumber \\
& & \mbox{\hglue0.6truecm} - \frac{17872N_c-34725}{576} d^{c8e} \{G^{ke},\{J^r,G^{r8}\}\} + \frac{17872N_c-34725}{1152} d^{c8e} \{G^{k8},\{J^r,G^{re}\}\} \nonumber \\
& & \mbox{\hglue0.6truecm} - \frac{17872N_c-34725}{1152} d^{88e} \{G^{kc},\{J^r,G^{re}\}\} + \frac{17872N_c-34725}{576} d^{88e} \{G^{ke},\{J^r,G^{rc}\}\} \nonumber \\
& & \mbox{\hglue0.6truecm} - \frac{71488N_c-2085N_f-130560}{9216} \epsilon^{kim} f^{c8e} \{T^e,\{J^i,G^{m8}\}\} - \frac{323928N_c-571207}{50688} i f^{c8e} d^{8eg} \mathcal{D}_4^{kg} \nonumber \\
& & \mbox{\hglue0.6truecm} - \frac{77}{192} i \epsilon^{kim} f^{c8e} f^{8eg} \{J^2,\{J^i,G^{mg}\}\} + \frac{N_c(24904N_c+49808N_f-323928)+571207}{25344N_f} i \epsilon^{kim} \delta^{c8} \{J^2,\{J^i,G^{m8}\}\} \nonumber \\
& & \mbox{\hglue0.6truecm} - \frac{196856N_c-65447}{50688} i f^{c8e} \{\mathcal{D}_2^{ke},\{J^r,G^{r8}\}\} + \frac{260392N_c-318327}{12672} i \epsilon^{kim} \{\{J^i,G^{m8}\},\{G^{r8},G^{rc}\}\} \nonumber \\
& & \mbox{\hglue0.6truecm} - \frac{196856N_c-65447}{25344} i \epsilon^{kim} \{\{J^i,G^{mc}\},\{G^{r8},G^{r8}\}\} + \frac{196856N_c-65447}{25344} i \epsilon^{rim} \{G^{k8},\{J^r,\{G^{ic},G^{m8}\}\}\} \nonumber \\
& & \mbox{\hglue0.6truecm} - \frac{323928N_c-571207}{50688} i \epsilon^{rim} d^{c8e} \{J^k,\{J^r,\{G^{i8},G^{me}\}\}\} - \frac{323928N_c-571207}{67584} i \epsilon^{kim} f^{cae} f^{8eb} \{\{J^i,G^{m8}\},\{T^a,T^b\}\} \nonumber \\
& & \mbox{\hglue0.6truecm} - \frac{696344N_c-971293}{50688} i f^{c8e} \{J^k,[\{J^i,G^{ie}\},\{J^r,G^{r8}\}]\} - \frac{696344N_c-971293}{50688} i f^{c8e} \{\{J^r,G^{re}\},[J^2,G^{k8}]\} \nonumber \\
& & \mbox{\hglue0.6truecm} + \frac{696344N_c-971293}{50688} i f^{c8e} \{\{J^r,G^{r8}\},[J^2,G^{ke}]\} + \frac{696344N_c-971293}{50688} i f^{c8e} \{J^2,[G^{ke},\{J^r,G^{r8}\}]\} \nonumber \\
& & \mbox{\hglue0.6truecm} - \frac{696344N_c-912317}{50688} i f^{c8e} \{J^2,[G^{k8},\{J^r,G^{re}\}]\} + \frac{323928N_c-571207}{50688} d^{c8e} \{J^2,[G^{ke},\{J^r,G^{r8}\}]\} \nonumber \\
& & \mbox{\hglue0.6truecm} - \frac{323928N_c-571207}{50688} d^{c8e} \{J^2,[G^{k8},\{J^r,G^{re}\}]\} - \frac{3971N_c-15805}{3168} [G^{kc},\{\{J^m,G^{m8}\},\{J^r,G^{r8}\}\}] \nonumber \\
& & \mbox{\hglue0.6truecm} - \frac{260392N_c-318327}{25344} [G^{k8},\{\{J^m,G^{m8}\},\{J^r,G^{rc}\}\}] + \frac{260392N_c-318327}{25344} \{\{J^m,G^{mc}\},[G^{k8},\{J^r,G^{r8}\}]\} \nonumber \\
& & \mbox{\hglue0.6truecm} + \frac{323928N_c-571207}{50688} i \epsilon^{kim} f^{cea} f^{e8b} \{\{J^i,G^{m8}\},\{G^{ra},G^{rb}\}\} + \frac{19}{48} f^{c8e} f^{8eg} \mathcal{D}_5^{kg} - \frac{283}{576} N_c i f^{c8e} d^{8eg} \mathcal{D}_5^{kg} \nonumber \\
& & \mbox{\hglue0.6truecm} - \frac{283}{576} N_c i d^{c8e} f^{8eg} \mathcal{D}_5^{kg} + \frac{29}{8} f^{c8e} f^{8eg} \mathcal{O}_5^{kg} + \frac{91}{24} d^{c8e} d^{8eg} \mathcal{O}_5^{kg} - \frac{343}{72} d^{ceg} d^{88e} \mathcal{O}_5^{kg} + \frac{12}{N_f} \delta^{c8} \mathcal{O}_5^{k8} \nonumber \\
& & \mbox{\hglue0.6truecm} + \frac{455}{576} \{J^2,\{G^{kc},\{G^{r8},G^{r8}\}\}\} - \frac{1607}{576} \{J^2,\{G^{k8},\{G^{rc},G^{r8}\}\}\} - \frac{19}{24} d^{c8e} \{J^2,\{J^k,\{G^{re},G^{r8}\}\}\} \nonumber \\
& & \mbox{\hglue0.6truecm} + \frac{19}{24} d^{88e} \{J^2,\{J^k,\{G^{rc},G^{re}\}\}\} - \frac{35744N_c-67123}{1152} d^{c8e} \{J^2,\{G^{ke},\{J^r,G^{r8}\}\}\} + \frac{53}{24} d^{c8e} \{J^2,\{G^{k8},\{J^r,G^{re}\}\}\} \nonumber \\
& & \mbox{\hglue0.6truecm} + \frac{343}{72} d^{88e} \{J^2,\{G^{kc},\{J^r,G^{re}\}\}\} - \frac{35744N_c-63507}{1152} d^{88e} \{J^2,\{G^{ke},\{J^r,G^{rc}\}\}\} \nonumber \\
& & \mbox{\hglue0.6truecm} + \frac{1}{48} (9N_f+55) \epsilon^{kim} f^{c8e} \{J^2,\{T^e,\{J^i,G^{m8}\}\}\} + \frac{17872N_c-34725}{576} \{G^{kc},\{\{J^m,G^{m8}\},\{J^r,G^{r8}\}\}\} \nonumber \\
& & \mbox{\hglue0.6truecm} - \frac{17872N_c-34725}{576} \{G^{k8},\{\{J^m,G^{m8}\},\{J^r,G^{rc}\}\}\} + \frac{457}{1152} \{J^k,\{\{J^m,G^{mc}\},\{G^{r8},G^{r8}\}\}\} \nonumber \\
& & \mbox{\hglue0.6truecm} + \frac{695}{1152} \{J^k,\{\{J^m,G^{m8}\},\{G^{r8},G^{rc}\}\}\} + \frac{35744N_c-68755}{2304} d^{c8e} \{\mathcal{D}_3^{ke},\{J^r,G^{r8}\}\} \nonumber \\
& & \mbox{\hglue0.6truecm} + \frac{35744N_c-69907}{2304} d^{88e} \{\mathcal{D}_3^{kc},\{J^r,G^{re}\}\} - \frac{695}{1152} \epsilon^{kim} f^{ab8} \{\{J^i,G^{m8}\},\{T^a,\{G^{rb},G^{rc}\}\}\} \nonumber \\
& & \mbox{\hglue0.6truecm} + \frac{283}{576} (N_c+N_f) i \epsilon^{kim} d^{c8e} \{J^2,\{T^e,\{J^i,G^{m8}\}\}\} + \frac{35744N_c-68755}{1152} i \epsilon^{kil} [\{J^i,G^{l8}\},\{\{J^m,G^{m8}\},\{J^r,G^{rc}\}\}] \nonumber
\end{eqnarray}
\begin{eqnarray}
& & \mbox{\hglue0.6truecm} + \frac{283}{288} i f^{c8e} d^{8eg} \mathcal{D}_6^{kg} + \frac{1843}{1584} d^{c8e} \{J^2,\{J^2,\{G^{ke},T^8\}\}\} + \frac{283}{144N_f} i \epsilon^{kim} \delta^{c8} \{J^2,\{J^2,\{J^i,G^{m8}\}\}\} \nonumber \\
& & \mbox{\hglue0.6truecm} - \frac{1843}{1584} d^{c8e} \{J^2,\{\mathcal{D}_2^{k8},\{J^r,G^{re}\}\}\} - \frac{77}{96} i f^{c8e} \{J^2,\{\mathcal{D}_2^{ke},\{J^r,G^{r8}\}\}\} - \frac{1843}{1584} \{J^2,\{\{J^r,G^{rc}\},\{G^{k8},T^8\}\}\} \nonumber \\
& & \mbox{\hglue0.6truecm} + \frac{1843}{1584} \{J^2,\{\{J^r,G^{r8}\},\{G^{kc},T^8\}\}\} - \frac{283}{576} i \epsilon^{kim} \{J^2,\{\{T^c,T^8\},\{J^i,G^{m8}\}\}\} - \frac{13}{36} i \epsilon^{kim} \{J^2,\{\{G^{rc},G^{r8}\},\{J^i,G^{m8}\}\}\} \nonumber \\
& & \mbox{\hglue0.6truecm} - \frac{77}{48} i \epsilon^{kim} \{J^2,\{\{G^{r8},G^{r8}\},\{J^i,G^{mc}\}\}\} + \frac{77}{48} i \epsilon^{rim} \{J^2,\{G^{k8},\{J^r,\{G^{ic},G^{m8}\}\}\}\} \nonumber \\
& & \mbox{\hglue0.6truecm} + \frac{283}{288} i \epsilon^{rim} d^{c8e} \{J^2,\{J^k,\{J^r,\{G^{i8},G^{me}\}\}\}\} + \frac{283}{288} i \epsilon^{kim} f^{cae} f^{8eb} \{J^2,\{\{J^i,G^{m8}\},\{T^a,T^b\}\}\} \nonumber \\
& & \mbox{\hglue0.6truecm} + \frac{73}{96} i f^{c8e} \{J^2,\{J^k,[\{J^i,G^{ie}\},\{J^r,G^{r8}\}]\}\} + \frac{73}{96} i f^{c8e} \{J^2,\{\{J^r,G^{re}\},[J^2,G^{k8}]\}\} \nonumber \\
& & \mbox{\hglue0.6truecm} - \frac{6095}{3168} i f^{c8e} \{J^2,\{\{J^r,G^{r8}\},[J^2,G^{ke}]\}\} - \frac{73}{96} i f^{c8e} \{J^2,\{J^2,[G^{ke},\{J^r,G^{r8}\}]\}\} + \frac{73}{96} i f^{c8e} \{J^2,\{J^2,[G^{k8},\{J^r,G^{re}\}]\}\} \nonumber \\
& & \mbox{\hglue0.6truecm} + \frac{1843}{1584} i \epsilon^{kim} [\{T^8,\{J^r,G^{r8}\}\},\{J^2,\{J^i,G^{mc}\}\}] - \frac{283}{288} d^{c8e} \{J^2,\{J^2,[G^{ke},\{J^r,G^{r8}\}]\}\} \nonumber \\
& & \mbox{\hglue0.6truecm} + \frac{283}{288} d^{c8e} \{J^2,\{J^2,[G^{k8},\{J^r,G^{re}\}]\}\} + \frac{257}{288} \{J^2,[G^{kc},\{\{J^m,G^{m8}\},\{J^r,G^{r8}\}\}]\} \nonumber \\
& & \mbox{\hglue0.6truecm} + \frac{13}{72} \{J^2,[G^{k8},\{\{J^m,G^{m8}\},\{J^r,G^{rc}\}\}]\} - \frac{13}{72} \{J^2,\{\{J^m,G^{mc}\},[G^{k8},\{J^r,G^{r8}\}]\}\} + \frac54 f^{c8e} f^{8eg} \mathcal{O}_7^{kg} \nonumber \\
& & \mbox{\hglue0.6truecm} + \frac54 d^{c8e} d^{8eg} \mathcal{O}_7^{kg} - d^{ceg} d^{88e} \mathcal{O}_7^{kg} + \frac{5}{N_f} \delta^{c8} \mathcal{O}_7^{k8} - \frac52 \{J^2,\{J^2,\{G^{kc},\{G^{r8},G^{r8}\}\}\}\} - \frac32 \{J^2,\{J^2,\{G^{k8},\{G^{rc},G^{r8}\}\}\}\} \nonumber \\
& & \mbox{\hglue0.6truecm} + \frac{13}{12} d^{c8e} \{J^2,\{J^2,\{G^{ke},\{J^r,G^{r8}\}\}\}\} + \frac54 d^{c8e} \{J^2,\{J^2,\{G^{k8},\{J^r,G^{re}\}\}\}\} + d^{88e} \{J^2,\{J^2,\{G^{kc},\{J^r,G^{re}\}\}\}\} \nonumber \\
& & \mbox{\hglue0.6truecm} + \frac43 d^{88e} \{J^2,\{J^2,\{G^{ke},\{J^r,G^{rc}\}\}\}\} + \frac58 \epsilon^{kim} f^{c8e} \{J^2,\{J^2,\{T^e,\{J^i,G^{m8}\}\}\}\} \nonumber \\
& & \mbox{\hglue0.6truecm} - \frac{13}{12} \{J^2,\{G^{kc},\{\{J^m,G^{m8}\},\{J^r,G^{r8}\}\}\}\} + \frac{19}{12} \{J^2,\{G^{k8},\{\{J^m,G^{m8}\},\{J^r,G^{rc}\}\}\}\} \nonumber \\
& & \mbox{\hglue0.6truecm} + \frac54 \{J^2,\{J^k,\{\{J^m,G^{mc}\},\{G^{r8},G^{r8}\}\}\}\} + \frac34 \{J^2,\{J^k,\{\{J^m,G^{m8}\},\{G^{r8},G^{rc}\}\}\}\} - \frac76 d^{c8e} \{J^2,\{\mathcal{D}_3^{ke},\{J^r,G^{r8}\}\}\} \nonumber \\
& & \mbox{\hglue0.6truecm} - \frac76 d^{88e} \{J^2,\{\mathcal{D}_3^{kc},\{J^r,G^{re}\}\}\} - \frac12 \epsilon^{kim} f^{ab8} \{J^2,\{\{J^i,G^{m8}\},\{T^a,\{G^{rb},G^{rc}\}\}\}\} \nonumber \\
& & \mbox{\hglue0.6truecm} + \frac{85}{36} i \epsilon^{kil} \{J^2,[\{J^i,G^{l8}\},\{\{J^m,G^{m8}\},\{J^r,G^{rc}\}\}]\} - \frac14 \{\mathcal{D}_3^{kc},\{\{J^m,G^{m8}\},\{J^r,G^{r8}\}\}\} \nonumber \\
& & \mbox{\hglue0.6truecm} - \frac{71}{18} i \epsilon^{kil} \{J^2,\{J^i,\{J^r,[G^{l8},\{G^{r8},\{J^m,G^{mc}\}\}]\}\}\}.
\end{eqnarray}

\section{\label{app:Loop2}Complete expressions from order $\mathcal{O}(m_q \ln m_q)$ corrections}

\subsection{\label{sec:Loop2ad}Figure \ref{fig:mmloop2}(a-d)}

The complete expressions for contributions from loop \ref{fig:mmloop2}(a-d) for $N_f=N_c=3$ can be organized as
\begin{eqnarray}
\delta \mu_{n}^{\mathrm{(loop\, 2ad)}} & = & \left[ \left( - \frac{7}{48} a_1^2 - \frac{1}{72} a_1b_2 - \frac{5}{216} a_1b_3 - \frac19 a_1c_3 - \frac{7}{432} b_2^2 - \frac{35}{648} b_2b_3 + \frac{2}{27} b_2c_3 - \frac{175}{3888} b_3^2 + \frac{10}{81} b_3c_3 - \frac{13}{108} c_3^2 \right) m_1 \right. \nonumber \\
& & \mbox{} + \left( \frac{35}{144} a_1^2 + \frac{5}{216} a_1b_2 + \frac{25}{648} a_1b_3 + \frac{5}{27} a_1c_3 + \frac{1}{432} b_2^2 + \frac{5}{648} b_2b_3 + \frac{25}{3888} b_3^2 + \frac{5}{108} c_3^2 \right) m_2 \nonumber \\
& & \mbox{} + \left( - \frac{7}{144} a_1^2 - \frac{35}{648} a_1b_2 - \frac{175}{1944} a_1b_3 + \frac{7}{81} a_1c_3 - \frac{7}{1296} b_2^2 - \frac{35}{1944} b_2b_3 - \frac{175}{11664} b_3^2 + \frac{7}{324} c_3^2 \right) m_3 \nonumber \\
& & \mbox{} + \left. \left( \frac{10}{27} a_1^2 + \frac{2}{27} a_1b_2 + \frac{10}{81} a_1b_3 + \frac{5}{27} a_1c_3 + \frac{1}{27} b_2c_3 + \frac{5}{81} b_3c_3 \right) m_4 \right] I_2(m_\pi,0,\mu) \nonumber \\
& & \mbox{} + \left[ \left( - \frac{7}{48} a_1^2 - \frac{1}{24} a_1b_2 - \frac{7}{72} a_1b_3 - \frac{23}{432} b_2^2 - \frac{25}{648} b_2b_3 + \frac{1}{27} b_2c_3 - \frac{95}{3888} b_3^2 + \frac{2}{81} b_3c_3 - \frac{1}{54} c_3^2 \right) m_1 \right. \nonumber \\
& & \mbox{} + \left( \frac{13}{144} a_1^2 + \frac{1}{216} a_1b_2 - \frac{1}{648} a_1b_3 + \frac{5}{54} a_1c_3 - \frac{1}{432} b_2^2 + \frac{1}{648} b_2b_3 - \frac{1}{3888} b_3^2 + \frac{5}{216} c_3^2 \right) m_2 \nonumber \\
& & \mbox{} + \left( \frac{11}{432} a_1^2 - \frac{25}{648} a_1b_2 - \frac{95}{1944} a_1b_3 + \frac{8}{81} a_1c_3 - \frac{23}{1296} b_2^2 - \frac{25}{1944} b_2b_3 - \frac{95}{11664} b_3^2 + \frac{2}{81} c_3^2 \right) m_3 \nonumber \\
& & \mbox{} + \left. \left( \frac{2}{27} a_1^2 + \frac{1}{27} a_1b_2 + \frac{2}{81} a_1b_3 + \frac{1}{27} a_1c_3 + \frac{1}{54} b_2c_3 + \frac{1}{81} b_3c_3 \right) m_4 \right] I_2(m_K,0,\mu) \nonumber \\
& & \mbox{} + \left[ \left( - \frac{1}{36} a_1^2 - \frac{1}{18} a_1b_2 - \frac{1}{54} a_1b_3 - \frac{1}{36} b_2^2 - \frac{1}{54} b_2b_3 - \frac{1}{324} b_3^2 \right) m_1 \right. \nonumber \\
& & \mbox{} + \left. \left( - \frac{1}{108} a_1^2 - \frac{1}{54} a_1b_2 - \frac{1}{162} a_1b_3 - \frac{1}{108} b_2^2 - \frac{1}{162} b_2b_3 - \frac{1}{972} b_3^2 \right) m_3 \right] I_2(m_\eta,0,\mu), \label{eq:mmnloop2ad}
\end{eqnarray}

\begin{eqnarray}
\delta \mu_{p}^{\mathrm{(loop\, 2ad)}} & = & \left[ \left( \frac{13}{48} a_1^2 + \frac{11}{72} a_1b_2 + \frac{55}{216} a_1b_3 - \frac19 a_1c_3 + \frac{13}{432} b_2^2 + \frac{65}{648} b_2b_3 - \frac{2}{27} b_2c_3 + \frac{325}{3888} b_3^2 - \frac{10}{81} b_3c_3 + \frac{7}{108} c_3^2 \right) m_1 \right. \nonumber \\
& & \mbox{} + \left( - \frac{17}{144} a_1^2 + \frac{25}{216} a_1b_2 + \frac{125}{648} a_1b_3 - \frac{11}{27} a_1c_3 + \frac{5}{432} b_2^2 + \frac{25}{648} b_2b_3 + \frac{125}{3888} b_3^2 - \frac{11}{108} c_3^2 \right) m_2 \nonumber \\
& & \mbox{} + \left( - \frac{281}{432} a_1^2 + \frac{65}{648} a_1b_2 + \frac{325}{1944} a_1b_3 - \frac{73}{81} a_1c_3 + \frac{13}{1296} b_2^2 + \frac{65}{1944} b_2b_3 + \frac{325}{11664} b_3^2 - \frac{73}{324} c_3^2 \right) m_3 \nonumber \\
& & \mbox{} + \left. \left( - \frac{10}{27} a_1^2 - \frac{2}{27} a_1b_2 - \frac{10}{81} a_1b_3 - \frac{5}{27} a_1c_3 - \frac{1}{27} b_2c_3 - \frac{5}{81} b_3c_3 \right) m_4 \right] I_2(m_\pi,0,\mu) \nonumber \\
& & \mbox{} + \left[ \left( \frac16 a_1^2 + \frac{5}{36} a_1b_2 + \frac{5}{54} a_1b_3 + \frac{1}{36} a_1c_3 + \frac{2}{27} b_2^2 + \frac{11}{324} b_2b_3 + \frac{1}{54} b_2c_3 + \frac{29}{972} b_3^2 - \frac{7}{162} b_3c_3 + \frac{17}{432} c_3^2 \right) m_1 \right. \nonumber \\
& & \mbox{} + \left( \frac{7}{72} a_1^2 + \frac{1}{27} a_1b_2 + \frac{23}{324} a_1b_3 - \frac{1}{108} a_1c_3 + \frac{5}{216} b_2^2 + \frac{1}{81} b_2b_3 + \frac{23}{1944} b_3^2 - \frac{1}{432} c_3^2 \right) m_2 \nonumber \\
& & \mbox{} + \left( \frac{1}{54} a_1^2 + \frac{11}{324} a_1b_2 + \frac{29}{486} a_1b_3 - \frac{23}{324} a_1c_3 + \frac{2}{81} b_2^2 + \frac{11}{972} b_2b_3 + \frac{29}{2916} b_3^2 - \frac{23}{1296} c_3^2 \right) m_3 \nonumber \\
& & \mbox{} + \left. \left( - \frac{7}{54} a_1^2 + \frac{1}{54} a_1b_2 - \frac{7}{162} a_1b_3 - \frac{7}{108} a_1c_3 + \frac{1}{108} b_2c_3 - \frac{7}{324} b_3c_3 \right) m_4 \right] I_2(m_K,0,\mu) \nonumber \\
& & \mbox{} + \left[ \left( \frac{1}{24} a_1^2 + \frac{1}{12} a_1b_2 + \frac{1}{36} a_1b_3 + \frac{1}{24} b_2^2 + \frac{1}{36} b_2b_3 + \frac{1}{216} b_3^2 \right) m_1 \right. \nonumber \\
& & \mbox{} + \left( \frac{1}{72} a_1^2 + \frac{1}{36} a_1b_2 + \frac{1}{108} a_1b_3 + \frac{1}{72} b_2^2 + \frac{1}{108} b_2b_3 + \frac{1}{648} b_3^2 \right) m_2 \nonumber \\
& & \mbox{} + \left. \left( \frac{1}{72} a_1^2 + \frac{1}{36} a_1b_2 + \frac{1}{108} a_1b_3 + \frac{1}{72} b_2^2 + \frac{1}{108} b_2b_3 + \frac{1}{648} b_3^2 \right) m_3 \right] I_2(m_\eta,0,\mu),
\end{eqnarray}

\begin{eqnarray}
\delta \mu_{\Sigma^-}^{\mathrm{(loop\, 2ad)}} & = & \left[ \left( - \frac{1}{12} a_1^2 - \frac{13}{108} a_1b_2 - \frac{5}{81} a_1b_3 + \frac{1}{108} a_1c_3 - \frac{1}{36} b_2^2 - \frac{1}{36} b_2b_3 - \frac{1}{54} b_2c_3 - \frac{1}{81} b_3^2 + \frac{1}{162} b_3c_3 - \frac{1}{432} c_3^2 \right) m_1 \right. \nonumber \\
& & \mbox{} + \left( - \frac{13}{72} a_1^2 - \frac{7}{54} a_1b_2 - \frac{37}{324} a_1b_3 - \frac{1}{108} a_1c_3 - \frac{7}{216} b_2^2 - \frac{7}{162} b_2b_3 - \frac{37}{1944} b_3^2 - \frac{1}{432} c_3^2 \right) m_2 \nonumber \\
& & \mbox{} + \left( \frac{7}{324} a_1^2 - \frac{1}{36} a_1b_2 - \frac{2}{81} a_1b_3 + \frac{19}{324} a_1c_3 - \frac{1}{108} b_2^2 - \frac{1}{108} b_2b_3 - \frac{1}{243} b_3^2 + \frac{19}{1296} c_3^2 \right) m_3 \nonumber \\
& & \mbox{} + \left. \left( \frac{1}{54} a_1^2 - \frac{1}{54} a_1b_2 + \frac{1}{162} a_1b_3 + \frac{1}{108} a_1c_3 - \frac{1}{108} b_2c_3 + \frac{1}{324} b_3c_3 \right) m_4 \right] I_2(m_\pi,0,\mu) \nonumber \\
& & \mbox{} + \left[ \left( - \frac{11}{144} a_1^2 - \frac{31}{216} a_1b_2 - \frac{89}{648} a_1b_3 + \frac{7}{54} a_1c_3 - \frac{1}{48} b_2^2 - \frac{5}{216} b_2b_3 - \frac{1}{27} b_2c_3 - \frac{35}{1296} b_3^2 + \frac{1}{81} b_3c_3 + \frac{5}{216} c_3^2 \right) m_1 \right. \nonumber \\
& & \mbox{} + \left( - \frac{7}{48} a_1^2 - \frac{17}{216} a_1b_2 - \frac{103}{648} a_1b_3 + \frac{5}{54} a_1c_3 - \frac{7}{432} b_2^2 - \frac{17}{648} b_2b_3 - \frac{103}{3888} b_3^2 + \frac{5}{216} c_3^2 \right) m_2 \nonumber \\
& & \mbox{} + \left( \frac{575}{1296} a_1^2 - \frac{5}{216} a_1b_2 - \frac{35}{648} a_1b_3 + \frac{85}{162} a_1c_3 - \frac{1}{144} b_2^2 - \frac{5}{648} b_2b_3 - \frac{35}{3888} b_3^2 + \frac{85}{648} c_3^2 \right) m_3 \nonumber \\
& & \mbox{} + \left. \left( \frac{1}{27} a_1^2 - \frac{1}{27} a_1b_2 + \frac{1}{81} a_1b_3 + \frac{1}{54} a_1c_3 - \frac{1}{54} b_2c_3 + \frac{1}{162} b_3c_3 \right) m_4 \right] I_2(m_K,0,\mu) \nonumber \\
& & \mbox{} + \left[ \left( - \frac{1}{27} a_1b_3 + \frac{1}{18} a_1c_3 - \frac{1}{162} b_3^2 + \frac{1}{72} c_3^2 \right) m_1 + \left( - \frac{1}{27} a_1b_3 + \frac{1}{18} a_1c_3 - \frac{1}{162} b_3^2 + \frac{1}{72} c_3^2 \right) m_2 \right. \nonumber \\
& & \mbox{} + \left. \left( \frac{5}{27} a_1^2 - \frac{1}{81} a_1b_3 + \frac{11}{54} a_1c_3 - \frac{1}{486} b_3^2 + \frac{11}{216} c_3^2 \right) m_3 \right] I_2(m_\eta,0,\mu),
\end{eqnarray}

\begin{eqnarray}
\delta \mu_{\Sigma^0}^{\mathrm{(loop\, 2ad)}} & = & \left[ \left( \frac{1}{12} a_1^2 + \frac{2}{27} a_1b_2 + \frac{11}{162} a_1b_3 - \frac{1}{54} a_1c_3 + \frac{1}{27} b_2^2 + \frac{4}{81} b_2b_3 - \frac{1}{27} b_2c_3 + \frac{19}{972} b_3^2 - \frac{2}{81} b_3c_3 + \frac{1}{72} c_3^2 \right) m_1 \right. \nonumber \\
& & \mbox{} + \left( \frac{25}{324} a_1^2 + \frac{4}{81} a_1b_2 + \frac{19}{486} a_1b_3 + \frac{1}{54} a_1c_3 + \frac{1}{81} b_2^2 + \frac{4}{243} b_2b_3 + \frac{19}{2916} b_3^2 + \frac{1}{216} c_3^2 \right) m_3 \nonumber \\
& & \mbox{} + \left. \left( - \frac{2}{27} a_1^2 - \frac{1}{27} a_1b_2 - \frac{2}{81} a_1b_3 - \frac{1}{27} a_1c_3 - \frac{1}{54} b_2c_3 - \frac{1}{81} b_3c_3 \right) m_4 \right] I_2(m_\pi,0,\mu) \nonumber \\
& & \mbox{} + \left[ \left( \frac{7}{144} a_1^2 - \frac{1}{54} a_1b_2 + \frac{1}{648} a_1b_3 + \frac{5}{108} a_1c_3 + \frac{5}{432} b_2^2 + \frac{1}{162} b_2b_3 - \frac{1}{54} b_2c_3 + \frac{41}{3888} b_3^2 - \frac{5}{162} b_3c_3 + \frac{5}{144} c_3^2 \right) m_1 \right. \nonumber \\
& & \mbox{} + \left( - \frac16 a_1^2 - \frac{1}{72} a_1b_2 - \frac{1}{54} a_1b_3 - \frac{5}{36} a_1c_3 - \frac{1}{216} b_2b_3 - \frac{1}{324} b_3^2 - \frac{5}{144} c_3^2 \right) m_2 \nonumber \\
& & \mbox{} + \left( - \frac{139}{1296} a_1^2 + \frac{1}{162} a_1b_2 + \frac{41}{1944} a_1b_3 - \frac{5}{36} a_1c_3 + \frac{5}{1296} b_2^2 + \frac{1}{486} b_2b_3 + \frac{41}{11664} b_3^2 - \frac{5}{144} c_3^2 \right) m_3 \nonumber \\
& & \mbox{} + \left. \left( - \frac{5}{54} a_1^2 - \frac{1}{54} a_1b_2 - \frac{5}{162} a_1b_3 - \frac{5}{108} a_1c_3 - \frac{1}{108} b_2c_3 - \frac{5}{324} b_3c_3 \right) m_4 \right] I_2(m_K,0,\mu) \nonumber \\
& & \mbox{} + \left[ \left( \frac{1}{36} a_1^2 + \frac{1}{36} a_1c_3 + \frac{1}{162} b_3^2 - \frac{1}{54} b_3c_3 + \frac{1}{48} c_3^2 \right) m_1 + \left( \frac{5}{108} a_1^2 + \frac{1}{81} a_1b_3 + \frac{1}{36} a_1c_3 + \frac{1}{486} b_3^2 + \frac{1}{144} c_3^2 \right) m_3 \right. \nonumber \\
& & \mbox{} + \left. \left( - \frac{1}{18} a_1^2 - \frac{1}{54} a_1b_3 - \frac{1}{36} a_1c_3 - \frac{1}{108} b_3c_3 \right) m_4 \right] I_2(m_\eta,0,\mu),
\end{eqnarray}

\begin{eqnarray}
\delta \mu_{\Sigma^+}^{\mathrm{(loop\, 2ad)}} & = & \left[ \left( \frac14 a_1^2 + \frac{29}{108} a_1b_2 + \frac{16}{81} a_1b_3 - \frac{5}{108} a_1c_3 + \frac{11}{108} b_2^2 + \frac{41}{324} b_2b_3 - \frac{1}{18} b_2c_3 + \frac{25}{486} b_3^2 - \frac{1}{18} b_3c_3 + \frac{13}{432} c_3^2 \right) m_1 \right. \nonumber \\
& & \mbox{} + \left( \frac{13}{72} a_1^2 + \frac{7}{54} a_1b_2 + \frac{37}{324} a_1b_3 + \frac{1}{108} a_1c_3 + \frac{7}{216} b_2^2 + \frac{7}{162} b_2b_3 + \frac{37}{1944} b_3^2 + \frac{1}{432} c_3^2 \right) m_2 \nonumber \\
& & \mbox{} + \left( \frac{43}{324} a_1^2 + \frac{41}{324} a_1b_2 + \frac{25}{243} a_1b_3 - \frac{7}{324} a_1c_3 + \frac{11}{324} b_2^2 + \frac{41}{972} b_2b_3 + \frac{25}{1458} b_3^2 - \frac{7}{1296} c_3^2 \right) m_3 \nonumber \\
& & \mbox{} + \left. \left( - \frac16 a_1^2 - \frac{1}{18} a_1b_2 - \frac{1}{18} a_1b_3 - \frac{1}{12} a_1c_3 - \frac{1}{36} b_2c_3 - \frac{1}{36} b_3c_3 \right) m_4 \right] I_2(m_\pi,0,\mu) \nonumber \\
& & \mbox{} + \left[ \left( \frac{25}{144} a_1^2 + \frac{23}{216} a_1b_2 + \frac{91}{648} a_1b_3 - \frac{1}{27} a_1c_3 + \frac{19}{432} b_2^2 + \frac{23}{648} b_2b_3 + \frac{187}{3888} b_3^2 - \frac{2}{27} b_3c_3 + \frac{5}{108} c_3^2 \right) m_1 \right. \nonumber \\
& & \mbox{} + \left( - \frac{3}{16} a_1^2 + \frac{11}{216} a_1b_2 + \frac{79}{648} a_1b_3 - \frac{10}{27} a_1c_3 + \frac{7}{432} b_2^2 + \frac{11}{648} b_2b_3 + \frac{79}{3888} b_3^2 - \frac{5}{54} c_3^2 \right) m_2 \nonumber \\
& & \mbox{} + \left( - \frac{853}{1296} a_1^2 + \frac{23}{648} a_1b_2 + \frac{187}{1944} a_1b_3 - \frac{65}{81} a_1c_3 + \frac{19}{1296} b_2^2 + \frac{23}{1944} b_2b_3 + \frac{187}{11664} b_3^2 - \frac{65}{324} c_3^2 \right) m_3 \nonumber \\
& & \mbox{} + \left. \left( - \frac29 a_1^2 - \frac{2}{27} a_1b_3 - \frac19 a_1c_3 - \frac{1}{27} b_3c_3 \right) m_4 \right] I_2(m_K,0,\mu) \nonumber \\
& & \mbox{} + \left[ \left( \frac{1}{18} a_1^2 + \frac{1}{27} a_1b_3 + \frac{1}{54} b_3^2 - \frac{1}{27} b_3c_3 + \frac{1}{36} c_3^2 \right) m_1 + \left( \frac{1}{27} a_1b_3 - \frac{1}{18} a_1c_3 + \frac{1}{162} b_3^2 - \frac{1}{72} c_3^2 \right) m_2 \right. \nonumber \\
& & \mbox{} + \left. \left( - \frac{5}{54} a_1^2 + \frac{1}{27} a_1b_3 - \frac{4}{27} a_1c_3 + \frac{1}{162} b_3^2 - \frac{1}{27} c_3^2 \right) m_3 + \left( - \frac19 a_1^2 - \frac{1}{27} a_1b_3 - \frac{1}{18} a_1c_3 - \frac{1}{54} b_3c_3 \right) m_4 \right] I_2(m_\eta,0,\mu),
\end{eqnarray}

\begin{eqnarray}
\delta \mu_{\Xi^-}^{\mathrm{(loop\, 2ad)}} & = & \left[ \left( \frac{1}{48} a_1^2 - \frac{1}{24} a_1b_2 + \frac{1}{72} a_1b_3 - \frac{7}{432} b_2^2 + \frac{7}{648} b_2b_3 - \frac{1}{27} b_2c_3 - \frac{7}{3888} b_3^2 + \frac{1}{81} b_3c_3 - \frac{1}{108} c_3^2 \right) m_1 \right. \nonumber \\
& & \mbox{} + \left( - \frac{7}{144} a_1^2 + \frac{5}{216} a_1b_2 - \frac{5}{648} a_1b_3 - \frac{1}{27} a_1c_3 - \frac{5}{432} b_2^2 + \frac{5}{648} b_2b_3 - \frac{5}{3888} b_3^2 - \frac{1}{108} c_3^2 \right) m_2 \nonumber \\
& & \mbox{} + \left( \frac{19}{432} a_1^2 + \frac{7}{648} a_1b_2 - \frac{7}{1944} a_1b_3 + \frac{4}{81} a_1c_3 - \frac{7}{1296} b_2^2 + \frac{7}{1944} b_2b_3 - \frac{7}{11664} b_3^2 + \frac{1}{81} c_3^2 \right) m_3 \nonumber \\
& & \mbox{} + \left. \left( \frac{1}{27} a_1^2 - \frac{1}{27} a_1b_2 + \frac{1}{81} a_1b_3 + \frac{1}{54} a_1c_3 - \frac{1}{54} b_2c_3 + \frac{1}{162} b_3c_3 \right) m_4 \right] I_2(m_\pi,0,\mu) \nonumber \\
& & \mbox{} + \left[ \left( - \frac19 a_1^2 - \frac{5}{36} a_1b_2 - \frac16 a_1b_3 + \frac{5}{36} a_1c_3 - \frac{1}{54} b_2^2 - \frac{11}{324} b_2b_3 - \frac{1}{54} b_2c_3 - \frac{29}{972} b_3^2 + \frac{1}{162} b_3c_3 + \frac{13}{432} c_3^2 \right) m_1 \right. \nonumber \\
& & \mbox{} + \left( - \frac{5}{24} a_1^2 - \frac{4}{27} a_1b_2 - \frac{71}{324} a_1b_3 + \frac{13}{108} a_1c_3 - \frac{5}{216} b_2^2 - \frac{4}{81} b_2b_3 - \frac{71}{1944} b_3^2 + \frac{13}{432} c_3^2 \right) m_2 \nonumber \\
& & \mbox{} + \left( \frac49 a_1^2 - \frac{11}{324} a_1b_2 - \frac{29}{486} a_1b_3 + \frac{173}{324} a_1c_3 - \frac{1}{162} b_2^2 - \frac{11}{972} b_2b_3 - \frac{29}{2916} b_3^2 + \frac{173}{1296} c_3^2 \right) m_3 \nonumber \\
& & \mbox{} + \left. \left( \frac{1}{54} a_1^2 - \frac{1}{54} a_1b_2 + \frac{1}{162} a_1b_3 + \frac{1}{108} a_1c_3 - \frac{1}{108} b_2c_3 + \frac{1}{324} b_3c_3 \right) m_4 \right] I_2(m_K,0,\mu) \nonumber \\
& & \mbox{} + \left[ \left( - \frac{5}{72} a_1^2 - \frac{1}{12} a_1b_2 - \frac{1}{12} a_1b_3 + \frac{1}{18} a_1c_3 - \frac{1}{72} b_2^2 - \frac{1}{36} b_2b_3 - \frac{1}{72} b_3^2 + \frac{1}{72} c_3^2 \right) m_1 \right. \nonumber \\
& & \mbox{} + \left( - \frac{5}{72} a_1^2 - \frac{1}{12} a_1b_2 - \frac{1}{12} a_1b_3 + \frac{1}{18} a_1c_3 - \frac{1}{72} b_2^2 - \frac{1}{36} b_2b_3 - \frac{1}{72} b_3^2 + \frac{1}{72} c_3^2 \right) m_2 \nonumber \\
& & \mbox{} + \left. \left( \frac{35}{216} a_1^2 - \frac{1}{36} a_1b_2 - \frac{1}{36} a_1b_3 + \frac{11}{54} a_1c_3 - \frac{1}{216} b_2^2 - \frac{1}{108} b_2b_3 - \frac{1}{216} b_3^2 + \frac{11}{216} c_3^2 \right) m_3 \right] I_2(m_\eta,0,\mu),
\end{eqnarray}

\begin{eqnarray}
\delta \mu_{\Xi^0}^{\mathrm{(loop\, 2ad)}} & = & \left[ \left( - \frac{1}{16} a_1^2 + \frac{1}{72} a_1b_2 - \frac{1}{216} a_1b_3 - \frac{1}{18} a_1c_3 - \frac{11}{432} b_2^2 + \frac{11}{648} b_2b_3 - \frac{1}{54} b_2c_3 - \frac{11}{3888} b_3^2 + \frac{1}{162} b_3c_3 - \frac{1}{54} c_3^2 \right) m_1 \right. \nonumber \\
& & \mbox{} + \left( \frac{13}{144} a_1^2 + \frac{1}{216} a_1b_2 - \frac{1}{648} a_1b_3 + \frac{5}{54} a_1c_3 - \frac{1}{432} b_2^2 + \frac{1}{648} b_2b_3 - \frac{1}{3888} b_3^2 + \frac{5}{216} c_3^2 \right) m_2 \nonumber \\
& & \mbox{} + \left( \frac{13}{144} a_1^2 + \frac{11}{648} a_1b_2 - \frac{11}{1944} a_1b_3 + \frac{8}{81} a_1c_3 - \frac{11}{1296} b_2^2 + \frac{11}{1944} b_2b_3 - \frac{11}{11664} b_3^2 + \frac{2}{81} c_3^2
 \right) m_3 \nonumber \\
& & \mbox{} + \left. \left( \frac{1}{54} a_1^2 - \frac{1}{54} a_1b_2 + \frac{1}{162} a_1b_3 + \frac{1}{108} a_1c_3 - \frac{1}{108} b_2c_3 + \frac{1}{324} b_3c_3 \right) m_4 \right] I_2(m_\pi,0,\mu) \nonumber \\
& & \mbox{} + \left[ \left( - \frac{25}{144} a_1^2 - \frac{5}{72} a_1b_2 - \frac{17}{216} a_1b_3 - \frac{1}{18} a_1c_3 - \frac{19}{432} b_2^2 - \frac{47}{648} b_2b_3 + \frac{2}{27} b_2c_3 - \frac{163}{3888} b_3^2 + \frac{7}{81} b_3c_3 - \frac{17}{216} c_3^2 \right) m_1 \right. \nonumber \\
& & \mbox{} + \left( \frac{35}{144} a_1^2 + \frac{5}{216} a_1b_2 + \frac{25}{648} a_1b_3 + \frac{5}{27} a_1c_3 + \frac{1}{432} b_2^2 + \frac{5}{648} b_2b_3 + \frac{25}{3888} b_3^2 + \frac{5}{108} c_3^2 \right) m_2 \nonumber \\
& & \mbox{} + \left( \frac{7}{432} a_1^2 - \frac{47}{648} a_1b_2 - \frac{163}{1944} a_1b_3 + \frac{23}{162} a_1c_3 - \frac{19}{1296} b_2^2 - \frac{47}{1944} b_2b_3 - \frac{163}{11664} b_3^2 + \frac{23}{648} c_3^2 \right) m_3 \nonumber \\
& & \mbox{} + \left. \left( \frac{7}{27} a_1^2 + \frac{2}{27} a_1b_2 + \frac{7}{81} a_1b_3 + \frac{7}{54} a_1c_3 + \frac{1}{27} b_2c_3 + \frac{7}{162} b_3c_3 \right) m_4 \right] I_2(m_K,0,\mu) \nonumber \\
& & \mbox{} + \left[ \left( - \frac{1}{12} a_1^2 - \frac{1}{18} a_1b_2 - \frac{1}{18} a_1b_3 - \frac{1}{36} b_2^2 - \frac{1}{18} b_2b_3 + \frac{1}{18} b_2c_3 - \frac{1}{36} b_3^2 + \frac{1}{18} b_3c_3 - \frac{1}{24} c_3^2
 \right) m_1 \right. \nonumber \\
& & \mbox{} + \left( - \frac{5}{36} a_1^2 - \frac{1}{18} a_1b_2 - \frac{1}{18} a_1b_3 - \frac{1}{18} a_1c_3 - \frac{1}{108} b_2^2 - \frac{1}{54} b_2b_3 - \frac{1}{108} b_3^2 - \frac{1}{72} c_3^2 \right) m_3 \nonumber \\
& & \mbox{} + \left. \left( \frac16 a_1^2 + \frac{1}{18} a_1b_2 + \frac{1}{18} a_1b_3 + \frac{1}{12} a_1c_3 + \frac{1}{36} b_2c_3 + \frac{1}{36} b_3c_3 \right) m_4 \right] I_2(m_\eta,0,\mu),
\end{eqnarray}

\begin{eqnarray}
\delta \mu_{\Lambda}^{\mathrm{(loop\, 2ad)}} & = & \left[ \left( \frac{1}{18} a_1b_3 - \frac{1}{12} a_1c_3 - \frac{1}{108} b_3^2 + \frac{1}{18} b_3c_3 - \frac{1}{16} c_3^2 \right) m_1 \right. \nonumber \\
& & \mbox{} + \left. \left( - \frac19 a_1^2 - \frac{1}{54} a_1b_3 - \frac{1}{12} a_1c_3 - \frac{1}{324} b_3^2 - \frac{1}{48} c_3^2 \right) m_3 \right] I_2(m_\pi,0,\mu) \nonumber \\
& & \mbox{} + \left[ \left( - \frac{5}{48} a_1^2 - \frac{1}{18} a_1b_2 - \frac{19}{216} a_1b_3 + \frac{1}{36} a_1c_3 - \frac{7}{144} b_2^2 - \frac{1}{18} b_2b_3 + \frac{1}{18} b_2c_3 - \frac{1}{48} b_3^2 + \frac{1}{54} b_3c_3 - \frac{1}{144} c_3^2 \right) m_1 \right. \nonumber \\
& & \mbox{} + \left( \frac16 a_1^2 + \frac{1}{72} a_1b_2 + \frac{1}{54} a_1b_3 + \frac{5}{36} a_1c_3 + \frac{1}{216} b_2b_3 + \frac{1}{324} b_3^2 + \frac{5}{144} c_3^2 \right) m_2 \nonumber \\
& & \mbox{} + \left( \frac{49}{432} a_1^2 - \frac{1}{18} a_1b_2 - \frac{1}{24} a_1b_3 + \frac{19}{108} a_1c_3 - \frac{7}{432} b_2^2 - \frac{1}{54} b_2b_3 - \frac{1}{144} b_3^2 + \frac{19}{432} c_3^2 \right) m_3 \nonumber \\
& & \mbox{} + \left. \left( \frac{1}{18} a_1^2 + \frac{1}{18} a_1b_2 + \frac{1}{54} a_1b_3 + \frac{1}{36} a_1c_3 + \frac{1}{36} b_2c_3 + \frac{1}{108} b_3c_3 \right) m_4 \right] I_2(m_K,0,\mu) \nonumber \\
& & \mbox{} + \left[ \left( - \frac{1}{18} a_1^2 - \frac{1}{27} a_1b_3 - \frac{1}{162} b_3^2 \right) m_1 + \left( - \frac{1}{54} a_1^2 - \frac{1}{81} a_1b_3 - \frac{1}{486} b_3^2 \right) m_3 \right] I_2(m_\eta,0,\mu),
\end{eqnarray}

\begin{eqnarray}
\delta \mu_{\Delta^{++}}^{\mathrm{(loop\, 2ad)}} & = & \left[ \left( \frac{9}{16} a_1^2 + \frac{23}{24} a_1b_2 + \frac{115}{72} a_1b_3 + \frac{1}{12} a_1c_3 + \frac{31}{48} b_2^2 + \frac{155}{72} b_2b_3 - \frac16 b_2c_3 + \frac{775}{432} b_3^2 - \frac{5}{18} b_3c_3 + \frac{1}{16} c_3^2 \right) m_1 \right. \nonumber \\
& & \mbox{} + \left( \frac{17}{16} a_1^2 + \frac{31}{24} a_1b_2 + \frac{155}{72} a_1b_3 + \frac{5}{12} a_1c_3 + \frac{31}{48} b_2^2 + \frac{155}{72} b_2b_3 + \frac{775}{432} b_3^2 + \frac{5}{48} c_3^2 \right) m_2 \nonumber \\
& & \mbox{} + \left( \frac{263}{144} a_1^2 + \frac{155}{72} a_1b_2 + \frac{775}{216} a_1b_3 + \frac34 a_1c_3 + \frac{155}{144} b_2^2 + \frac{775}{216} b_2b_3 + \frac{3875}{1296} b_3^2 + \frac{3}{16} c_3^2 \right) m_3 \nonumber \\
& & \mbox{} + \left. \left( - \frac16 a_1^2 - \frac16 a_1b_2 - \frac{5}{18} a_1b_3 - \frac{1}{12} a_1c_3 - \frac{1}{12} b_2c_3 - \frac{5}{36} b_3c_3 \right) m_4 \right] I_2(m_\pi,0,\mu) \nonumber \\
& & \mbox{} + \left[ \left( \frac{5}{16} a_1^2 + \frac{11}{24} a_1b_2 + \frac{55}{72} a_1b_3 + \frac{1}{12} a_1c_3 + \frac{19}{48} b_2^2 + \frac{95}{72} b_2b_3 - \frac16 b_2c_3 + \frac{475}{432} b_3^2 - \frac{5}{18} b_3c_3 + \frac{1}{16} c_3^2 \right) m_1 \right. \nonumber \\
& & \mbox{} + \left( \frac{13}{16} a_1^2 + \frac{19}{24} a_1b_2 + \frac{95}{72} a_1b_3 + \frac{5}{12} a_1c_3 + \frac{19}{48} b_2^2 + \frac{95}{72} b_2b_3 + \frac{475}{432} b_3^2 + \frac{5}{48} c_3^2 \right) m_2 \nonumber \\
& & \mbox{} + \left( \frac{203}{144} a_1^2 + \frac{95}{72} a_1b_2 + \frac{475}{216} a_1b_3 + \frac34 a_1c_3 + \frac{95}{144} b_2^2 + \frac{475}{216} b_2b_3 + \frac{2375}{1296} b_3^2 + \frac{3}{16} c_3^2 \right) m_3 \nonumber \\
& & \mbox{} + \left. \left( - \frac16 a_1^2 - \frac16 a_1b_2 - \frac{5}{18} a_1b_3 - \frac{1}{12} a_1c_3 - \frac{1}{12} b_2c_3 - \frac{5}{36} b_3c_3 \right) m_4 \right] I_2(m_K,0,\mu) \nonumber \\
& & \mbox{} + \left[ \left( \frac{1}{12} a_1^2 + \frac16 a_1b_2 + \frac{5}{18} a_1b_3 + \frac{1}{12} b_2^2 + \frac{5}{18} b_2b_3 + \frac{25}{108} b_3^2 \right) m_1 \right. \nonumber \\
& & \mbox{} + \left( \frac{1}{12} a_1^2 + \frac16 a_1b_2 + \frac{5}{18} a_1b_3 + \frac{1}{12} b_2^2 + \frac{5}{18} b_2b_3 + \frac{25}{108} b_3^2 \right) m_2 \nonumber \\
& & \mbox{} + \left. \left( \frac{5}{36} a_1^2 + \frac{5}{18} a_1b_2 + \frac{25}{54} a_1b_3 + \frac{5}{36} b_2^2 + \frac{25}{54} b_2b_3 + \frac{125}{324} b_3^2 \right) m_3 \right] I_2(m_\eta,0,\mu),
\end{eqnarray}

\begin{eqnarray}
\delta \mu_{\Delta^+}^{\mathrm{(loop\, 2ad)}} & = & \left[ \left( \frac{5}{16} a_1^2 + \frac{11}{24} a_1b_2 + \frac{55}{72} a_1b_3 + \frac{1}{12} a_1c_3 + \frac{41}{144} b_2^2 + \frac{205}{216} b_2b_3 - \frac{1}{18} b_2c_3 + \frac{1025}{1296} b_3^2 - \frac{5}{54} b_3c_3 + \frac{5}{144} c_3^2 \right) m_1 \right. \nonumber \\
& & \mbox{} + \left( \frac{23}{48} a_1^2 + \frac{41}{72} a_1b_2 + \frac{205}{216} a_1b_3 + \frac{7}{36} a_1c_3 + \frac{41}{144} b_2^2 + \frac{205}{216} b_2b_3 + \frac{1025}{1296} b_3^2 + \frac{7}{144} c_3^2 \right) m_2 \nonumber \\
& & \mbox{} + \left( \frac{41}{48} a_1^2 + \frac{205}{216} a_1b_2 + \frac{1025}{648} a_1b_3 + \frac{41}{108} a_1c_3 + \frac{205}{432} b_2^2 + \frac{1025}{648} b_2b_3 + \frac{5125}{3888} b_3^2 + \frac{41}{432} c_3^2 \right) m_3 \nonumber \\
& & \mbox{} + \left. \left( - \frac{1}{18} a_1^2 - \frac{1}{18} a_1b_2 - \frac{5}{54} a_1b_3 - \frac{1}{36} a_1c_3 - \frac{1}{36} b_2c_3 - \frac{5}{108} b_3c_3 \right) m_4 \right] I_2(m_\pi,0,\mu) \nonumber \\
& & \mbox{} + \left[ \left( \frac18 a_1^2 + \frac14 a_1b_2 + \frac{5}{12} a_1b_3 + \frac{17}{72} b_2^2 + \frac{85}{108} b_2b_3 - \frac19 b_2c_3 + \frac{425}{648} b_3^2 - \frac{5}{27} b_3c_3 + \frac{1}{36} c_3^2 \right) m_1 \right. \nonumber \\
& & \mbox{} + \left( \frac{11}{24} a_1^2 + \frac{17}{36} a_1b_2 + \frac{85}{108} a_1b_3 + \frac29 a_1c_3 + \frac{17}{72} b_2^2 + \frac{85}{108} b_2b_3 + \frac{425}{648} b_3^2 + \frac{1}{18} c_3^2 \right) m_2 \nonumber \\
& & \mbox{} + \left( \frac{55}{72} a_1^2 + \frac{85}{108} a_1b_2 + \frac{425}{324} a_1b_3 + \frac{10}{27} a_1c_3 + \frac{85}{216} b_2^2 + \frac{425}{324} b_2b_3 + \frac{2125}{1944} b_3^2 + \frac{5}{54} c_3^2 \right) m_3 \nonumber \\
& & \mbox{} + \left. \left( - \frac19 a_1^2 - \frac19 a_1b_2 - \frac{5}{27} a_1b_3 - \frac{1}{18} a_1c_3 - \frac{1}{18} b_2c_3 - \frac{5}{54} b_3c_3 \right) m_4 \right] I_2(m_K,0,\mu) \nonumber \\
& & \mbox{} + \left[ \left( \frac{1}{24} a_1^2 + \frac{1}{12} a_1b_2 + \frac{5}{36} a_1b_3 + \frac{1}{24} b_2^2 + \frac{5}{36} b_2b_3 + \frac{25}{216} b_3^2 \right) m_1 \right. \nonumber \\
& & \mbox{} + \left( \frac{1}{24} a_1^2 + \frac{1}{12} a_1b_2 + \frac{5}{36} a_1b_3 + \frac{1}{24} b_2^2 + \frac{5}{36} b_2b_3 + \frac{25}{216} b_3^2 \right) m_2 \nonumber \\
& & \mbox{} + \left. \left( \frac{5}{72} a_1^2 + \frac{5}{36} a_1b_2 + \frac{25}{108} a_1b_3 + \frac{5}{72} b_2^2 + \frac{25}{108} b_2b_3 + \frac{125}{648} b_3^2 \right) m_3 \right] I_2(m_\eta,0,\mu),
\end{eqnarray}

\begin{eqnarray}
\delta \mu_{\Delta^0}^{\mathrm{(loop\, 2ad)}} & = & \left[ \left( \frac{1}{16} a_1^2 - \frac{1}{24} a_1b_2 - \frac{5}{72} a_1b_3 + \frac{1}{12} a_1c_3 - \frac{11}{144} b_2^2 - \frac{55}{216} b_2b_3 + \frac{1}{18} b_2c_3 - \frac{275}{1296} b_3^2 + \frac{5}{54} b_3c_3 + \frac{1}{144} c_3^2 \right) m_1 \right. \nonumber \\
& & \mbox{} + \left( - \frac{5}{48} a_1^2 - \frac{11}{72} a_1b_2 - \frac{55}{216} a_1b_3 - \frac{1}{36} a_1c_3 - \frac{11}{144} b_2^2 - \frac{55}{216} b_2b_3 - \frac{275}{1296} b_3^2 - \frac{1}{144} c_3^2 \right) m_2 \nonumber \\
& & \mbox{} + \left( - \frac{17}{144} a_1^2 - \frac{55}{216} a_1b_2 - \frac{275}{648} a_1b_3 + \frac{1}{108} a_1c_3 - \frac{55}{432} b_2^2 - \frac{275}{648} b_2b_3 - \frac{1375}{3888} b_3^2 + \frac{1}{432} c_3^2 \right) m_3 \nonumber \\
& & \mbox{} + \left. \left( \frac{1}{18} a_1^2 + \frac{1}{18} a_1b_2 + \frac{5}{54} a_1b_3 + \frac{1}{36} a_1c_3 + \frac{1}{36} b_2c_3 + \frac{5}{108} b_3c_3 \right) m_4 \right] I_2(m_\pi,0,\mu) \nonumber \\
& & \mbox{} + \left[ \left( - \frac{1}{16} a_1^2 + \frac{1}{24} a_1b_2 + \frac{5}{72} a_1b_3 - \frac{1}{12} a_1c_3 + \frac{11}{144} b_2^2 + \frac{55}{216} b_2b_3 - \frac{1}{18} b_2c_3 + \frac{275}{1296} b_3^2 - \frac{5}{54} b_3c_3 - \frac{1}{144} c_3^2 \right) m_1 \right. \nonumber \\
& & \mbox{} + \left( \frac{5}{48} a_1^2 + \frac{11}{72} a_1b_2 + \frac{55}{216} a_1b_3 + \frac{1}{36} a_1c_3 + \frac{11}{144} b_2^2 + \frac{55}{216} b_2b_3 + \frac{275}{1296} b_3^2 + \frac{1}{144} c_3^2 \right) m_2 \nonumber \\
& & \mbox{} + \left( \frac{17}{144} a_1^2 + \frac{55}{216} a_1b_2 + \frac{275}{648} a_1b_3 - \frac{1}{108} a_1c_3 + \frac{55}{432} b_2^2 + \frac{275}{648} b_2b_3 + \frac{1375}{3888} b_3^2 - \frac{1}{432} c_3^2 \right) m_3 \nonumber \\
& & \mbox{} + \left. \left( - \frac{1}{18} a_1^2 - \frac{1}{18} a_1b_2 - \frac{5}{54} a_1b_3 - \frac{1}{36} a_1c_3 - \frac{1}{36} b_2c_3 - \frac{5}{108} b_3c_3 \right) m_4 \right] I_2(m_K,0,\mu),
\end{eqnarray}

\begin{eqnarray}
\delta \mu_{\Delta^-}^{\mathrm{(loop\, 2ad)}} & = & \left[ \left( - \frac{3}{16} a_1^2 - \frac{13}{24} a_1b_2 - \frac{65}{72} a_1b_3 + \frac{1}{12} a_1c_3 - \frac{7}{16} b_2^2 - \frac{35}{24} b_2b_3 + \frac16 b_2c_3 - \frac{175}{144} b_3^2 + \frac{5}{18} b_3c_3 - \frac{1}{48} c_3^2 \right) m_1 \right. \nonumber \\
& & \mbox{} + \left( - \frac{11}{16} a_1^2 - \frac78 a_1b_2 - \frac{35}{24} a_1b_3 - \frac14 a_1c_3 - \frac{7}{16} b_2^2 - \frac{35}{24} b_2b_3 - \frac{175}{144} b_3^2 - \frac{1}{16} c_3^2 \right) m_2 \nonumber \\
& & \mbox{} + \left( - \frac{157}{144} a_1^2 - \frac{35}{24} a_1b_2 - \frac{175}{72} a_1b_3 - \frac{13}{36} a_1c_3 - \frac{35}{48} b_2^2 - \frac{175}{72} b_2b_3 - \frac{875}{432} b_3^2 - \frac{13}{144} c_3^2 \right) m_3 \nonumber \\
& & \mbox{} + \left. \left( \frac16 a_1^2 + \frac16 a_1b_2 + \frac{5}{18} a_1b_3 + \frac{1}{12} a_1c_3 + \frac{1}{12} b_2c_3 + \frac{5}{36} b_3c_3 \right) m_4 \right] I_2(m_\pi,0,\mu) \nonumber \\
& & \mbox{} + \left[ \left( - \frac14 a_1^2 - \frac16 a_1b_2 - \frac{5}{18} a_1b_3 - \frac16 a_1c_3 - \frac{1}{12} b_2^2 - \frac{5}{18} b_2b_3 - \frac{25}{108} b_3^2 - \frac{1}{24} c_3^2 \right) m_1 \right. \nonumber \\
& & \mbox{} + \left( - \frac14 a_1^2 - \frac16 a_1b_2 - \frac{5}{18} a_1b_3 - \frac16 a_1c_3 - \frac{1}{12} b_2^2 - \frac{5}{18} b_2b_3 - \frac{25}{108} b_3^2 - \frac{1}{24} c_3^2 \right) m_2 \nonumber \\
& & \mbox{} + \left. \left( - \frac{19}{36} a_1^2 - \frac{5}{18} a_1b_2 - \frac{25}{54} a_1b_3 - \frac{7}{18} a_1c_3 - \frac{5}{36} b_2^2 - \frac{25}{54} b_2b_3 - \frac{125}{324} b_3^2 - \frac{7}{72} c_3^2 \right) m_3 \right] I_2(m_K,0,\mu) \nonumber \\
& & \mbox{} + \left[ \left( - \frac{1}{24} a_1^2 - \frac{1}{12} a_1b_2 - \frac{5}{36} a_1b_3 - \frac{1}{24} b_2^2 - \frac{5}{36} b_2b_3 - \frac{25}{216} b_3^2 \right) m_1 \right. \nonumber \\
& & \mbox{} + \left( - \frac{1}{24} a_1^2 - \frac{1}{12} a_1b_2 - \frac{5}{36} a_1b_3 - \frac{1}{24} b_2^2 - \frac{5}{36} b_2b_3 - \frac{25}{216} b_3^2 \right) m_2 \nonumber \\
& & \mbox{} + \left. \left( - \frac{5}{72} a_1^2 - \frac{5}{36} a_1b_2 - \frac{25}{108} a_1b_3 - \frac{5}{72} b_2^2 - \frac{25}{108} b_2b_3 - \frac{125}{648} b_3^2 \right) m_3 \right] I_2(m_\eta,0,\mu),
\end{eqnarray}

\begin{eqnarray}
\delta \mu_{{\Sigma^*}^+}^{\mathrm{(loop\, 2ad)}} & = & \left[ \left( \frac38 a_1^2 + \frac{19}{36} a_1b_2 + \frac{95}{108} a_1b_3 + \frac19 a_1c_3 + \frac{19}{72} b_2^2 + \frac{95}{108} b_2b_3 + \frac{475}{648} b_3^2 + \frac{1}{36} c_3^2 \right) m_1 \right. \nonumber \\
& & \mbox{} + \left( \frac{11}{24} a_1^2 + \frac{19}{36} a_1b_2 + \frac{95}{108} a_1b_3 + \frac{7}{36} a_1c_3 + \frac{19}{72} b_2^2 + \frac{95}{108} b_2b_3 + \frac{475}{648} b_3^2 + \frac{7}{144} c_3^2 \right) m_2 \nonumber \\
& & \mbox{} + \left. \left( \frac{163}{216} a_1^2 + \frac{95}{108} a_1b_2 + \frac{475}{324} a_1b_3 + \frac{17}{54} a_1c_3 + \frac{95}{216} b_2^2 + \frac{475}{324} b_2b_3 + \frac{2375}{1944} b_3^2 + \frac{17}{216} c_3^2 \right) m_3 \right] I_2(m_\pi,0,\mu) \nonumber \\
& & \mbox{} + \left[ \left( \frac{5}{48} a_1^2 + \frac{19}{72} a_1b_2 + \frac{95}{216} a_1b_3 - \frac{1}{36} a_1c_3 + \frac{43}{144} b_2^2 + \frac{215}{216} b_2b_3 - \frac16 b_2c_3 + \frac{1075}{1296} b_3^2 - \frac{5}{18} b_3c_3 + \frac{5}{144} c_3^2 \right) m_1 \right. \nonumber \\
& & \mbox{} + \left( \frac{7}{16} a_1^2 + \frac{43}{72} a_1b_2 + \frac{215}{216} a_1b_3 + \frac{5}{36} a_1c_3 + \frac{43}{144} b_2^2 + \frac{215}{216} b_2b_3 + \frac{1075}{1296} b_3^2 + \frac{5}{144} c_3^2 \right) m_2 \nonumber \\
& & \mbox{} + \left( \frac{331}{432} a_1^2 + \frac{215}{216} a_1b_2 + \frac{1075}{648} a_1b_3 + \frac{29}{108} a_1c_3 + \frac{215}{432} b_2^2 + \frac{1075}{648} b_2b_3 + \frac{5375}{3888} b_3^2 + \frac{29}{432} c_3^2 \right) m_3 \nonumber \\
& & \mbox{} + \left. \left( - \frac16 a_1^2 - \frac16 a_1b_2 - \frac{5}{18} a_1b_3 - \frac{1}{12} a_1c_3 - \frac{1}{12} b_2c_3 - \frac{5}{36} b_3c_3 \right) m_4 \right] I_2(m_K,0,\mu) \nonumber \\
& & \mbox{} + \left[ \left( \frac{1}{12} a_1^2 + \frac{1}{12} a_1c_3 + \frac{1}{48} c_3^2 \right) m_2 + \left( \frac16 a_1^2 + \frac16 a_1c_3 + \frac{1}{24} c_3^2 \right) m_3 \right] I_2(m_\eta,0,\mu),
\end{eqnarray}

\begin{eqnarray}
\delta \mu_{{\Sigma^*}^0}^{\mathrm{(loop\, 2ad)}} & = & \left[ \left( \frac18 a_1^2 + \frac19 a_1b_2 + \frac{5}{27} a_1b_3 + \frac{5}{72} a_1c_3 + \frac{1}{18} b_2c_3 + \frac{5}{54} b_3c_3 + \frac{1}{288} c_3^2 \right) m_1 + \left( \frac{1}{216} a_1^2 + \frac{1}{216} a_1c_3 + \frac{1}{864} c_3^2 \right) m_3 \right. \nonumber \\
& & \mbox{} + \left. \left( \frac{1}{18} a_1^2 + \frac{1}{18} a_1b_2 + \frac{5}{54} a_1b_3 + \frac{1}{36} a_1c_3 + \frac{1}{36} b_2c_3 + \frac{5}{108} b_3c_3 \right) m_4 \right] I_2(m_\pi,0,\mu) \nonumber \\
& & \mbox{} + \left[ \left( - \frac{1}{12} a_1^2 - \frac19 a_1b_2 - \frac{5}{27} a_1b_3 - \frac{1}{36} a_1c_3 - \frac{1}{18} b_2c_3 - \frac{5}{54} b_3c_3 + \frac{1}{144} c_3^2 \right) m_1 \right. \nonumber \\
& & \mbox{} + \left( \frac{1}{108} a_1^2 + \frac{1}{108} a_1c_3 + \frac{1}{432} c_3^2 \right) m_3 \nonumber \\
& & \mbox{} + \left. \left( - \frac{1}{18} a_1^2 - \frac{1}{18} a_1b_2 - \frac{5}{54} a_1b_3 - \frac{1}{36} a_1c_3 - \frac{1}{36} b_2c_3 - \frac{5}{108} b_3c_3 \right) m_4 \right] I_2(m_K,0,\mu) \nonumber \\
& & \mbox{} + \left[ \left( - \frac{1}{24} a_1^2 - \frac{1}{24} a_1c_3 - \frac{1}{96} c_3^2 \right) m_1 + \left( - \frac{1}{72} a_1^2 - \frac{1}{72} a_1c_3 - \frac{1}{288} c_3^2 \right) m_3 \right] I_2(m_\eta,0,\mu),
\end{eqnarray}

\begin{eqnarray}
\delta \mu_{{\Sigma^*}^-}^{\mathrm{(loop\, 2ad)}} & = & \left[ \left( - \frac18 a_1^2 - \frac{11}{36} a_1b_2 - \frac{55}{108} a_1b_3 + \frac{1}{36} a_1c_3 - \frac{19}{72} b_2^2 - \frac{95}{108} b_2b_3 + \frac19 b_2c_3 - \frac{475}{648} b_3^2 + \frac{5}{27} b_3c_3 - \frac{1}{48} c_3^2 \right) m_1 \right. \nonumber \\
& & \mbox{} + \left( - \frac{11}{24} a_1^2 - \frac{19}{36} a_1b_2 - \frac{95}{108} a_1b_3 - \frac{7}{36} a_1c_3 - \frac{19}{72} b_2^2 - \frac{95}{108} b_2b_3 - \frac{475}{648} b_3^2 - \frac{7}{144} c_3^2 \right) m_2 \nonumber \\
& & \mbox{} + \left( - \frac{161}{216} a_1^2 - \frac{95}{108} a_1b_2 - \frac{475}{324} a_1b_3 - \frac{11}{36} a_1c_3 - \frac{95}{216} b_2^2 - \frac{475}{324} b_2b_3 - \frac{2375}{1944} b_3^2 - \frac{11}{144} c_3^2 \right) m_3 \nonumber \\
& & \mbox{} + \left. \left( \frac19 a_1^2 + \frac19 a_1b_2 + \frac{5}{27} a_1b_3 + \frac{1}{18} a_1c_3 + \frac{1}{18} b_2c_3 + \frac{5}{54} b_3c_3 \right) m_4 \right] I_2(m_\pi,0,\mu) \nonumber \\
& & \mbox{} + \left[ \left( - \frac{13}{48} a_1^2 - \frac{35}{72} a_1b_2 - \frac{175}{216} a_1b_3 - \frac{1}{36} a_1c_3 - \frac{43}{144} b_2^2 - \frac{215}{216} b_2b_3 + \frac{1}{18} b_2c_3 - \frac{1075}{1296} b_3^2 + \frac{5}{54} b_3c_3 - \frac{1}{48} c_3^2 \right) m_1 \right. \nonumber \\
& & \mbox{} + \left( - \frac{7}{16} a_1^2 - \frac{43}{72} a_1b_2 - \frac{215}{216} a_1b_3 - \frac{5}{36} a_1c_3 - \frac{43}{144} b_2^2 - \frac{215}{216} b_2b_3 - \frac{1075}{1296} b_3^2 - \frac{5}{144} c_3^2 \right) m_2 \nonumber \\
& & \mbox{} + \left( - \frac{323}{432} a_1^2 - \frac{215}{216} a_1b_2 - \frac{1075}{648} a_1b_3 - \frac14 a_1c_3 - \frac{215}{432} b_2^2 - \frac{1075}{648} b_2b_3 - \frac{5375}{3888} b_3^2 - \frac{1}{16} c_3^2 \right) m_3 \nonumber \\
& & \mbox{} + \left. \left( \frac{1}{18} a_1^2 + \frac{1}{18} a_1b_2 + \frac{5}{54} a_1b_3 + \frac{1}{36} a_1c_3 + \frac{1}{36} b_2c_3 + \frac{5}{108} b_3c_3 \right) m_4 \right] I_2(m_K,0,\mu) \nonumber \\
& & \mbox{} + \left[ \left( - \frac{1}{12} a_1^2 - \frac{1}{12} a_1c_3 - \frac{1}{48} c_3^2 \right) m_1 + \left( - \frac{1}{12} a_1^2 - \frac{1}{12} a_1c_3 - \frac{1}{48} c_3^2 \right) m_2 \right. \nonumber \\
& & \mbox{} + \left. \left( - \frac{7}{36} a_1^2 - \frac{7}{36} a_1c_3 - \frac{7}{144} c_3^2 \right) m_3 \right] I_2(m_\eta,0,\mu),
\end{eqnarray}

\begin{eqnarray}
\delta \mu_{{\Xi^*}^0}^{\mathrm{(loop\, 2ad)}} & = & \left[ \left( \frac{3}{16} a_1^2 + \frac{5}{24} a_1b_2 + \frac{25}{72} a_1b_3 + \frac{1}{12} a_1c_3 + \frac{11}{144} b_2^2 + \frac{55}{216} b_2b_3 + \frac{1}{36} b_2c_3 + \frac{275}{1296} b_3^2 + \frac{5}{108} b_3c_3 + \frac{1}{72} c_3^2 \right) m_1 \right. \nonumber \\
& & \mbox{} + \left( \frac{5}{48} a_1^2 + \frac{11}{72} a_1b_2 + \frac{55}{216} a_1b_3 + \frac{1}{36} a_1c_3 + \frac{11}{144} b_2^2 + \frac{55}{216} b_2b_3 + \frac{275}{1296} b_3^2 + \frac{1}{144} c_3^2 \right) m_2 \nonumber \\
& & \mbox{} + \left( \frac{7}{48} a_1^2 + \frac{55}{216} a_1b_2 + \frac{275}{648} a_1b_3 + \frac{1}{54} a_1c_3 + \frac{55}{432} b_2^2 + \frac{275}{648} b_2b_3 + \frac{1375}{3888} b_3^2 + \frac{1}{216} c_3^2 \right) m_3 \nonumber \\
& & \mbox{} + \left. \left( \frac{1}{36} a_1^2 + \frac{1}{36} a_1b_2 + \frac{5}{108} a_1b_3 + \frac{1}{72} a_1c_3 + \frac{1}{72} b_2c_3 + \frac{5}{216} b_3c_3 \right) m_4 \right] I_2(m_\pi,0,\mu) \nonumber \\
& & \mbox{} + \left[ \left( - \frac{5}{48} a_1^2 - \frac{1}{24} a_1b_2 - \frac{5}{72} a_1b_3 - \frac{1}{12} a_1c_3 - \frac{11}{144} b_2^2 - \frac{55}{216} b_2b_3 + \frac{1}{18} b_2c_3 - \frac{275}{1296} b_3^2 + \frac{5}{54} b_3c_3 - \frac{5}{144} c_3^2 \right) m_1 \right. \nonumber \\
& & \mbox{} + \left( - \frac{5}{48} a_1^2 - \frac{11}{72} a_1b_2 - \frac{55}{216} a_1b_3 - \frac{1}{36} a_1c_3 - \frac{11}{144} b_2^2 - \frac{55}{216} b_2b_3 - \frac{275}{1296} b_3^2 - \frac{1}{144} c_3^2 \right) m_2 \nonumber \\
& & \mbox{} + \left( - \frac{25}{144} a_1^2 - \frac{55}{216} a_1b_2 - \frac{275}{648} a_1b_3 - \frac{5}{108} a_1c_3 - \frac{55}{432} b_2^2 - \frac{275}{648} b_2b_3 - \frac{1375}{3888} b_3^2 - \frac{5}{432} c_3^2 \right) m_3 \nonumber \\
& & \mbox{} + \left. \left( \frac{1}{18} a_1^2 + \frac{1}{18} a_1b_2 + \frac{5}{54} a_1b_3 + \frac{1}{36} a_1c_3 + \frac{1}{36} b_2c_3 + \frac{5}{108} b_3c_3 \right) m_4 \right] I_2(m_K,0,\mu) \nonumber \\
& & \mbox{} + \left[ \left( - \frac{1}{12} a_1^2 - \frac16 a_1b_2 - \frac{5}{18} a_1b_3 - \frac{1}{12} b_2c_3 - \frac{5}{36} b_3c_3 + \frac{1}{48} c_3^2 \right) m_1 + \left( \frac{1}{36} a_1^2 + \frac{1}{36} a_1c_3 + \frac{1}{144} c_3^2 \right) m_3 \right. \nonumber \\
& & \mbox{} + \left. \left( - \frac{1}{12} a_1^2 - \frac{1}{12} a_1b_2 - \frac{5}{36} a_1b_3 - \frac{1}{24} a_1c_3 - \frac{1}{24} b_2c_3 - \frac{5}{72} b_3c_3 \right) m_4 \right] I_2(m_\eta,0,\mu),
\end{eqnarray}

\begin{eqnarray}
\delta \mu_{{\Xi^*}^-}^{\mathrm{(loop\, 2ad)}} & = & \left[ \left( - \frac{1}{16} a_1^2 - \frac18 a_1b_2 - \frac{5}{24} a_1b_3 - \frac{17}{144} b_2^2 - \frac{85}{216} b_2b_3 + \frac{1}{18} b_2c_3 - \frac{425}{1296} b_3^2 + \frac{5}{54} b_3c_3 - \frac{1}{72} c_3^2 \right) m_1 \right. \nonumber \\
& & \mbox{} + \left( - \frac{11}{48} a_1^2 - \frac{17}{72} a_1b_2 - \frac{85}{216} a_1b_3 - \frac19 a_1c_3 - \frac{17}{144} b_2^2 - \frac{85}{216} b_2b_3 - \frac{425}{1296} b_3^2 - \frac{1}{36} c_3^2 \right) m_2 \nonumber \\
& & \mbox{} + \left( - \frac{55}{144} a_1^2 - \frac{85}{216} a_1b_2 - \frac{425}{648} a_1b_3 - \frac{5}{27} a_1c_3 - \frac{85}{432} b_2^2 - \frac{425}{648} b_2b_3 - \frac{2125}{3888} b_3^2 - \frac{5}{108} c_3^2 \right) m_3 \nonumber \\
& & \mbox{} + \left. \left( \frac{1}{18} a_1^2 + \frac{1}{18} a_1b_2 + \frac{5}{54} a_1b_3 + \frac{1}{36} a_1c_3 + \frac{1}{36} b_2c_3 + \frac{5}{108} b_3c_3 \right) m_4 \right] I_2(m_\pi,0,\mu) \nonumber \\
& & \mbox{} + \left[ \left( - \frac{7}{24} a_1^2 - \frac{7}{12} a_1b_2 - \frac{35}{36} a_1b_3 - \frac{29}{72} b_2^2 - \frac{145}{108} b_2b_3 + \frac19 b_2c_3 - \frac{725}{648} b_3^2 + \frac{5}{27} b_3c_3 - \frac{1}{36} c_3^2 \right) m_1 \right. \nonumber \\
& & \mbox{} + \left( - \frac58 a_1^2 - \frac{29}{36} a_1b_2 - \frac{145}{108} a_1b_3 - \frac29 a_1c_3 - \frac{29}{72} b_2^2 - \frac{145}{108} b_2b_3 - \frac{725}{648} b_3^2 - \frac{1}{18} c_3^2 \right) m_2 \nonumber \\
& & \mbox{} + \left( - \frac{25}{24} a_1^2 - \frac{145}{108} a_1b_2 - \frac{725}{324} a_1b_3 - \frac{10}{27} a_1c_3 - \frac{145}{216} b_2^2 - \frac{725}{324} b_2b_3 - \frac{3625}{1944} b_3^2 - \frac{5}{54} c_3^2 \right) m_3 \nonumber \\
& & \mbox{} + \left. \left( \frac19 a_1^2 + \frac19 a_1b_2 + \frac{5}{27} a_1b_3 + \frac{1}{18} a_1c_3 + \frac{1}{18} b_2c_3 + \frac{5}{54} b_3c_3 \right) m_4 \right] I_2(m_K,0,\mu) \nonumber \\
& & \mbox{} + \left[ \left( - \frac18 a_1^2 - \frac{1}{12} a_1b_2 - \frac{5}{36} a_1b_3 - \frac{1}{12} a_1c_3 - \frac{1}{24} b_2^2 - \frac{5}{36} b_2b_3 - \frac{25}{216} b_3^2 - \frac{1}{48} c_3^2 \right) m_1 \right. \nonumber \\
& & \mbox{} + \left( - \frac18 a_1^2 - \frac{1}{12} a_1b_2 - \frac{5}{36} a_1b_3 - \frac{1}{12} a_1c_3 - \frac{1}{24} b_2^2 - \frac{5}{36} b_2b_3 - \frac{25}{216} b_3^2 - \frac{1}{48} c_3^2 \right) m_2 \nonumber \\
& & \mbox{} + \left. \left( - \frac{19}{72} a_1^2 - \frac{5}{36} a_1b_2 - \frac{25}{108} a_1b_3 - \frac{7}{36} a_1c_3 - \frac{5}{72} b_2^2 - \frac{25}{108} b_2b_3 - \frac{125}{648} b_3^2 - \frac{7}{144} c_3^2 \right) m_3 \right] I_2(m_\eta,0,\mu),
\end{eqnarray}

\begin{eqnarray}
\delta \mu_{\Omega^-}^{\mathrm{(loop\, 2ad)}} & = & \left[ \left( - \frac{5}{16} a_1^2 - \frac{11}{24} a_1b_2 - \frac{55}{72} a_1b_3 - \frac{1}{12} a_1c_3 - \frac{19}{48} b_2^2 - \frac{95}{72} b_2b_3 + \frac16 b_2c_3 - \frac{475}{432} b_3^2 + \frac{5}{18} b_3c_3 - \frac{1}{16} c_3^2 \right) m_1 \right. \nonumber \\
& & \mbox{} + \left( - \frac{13}{16} a_1^2 - \frac{19}{24} a_1b_2 - \frac{95}{72} a_1b_3 - \frac{5}{12} a_1c_3 - \frac{19}{48} b_2^2 - \frac{95}{72} b_2b_3 - \frac{475}{432} b_3^2 - \frac{5}{48} c_3^2 \right) m_2 \nonumber \\
& & \mbox{} + \left( - \frac{203}{144} a_1^2 - \frac{95}{72} a_1b_2 - \frac{475}{216} a_1b_3 - \frac34 a_1c_3 - \frac{95}{144} b_2^2 - \frac{475}{216} b_2b_3 - \frac{2375}{1296} b_3^2 - \frac{3}{16} c_3^2 \right) m_3 \nonumber \\
& & \mbox{} + \left. \left( \frac16 a_1^2 + \frac16 a_1b_2 + \frac{5}{18} a_1b_3 + \frac{1}{12} a_1c_3 + \frac{1}{12} b_2c_3 + \frac{5}{36} b_3c_3 \right) m_4 \right] I_2(m_K,0,\mu) \nonumber \\
& & \mbox{} + \left[ \left( - \frac16 a_1^2 - \frac13 a_1b_2 - \frac59 a_1b_3 - \frac16 b_2^2 - \frac59 b_2b_3 - \frac{25}{54} b_3^2 \right) m_1 \right. \nonumber \\
& & \mbox{} + \left( - \frac16 a_1^2 - \frac13 a_1b_2 - \frac59 a_1b_3 - \frac16 b_2^2 - \frac59 b_2b_3 - \frac{25}{54} b_3^2 \right) m_2 \nonumber \\
& & \mbox{} + \left. \left( - \frac{5}{18} a_1^2 - \frac59 a_1b_2 - \frac{25}{27} a_1b_3 - \frac{5}{18} b_2^2 - \frac{25}{27} b_2b_3 - \frac{125}{162} b_3^2 \right) m_3 \right] I_2(m_\eta,0,\mu),
\end{eqnarray}

\begin{eqnarray}
\sqrt{3} \delta \mu_{\Sigma^0\Lambda}^{\mathrm{(loop\, 2ad)}} & = & \left[ \left( \frac14 a_1^2 + \frac{5}{36} a_1b_3 + \frac{1}{24} a_1c_3 + \frac{1}{24} b_2^2 + \frac{1}{27} b_2b_3 - \frac{1}{18} b_2c_3 + \frac{25}{648} b_3^2 - \frac{5}{108} b_3c_3 + \frac{13}{288} c_3^2 \right) m_1 \right. \nonumber \\
& & \mbox{} + \left( - \frac13 a_1^2 - \frac{1}{36} a_1b_2 - \frac{1}{27} a_1b_3 - \frac{5}{18} a_1c_3 - \frac{1}{108} b_2b_3 - \frac{1}{162} b_3^2 - \frac{5}{72} c_3^2 \right) m_2 \nonumber \\
& & \mbox{} + \left( - \frac{7}{36} a_1^2 + \frac{1}{27} a_1b_2 + \frac{25}{324} a_1b_3 - \frac{67}{216} a_1c_3 + \frac{1}{72} b_2^2 + \frac{1}{81} b_2b_3 + \frac{25}{1944} b_3^2 - \frac{67}{864} c_3^2 \right) m_3 \nonumber \\
& & \mbox{} + \left. \left( - \frac{5}{36} a_1^2 - \frac{1}{18} a_1b_2 - \frac{5}{108} a_1b_3 - \frac{5}{72} a_1c_3 - \frac{1}{36} b_2c_3 - \frac{5}{216} b_3c_3 \right) m_4 \right] I_2(m_\pi,0,\mu) \nonumber \\
& & \mbox{} + \left[ \left( \frac{3}{16} a_1^2 + \frac16 a_1b_2 + \frac{5}{72} a_1b_3 + \frac{1}{12} a_1c_3 + \frac{5}{48} b_2^2 + \frac{7}{54} b_2b_3 - \frac19 b_2c_3 + \frac{79}{1296} b_3^2 - \frac{4}{27} b_3c_3 + \frac{19}{144} c_3^2 \right) m_1 \right. \nonumber \\
& & \mbox{} + \left( - \frac16 a_1^2 - \frac{1}{72} a_1b_2 - \frac{1}{54} a_1b_3 - \frac{5}{36} a_1c_3 - \frac{1}{216} b_2b_3 - \frac{1}{324} b_3^2 - \frac{5}{144} c_3^2 \right) m_2 \nonumber \\
& & \mbox{} + \left( \frac{25}{144} a_1^2 + \frac{7}{54} a_1b_2 + \frac{79}{648} a_1b_3 - \frac{1}{108} a_1c_3 + \frac{5}{144} b_2^2 + \frac{7}{162} b_2b_3 + \frac{79}{3888} b_3^2 - \frac{1}{432} c_3^2 \right) m_3 \nonumber \\
& & \mbox{} + \left. \left( - \frac49 a_1^2 - \frac19 a_1b_2 - \frac{4}{27} a_1b_3 - \frac29 a_1c_3 - \frac{1}{18} b_2c_3 - \frac{2}{27} b_3c_3 \right) m_4 \right] I_2(m_K,0,\mu) \nonumber \\
& & \mbox{} + \left[ \left( \frac{1}{24} a_1^2 + \frac{1}{24} a_1c_3 + \frac{1}{108} b_3^2 - \frac{1}{36} b_3c_3 + \frac{1}{32} c_3^2 \right) m_1 + \left( \frac{5}{72} a_1^2 + \frac{1}{54} a_1b_3 + \frac{1}{24} a_1c_3 + \frac{1}{324} b_3^2 + \frac{1}{96} c_3^2 \right) m_3 \right. \nonumber \\
& & \mbox{} + \left. \left( - \frac{1}{12} a_1^2 - \frac{1}{36} a_1b_3 - \frac{1}{24} a_1c_3 - \frac{1}{72} b_3c_3 \right) m_4 \right] I_2(m_\eta,0,\mu),
\end{eqnarray}

\begin{eqnarray}
\sqrt{2} \delta \mu_{\Delta^+ p}^{\mathrm{(loop\, 2ad)}} & = & \left[ \left( \frac13 a_1^2 + \frac13 a_1c_3 + \frac{23}{54} b_2^2 + \frac{115}{81} b_2b_3 - \frac{5}{18} b_2c_3 + \frac{575}{486} b_3^2 - \frac{25}{54} b_3c_3 + \frac{43}{216} c_3^2 \right) m_1 \right. \nonumber \\
& & \mbox{} + \left( - \frac59 a_1^2 - \frac{13}{27} a_1b_2 - \frac{65}{81} a_1b_3 - \frac{5}{18} a_1c_3 - \frac{13}{54} b_2c_3 - \frac{65}{162} b_3c_3 \right) m_2 \nonumber \\
& & \mbox{} + \left( - \frac{25}{27} a_1^2 - \frac{65}{81} a_1b_2 - \frac{325}{243} a_1b_3 - \frac{25}{54} a_1c_3 - \frac{65}{162} b_2c_3 - \frac{325}{486} b_3c_3 \right) m_3 \nonumber \\
& & \mbox{} + \left. \left( \frac{17}{27} a_1^2 + \frac{5}{18} a_1b_2 + \frac{25}{54} a_1b_3 + \frac{43}{108} a_1c_3 + \frac{23}{108} b_2^2 + \frac{115}{162} b_2b_3 + \frac{575}{972} b_3^2 + \frac{43}{432} c_3^2 \right) m_4 \right] I_2(m_\pi,0,\mu) \nonumber \\
& & \mbox{} + \left[ \left( \frac14 a_1^2 + \frac19 a_1b_2 + \frac{13}{54} a_1b_3 + \frac16 a_1c_3 + \frac{17}{108} b_2^2 + \frac{41}{81} b_2b_3 - \frac{1}{18} b_2c_3 + \frac{485}{972} b_3^2 - \frac19 b_3c_3 + \frac{17}{216} c_3^2 \right) m_1 \right. \nonumber \\
& & \mbox{} + \left( - \frac19 a_1^2 - \frac{2}{27} a_1b_2 - \frac{13}{81} a_1b_3 - \frac{1}{18} a_1c_3 - \frac{1}{27} b_2c_3 - \frac{13}{162} b_3c_3 \right) m_2 \nonumber \\
& & \mbox{} + \left( - \frac29 a_1^2 - \frac{13}{81} a_1b_2 - \frac{68}{243} a_1b_3 - \frac19 a_1c_3 - \frac{13}{162} b_2c_3 - \frac{34}{243} b_3c_3 \right) m_3 \nonumber \\
& & \mbox{} + \left. \left( \frac{59}{216} a_1^2 + \frac19 a_1b_2 + \frac{25}{108} a_1b_3 + \frac{17}{108} a_1c_3 + \frac{17}{216} b_2^2 + \frac{41}{162} b_2b_3 + \frac{485}{1944} b_3^2 + \frac{17}{432} c_3^2 \right) m_4 \right] I_2(m_K,0,\mu) \nonumber \\
& & \mbox{} + \left[ \left( \frac{1}{18} a_1^2 + \frac19 a_1b_2 + \frac29 a_1b_3 + \frac{1}{18} b_2^2 + \frac29 b_2b_3 + \frac{41}{162} b_3^2 \right) m_1 \right. \nonumber \\
& & \mbox{} + \left. \left( \frac{1}{36} a_1^2 + \frac{1}{18} a_1b_2 + \frac19 a_1b_3 + \frac{1}{36} b_2^2 + \frac19 b_2b_3 + \frac{41}{324} b_3^2 \right) m_4 \right] I_2(m_\eta,0,\mu),
\end{eqnarray}

\begin{eqnarray}
\sqrt{2} \delta \mu_{\Delta^0 n}^{\mathrm{(loop\, 2ad)}} & = & \left[ \left( \frac13 a_1^2 + \frac13 a_1c_3 + \frac{23}{54} b_2^2 + \frac{115}{81} b_2b_3 - \frac{5}{18} b_2c_3 + \frac{575}{486} b_3^2 - \frac{25}{54} b_3c_3 + \frac{43}{216} c_3^2 \right) m_1 \right. \nonumber \\
& & \mbox{} + \left( - \frac59 a_1^2 - \frac{13}{27} a_1b_2 - \frac{65}{81} a_1b_3 - \frac{5}{18} a_1c_3 - \frac{13}{54} b_2c_3 - \frac{65}{162} b_3c_3 \right) m_2 \nonumber \\
& & \mbox{} + \left( - \frac{25}{27} a_1^2 - \frac{65}{81} a_1b_2 - \frac{325}{243} a_1b_3 - \frac{25}{54} a_1c_3 - \frac{65}{162} b_2c_3 - \frac{325}{486} b_3c_3 \right) m_3 \nonumber \\
& & \mbox{} + \left. \left( \frac{17}{27} a_1^2 + \frac{5}{18} a_1b_2 + \frac{25}{54} a_1b_3 + \frac{43}{108} a_1c_3 + \frac{23}{108} b_2^2 + \frac{115}{162} b_2b_3 + \frac{575}{972} b_3^2 + \frac{43}{432} c_3^2 \right) m_4 \right] I_2(m_\pi,0,\mu) \nonumber \\
& & \mbox{} + \left[ \left( \frac14 a_1^2 + \frac19 a_1b_2 + \frac{13}{54} a_1b_3 + \frac16 a_1c_3 + \frac{17}{108} b_2^2 + \frac{41}{81} b_2b_3 - \frac{1}{18} b_2c_3 + \frac{485}{972} b_3^2 - \frac19 b_3c_3 + \frac{17}{216} c_3^2 \right) m_1 \right. \nonumber \\
& & \mbox{} + \left( - \frac19 a_1^2 - \frac{2}{27} a_1b_2 - \frac{13}{81} a_1b_3 - \frac{1}{18} a_1c_3 - \frac{1}{27} b_2c_3 - \frac{13}{162} b_3c_3 \right) m_2 \nonumber \\
& & \mbox{} + \left( - \frac29 a_1^2 - \frac{13}{81} a_1b_2 - \frac{68}{243} a_1b_3 - \frac19 a_1c_3 - \frac{13}{162} b_2c_3 - \frac{34}{243} b_3c_3 \right) m_3 \nonumber \\
& & \mbox{} + \left. \left( \frac{59}{216} a_1^2 + \frac19 a_1b_2 + \frac{25}{108} a_1b_3 + \frac{17}{108} a_1c_3 + \frac{17}{216} b_2^2 + \frac{41}{162} b_2b_3 + \frac{485}{1944} b_3^2 + \frac{17}{432} c_3^2 \right) m_4 \right] I_2(m_K,0,\mu) \nonumber \\
& & \mbox{} + \left[ \left( \frac{1}{18} a_1^2 + \frac19 a_1b_2 + \frac29 a_1b_3 + \frac{1}{18} b_2^2 + \frac29 b_2b_3 + \frac{41}{162} b_3^2 \right) m_1 \right. \nonumber \\
& & \mbox{} + \left. \left( \frac{1}{36} a_1^2 + \frac{1}{18} a_1b_2 + \frac19 a_1b_3 + \frac{1}{36} b_2^2 + \frac19 b_2b_3 + \frac{41}{324} b_3^2 \right) m_4 \right] I_2(m_\eta,0,\mu),
\end{eqnarray}

\begin{eqnarray}
\sqrt{6} \delta \mu_{{\Sigma^*}^0\Lambda}^{\mathrm{(loop\, 2ad)}} & = & \left[ \left( \frac12 a_1^2 + \frac{1}{36} a_1b_3 + \frac{11}{24} a_1c_3 + \frac{5}{12} b_2^2 + \frac{35}{27} b_2b_3 - \frac{5}{18} b_2c_3 + \frac{677}{648} b_3^2 - \frac{19}{36} b_3c_3 + \frac{67}{288} c_3^2 \right) m_1 \right. \nonumber \\
& & \mbox{} + \left( - \frac23 a_1^2 - \frac59 a_1b_2 - \frac{26}{27} a_1b_3 - \frac13 a_1c_3 - \frac{5}{18} b_2c_3 - \frac{13}{27} b_3c_3 \right) m_2 \nonumber \\
& & \mbox{} + \left( - \frac{19}{18} a_1^2 - \frac{25}{27} a_1b_2 - \frac{257}{162} a_1b_3 - \frac{19}{36} a_1c_3 - \frac{25}{54} b_2c_3 - \frac{257}{324} b_3c_3 \right) m_3 \nonumber \\
& & \mbox{} + \left. \left( \frac{13}{18} a_1^2 + \frac{5}{18} a_1b_2 + \frac{13}{24} a_1b_3 + \frac{67}{144} a_1c_3 + \frac{5}{24} b_2^2 + \frac{35}{54} b_2b_3 + \frac{677}{1296} b_3^2 + \frac{67}{576} c_3^2 \right) m_4 \right] I_2(m_\pi,0,\mu) \nonumber \\
& & \mbox{} + \left[ \left( \frac38 a_1^2 + \frac13 a_1b_2 + \frac{23}{36} a_1b_3 + \frac14 a_1c_3 + \frac{13}{24} b_2^2 + \frac{52}{27} b_2b_3 - \frac29 b_2c_3 + \frac{1195}{648} b_3^2 - \frac{11}{36} b_3c_3 + \frac{11}{72} c_3^2 \right) m_1 \right. \nonumber \\
& & \mbox{} + \left( - \frac13 a_1^2 - \frac{5}{18} a_1b_2 - \frac{13}{27} a_1b_3 - \frac16 a_1c_3 - \frac{5}{36} b_2c_3 - \frac{13}{54} b_3c_3 \right) m_2 \nonumber \\
& & \mbox{} + \left( - \frac{11}{18} a_1^2 - \frac{14}{27} a_1b_2 - \frac{133}{162} a_1b_3 - \frac{11}{36} a_1c_3 - \frac{7}{27} b_2c_3 - \frac{133}{324} b_3c_3 \right) m_3 \nonumber \\
& & \mbox{} + \left. \left( \frac{79}{144} a_1^2 + \frac{7}{18} a_1b_2 + \frac58 a_1b_3 + \frac{11}{36} a_1c_3 + \frac{13}{48} b_2^2 + \frac{26}{27} b_2b_3 + \frac{1195}{1296} b_3^2 + \frac{11}{144} c_3^2 \right) m_4 \right] I_2(m_K,0,\mu) \nonumber \\
& & \mbox{} + \left[ \left( \frac{1}{12} a_1^2 + \frac{1}{36} a_1b_3 + \frac{1}{24} a_1c_3 + \frac{1}{72} b_3^2 - \frac{1}{36} b_3c_3 + \frac{1}{32} c_3^2 \right) m_1 \right. \nonumber \\
& & \mbox{} + \left( - \frac{1}{18} a_1^2 - \frac{1}{54} a_1b_3 - \frac{1}{36} a_1c_3 - \frac{1}{108} b_3c_3 \right) m_3 \nonumber \\
& & \mbox{} + \left. \left( \frac18 a_1^2 + \frac{1}{24} a_1b_3 + \frac{1}{16} a_1c_3 + \frac{1}{144} b_3^2 + \frac{1}{64} c_3^2 \right) m_4 \right] I_2(m_\eta,0,\mu),
\end{eqnarray}

\begin{eqnarray}
\sqrt{2} \delta \mu_{{\Sigma^*}^0\Sigma^0}^{\mathrm{(loop\, 2ad)}} & = & \left[ \left( \frac{1}{12} a_1^2 + \frac{2}{27} a_1b_2 + \frac{49}{324} a_1b_3 + \frac{1}{24} a_1c_3 + \frac{2}{27} b_2^2 + \frac{23}{81} b_2b_3 - \frac{1}{54} b_2c_3 + \frac{583}{1944} b_3^2 - \frac{7}{324} b_3c_3 + \frac{23}{864} c_3^2 \right) m_1 \right. \nonumber \\
& & \mbox{} + \left( - \frac{7}{162} a_1^2 - \frac{1}{81} a_1b_2 - \frac{7}{486} a_1b_3 - \frac{7}{324} a_1c_3 - \frac{1}{162} b_2c_3 - \frac{7}{972} b_3c_3 \right) m_3 \nonumber \\
& & \mbox{} + \left. \left( \frac{23}{216} a_1^2 + \frac{1}{18} a_1b_2 + \frac{7}{72} a_1b_3 + \frac{23}{432} a_1c_3 + \frac{1}{27} b_2^2 + \frac{23}{162} b_2b_3 + \frac{583}{3888} b_3^2 + \frac{23}{1728} c_3^2 \right) m_4 \right] I_2(m_\pi,0,\mu) \nonumber \\
& & \mbox{} + \left[ \left( \frac{11}{72} a_1^2 + \frac{1}{27} a_1b_2 + \frac{23}{324} a_1b_3 + \frac{5}{36} a_1c_3 + \frac{53}{216} b_2^2 + \frac{64}{81} b_2b_3 - \frac{4}{27} b_2c_3 + \frac{1289}{1944} b_3^2 - \frac{83}{324} b_3c_3 + \frac{19}{216} c_3^2 \right) m_1 \right. \nonumber \\
& & \mbox{} + \left( - \frac13 a_1^2 - \frac{5}{18} a_1b_2 - \frac{13}{27} a_1b_3 - \frac16 a_1c_3 - \frac{5}{36} b_2c_3 - \frac{13}{54} b_3c_3 \right) m_2 \nonumber \\
& & \mbox{} + \left( - \frac{83}{162} a_1^2 - \frac{38}{81} a_1b_2 - \frac{383}{486} a_1b_3 - \frac{83}{324} a_1c_3 - \frac{19}{81} b_2c_3 - \frac{383}{972} b_3c_3 \right) m_3 \nonumber \\
& & \mbox{} + \left. \left( \frac{125}{432} a_1^2 + \frac16 a_1b_2 + \frac{7}{24} a_1b_3 + \frac{19}{108} a_1c_3 + \frac{53}{432} b_2^2 + \frac{32}{81} b_2b_3 + \frac{1289}{3888} b_3^2 + \frac{19}{432} c_3^2 \right) m_4 \right] I_2(m_K,0,\mu) \nonumber \\
& & \mbox{} + \left[ \left( \frac{1}{12} a_1^2 + \frac{1}{108} a_1b_3 + \frac{5}{72} a_1c_3 + \frac{1}{216} b_3^2 - \frac{1}{108} b_3c_3 + \frac{7}{288} c_3^2 \right) m_1 \right. \nonumber \\
& & \mbox{} + \left( - \frac{1}{54} a_1^2 - \frac{1}{162} a_1b_3 - \frac{1}{108} a_1c_3 - \frac{1}{324} b_3c_3 \right) m_3 \nonumber \\
& & \mbox{} + \left. \left( \frac{5}{72} a_1^2 + \frac{1}{72} a_1b_3 + \frac{7}{144} a_1c_3 + \frac{1}{432} b_3^2 + \frac{7}{576} c_3^2 \right) m_4 \right] I_2(m_\eta,0,\mu),
\end{eqnarray}

\begin{eqnarray}
\sqrt{2} \delta \mu_{{\Sigma^*}^+\Sigma^+}^{\mathrm{(loop\, 2ad)}} & = & \left[ \left( \frac14 a_1^2 + \frac{5}{27} a_1b_2 + \frac{59}{162} a_1b_3 + \frac16 a_1c_3 + \frac{7}{36} b_2^2 + \frac{17}{27} b_2b_3 - \frac{1}{54} b_2c_3 + \frac{31}{54} b_3^2 + \frac{11}{324} b_3c_3 + \frac{1}{16} c_3^2 \right) m_1 \right. \nonumber \\
& & \mbox{} + \left( \frac{1}{18} a_1^2 + \frac{2}{27} a_1b_2 + \frac{23}{162} a_1b_3 + \frac{1}{36} a_1c_3 + \frac{1}{27} b_2c_3 + \frac{23}{324} b_3c_3 \right) m_2 \nonumber \\
& & \mbox{} + \left( \frac{11}{162} a_1^2 + \frac19 a_1b_2 + \frac{37}{162} a_1b_3 + \frac{11}{324} a_1c_3 + \frac{1}{18} b_2c_3 + \frac{37}{324} b_3c_3 \right) m_3 \nonumber \\
& & \mbox{} + \left. \left( \frac{5}{24} a_1^2 + \frac19 a_1b_2 + \frac{4}{27} a_1b_3 + \frac18 a_1c_3 + \frac{7}{72} b_2^2 + \frac{17}{54} b_2b_3 + \frac{31}{108} b_3^2 + \frac{1}{32} c_3^2 \right) m_4 \right] I_2(m_\pi,0,\mu) \nonumber \\
& & \mbox{} + \left[ \left( \frac{5}{18} a_1^2 + \frac{1}{27} a_1b_2 + \frac{8}{81} a_1b_3 + \frac29 a_1c_3 + \frac49 b_2^2 + \frac{41}{27} b_2b_3 - \frac{17}{54} b_2c_3 + \frac{73}{54} b_3^2 - \frac{47}{81} b_3c_3 + \frac16 c_3^2 \right) m_1 \right. \nonumber \\
& & \mbox{} + \left( - \frac23 a_1^2 - \frac{17}{27} a_1b_2 - \frac{88}{81} a_1b_3 - \frac13 a_1c_3 - \frac{17}{54} b_2c_3 - \frac{44}{81} b_3c_3 \right) m_2 \nonumber \\
& & \mbox{} + \left( - \frac{94}{81} a_1^2 - \frac{29}{27} a_1b_2 - \frac{148}{81} a_1b_3 - \frac{47}{81} a_1c_3 - \frac{29}{54} b_2c_3 - \frac{74}{81} b_3c_3 \right) m_3 \nonumber \\
& & \mbox{} + \left. \left( \frac{7}{12} a_1^2 + \frac13 a_1b_2 + \frac{17}{27} a_1b_3 + \frac13 a_1c_3 + \frac29 b_2^2 + \frac{41}{54} b_2b_3 + \frac{73}{108} b_3^2 + \frac{1}{12} c_3^2 \right) m_4 \right] I_2(m_K,0,\mu) \nonumber \\
& & \mbox{} + \left[ \left( \frac19 a_1^2 + \frac19 a_1c_3 + \frac{1}{108} b_3^2 - \frac{1}{36} b_3c_3 + \frac{7}{144} c_3^2 \right) m_1 + \left( - \frac{1}{18} a_1^2 - \frac{1}{54} a_1b_3 - \frac{1}{36} a_1c_3 - \frac{1}{108} b_3c_3 \right) m_2 \right. \nonumber \\
& & \mbox{} + \left( - \frac{1}{18} a_1^2 - \frac{1}{54} a_1b_3 - \frac{1}{36} a_1c_3 - \frac{1}{108} b_3c_3 \right) m_3 \nonumber \\
& & \mbox{} + \left. \left( \frac{5}{36} a_1^2 + \frac{1}{36} a_1b_3 + \frac{7}{72} a_1c_3 + \frac{1}{216} b_3^2 + \frac{7}{288} c_3^2 \right) m_4 \right] I_2(m_\eta,0,\mu),
\end{eqnarray}

\begin{eqnarray}
\sqrt{2} \delta \mu_{{\Sigma^*}^-\Sigma^-}^{\mathrm{(loop\, 2ad)}} & = & \left[ \left( - \frac{1}{12} a_1^2 - \frac{1}{27} a_1b_2 - \frac{5}{81} a_1b_3 - \frac{1}{12} a_1c_3 - \frac{5}{108} b_2^2 - \frac{5}{81} b_2b_3 - \frac{1}{54} b_2c_3 + \frac{25}{972} b_3^2 - \frac{25}{324} b_3c_3 - \frac{1}{108} c_3^2 \right) m_1 \right. \nonumber \\
& & \mbox{} + \left( - \frac{1}{18} a_1^2 - \frac{2}{27} a_1b_2 - \frac{23}{162} a_1b_3 - \frac{1}{36} a_1c_3 - \frac{1}{27} b_2c_3 - \frac{23}{324} b_3c_3 \right) m_2 \nonumber \\
& & \mbox{} + \left( - \frac{25}{162} a_1^2 - \frac{11}{81} a_1b_2 - \frac{125}{486} a_1b_3 - \frac{25}{324} a_1c_3 - \frac{11}{162} b_2c_3 - \frac{125}{972} b_3c_3 \right) m_3 \nonumber \\
& & \mbox{} + \left. \left( \frac{1}{216} a_1^2 + \frac{5}{108} a_1b_3 - \frac{1}{54} a_1c_3 - \frac{5}{216} b_2^2 - \frac{5}{162} b_2b_3 + \frac{25}{1944} b_3^2 - \frac{1}{216} c_3^2 \right) m_4 \right] I_2(m_\pi,0,\mu) \nonumber \\
& & \mbox{} + \left[ \left( \frac{1}{36} a_1^2 + \frac{1}{27} a_1b_2 + \frac{7}{162} a_1b_3 + \frac{1}{18} a_1c_3 + \frac{5}{108} b_2^2 + \frac{5}{81} b_2b_3 + \frac{1}{54} b_2c_3 - \frac{25}{972} b_3^2 + \frac{11}{162} b_3c_3 + \frac{1}{108} c_3^2 \right) m_1 \right. \nonumber \\
& & \mbox{} + \left( \frac{2}{27} a_1b_2 + \frac{10}{81} a_1b_3 + \frac{1}{27} b_2c_3 + \frac{5}{81} b_3c_3 \right) m_2 \nonumber \\
& & \mbox{} + \left( \frac{11}{81} a_1^2 + \frac{11}{81} a_1b_2 + \frac{61}{243} a_1b_3 + \frac{11}{162} a_1c_3 + \frac{11}{162} b_2c_3 + \frac{61}{486} b_3c_3 \right) m_3 \nonumber \\
& & \mbox{} + \left. \left( - \frac{1}{216} a_1^2 - \frac{5}{108} a_1b_3 + \frac{1}{54} a_1c_3 + \frac{5}{216} b_2^2 + \frac{5}{162} b_2b_3 - \frac{25}{1944} b_3^2 + \frac{1}{216} c_3^2 \right) m_4 \right] I_2(m_K,0,\mu) \nonumber \\
& & \mbox{} + \left[ \left( \frac{1}{18} a_1^2 + \frac{1}{54} a_1b_3 + \frac{1}{36} a_1c_3 + \frac{1}{108} b_3c_3 \right) m_1 + \left( \frac{1}{18} a_1^2 + \frac{1}{54} a_1b_3 + \frac{1}{36} a_1c_3 + \frac{1}{108} b_3c_3 \right) m_2 \right. \nonumber \\
& & \mbox{} + \left. \left( \frac{1}{54} a_1^2 + \frac{1}{162} a_1b_3 + \frac{1}{108} a_1c_3 + \frac{1}{324} b_3c_3 \right) m_3 \right] I_2(m_\eta,0,\mu),
\end{eqnarray}

\begin{eqnarray}
\sqrt{2} \delta \mu_{{\Xi^*}^0\Xi^0}^{\mathrm{(loop\, 2ad)}} & = & \left[ \left( \frac14 a_1^2 + \frac19 a_1b_2 + \frac{13}{54} a_1b_3 + \frac16 a_1c_3 + \frac{11}{108} b_2^2 + \frac{49}{162} b_2b_3 - \frac{1}{36} b_2c_3 + \frac{74}{243} b_3^2 - \frac{1}{12} b_3c_3 + \frac{25}{432} c_3^2 \right) m_1 \right. \nonumber \\
& & \mbox{} + \left( - \frac19 a_1^2 - \frac{2}{27} a_1b_2 - \frac{13}{81} a_1b_3 - \frac{1}{18} a_1c_3 - \frac{1}{27} b_2c_3 - \frac{13}{162} b_3c_3 \right) m_2 \nonumber \\
& & \mbox{} + \left( - \frac16 a_1^2 - \frac{23}{162} a_1b_2 - \frac{127}{486} a_1b_3 - \frac{1}{12} a_1c_3 - \frac{23}{324} b_2c_3 - \frac{127}{972} b_3c_3 \right) m_3 \nonumber \\
& & \mbox{} + \left. \left( \frac{41}{216} a_1^2 + \frac{1}{12} a_1b_2 + \frac{11}{54} a_1b_3 + \frac{25}{216} a_1c_3 + \frac{11}{216} b_2^2 + \frac{49}{324} b_2b_3 + \frac{37}{243} b_3^2 + \frac{25}{864} c_3^2 \right) m_4 \right] I_2(m_\pi,0,\mu) \nonumber \\
& & \mbox{} + \left[ \left( \frac{5}{18} a_1^2 + \frac19 a_1b_2 + \frac29 a_1b_3 + \frac29 a_1c_3 + \frac{13}{27} b_2^2 + \frac{133}{81} b_2b_3 - \frac{5}{18} b_2c_3 + \frac{349}{243} b_3^2 - \frac{25}{54} b_3c_3 + \frac{37}{216} c_3^2 \right) m_1 \right. \nonumber \\
& & \mbox{} + \left( - \frac59 a_1^2 - \frac{13}{27} a_1b_2 - \frac{65}{81} a_1b_3 - \frac{5}{18} a_1c_3 - \frac{13}{54} b_2c_3 - \frac{65}{162} b_3c_3 \right) m_2 \nonumber \\
& & \mbox{} + \left( - \frac{25}{27} a_1^2 - \frac{65}{81} a_1b_2 - \frac{325}{243} a_1b_3 - \frac{25}{54} a_1c_3 - \frac{65}{162} b_2c_3 - \frac{325}{486} b_3c_3 \right) m_3 \nonumber \\
& & \mbox{} + \left. \left( \frac{65}{108} a_1^2 + \frac13 a_1b_2 + \frac{31}{54} a_1b_3 + \frac{37}{108} a_1c_3 + \frac{13}{54} b_2^2 + \frac{133}{162} b_2b_3 + \frac{349}{486} b_3^2 + \frac{37}{432} c_3^2 \right) m_4 \right] I_2(m_K,0,\mu) \nonumber \\
& & \mbox{} + \left[ \left( \frac19 a_1^2 + \frac19 a_1c_3 + \frac{1}{18} b_2^2 + \frac{11}{54} b_2b_3 - \frac{1}{36} b_2c_3 + \frac{7}{36} b_3^2 - \frac{1}{36} b_3c_3 + \frac{7}{144} c_3^2 \right) m_1 \right. \nonumber \\
& & \mbox{} + \left( - \frac{1}{18} a_1^2 - \frac{1}{54} a_1b_2 - \frac{1}{54} a_1b_3 - \frac{1}{36} a_1c_3 - \frac{1}{108} b_2c_3 - \frac{1}{108} b_3c_3 \right) m_3 \nonumber \\
& & \mbox{} + \left. \left( \frac{5}{36} a_1^2 + \frac{1}{36} a_1b_2 + \frac{1}{36} a_1b_3 + \frac{7}{72} a_1c_3 + \frac{1}{36} b_2^2 + \frac{11}{108} b_2b_3 + \frac{7}{72} b_3^2 + \frac{7}{288} c_3^2 \right) m_4 \right] I_2(m_\eta,0,\mu),
\end{eqnarray}

\begin{eqnarray}
\sqrt{2} \delta \mu_{{\Xi^*}^-\Xi^-}^{\mathrm{(loop\, 2ad)}} & = & \left[ \left( - \frac{1}{12} a_1^2 - \frac{1}{12} a_1c_3 - \frac{5}{108} b_2^2 - \frac{5}{81} b_2b_3 + \frac{25}{972} b_3^2 - \frac{5}{108} b_3c_3 - \frac{1}{108} c_3^2 \right) m_1 \right. \nonumber \\
& & \mbox{} + \left( - \frac{1}{18} a_1^2 - \frac{1}{27} a_1b_2 - \frac{13}{162} a_1b_3 - \frac{1}{36} a_1c_3 - \frac{1}{54} b_2c_3 - \frac{13}{324} b_3c_3 \right) m_2 \nonumber \\
& & \mbox{} + \left( - \frac{5}{54} a_1^2 - \frac{5}{81} a_1b_2 - \frac{65}{486} a_1b_3 - \frac{5}{108} a_1c_3 - \frac{5}{162} b_2c_3 - \frac{65}{972} b_3c_3 \right) m_3 \nonumber \\
& & \mbox{} + \left. \left( \frac{1}{216} a_1^2 + \frac{5}{108} a_1b_3 - \frac{1}{54} a_1c_3 - \frac{5}{216} b_2^2 - \frac{5}{162} b_2b_3 + \frac{25}{1944} b_3^2 - \frac{1}{216} c_3^2 \right) m_4 \right] I_2(m_\pi,0,\mu) \nonumber \\
& & \mbox{} + \left[ \left( \frac{1}{36} a_1^2 - \frac19 a_1b_2 - \frac{11}{54} a_1b_3 + \frac{1}{18} a_1c_3 + \frac{5}{108} b_2^2 + \frac{5}{81} b_2b_3 - \frac{1}{18} b_2c_3 - \frac{25}{972} b_3^2 - \frac{1}{18} b_3c_3 + \frac{1}{108} c_3^2 \right) m_1 \right. \nonumber \\
& & \mbox{} + \left( - \frac{2}{27} a_1b_2 - \frac{10}{81} a_1b_3 - \frac{1}{27} b_2c_3 - \frac{5}{81} b_3c_3 \right) m_2 \nonumber \\
& & \mbox{} + \left( - \frac19 a_1^2 - \frac{13}{81} a_1b_2 - \frac{59}{243} a_1b_3 - \frac{1}{18} a_1c_3 - \frac{13}{162} b_2c_3 - \frac{59}{486} b_3c_3 \right) m_3 \nonumber \\
& & \mbox{} + \left. \left( - \frac{1}{216} a_1^2 - \frac{5}{108} a_1b_3 + \frac{1}{54} a_1c_3 + \frac{5}{216} b_2^2 + \frac{5}{162} b_2b_3 - \frac{25}{1944} b_3^2 + \frac{1}{216} c_3^2 \right) m_4 \right] I_2(m_K,0,\mu) \nonumber \\
& & \mbox{} + \left[ \left( \frac{1}{18} a_1^2 + \frac19 a_1b_2 + \frac{11}{54} a_1b_3 + \frac{1}{36} a_1c_3 + \frac{1}{18} b_2c_3 + \frac{11}{108} b_3c_3 \right) m_1 \right. \nonumber \\
& & \mbox{} + \left( \frac{1}{18} a_1^2 + \frac19 a_1b_2 + \frac{11}{54} a_1b_3 + \frac{1}{36} a_1c_3 + \frac{1}{18} b_2c_3 + \frac{11}{108} b_3c_3 \right) m_2 \nonumber \\
& & \mbox{} + \left. \left( \frac{11}{54} a_1^2 + \frac29 a_1b_2 + \frac{61}{162} a_1b_3 + \frac{11}{108} a_1c_3 + \frac19 b_2c_3 + \frac{61}{324} b_3c_3 \right) m_3 \right] I_2(m_\eta,0,\mu). \label{eq:mmxsmxmloop2ad}
\end{eqnarray}

Using relations (\ref{eq:su3inv}) and (\ref{eq:rel1inv}) yields the magnetic moments expressed in terms of the $SU(3)$ invariants $\mu_D$, $\mu_F$, $\mu_C$, $\mu_T$, $D$, $F$, $\mathcal{C}$, and $\mathcal{H}$. These expressions read,
\begin{eqnarray}
\delta \mu_{n}^{\mathrm{(loop\,2ad)}} & = & \left[ \left( -\frac32 (D + F)^2 - \frac43 \mathcal{C}^2 \right) \mu_ D + \frac12 (D + F)^2 \mu_ F + \frac{10}{27} \mathcal{C}^2 \mu_C + \frac49 (D + F) \mathcal{C} \mu_T \right] I_2(m_\pi,0,\mu) \nonumber \\
& & \mbox{} + \left[ \left( - \frac{23}{18} D^2 + \frac53 D F - \frac72 F^2 - \frac13 \mathcal{C}^2 \right) \mu_ D - \frac12 (D - F)^2 \mu_ F + \frac{5}{27} \mathcal{C}^2 \mu_C + \frac29 F \mathcal{C} \mu_T \right] I_2(m_K,0,\mu) \nonumber \\
& & \mbox{} + \left[ \left( - \frac29 D^2 + \frac43 D F - 2 F^2 \right) \mu_ D \right] I_2(m_\eta,0,\mu), \label{eq:mmnloop2adch}
\end{eqnarray}

\begin{eqnarray}
\delta \mu_{p}^{\mathrm{(loop\, 2ad)}} & = & \left[ \left( \frac12 (D + F)^2 + \frac23 \mathcal{C}^2 \right) \mu_ D + \left( \frac52 (D + F)^2 + 2 \mathcal{C}^2 \right) \mu_ F - \frac{40}{27} \mathcal{C}^2 \mu_C - \frac49 (D + F) \mathcal{C} \mu_T \right] I_2(m_\pi,0,\mu) \nonumber \\
& & \mbox{} + \left[ \left( \frac89 D^2 - 2 D F + 2 F^2 + \frac16 \mathcal{C}^2 \right) \mu_ D + \left( 3 D^2 - 4 D F + 5 F^2 + \frac12 \mathcal{C}^2 \right) \mu_ F \right. \nonumber \\
& & \mbox{} - \left. \frac{5}{27} \mathcal{C}^2 \mu_C - \frac19 (3 D - F) \mathcal{C} \mu_T \right] I_2(m_K,0,\mu) \nonumber \\
& & \mbox{} + \left[ \left( \frac19 D^2 - \frac23 D F + F^2 \right) \mu_ D + \left( \frac13 D^2 - 2 D F + 3 F^2 \right) \mu_ F \right] I_2(m_\eta,0,\mu),
\end{eqnarray}

\begin{eqnarray}
\delta \mu_{\Sigma^-}^{\mathrm{(loop\, 2ad)}} & = & \left[ \left( \frac29 D^2 + \frac23 D F + \frac83 F^2 + \frac19 \mathcal{C}^2 \right) \mu_ D + \left( - D^2 - 7 F^2 - \frac13 \mathcal{C}^2 \right) \mu_ F + \frac{5}{54} \mathcal{C}^2 \mu_C + \frac19 (D - F) \mathcal{C} \mu_T \right] I_2(m_\pi,0,\mu) \nonumber \\
& & \mbox{} + \left[ \left( \frac56 D^2 + D F + \frac56 F^2 + \frac59 \mathcal{C}^2 \right) \mu_ D + \left( - \frac72 D^2 - D F - \frac72 F^2 - \frac53 \mathcal{C}^2 \right) \mu_ F \right. \nonumber \\
& & \mbox{} + \left. \frac{20}{27} \mathcal{C}^2 \mu_C + \frac29 (D - F) \mathcal{C} \mu_T \right] I_2(m_K,0,\mu) \nonumber \\
& & \mbox{} + \left[ \left( \frac49 D^2 + \frac16 \mathcal{C}^2 \right) \mu_ D + \left( - \frac43 D^2 - \frac12 \mathcal{C}^2 \right) \mu_ F + \frac{5}{18} \mathcal{C}^2 \mu_C \right] I_2(m_\eta,0,\mu),
\end{eqnarray}

\begin{eqnarray}
\delta \mu_{\Sigma^0}^{\mathrm{(loop\, 2ad)}} & = & \left[ \left( \frac29 D^2 + \frac83 F^2 + \frac19 \mathcal{C}^2 \right) \mu_ D - \frac29 F \mathcal{C} \mu_T \right] I_2(m_\pi,0,\mu) \nonumber \\
& & \mbox{} + \left[ \left( \frac56 D^2 + \frac56 F^2 + \frac59 \mathcal{C}^2 \right) \mu_ D - D F \mu_ F - \frac{5}{18} \mathcal{C}^2 \mu_C - \frac19 (D + F) \mathcal{C} \mu_T \right] I_2(m_K,0,\mu) \nonumber \\
& & \mbox{} + \left[ \left( \frac49 D^2 + \frac16 \mathcal{C}^2 \right) \mu_ D - \frac19 D \mathcal{C} \mu_T \right] I_2(m_\eta,0,\mu),
\end{eqnarray}

\begin{eqnarray}
\delta \mu_{\Sigma^+}^{\mathrm{(loop\, 2ad)}} & = & \left[ \left( \frac29 D^2 - \frac23 D F + \frac83 F^2 + \frac19 \mathcal{C}^2 \right) \mu_ D + \left( D^2 + 7 F^2 + \frac13 \mathcal{C}^2 \right) \mu_ F - \frac{5}{54} \mathcal{C}^2 \mu_C - \frac19 (D + 3 F) \mathcal{C} \mu_T \right] I_2(m_\pi,0,\mu) \nonumber \\
& & \mbox{} + \left[ \left( \frac56 D^2 - D F + \frac56 F^2 + \frac59 \mathcal{C}^2 \right) \mu_ D + \left( \frac72 D^2 - D F + \frac72 F^2 + \frac53 \mathcal{C}^2 \right) \mu_ F - \frac{35}{27} \mathcal{C}^2 \mu_C - \frac49 D \mathcal{C} \mu_T \right] I_2(m_K,0,\mu) \nonumber \\
& & \mbox{} + \left[ \left( \frac49 D^2 + \frac16 \mathcal{C}^2 \right) \mu_ D + \left( \frac43 D^2 + \frac12 \mathcal{C}^2 \right) \mu_ F - \frac{5}{18} \mathcal{C}^2 \mu_C - \frac29 D \mathcal{C} \mu_T \right] I_2(m_\eta,0,\mu),
\end{eqnarray}

\begin{eqnarray}
\delta \mu_{\Xi^-}^{\mathrm{(loop\, 2ad)}} & = & \left[ \left( \frac12 (D - F)^2 + \frac16 \mathcal{C}^2 \right) \mu_ D + \left( - \frac52 (D - F)^2 - \frac12 \mathcal{C}^2 \right) \mu_ F + \frac{5}{54} \mathcal{C}^2 \mu_C + \frac29 (D - F) \mathcal{C} \mu_T \right] I_2(m_\pi,0,\mu) \nonumber \\
& & \mbox{} + \left[ \left( \frac89 D^2 + 2 D F + 2 F^2 + \frac12 \mathcal{C}^2 \right) \mu_ D + \left( - 3 D^2 - 4 D F - 5 F^2 - \frac32 \mathcal{C}^2 \right) \mu_ F \right. \nonumber \\
& & \mbox{} + \left. \frac{20}{27} \mathcal{C}^2 \mu_C + \frac19 (D - F) \mathcal{C} \mu_T \right] I_2(m_K,0,\mu) \nonumber \\
& & \mbox{} + \left[ \left( \frac19 D^2 + \frac23 D F + F^2 + \frac16 \mathcal{C}^2 \right) \mu_ D + \left( - \frac13 D^2 - 2 D F - 3 F^2 - \frac12 \mathcal{C}^2 \right) \mu_ F + \frac{5}{18} \mathcal{C}^2 \mu_C \right] I_2(m_\eta,0,\mu),
\end{eqnarray}

\begin{eqnarray}
\delta \mu_{\Xi^0}^{\mathrm{(loop\, 2ad)}} & = & \left[ \left( - \frac32 (D - F)^2 - \frac13 \mathcal{C}^2 \right) \mu_ D - \frac12 (D - F)^2 \mu_ F + \frac{5}{27} \mathcal{C}^2 \mu_C + \frac19 (D - F) \mathcal{C} \mu_T \right] I_2(m_\pi,0,\mu) \nonumber \\
& & \mbox{} + \left[ \left( - \frac{23}{18} D^2 - \frac53 D F - \frac72 F^2 - \mathcal{C}^2 \right) \mu_ D + \frac12 (D + F)^2 \mu_ F + \frac{10}{27} \mathcal{C}^2 \mu_C + \frac29 (D + 2 F) \mathcal{C} \mu_T \right] I_2(m_K,0,\mu) \nonumber \\
& & \mbox{} + \left[ \left( - \frac29 D^2 - \frac43 D F - 2 F^2 - \frac13 \mathcal{C}^2 \right) \mu_ D + \frac19 (D + 3 F) \mathcal{C} \mu_T \right] I_2(m_\eta,0,\mu),
\end{eqnarray}

\begin{eqnarray}
\delta \mu_{\Lambda}^{\mathrm{(loop\, 2ad)}} & = & \left[ \left( - \frac23 D^2 - \frac12 \mathcal{C}^2 \right) \mu_ D \right] I_2(m_\pi,0,\mu) \nonumber \\
& & \mbox{} + \left[ \left( - \frac{7}{18} D^2 - \frac72 F^2 - \frac13 \mathcal{C}^2 \right) \mu_ D + D F \mu_ F + \frac{5}{18} \mathcal{C}^2 \mu_C - \frac19 (D - 3 F) \mathcal{C} \mu_T \right] I_2(m_K,0,\mu) \nonumber \\
& & \mbox{} + \left[ - \frac49 D^2 \mu_ D \right] I_2(m_\eta,0,\mu),
\end{eqnarray}

\begin{eqnarray}
\delta \mu_{\Delta^{++}}^{\mathrm{(loop\, 2ad)}} & = & \left[ - \frac16 \mathcal{C}^2 \mu_ D - \frac12 \mathcal{C}^2 \mu_ F + \left( \mathcal{C}^2 + \frac{31}{54} \mathcal{H}^2 \right) \mu_C + \frac19 \mathcal{C} \mathcal{H} \mu_T \right] I_2(m_\pi,0,\mu) \nonumber \\
& & \mbox{} + \left[ - \frac16 \mathcal{C}^2 \mu_ D - \frac12 \mathcal{C}^2 \mu_ F + \left( \mathcal{C}^2 + \frac{19}{54} \mathcal{H}^2 \right) \mu_C + \frac19 \mathcal{C} \mathcal{H} \mu_T \right] I_2(m_K,0,\mu) + \left[ \frac{2}{27} \mathcal{H}^2 \mu_C \right] I_2(m_\eta,0,\mu),
\end{eqnarray}

\begin{eqnarray}
\delta \mu_{\Delta^+}^{\mathrm{(loop\, 2ad)}} & = & \left[ - \frac13 \mathcal{C}^2 \mu_ F + \left( \frac12 \mathcal{C}^2 + \frac{41}{162} \mathcal{H}^2 \right) \mu_C + \frac{1}{27} \mathcal{C} \mathcal{H} \mu_T \right] I_2(m_\pi,0,\mu) \nonumber \\
& & \mbox{} + \left[ - \frac16 \mathcal{C}^2 \mu_ D - \frac16 \mathcal{C}^2 \mu_ F + \left( \frac12 \mathcal{C}^2 + \frac{17}{81} \mathcal{H}^2 \right) \mu_C + \frac{2}{27} \mathcal{C} \mathcal{H} \mu_T \right] I_2(m_K,0,\mu) + \left[ \frac{1}{27} \mathcal{H}^2 \mu_C \right] I_2(m_\eta,0,\mu),
\end{eqnarray}

\begin{eqnarray}
\delta \mu_{\Delta^0}^{\mathrm{(loop\, 2ad)}} & = & \left[ \frac16 \mathcal{C}^2 \mu_ D - \frac16 \mathcal{C}^2 \mu_ F - \frac{11}{162} \mathcal{H}^2 \mu_C - \frac{1}{27} \mathcal{C} \mathcal{H} \mu_T \right] I_2(m_\pi,0,\mu) \nonumber \\
& & \mbox{} + \left[ - \frac16 \mathcal{C}^2 \mu_ D + \frac16 \mathcal{C}^2 \mu_ F + \frac{11}{162} \mathcal{H}^2 \mu_C + \frac{1}{27} \mathcal{C} \mathcal{H} \mu_T \right] I_2(m_K,0,\mu),
\end{eqnarray}

\begin{eqnarray}
\delta \mu_{\Delta^-}^{\mathrm{(loop\, 2ad)}} & = & \left[ \frac13 \mathcal{C}^2 \mu_ D + \left( - \frac12 \mathcal{C}^2 - \frac{7}{18} \mathcal{H}^2 \right) \mu_C - \frac19 \mathcal{C} \mathcal{H} \mu_T \right] I_2(m_\pi,0,\mu) \nonumber \\
& & \mbox{} + \left[ - \frac16 \mathcal{C}^2 \mu_ D + \frac12 \mathcal{C}^2 \mu_ F + \left( - \frac12 \mathcal{C}^2 - \frac{2}{27} \mathcal{H}^2 \right) \mu_C \right] I_2(m_K,0,\mu) + \left[ - \frac{1}{27} \mathcal{H}^2 \mu_C \right] I_2(m_\eta,0,\mu),
\end{eqnarray}

\begin{eqnarray}
\delta \mu_{{\Sigma^*}^+}^{\mathrm{(loop\, 2ad)}} & = & \left[ - \frac{5}{36} \mathcal{C}^2 \mu_ D - \frac{1}{12} \mathcal{C}^2 \mu_ F + \left( \frac{5}{12} \mathcal{C}^2 + \frac{19}{81} \mathcal{H}^2 \right) \mu_C \right] I_2(m_\pi,0,\mu) \nonumber \\
& & \mbox{} + \left[ \frac{1}{18} \mathcal{C}^2 \mu_ D - \frac16 \mathcal{C}^2 \mu_ F + \left( \frac13 \mathcal{C}^2 + \frac{43}{162} \mathcal{H}^2 \right) \mu_C + \frac19 \mathcal{C} \mathcal{H} \mu_T \right] I_2(m_K,0,\mu) \nonumber \\
& & \mbox{}
 + \left[ - \frac{1}{12} \mathcal{C}^2 \mu_ D - \frac14 \mathcal{C}^2 \mu_ F + \frac14 \mathcal{C}^2 \mu_C \right] I_2(m_\eta,0,\mu),
\end{eqnarray}

\begin{equation}
\delta \mu_{{\Sigma^*}^0}^{\mathrm{(loop\, 2ad)}} = \left[ \frac{1}{36} \mathcal{C}^2 \mu_ D - \frac{1}{27} \mathcal{C} \mathcal{H} \mu_T \right] I_2(m_\pi,0,\mu) + \left[ \frac{1}{18} \mathcal{C}^2 \mu_ D + \frac{1}{27} \mathcal{C} \mathcal{H} \mu_T \right] I_2(m_K,0,\mu) + \left[ - \frac{1}{12} \mathcal{C}^2 \mu_ D \right] I_2(m_\eta,0,\mu),
\end{equation}

\begin{eqnarray}
\delta \mu_{{\Sigma^*}^-}^{\mathrm{(loop\, 2ad)}} & = & \left[ \frac{7}{36} \mathcal{C}^2 \mu_ D + \frac{1}{12} \mathcal{C}^2 \mu_ F + \left( - \frac{5}{12} \mathcal{C}^2 - \frac{19}{81} \mathcal{H}^2 \right) \mu_C - \frac{2}{27} \mathcal{C} \mathcal{H} \mu_T \right] I_2(m_\pi,0,\mu) \nonumber \\
& & \mbox{} + \left[ \frac{1}{18} \mathcal{C}^2 \mu_ D + \frac16 \mathcal{C}^2 \mu_ F + \left( - \frac13 \mathcal{C}^2 - \frac{43}{162} \mathcal{H}^2 \right) \mu_C - \frac{1}{27} \mathcal{C} \mathcal{H} \mu_T \right] I_2(m_K,0,\mu) \nonumber \\
& & \mbox{} + \left[ - \frac{1}{12} \mathcal{C}^2 \mu_ D + \frac14 \mathcal{C}^2 \mu_ F - \frac14 \mathcal{C}^2 \mu_C \right] I_2(m_\eta,0,\mu),
\end{eqnarray}

\begin{eqnarray}
\delta \mu_{{\Xi^*}^0}^{\mathrm{(loop\, 2ad)}} & = & \left[ \frac16 \mathcal{C}^2 \mu_ F + \frac{11}{162} \mathcal{H}^2 \mu_C - \frac{1}{54} \mathcal{C} \mathcal{H} \mu_T \right] I_2(m_\pi,0,\mu) \nonumber \\
& & \mbox{} + \left[ - \frac16 \mathcal{C}^2 \mu_ D - \frac16 \mathcal{C}^2 \mu_ F - \frac{11}{162} \mathcal{H}^2 \mu_C - \frac{1}{27} \mathcal{C} \mathcal{H} \mu_T \right] I_2(m_K,0,\mu) + \left[ \frac16 \mathcal{C}^2 \mu_ D + \frac{1}{18} \mathcal{C} \mathcal{H} \mu_T \right] I_2(m_\eta,0,\mu),
\end{eqnarray}

\begin{eqnarray}
\delta \mu_{{\Xi^*}^-}^{\mathrm{(loop\, 2ad)}} & = & \left[ \frac{1}{12} \mathcal{C}^2 \mu_ D + \frac{1}{12} \mathcal{C}^2 \mu_ F + \left( - \frac14 \mathcal{C}^2 - \frac{17}{162} \mathcal{H}^2 \right) \mu_C - \frac{1}{27} \mathcal{C} \mathcal{H} \mu_T \right] I_2(m_\pi,0,\mu) \nonumber \\
& & \mbox{} + \left[ \frac16 \mathcal{C}^2 \mu_ D + \frac16 \mathcal{C}^2 \mu_ F + \left( - \frac12 \mathcal{C}^2 - \frac{29}{81} \mathcal{H}^2 \right) \mu_C - \frac{2}{27} \mathcal{C} \mathcal{H} \mu_T \right] I_2(m_K,0,\mu) \nonumber \\
& & \mbox{} + \left[ - \frac{1}{12} \mathcal{C}^2 \mu_ D + \frac14 \mathcal{C}^2 \mu_ F - \frac14 \mathcal{C}^2 - \frac{1}{27} \mathcal{H}^2 \mu_C \right] I_2(m_\eta,0,\mu),
\end{eqnarray}

\begin{equation}
\delta \mu_{\Omega^-}^{\mathrm{(loop\, 2ad)}} = \left[ \frac16 \mathcal{C}^2 \mu_ D + \frac12 \mathcal{C}^2 \mu_ F + \left( - \mathcal{C}^2 - \frac{19}{54} \mathcal{H}^2 \right) \mu_C - \frac19 \mathcal{C} \mathcal{H} \mu_T \right] I_2(m_K,0,\mu)
 + \left[ - \frac{4}{27} \mathcal{H}^2 \mu_C \right] I_2(m_\eta,0,\mu),
\end{equation}

\begin{eqnarray}
\sqrt{3} \delta \mu_{\Sigma^0\Lambda}^{\mathrm{(loop\, 2ad)}} & = & \left[ \left( \frac73 D^2 + 3 F^2 + \frac{11}{12} \mathcal{C}^2 \right) \mu_ D - 2 D F \mu_ F - \frac59 \mathcal{C}^2 \mu_C - \frac{1}{18} (D + 6 F) \mathcal{C} \mu_T \right] I_2(m_\pi,0,\mu) \nonumber \\
& & \mbox{} + \left[ \left( \frac32 D^2 + \frac{15}{2} F^2 + \frac43 \mathcal{C}^2 \right) \mu_ D - D F \mu_ F - \frac{5}{18} \mathcal{C}^2 \mu_C + \left( - \frac49 D - \frac23 F \right) \mathcal{C} \mu_T \right] I_2(m_K,0,\mu) \nonumber \\
& & \mbox{} + \left[ \left( \frac23 D^2 + \frac14 \mathcal{C}^2 \right) \mu_ D - \frac16 D \mathcal{C} \mu_T \right] I_2(m_\eta,0,\mu),
\end{eqnarray}

\begin{eqnarray}
\sqrt{2} \delta \mu_{\Delta^+ p}^{\mathrm{(loop\, 2ad)}} & = & \left[ \frac23 (D + F) \mathcal{C} \mu_ D + \frac23 (D + F) \mathcal{C} \mu_ F - \frac{50}{81} \mathcal{C} \mathcal{H} \mu_C \right. \nonumber \\
& & \mbox{} + \left. \left( - \frac38 (D + F)^2 - \frac{25}{108} (D + F) \mathcal{H} - \frac{43}{108} \mathcal{C}^2 - \frac{25}{216} \mathcal{H}^2 \right) \mu_T \right] I_2(m_\pi,0,\mu) \nonumber \\
& & \mbox{} + \left[ \frac29 (D + 3 F) \mathcal{C} \mu_ D + \frac23 (D - F) \mathcal{C} \mu_ F - \frac{10}{81} \mathcal{C} \mathcal{H} \mu_C \right. \nonumber \\
& & \mbox{} + \left. \left( - \frac{5}{12} D^2 + \frac12 D F - \frac34 F^2 - \frac{5}{27} F \mathcal{H} - \frac{17}{108} \mathcal{C}^2 - \frac{5}{108} \mathcal{H}^2 \right) \mu_T \right] I_2(m_K,0,\mu) \nonumber \\
& & \mbox{} + \left[ \left( - \frac{1}{24} D^2 + \frac14 D F - \frac38 F^2 + \frac{5}{108} (D - 3 F) \mathcal{H} - \frac{5}{216} \mathcal{H}^2 \right) \mu_T \right] I_2(m_\eta,0,\mu),
\end{eqnarray}

\begin{eqnarray}
\sqrt{2} \delta \mu_{\Delta^0 n}^{\mathrm{(loop\, 2ad)}} & = & \left[ \frac23 (D + F) \mathcal{C} \mu_ D + \frac23 (D + F) \mathcal{C} \mu_ F - \frac{50}{81} \mathcal{C} \mathcal{H} \mu_C \right. \nonumber \\
& & \mbox{} + \left. \left( - \frac38 (D + F)^2 - \frac{25}{108} (D + F) \mathcal{H} - \frac{43}{108} \mathcal{C}^2 - \frac{25}{216} \mathcal{H}^2 \right) \mu_T \right] I_2(m_\pi,0,\mu) \nonumber \\
& & \mbox{} + \left[ \frac29 (D + 3 F) \mathcal{C} \mu_ D + \frac23 (D - F) \mathcal{C} \mu_ F - \frac{10}{81} \mathcal{C} \mathcal{H} \mu_C \right. \nonumber \\
& & \mbox{} + \left. \left( - \frac{5}{12} D^2 + \frac12 D F - \frac34 F^2 - \frac{5}{27} F \mathcal{H} - \frac{17}{108} \mathcal{C}^2 - \frac{5}{108} \mathcal{H}^2 \right) \mu_T \right] I_2(m_K,0,\mu) \nonumber \\
& & \mbox{} + \left[ \left( - \frac{1}{24} D^2 + \frac14 D F - \frac38 F^2 + \frac{5}{108} (D - 3 F) \mathcal{H} - \frac{5}{216} \mathcal{H}^2 \right) \mu_T \right] I_2(m_\eta,0,\mu),
\end{eqnarray}

\begin{eqnarray}
\sqrt{6} \delta \mu_{{\Sigma^*}^0\Lambda}^{\mathrm{(loop\, 2ad)}} & = & \left[ \frac23 D \mathcal{C} \mu_ D + \frac43 D \mathcal{C} \mu_ F - \frac{20}{27} \mathcal{C} \mathcal{H} \mu_C + \left( - \frac34 D^2 - \frac{5}{27} D \mathcal{H} - \frac{67}{144} \mathcal{C}^2 - \frac{5}{54} \mathcal{H}^2 \right) \mu_T \right] I_2(m_\pi,0,\mu) \nonumber \\
& & \mbox{} + \left[ 2 F \mathcal{C} \mu_ D + \frac23 D \mathcal{C} \mu_ F - \frac{10}{27} \mathcal{C} \mathcal{H} \mu_C + \left( - \frac14 (D^2 + 9 F^2) - \frac{5}{54} (D + 9 F) \mathcal{H} - \frac{11}{36} \mathcal{C}^2 - \frac{5}{27} \mathcal{H}^2 \right) \mu_T \right] I_2(m_K,0,\mu) \nonumber \\
& & \mbox{} + \left[ \frac23 D \mathcal{C} \mu_ D + \left( - \frac14 D^2 - \frac{1}{16} \mathcal{C}^2 \right) \mu_T \right] I_2(m_\eta,0,\mu),
\end{eqnarray}

\begin{eqnarray}
\sqrt{2} \delta \mu_{{\Sigma^*}^0\Sigma^0}^{\mathrm{(loop\, 2ad)}} & = & \left[ \frac29 (D + 2 F) \mathcal{C} \mu_ D + \left( - \frac{1}{12} D^2 - \frac12 F^2 - \frac{5}{27} F \mathcal{H} - \frac{23}{432} \mathcal{C}^2 - \frac{5}{162} \mathcal{H}^2 \right) \mu_T \right] I_2(m_\pi,0,\mu) \nonumber \\
& & \mbox{} + \left[ \frac29 F \mathcal{C} \mu_ D + \frac23 D \mathcal{C} \mu_ F - \frac{10}{27} \mathcal{C} \mathcal{H} \mu_C +
\left( - \frac14 (D^2 + F^2) - \frac{5}{54} (D + F) \mathcal{H} - \frac{19}{108} \mathcal{C}^2 - \frac{5}{81} \mathcal{H}^2 \right) \mu_T \right] I_2(m_K,0,\mu) \nonumber \\
& & \mbox{} + \left[ \frac29 D \mathcal{C} \mu_ D + \left( - \frac{1}{12} D^2 - \frac{7}{144} \mathcal{C}^2 \right) \mu_T \right] I_2(m_\eta,0,\mu),
\end{eqnarray}

\begin{eqnarray}
\sqrt{2} \delta \mu_{{\Sigma^*}^+\Sigma^+}^{\mathrm{(loop\, 2ad)}} & = & \left[ \frac{10}{9} F \mathcal{C} \mu_ D + \frac23 F \mathcal{C} \mu_ F + \frac{10}{81} \mathcal{C} \mathcal{H} \mu_C + \left( - \frac16 (D^2 + 6 F^2) - \frac{5}{54} (D + 3 F) \mathcal{H} - \frac18 \mathcal{C}^2 - \frac{5}{81} \mathcal{H}^2 \right) \mu_T \right] I_2(m_\pi,0,\mu) \nonumber \\
& & \mbox{} + \left[ \frac29 (3 D + F) \mathcal{C} \mu_ D + \frac23 (D - F) \mathcal{C} \mu_ F - \frac{70}{81} \mathcal{C} \mathcal{H} \mu_C \right. \nonumber \\
& & \mbox{} + \left. \left( - \frac12 (D^2 + F^2) - \frac{5}{54} (D + 3 F) \mathcal{H} - \frac13 \mathcal{C}^2 - \frac{10}{81} \mathcal{H}^2 \right) \mu_T \right] I_2(m_K,0,\mu) \nonumber \\
& & \mbox{} + \left[ \frac29 D \mathcal{C} \mu_ D + \frac23 D \mathcal{C} \mu_ F + \left( - \frac16 D^2 - \frac{7}{72} \mathcal{C}^2 \right) \mu_T \right] I_2(m_\eta,0,\mu),
\end{eqnarray}

\begin{eqnarray}
\sqrt{2} \delta \mu_{{\Sigma^*}^-\Sigma^-}^{\mathrm{(loop\, 2ad)}} & = & \left[ \frac49 (D - 2 F) \mathcal{C} \mu_ D - \frac23 F \mathcal{C} \mu_ F - \frac{10}{81} \mathcal{C} \mathcal{H} \mu_C + \left( \frac{5}{54} (D - F) \mathcal{H} + \frac{1}{54} \mathcal{C}^2 \right) \mu_T \right] I_2(m_\pi,0,\mu) \nonumber \\
& & \mbox{} + \left[ - \frac29(3 D - F) \mathcal{C} \mu_ D + \frac23 (D + F) \mathcal{C} \mu_ F + \frac{10}{81} \mathcal{C} \mathcal{H} \mu_C + \left( - \frac{5}{54} (D - F) \mathcal{H} - \frac{1}{54} \mathcal{C}^2 \right) \mu_T \right] I_2(m_K,0,\mu) \nonumber \\
& & \mbox{} + \left[ \frac29 D \mathcal{C} \mu_ D - \frac23 D \mathcal{C} \mu_ F \right] I_2(m_\eta,0,\mu),
\end{eqnarray}

\begin{eqnarray}
\sqrt{2} \delta \mu_{{\Xi^*}^0\Xi^0}^{\mathrm{(loop\, 2ad)}} & = & \left[ \frac23 (D - F) \mathcal{C} \mu_ F - \frac{10}{81} \mathcal{C} \mathcal{H} \mu_C + \left( - \frac38 (D - F)^2 + \frac{5}{108} (D - F) \mathcal{H} - \frac{25}{216} \mathcal{C}^2 - \frac{5}{216} \mathcal{H}^2 \right) \mu_T \right] I_2(m_\pi,0,\mu) \nonumber \\
& & \mbox{} + \left[ \frac23 (D + F) \mathcal{C} \mu_ D + \frac23 (D + F) \mathcal{C} \mu_ F - \frac{50}{81} \mathcal{C} \mathcal{H} \mu_C \right. \nonumber \\
& & \mbox{} + \left. \left( - \frac{5}{12} D^2 - \frac12 D F - \frac34 F^2 - \frac{5}{27} (D + 2 F) \mathcal{H} - \frac{37}{108} \mathcal{C}^2 - \frac{5}{36} \mathcal{H}^2 \right) \mu_T \right] I_2(m_K,0,\mu) \nonumber \\
& & \mbox{} + \left[ \frac29 (D + 3 F) \mathcal{C} \mu_ D + \left( - \frac{1}{24} D^2 - \frac14 D F - \frac38 F^2 - \frac{5}{108} (D + 3 F) \mathcal{H} - \frac{7}{72} \mathcal{C}^2 - \frac{5}{216} \mathcal{H}^2 \right) \mu_T \right] I_2(m_\eta,0,\mu),
\end{eqnarray}

\begin{eqnarray}
\sqrt{2} \delta \mu_{{\Xi^*}^-\Xi^-}^{\mathrm{(loop\, 2ad)}} & = &
\left[ \frac13 (D - F) \mathcal{C} \mu_ D + \frac13 (D - F) \mathcal{C} \mu_ F - \frac{5}{81} \mathcal{C} \mathcal{H} \mu_C + \left( \frac{5}{54} (D - F) \mathcal{H} + \frac{1}{54} \mathcal{C}^2 \right) \mu_T \right] I_2(m_\pi,0,\mu) \nonumber \\
& & \mbox{} + \left[ - \frac29 (D - 3 F) \mathcal{C} \mu_ D - \frac23 (D + F) \mathcal{C} \mu_ F - \frac{10}{81} \mathcal{C} \mathcal{H} \mu_C + \left( - \frac{5}{54} (D - F) \mathcal{H} - \frac{1}{54} \mathcal{C}^2 \right) \mu_T \right] I_2(m_K,0,\mu) \nonumber \\
& & \mbox{} + \left[ - \frac19 (D + 3 F) \mathcal{C} \mu_ D + \frac13 (D + 3 F) \mathcal{C} \mu_ F + \frac{5}{27} \mathcal{C} \mathcal{H} \mu_C \right] I_2(m_\eta,0,\mu). \label{eq:mmxsmxmloop2adch}
\end{eqnarray}

\subsection{\label{sec:Loop2e}Figure \ref{fig:mmloop2}(e)}

The final expressions for the loop contribution Fig.~\ref{fig:mmloop2}(e) simply read
\begin{equation}
\delta \mu_{n}^{\mathrm{(loop\, 2e)}} = \left[ \frac{5}{12} m_1 + \frac{1}{12} m_2 + \frac{5}{36} m_3 \right] I_2(m_\pi,0,\mu) + \left[ \frac{1}{12} m_1 - \frac{1}{12} m_2 + \frac{1}{36} m_3 \right] I_2(m_K,0,\mu), \label{eq:mmnloop2e}
\end{equation}

\begin{equation}
\delta \mu_{p}^{\mathrm{(loop\, 2e)}} = \left[ - \frac{5}{12} m_1 - \frac{1}{12} m_2 - \frac{5}{36} m_3 \right] I_2(m_\pi,0,\mu) + \left[ - \frac13 m_1 - \frac16 m_2 - \frac19 m_3 \right] I_2(m_K,0,\mu),
\end{equation}

\begin{equation}
\delta \mu_{\Sigma^-}^{\mathrm{(loop\, 2e)}} = \left[ \frac13 m_1 + \frac16 m_2 + \frac19 m_3 \right] I_2(m_\pi,0,\mu) + \left[ - \frac{1}{12} m_1 + \frac{1}{12} m_2 - \frac{1}{36} m_3 \right] I_2(m_K,0,\mu),
\end{equation}

\begin{equation}
\delta \mu_{\Sigma^0}^{\mathrm{(loop\, 2e)}} = \left[ - \frac14 m_1 - \frac{1}{12} m_3 \right] I_2(m_K,0,\mu),
\end{equation}

\begin{equation}
\delta \mu_{\Sigma^+}^{\mathrm{(loop\, 2e)}} = \left[ - \frac13 m_1 - \frac16 m_2 - \frac19 m_3 \right] I_2(m_\pi,0,\mu) + \left[ - \frac{5}{12} m_1 - \frac{1}{12} m_2 - \frac{5}{36} m_3 \right] I_2(m_K,0,\mu),
\end{equation}

\begin{equation}
\delta \mu_{\Xi^-}^{\mathrm{(loop\, 2e)}} = \left[ - \frac{1}{12} m_1 + \frac{1}{12} m_2 - \frac{1}{36} m_3 \right] I_2(m_\pi,0,\mu) + \left[ \frac13 m_1 + \frac16 m_2 + \frac19 m_3 \right] I_2(m_K,0,\mu),
\end{equation}

\begin{equation}
\delta \mu_{\Xi^0}^{\mathrm{(loop\, 2e)}} = \left[ \frac{1}{12} m_1 - \frac{1}{12} m_2 + \frac{1}{36} m_3 \right] I_2(m_\pi,0,\mu) + \left[ \frac{5}{12} m_1 + \frac{1}{12} m_2 + \frac{5}{36} m_3 \right] I_2(m_K,0,\mu),
\end{equation}

\begin{equation}
\delta \mu_{\Lambda}^{\mathrm{(loop\, 2e)}} = \left[ \frac14 m_1 + \frac{1}{12} m_3 \right] I_2(m_K,0,\mu),
\end{equation}

\begin{equation}
\sqrt{3} \delta \mu_{\Sigma^0 \Lambda}^{\mathrm{(loop\, 2e)}} = \left[ - \frac12 m_1 - \frac16 m_3 \right] I_2(m_\pi,0,\mu) + \left[ - \frac14 m_1 - \frac{1}{12} m_3 \right] I_2(m_K,0,\mu),
\end{equation}

\begin{equation}
\delta \mu_{\Delta^{++}}^{\mathrm{(loop\, 2e)}} = \left[ - \frac34 m_1 - \frac34 m_2 - \frac54 m_3 \right] I_2(m_\pi,0,\mu) + \left[ - \frac34 m_1 - \frac34 m_2 - \frac54 m_3 \right] I_2(m_K,0,\mu),
\end{equation}

\begin{equation}
\delta \mu_{\Delta^+}^{\mathrm{(loop\, 2e)}} = \left[ - \frac14 m_1 - \frac14 m_2 - \frac{5}{12} m_3 \right] I_2(m_\pi,0,\mu) + \left[ - \frac12 m_1 - \frac12 m_2 - \frac56 m_3 \right] I_2(m_K,0,\mu),
\end{equation}

\begin{equation}
\delta \mu_{\Delta^0}^{\mathrm{(loop\, 2e)}} = \left[ \frac14 m_1 + \frac14 m_2 + \frac{5}{12} m_3 \right] I_2(m_\pi,0,\mu) + \left[ - \frac14 m_1 - \frac14 m_2 - \frac{5}{12} m_3 \right] I_2(m_K,0,\mu),
\end{equation}

\begin{equation}
\delta \mu_{\Delta^-}^{\mathrm{(loop\, 2e)}} = \left[ \frac34 m_1 + \frac34 m_2 + \frac54 m_3 \right] I_2(m_\pi,0,\mu),
\end{equation}

\begin{equation}
\delta \mu_{{\Sigma^*}^+}^{\mathrm{(loop\, 2e)}} = \left[ - \frac12 m_1 - \frac12 m_2 - \frac56 m_3 \right] I_2(m_\pi,0,\mu) + \left[ - \frac14 m_1 - \frac14 m_2 - \frac{5}{12} m_3 \right] I_2(m_K,0,\mu),
\end{equation}

\begin{equation}
\delta \mu_{{\Sigma^*}^0}^{\mathrm{(loop\, 2e)}} = 0,
\end{equation}

\begin{equation}
\delta \mu_{{\Sigma^*}^-}^{\mathrm{(loop\, 2e)}} = \left[ \frac12 m_1 + \frac12 m_2 + \frac56 m_3 \right] I_2(m_\pi,0,\mu) + \left[ \frac14 m_1 + \frac14 m_2 + \frac{5}{12} m_3 \right] I_2(m_K,0,\mu),
\end{equation}

\begin{equation}
\delta \mu_{{\Xi^*}^0}^{\mathrm{(loop\, 2e)}} = \left[ - \frac14 m_1 - \frac14 m_2 - \frac{5}{12} m_3 \right] I_2(m_\pi,0,\mu) + \left[ \frac14 m_1 + \frac14 m_2 + \frac{5}{12} m_3 \right] I_2(m_K,0,\mu),
\end{equation}

\begin{equation}
\delta \mu_{{\Xi^*}^-}^{\mathrm{(loop\, 2e)}} = \left[ \frac14 m_1 + \frac14 m_2 + \frac{5}{12} m_3 \right] I_2(m_\pi,0,\mu) + \left[ \frac12 m_1 + \frac12 m_2 + \frac56 m_3 \right] I_2(m_K,0,\mu),
\end{equation}

\begin{equation}
\delta \mu_{\Omega^-}^{\mathrm{(loop\, 2e)}} = \left[ \frac34 m_1 + \frac34 m_2 + \frac54 m_3 \right] I_2(m_K,0,\mu),
\end{equation}

\begin{equation}
\sqrt{2} \delta \mu_{\Delta^+ p}^{\mathrm{(loop\, 2e)}} = \left[ - \frac23 m_1 - \frac13 m_4 \right] I_2(m_\pi,0,\mu) + \left[ - \frac13 m_1 - \frac16 m_4 \right] I_2(m_K,0,\mu),
\end{equation}

\begin{equation}
\sqrt{2} \delta \mu_{\Delta^0 n}^{\mathrm{(loop\, 2e)}} = \left[ - \frac23 m_1 - \frac13 m_4 \right] I_2(m_\pi,0,\mu) + \left[ - \frac13 m_1 - \frac16 m_4 \right] I_2(m_K,0,\mu),
\end{equation}

\begin{equation}
\sqrt{6} \delta \mu_{{\Sigma^*}^0 \Lambda}^{\mathrm{(loop\, 2e)}} = \left[ - m_1 - \frac12 m_4 \right] I_2(m_\pi,0,\mu) + \left[ - \frac12 m_1 - \frac14 m_4 \right] I_2(m_K,0,\mu),
\end{equation}

\begin{equation}
\sqrt{2} \delta \mu_{{\Sigma^*}^0 \Sigma^0}^{\mathrm{(loop\, 2e)}} = \left[ - \frac12 m_1 - \frac14 m_4 \right] I_2(m_K,0,\mu),
\end{equation}

\begin{equation}
\sqrt{2} \delta \mu_{{\Sigma^*}^+ \Sigma^+}^{\mathrm{(loop\, 2e)}} = \left[ - \frac13 m_1 - \frac16 m_4 \right] I_2(m_\pi,0,\mu) + \left[ - \frac23 m_1 - \frac13 m_4 \right] I_2(m_K,0,\mu),
\end{equation}

\begin{equation}
\sqrt{2} \delta \mu_{{\Sigma^*}^- \Sigma^-}^{\mathrm{(loop\, 2e)}} = \left[ \frac13 m_1 + \frac16 m_4 \right] I_2(m_\pi,0,\mu) + \left[ - \frac13 m_1 - \frac16 m_4 \right] I_2(m_K,0,\mu),
\end{equation}

\begin{equation}
\sqrt{2} \delta \mu_{{\Xi^*}^0 \Xi^0}^{\mathrm{(loop\, 2e)}} = \left[ - \frac13 m_1 - \frac16 m_4 \right] I_2(m_\pi,0,\mu) + \left[ - \frac23 m_1 - \frac13 m_4 \right] I_2(m_K,0,\mu),
\end{equation}

\begin{equation}
\sqrt{2} \delta \mu_{{\Xi^*}^- \Xi^-}^{\mathrm{(loop\, 2e)}} = \left[ \frac13 m_1 + \frac16 m_4 \right] I_2(m_\pi,0,\mu) + \left[ - \frac13 m_1 - \frac16 m_4 \right] I_2(m_K,0,\mu). \label{eq:mmxsmxmloop2e}
\end{equation}

The use of relations (\ref{eq:su3inv}) to rewrite the above expressions in terms of the $SU(3)$ invariants $\mu_D$, $\mu_F$, $\mu_C$, and $\mu_T$ yields,
\begin{equation}
\delta \mu_{n}^{\mathrm{(loop\, 2e)}} = \frac12 (\mu_D + \mu_F) I_2(m_\pi,0,\mu) + \frac12 (\mu_D - \mu_F) I_2(m_K,0,\mu), \label{eq:mmnloop2ech}
\end{equation}

\begin{equation}
\delta \mu_{p}^{\mathrm{(loop\, 2e)}} = - \frac12 (\mu_D + \mu_F) I_2(m_\pi,0,\mu) - \mu_F I_2(m_K,0,\mu),
\end{equation}

\begin{equation}
\delta \mu_{\Sigma^-}^{\mathrm{(loop\, 2e)}} = \mu_F I_2(m_\pi,0,\mu) - \frac12 (\mu_D - \mu_F ) I_2(m_K,0,\mu),
\end{equation}

\begin{equation}
\delta \mu_{\Sigma^0}^{\mathrm{(loop\, 2e)}} = - \frac12 \mu_D I_2(m_K,0,\mu),
\end{equation}

\begin{equation}
\delta \mu_{\Sigma^+}^{\mathrm{(loop\, 2e)}} = - \mu_F I_2(m_\pi,0,\mu) - \frac12 (\mu_D + \mu_F) I_2(m_K,0,\mu),
\end{equation}

\begin{equation}
\delta \mu_{\Xi^-}^{\mathrm{(loop\, 2e)}} = - \frac12 (\mu_D - \mu_F ) I_2(m_\pi,0,\mu) + \mu_F I_2(m_K,0,\mu),
\end{equation}

\begin{equation}
\delta \mu_{\Xi^0}^{\mathrm{(loop\, 2e)}} = \frac12 (\mu_D - \mu_F) I_2(m_\pi,0,\mu) + \frac12 (\mu_D + \mu_F) I_2(m_K,0,\mu),
\end{equation}

\begin{equation}
\delta \mu_{\Lambda}^{\mathrm{(loop\, 2e)}} = \frac12 \mu_D I_2(m_K,0,\mu),
\end{equation}

\begin{equation}
\sqrt{3} \delta \mu_{\Sigma^0 \Lambda}^{\mathrm{(loop\, 2e)}} = - \mu_D I_2(m_\pi,0,\mu) - \frac12 \mu_D I_2(m_K,0,\mu),
\end{equation}

\begin{equation}
\delta \mu_{\Delta^{++}}^{\mathrm{(loop\, 2e)}} = - \frac32 \mu_C I_2(m_\pi,0,\mu) - \frac32 \mu_C I_2(m_K,0,\mu),
\end{equation}

\begin{equation}
\delta \mu_{\Delta^+}^{\mathrm{(loop\, 2e)}} = - \frac12 \mu_C I_2(m_\pi,0,\mu) - \mu_C I_2(m_K,0,\mu),
\end{equation}

\begin{equation}
\delta \mu_{\Delta^0}^{\mathrm{(loop\, 2e)}} = \frac12 \mu_C I_2(m_\pi,0,\mu) - \frac12 \mu_C I_2(m_K,0,\mu),
\end{equation}

\begin{equation}
\delta \mu_{\Delta^-}^{\mathrm{(loop\, 2e)}} = \frac32 \mu_C I_2(m_\pi,0,\mu),
\end{equation}

\begin{equation}
\delta \mu_{{\Sigma^*}^+}^{\mathrm{(loop\, 2e)}} = - \mu_C I_2(m_\pi,0,\mu) - \frac12 \mu_C I_2(m_K,0,\mu),
\end{equation}

\begin{equation}
\delta \mu_{{\Sigma^*}^0}^{\mathrm{(loop\, 2e)}} = 0,
\end{equation}

\begin{equation}
\delta \mu_{{\Sigma^*}^-}^{\mathrm{(loop\, 2e)}} = \mu_C I_2(m_\pi,0,\mu) + \frac12 \mu_C I_2(m_K,0,\mu),
\end{equation}

\begin{equation}
\delta \mu_{{\Xi^*}^0}^{\mathrm{(loop\, 2e)}} = - \frac12 \mu_C I_2(m_\pi,0,\mu) + \frac12 \mu_C I_2(m_K,0,\mu),
\end{equation}

\begin{equation}
\delta \mu_{{\Xi^*}^-}^{\mathrm{(loop\, 2e)}} = \frac12 \mu_C I_2(m_\pi,0,\mu) + \mu_C I_2(m_K,0,\mu),
\end{equation}

\begin{equation}
\delta \mu_{\Omega^-}^{\mathrm{(loop\, 2e)}} = \frac32 \mu_C I_2(m_K,0,\mu),
\end{equation}

\begin{equation}
\sqrt{2} \delta \mu_{\Delta^+ p}^{\mathrm{(loop\, 2e)}} = \frac13 \mu_T I_2(m_\pi,0,\mu) + \frac16 \mu_T I_2(m_K,0,\mu),
\end{equation}

\begin{equation}
\sqrt{2} \delta \mu_{\Delta^0 n}^{\mathrm{(loop\, 2e)}} = \frac13 \mu_T I_2(m_\pi,0,\mu) + \frac16 \mu_T I_2(m_K,0,\mu),
\end{equation}

\begin{equation}
\sqrt{6} \delta \mu_{{\Sigma^*}^0 \Lambda}^{\mathrm{(loop\, 2e)}} = \frac12 \mu_T I_2(m_\pi,0,\mu) + \frac14 \mu_T I_2(m_K,0,\mu),
\end{equation}

\begin{equation}
\sqrt{2} \delta \mu_{{\Sigma^*}^0 \Sigma^0}^{\mathrm{(loop\, 2e)}} = \frac14 \mu_T I_2(m_K,0,\mu),
\end{equation}

\begin{equation}
\sqrt{2} \delta \mu_{{\Sigma^*}^+ \Sigma^+}^{\mathrm{(loop\, 2e)}} = \frac16 \mu_T I_2(m_\pi,0,\mu) + \frac13 \mu_T I_2(m_K,0,\mu),
\end{equation}

\begin{equation}
\sqrt{2}\delta \mu_{{\Sigma^*}^- \Sigma^-}^{\mathrm{(loop\, 2e)}} = - \frac16 \mu_T I_2(m_\pi,0,\mu) + \frac16 \mu_T I_2(m_K,0,\mu),
\end{equation}

\begin{equation}
\sqrt{2} \delta \mu_{{\Xi^*}^0 \Xi^0}^{\mathrm{(loop\, 2e)}} = \frac16 \mu_T I_2(m_\pi,0,\mu) + \frac13 \mu_T I_2(m_K,0,\mu),
\end{equation}

\begin{equation}
\sqrt{2} \delta \mu_{{\Xi^*}^- \Xi^-}^{\mathrm{(loop\, 2e)}} = - \frac16 \mu_T I_2(m_\pi,0,\mu) + \frac16 \mu_T I_2(m_K,0,\mu). \label{eq:mmxsmxmloop2ech}
\end{equation}

\section{\label{app:SB}Complete expressions from explicit symmetry breaking corrections}

Contributions to baryon magnetic moments due to explicit SB for $N_f=N_c=3$ read
\begin{equation}
\sqrt{3} \delta \mu_{n}^{\mathrm{SB}} = \frac12 m_1^{1,\mathbf{1}} + \frac{1}{12} m_3^{1,\mathbf{1}} - \frac12 n_1^{1,\mathbf{8}} - \frac16 n_2^{1,\mathbf{8}} - \frac16 n_3^{1,\mathbf{8}} - \frac13 m_2^{1,\mathbf{10}+\overline{\mathbf{10}}} - \frac13 m_2^{1,\mathbf{27}} - \frac19 \bar{c}_3^{1,\mathbf{27}},
\end{equation}

\begin{equation}
\sqrt{3} \delta \mu_{p}^{\mathrm{SB}} = \frac12 m_1^{1,\mathbf{1}} + \frac{1}{12} m_3^{1,\mathbf{1}} + \frac13 n_1^{1,\mathbf{8}} + \frac19 n_3^{1,\mathbf{8}} + \frac13 m_2^{1,\mathbf{10}+\overline{\mathbf{10}}} + \frac23 m_2^{1,\mathbf{27}} + \frac13 m_3^{1,\mathbf{27}} + \frac16 \bar{c}_3^{1,\mathbf{27}},
\end{equation}

\begin{equation}
\sqrt{3} \delta \mu_{\Lambda}^{\mathrm{SB}} = \frac12 m_1^{1,\mathbf{1}} + \frac{1}{12} m_3^{1,\mathbf{1}} + \frac16 n_1^{1,\mathbf{8}} + \frac{1}{18} n_3^{1,\mathbf{8}} + \frac19 \bar{c}_3^{1,\mathbf{27}},
\end{equation}

\begin{equation}
\sqrt{3} \delta \mu_{\Sigma^0}^{\mathrm{SB}} = \frac12 m_1^{1,\mathbf{1}} + \frac{1}{12} m_3^{1,\mathbf{1}} - \frac16 n_1^{1,\mathbf{8}} - \frac{1}{18} n_3^{1,\mathbf{8}} + \frac19 \bar{c}_3^{1,\mathbf{27}},
\end{equation}

\begin{equation}
\sqrt{3} \delta \mu_{\Sigma^+}^{\mathrm{SB}} = \frac12 m_1^{1,\mathbf{1}} + \frac{1}{12} m_3^{1,\mathbf{1}} + \frac16 n_1^{1,\mathbf{8}} + \frac16 n_2^{1,\mathbf{8}} + \frac{1}{18} n_3^{1,\mathbf{8}} - \frac13 m_2^{1,\mathbf{10}+\overline{\mathbf{10}}} + \frac13 m_2^{1,\mathbf{27}} + \frac13 \bar{c}_3^{1,\mathbf{27}},
\end{equation}

\begin{equation}
\sqrt{3} \delta \mu_{\Sigma^-}^{\mathrm{SB}} = \frac12 m_1^{1,\mathbf{1}} + \frac{1}{12} m_3^{1,\mathbf{1}} - \frac12 n_1^{1,\mathbf{8}} - \frac16 n_2^{1,\mathbf{8}} - \frac16 n_3^{1,\mathbf{8}} + \frac13 m_2^{1,\mathbf{10}+\overline{\mathbf{10}}} - \frac13 m_2^{1,\mathbf{27}} - \frac19 \bar{c}_3^{1,\mathbf{27}},
\end{equation}

\begin{equation}
\sqrt{3} \delta \mu_{\Xi^0}^{\mathrm{SB}} = \frac12 m_1^{1,\mathbf{1}} + \frac{1}{12} m_3^{1,\mathbf{1}} + \frac16 n_1^{1,\mathbf{8}} + \frac16 n_2^{1,\mathbf{8}} + \frac{1}{18} n_3^{1,\mathbf{8}} + \frac13 m_2^{1,\mathbf{10}+\overline{\mathbf{10}}} + \frac13 m_2^{1,\mathbf{27}} + \frac13 \bar{c}_3^{1,\mathbf{27}},
\end{equation}

\begin{equation}
\sqrt{3} \delta \mu_{\Xi^-}^{\mathrm{SB}} = \frac12 m_1^{1,\mathbf{1}} + \frac{1}{12} m_3^{1,\mathbf{1}} + \frac13 n_1^{1,\mathbf{8}} + \frac19 n_3^{1,\mathbf{8}} - \frac13 m_2^{1,\mathbf{10}+\overline{\mathbf{10}}} + \frac23 m_2^{1,\mathbf{27}} + \frac13 m_3^{1,\mathbf{27}} + \frac16 \bar{c}_3^{1,\mathbf{27}},
\end{equation}

\begin{equation}
 \delta \mu_{\Sigma^0\Lambda}^{\mathrm{SB}} = \frac16 n_1^{1,\mathbf{8}} + \frac{1}{18} n_3^{1,\mathbf{8}},
\end{equation}

\begin{equation}
\sqrt{3} \delta \mu_{\Delta^{++}}^{\mathrm{SB}} = \frac32 m_1^{1,\mathbf{1}} + \frac54 m_3^{1,\mathbf{1}} + \frac12 n_1^{1,\mathbf{8}} + \frac12 n_2^{1,\mathbf{8}} + \frac56 n_3^{1,\mathbf{8}} + 2 m_2^{1,\mathbf{27}} + 2 m_3^{1,\mathbf{27}} + \frac53 \bar{c}_3^{1,\mathbf{27}},
\end{equation}

\begin{equation}
\sqrt{3} \delta \mu_{\Delta^+}^{\mathrm{SB}} = \frac32 m_1^{1,\mathbf{1}} + \frac54 m_3^{1,\mathbf{1}} + m_2^{1,\mathbf{27}} + m_3^{1,\mathbf{27}} + \frac56 \bar{c}_3^{1,\mathbf{27}},
\end{equation}

\begin{equation}
\sqrt{3} \delta \mu_{\Delta^0}^{\mathrm{SB}} = \frac32 m_1^{1,\mathbf{1}} + \frac54 m_3^{1,\mathbf{1}} - \frac12 n_1^{1,\mathbf{8}} - \frac12 n_2^{1,\mathbf{8}} - \frac56 n_3^{1,\mathbf{8}},
\end{equation}

\begin{equation}
\sqrt{3} \delta \mu_{\Delta^-}^{\mathrm{SB}} = \frac32 m_1^{1,\mathbf{1}} + \frac54 m_3^{1,\mathbf{1}} - n_1^{1,\mathbf{8}} - n_2^{1,\mathbf{8}} - \frac53 n_3^{1,\mathbf{8}} - m_2^{1,\mathbf{27}} - m_3^{1,\mathbf{27}} - \frac56 \bar{c}_3^{1,\mathbf{27}},
\end{equation}

\begin{equation}
\sqrt{3} \delta \mu_{{\Sigma^*}^+}^{\mathrm{SB}} = \frac32 m_1^{1,\mathbf{1}} + \frac54 m_3^{1,\mathbf{1}} + \frac12 n_1^{1,\mathbf{8}} + \frac12 n_2^{1,\mathbf{8}} + \frac56 n_3^{1,\mathbf{8}},
\end{equation}

\begin{equation}
\sqrt{3} \delta \mu_{{\Sigma^*}^-}^{\mathrm{SB}} = \frac32 m_1^{1,\mathbf{1}} + \frac54 m_3^{1,\mathbf{1}} - \frac12 n_1^{1,\mathbf{8}} - \frac12 n_2^{1,\mathbf{8}} - \frac56 n_3^{1,\mathbf{8}},
\end{equation}

\begin{equation}
\sqrt{3} \delta \mu_{{\Sigma^*}^0}^{\mathrm{SB}} = \frac32 m_1^{1,\mathbf{1}} + \frac54 m_3^{1,\mathbf{1}},
\end{equation}

\begin{equation}
\sqrt{3} \delta \mu_{{\Xi^*}^0}^{\mathrm{SB}} = \frac32 m_1^{1,\mathbf{1}} + \frac54 m_3^{1,\mathbf{1}} + \frac12 n_1^{1,\mathbf{8}} +
 \frac12 n_2^{1,\mathbf{8}} + \frac56 n_3^{1,\mathbf{8}},
\end{equation}

\begin{equation}
\sqrt{3} \delta \mu_{{\Xi^*}^-}^{\mathrm{SB}} = \frac32 m_1^{1,\mathbf{1}} + \frac54 m_3^{1,\mathbf{1}} + m_2^{1,\mathbf{27}} + m_3^{1,\mathbf{27}} + \frac56 \bar{c}_3^{1,\mathbf{27}},
\end{equation}

\begin{equation}
\sqrt{3} \delta \mu_{\Omega^-}^{\mathrm{SB}} = \frac32 m_1^{1,\mathbf{1}} + \frac54 m_3^{1,\mathbf{1}} + \frac12 n_1^{1,\mathbf{8}} + \frac12 n_2^{1,\mathbf{8}} + \frac56 n_3^{1,\mathbf{8}} + 2 m_2^{1,\mathbf{27}} + 2 m_3^{1,\mathbf{27}} + \frac53 \bar{c}_3^{1,\mathbf{27}},
\end{equation}

\begin{equation}
\sqrt{6} \delta \mu_{\Delta^+p}^{\mathrm{SB}} = \frac23 n_1^{1,\mathbf{8}} + \frac13 \bar{n}_3^{1,\mathbf{8}} + \frac23 m_2^{1,\mathbf{10}+\overline{\mathbf{10}}} + \frac13 m_3^{1,\mathbf{10}+\overline{\mathbf{10}}} + \frac23 m_2^{1,\mathbf{27}} + \frac13 \bar{c}_3^{1,\mathbf{27}},
\end{equation}

\begin{equation}
\sqrt{6} \delta \mu_{\Delta^0n}^{\mathrm{SB}} = \frac23 n_1^{1,\mathbf{8}} + \frac13 \bar{n}_3^{1,\mathbf{8}} + \frac23 m_2^{1,\mathbf{10}+\overline{\mathbf{10}}} + \frac13 m_3^{1,\mathbf{10}+\overline{\mathbf{10}}} + \frac23 m_2^{1,\mathbf{27}} + \frac13 \bar{c}_3^{1,\mathbf{27}},
\end{equation}

\begin{equation}
\sqrt{2} \delta \mu_{{\Sigma^*}^0\Lambda}^{\mathrm{SB}} = \frac13 n_1^{1,\mathbf{8}} + \frac16 \bar{n}_3^{1,\mathbf{8}} - \frac19 \bar{c}_3^{1,\mathbf{27}},
\end{equation}

\begin{equation}
\sqrt{6} \delta \mu_{{\Sigma^*}^0\Sigma^0}^{\mathrm{SB}} = - \frac13 n_1^{1,\mathbf{8}} - \frac16 \bar{n}_3^{1,\mathbf{8}} + \frac19 \bar{c}_3^{1,\mathbf{27}},
\end{equation}

\begin{equation}
\sqrt{6} \delta \mu_{{\Sigma^*}^+\Sigma^+}^{\mathrm{SB}} = - \frac23 m_2^{1,\mathbf{10}+\overline{\mathbf{10}}} - \frac13 m_3^{1,\mathbf{10}+\overline{\mathbf{10}}} + \frac23 m_2^{1,\mathbf{27}} + \frac59 \bar{c}_3^{1,\mathbf{27}},
\end{equation}

\begin{equation}
\sqrt{6} \delta \mu_{{\Sigma^*}^-\Sigma^-}^{\mathrm{SB}} = - \frac23 n_1^{1,\mathbf{8}} - \frac13 \bar{n}_3^{1,\mathbf{8}} + \frac23 m_2^{1,\mathbf{10}+\overline{\mathbf{10}}} + \frac13 m_3^{1,\mathbf{10}+\overline{\mathbf{10}}} - \frac23 m_2^{1,\mathbf{27}} - \frac13 \bar{c}_3^{1,\mathbf{27}},
\end{equation}

\begin{equation}
\sqrt{6} \delta \mu_{{\Xi^*}^0\Xi^0}^{\mathrm{SB}} = - \frac23 m_2^{1,\mathbf{10}+\overline{\mathbf{10}}} - \frac13 m_3^{1,\mathbf{10}+\overline{\mathbf{10}}} - \frac23 m_2^{1,\mathbf{27}} - \frac59 \bar{c}_3^{1,\mathbf{27}},
\end{equation}

\begin{equation}
\sqrt{6} \delta \mu_{{\Xi^*}^-\Xi^-}^{\mathrm{SB}} = - \frac23 n_1^{1,\mathbf{8}} - \frac13 \bar{n}_3^{1,\mathbf{8}} + \frac23 m_2^{1,\mathbf{10}+\overline{\mathbf{10}}} + \frac13 m_3^{1,\mathbf{10}+\overline{\mathbf{10}}} - \frac23 m_2^{1,\mathbf{27}} - \frac13 \bar{c}_3^{1,\mathbf{27}}.
\end{equation}

\end{document}